\newcounter{saveeqn}%
\newcommand{\alpheqn}{\setcounter{saveeqn}{\value{equation}}%
\stepcounter{saveeqn}\setcounter{equation}{0}%
\renewcommand{\theequation}
    {\mbox{\arabic{saveeqn}-\arabic{equation}}}}%
\newcommand{\reseteqn}{\setcounter{equation}{\value{saveeqn}}%
\renewcommand{\theequation}{\arabic{equation}}}%

\newcounter{savesec}%
\newcommand{\resec}{\setcounter{savesec}{\value{section}}%
\stepcounter{savesec}\setcounter{section}{0}%
\renewcommand{\thesection}
    {\mbox{\arabic{savesec}.\arabic{section}}}}%
\newcommand{\resetsec}{\setcounter{section}{\value{savesec}}%
\renewcommand{\thesection}{\arabic{section}}}%

\newcounter{savefig}%
\newcommand{\refig}{\setcounter{savefig}{\value{figure}}%
\stepcounter{savefig}\setcounter{figure}{0}%
\renewcommand{\thefigure}
    {\mbox{\arabic{savefig}.\arabic{figure}}}}%
\newcommand{\resetfig}{\setcounter{figure}{\value{savefig}}%
\renewcommand{\thefigure}{\arabic{section}}}%

\newcounter{savetab}%
\newcommand{\retab}{\setcounter{savetab}{\value{table}}%
\stepcounter{savetab}\setcounter{table}{0}%
\renewcommand{\thetable}
    {\mbox{\arabic{savetab}.\arabic{table}}}}%
\newcommand{\resettab}{\setcounter{table}{\value{savetab}}%
\renewcommand{\thetable}{\arabic{table}}}%

\def\ahat{{\mathaccent "7E {\cal A}}}
\def\bhat{{\mathaccent "7E {\cal B}}}
\def\chat{{\mathaccent "7E {\cal F}}}
\def\dhat{{\mathaccent "7E {\cal D}}}
\def\shat{{\mathaccent "7E {\cal S}}}
\def\proj{{\cal P}}
\def\ket{\vert \vert  \{ \emptyset \} \rangle}
\def\ket2{\vert \vert \otimes \{ R \} \rangle}
\def\sqr{$^{2}$}
\def\T{$T$}
\def\P{$P$}
\def\Q{$Q$}
\def\G{$G$}
\def\M{{\bf M}}
\def\H{{\bf H}}
\def\I{{\bf I}}
\def\Pr{{\bf P}}
\def\Tr{{\bf T}}
\def\eq{\enskip =\enskip}
\def\pls{\enskip+\enskip}
\def\mns{\enskip -\enskip}
\def\oper{O^{\bf k}_{R}}
\def\aug{\tilde{\cal H}}
\def\envfn{\chi^{\alpha}_{RL}}
\def\struc{S^{\alpha}_{RL,R^{\prime}L^{\prime}}}
\def\rR{r_{R}}
\def\aug{\vert R,L,\{\emptyset\}\rangle }
  \def\ket{\vert \vert	\{ \emptyset \} \rangle}
  \def\ket2{\vert \vert \otimes \{ R \} \rangle}
  \def\sqr{$^{2}$}
\def\dpr{\prime\prime}
\def\tpr{\prime\prime\prime}
\def\pr#1{ Phys.Rev. {\bf B#1}}
\def\pj#1{\proj_{{\cal #1}}}
\def\barr#1{{\overline{#1}}}
\def\jpc#1{J.Phys. Condensed Matter {\bf #1}}
\def\prl#1{ Phys. Rev. Lett. {\bf #1}}
\def\.#1{\mathaccent 95#1}
\def\^#1{\mathaccent 94 #1}
\def\~#1{\mathaccent "7E #1}
\def\Ir{{\mbox{I}}}
\def\Mr{{\mbox{M}}}
\def\Hr{{\mbox{H}}}
\def\sund{\mathaccent 22{\sigma}}
\def\equal{\enskip =\enskip}
\def\plus{\enskip +\enskip}
\def\minus{\enskip -\enskip}
\def\eq{\enskip =\enskip}
\def\pls{\enskip +\enskip}
\def\mns{\enskip -\enskip}
\def\Gund{\mathaccent 22 {G}}
\def\ul#1{\underline{#1}}
\def\ac#1{\mathaccent 95#1}
\def\td#1{\mathaccent "7E#1}
\def\un#1{\underline{#1}}
\def\nbox{\raisebox{.6ex}{\fbox{{\scriptsize{\phantom{$\sqrt{}$}}}}}$\:$}
\def\ybox{\raisebox{.6ex}{\fbox{{\scriptsize{$\sqrt{}$}}}}$\:$}
\def\pbox#1{\raisebox{.6ex}{\fbox{{#1}}}$\:$}
\def\c#1{\mbox{\bf #1}}
\def\und#1{$\underline{\mbox{\bf #1}}\:$}
\def\unit{{\cal I}}
\def\trans{{\cal T}}
\def\proj{{\cal P}}
\def\T{$T$}
\def\P{$P$}
\def\Q{$Q$}
\def\G{$G$}
\def\M{{\bf M}}
\def\H{{\bf H}}
\def\I{{\bf I}}
\def\Pr{{\bf P}}
\def\Tra{{\bf T}}
\def\diag{\varepsilon_{i}}
  \def\proj{{\cal P}}
  \def\trans{{\cal T}}
  \def\ket{\vert \vert	\{ \emptyset \} \rangle}
  \def\ket2{\vert \vert \otimes \{ R \} \rangle}
  \def\sqr{$^{2}$}
\def\k{{\bf k}}
  \def\ahat{{\mathaccent "7E  A}}
  \def\bhat{{\mathaccent "7E  B}}
  \def\chat{{\mathaccent "7E  C}}
  \def\fhat{{\mathaccent "7E  F}}
  \def\dhat{{\mathaccent "7E  D}}
  \def\shat{{\mathaccent "7E  S}}
  \def\phat{{\mathaccent "7E  P}}
  \def\jhat{{\mathaccent "7E  J}}
  \def\khat{{\mathaccent "7E  K}}
  \def\ohat{{\mathaccent "7E o}}
\def\ve{\varepsilon}
\def\car{\{{\cal C}\}}
\def\gt{\; > \;}
\def\lt{\: < \:}
\def\dpr{\prime\prime}
\def\tpr{\prime\prime\prime}
\def\pr#1{ Phys.Rev. {\bf B#1}}
\def\jpc#1{J.Phys. Condensed Matter {\bf #1}}
\def\prl#1{ Phys. Rev. Lett. {\bf #1}}
\def\.#1{\mathaccent 95#1}
\def\^#1{\mathaccent 94 #1}
\def\~#1{\mathaccent "7E #1}
\def\sund{\mathaccent 22{\sigma}}
\def\equal{\enskip =\enskip}
\def\plus{\enskip +\enskip}
\def\minus{\enskip -\enskip}
\def\eq{\enskip =\enskip}
\def\pls{\enskip +\enskip}
\def\mns{\enskip -\enskip}
\def\Gund{\mathaccent 22 {G}}
\def\ul#1{\underline{#1}}
\def\ac#1{\mathaccent 95#1}
\def\td#1{\mathaccent "7E#1}
\def\un#1{\underline{#1}}
\def\nbox{\raisebox{.6ex}{\fbox{{\scriptsize{\phantom{$\sqrt{}$}}}}}$\:$}
\def\ybox{\raisebox{.6ex}{\fbox{{\scriptsize{$\sqrt{}$}}}}$\:$}
\def\pbox#1{\raisebox{.6ex}{\fbox{{#1}}}$\:$}
\def\und#1{$\underline{\mbox{\bf #1}}\:$}
\def\unit{{\cal I}}
\def\trans{{\cal T}}
\def\proj{{\cal P}}
\def\T{$T$}
\def\P{$P$}
\def\Q{$Q$}
\def\G{$G$}
\def\M{{\bf M}}
\def\H{{\bf H}}
\def\I{{\bf I}}
\def\Pr{\tilde{P}}
\def\Tr{\tilde{T}}
\def\aug{\vert R,L,\{\emptyset\}\rangle }
  \def\ket{\vert \vert	\{ \emptyset \} \rangle}
  \def\ket2{\vert \vert \otimes \{ R \} \rangle}
  \def\sqr{$^{2}$}
\def\dpr{\prime\prime}
\def\tpr{\prime\prime\prime}
\def\pr#1{ Phys.Rev. {\bf B #1}}
\def\pj#1{\proj_{{\cal #1}}}
\def\barr#1{{\overline{#1}}}
\def\jpc#1{J.Phys. Condensed Matter {\bf #1}}
\def\prl#1{ Phys. Rev. Lett. {\bf #1}}
\def\.#1{\mathaccent 95#1}
\def\^#1{\mathaccent 94 #1}
\def\~#1{\mathaccent "7E #1}
\def\Ir{{\mbox{I}}}
\def\Mr{{\mbox{M}}}
\def\Hr{{\mbox{H}}}
\def\sund{\mathaccent 22{\sigma}}
\def\equal{\enskip =\enskip}
\def\plus{\enskip +\enskip}
\def\minus{\enskip -\enskip}
\def\eq{\enskip =\enskip}
\def\pls{\enskip +\enskip}
\def\mns{\enskip -\enskip}
\def\Gund{\mathaccent 22 {G}}
\def\ul#1{\underline{#1}}
\def\ac#1{\mathaccent 95#1}
\def\td#1{\mathaccent "7E#1}
\def\un#1{\underline{#1}}
\def\nbox{\raisebox{.6ex}{\fbox{{\scriptsize{\phantom{$\sqrt{}$}}}}}$\:$}
\def\ybox{\raisebox{.6ex}{\fbox{{\scriptsize{$\sqrt{}$}}}}$\:$}
\def\pbox#1{\raisebox{.6ex}{\fbox{{#1}}}$\:$}
\def\c#1{\mbox{\bf #1}}
\def\und#1{$\underline{\mbox{\bf #1}}\:$}
\def\unit{{\cal I}}
\def\trans{{\cal T}}
\def\proj{{\cal P}}
\def\T{$T$}
\def\P{$P$}
\def\Q{$Q$}
\def\G{$G$}
\def\M{{\bf M}}
\def\H{{\bf H}}
\def\I{{\bf I}}
\def\Pr{{\bf P}}
\def\Tra{{\bf T}}
\def\diag{\varepsilon_{i}}
  \def\proj{{\cal P}}
  \def\trans{{\cal T}}
  \def\ket{\vert \vert	\{ \emptyset \} \rangle}
  \def\ket2{\vert \vert \otimes \{ R \} \rangle}
  \def\sqr{$^{2}$}
\def\k{{ ( k}}
  \def\ahat{{\mathaccent "7E  A}}
  \def\bhat{{\mathaccent "7E  B}}
  \def\chat{{\mathaccent "7E  C}}
  \def\fhat{{\mathaccent "7E  F}}
  \def\dhat{{\mathaccent "7E  D}}
  \def\shat{{\mathaccent "7E  S}}
  \def\phat{{\mathaccent "7E  P}}
  \def\jhat{{\mathaccent "7E  J}}
  \def\khat{{\mathaccent "7E  K}}
  \def\ohat{{\mathaccent "7E o}}
\def\ve{\varepsilon}
\def\car{\{{\cal C}\}}
\def\gt{\; > \;}
\def\lt{\: < \:}

\def\be{\begin{equation}}
\def\ee{\end{equation}}

 3

\def\plus{\enskip +\enskip}
\def\equal{\enskip =\enskip}
\def\k{{\bf k}}
\def\aug{{\tilde{\cal H}}}

\documentclass[12pt]{thesis}
\usepackage{epsfig,amssymb,amsmath,graphicx}

\usepackage[round,comma]{authyear}
\def\lsim{\lower.5ex\hbox{$\; \buildrel < \over \sim \;$}}
\def\gsim{\lower.5ex\hbox{$\; \buildrel > \over \sim \;$}}

\usepackage{makeidx}
\makeindex

\begin{document}
\setcounter{section}{0}
\setcounter{figure}{0}
\setcounter{table}{0}
\vskip 3cm
\thispagestyle{empty}
%{\baselineskip 25pt
\centerline{\Huge\bf X-ray properties of} 
\vskip 0.5cm
\centerline{\Huge\bf the Sun and some compact}
\vskip 0.5cm
\centerline{\Huge\bf objects of our Galaxy}

\vskip 3cm
\begin{center}
{\large\bf  Ph.D. Thesis under the supervision of\\
            Prof. Sandip K. Chakrabarti\\ }
\end{center}

\vskip 1cm
\begin{center}
{\large\bf  Thesis submitted for the degree of\\
            Doctor of Philosophy (Science)\\
            in Physics (Theoretical) of the \\
            Calcutta University\\ }
\end{center}

%\vskip 2.0cm
%\centerline{\large\bf (Corrected Copy)}
\vfill
\begin{figure}[h]
\vspace{1.5cm}
\centering
\includegraphics[height=0.7in,width=1.0in,angle=0]{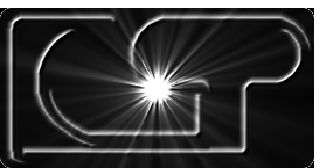}
\end{figure}
\centerline{\large\bf Dipak Debnath}
\vskip 0.2cm
{\baselineskip 15pt
\centerline{\large Indian Centre for Space Physics}
\centerline{\large 43-Chalantika, Garia Station Road}
\centerline{\large Kolkata-700084, India}}
\vskip 0.5cm
\centerline{\large 2010}

%Next 3 lines to add an extra empty page
%\newpage
%\thispagestyle{empty}
%\mbox{}

\newpage
\thispagestyle{empty}
\vfill
\vskip 25cm
\rightline{\large \bf I want to dedicate this thesis to parents.}

\newpage
\pagestyle{newheadings}
\pagenumbering{roman}
\setcounter{page}{1}
%\vskip 2cm
%\input{letter.tex}

\newpage
\vskip 2cm
\centerline{\underline{\bf ABSTRACT}}

\vskip 0.4cm
Sun is the closest and the brightest star from us. Even though its surface 
temperature is  only $\sim 6000$K, it emits X-rays and $\gamma$-rays 
up to a few Mev. This is primarily because of many non-thermal processes
and rapid magnetic reconnections which 
may produce energetic solar activities. Apart from the thermal electrons which 
obey Maxwell-Boltzmann distribution, the charged particles, especially 
electrons  are accelerated by shocks and acquire a non-thermal (power-law) 
distribution. These non-thermal electrons emit energetic synchrotron 
emissions. With time, the energy is shifted from one wavelength to another.
One of our goals is to understand the energy transport processes on the 
solar surface through detailed observation. It is due to plasma oscillations, 
or pinching or sausage instabilities in the magnetic field. For the study of temporal 
and spectral behaviours of the Sun, three detector payloads (two phoswich type payloads 
(RT-2/S, RT-2/G) and one solid state imaging type payload (RT-2/CZT) with 
detectors CZT \& CMOS) and one control electronic payload (RT-2/E) developed 
by Indian scientists. As a whole, these four Indian payloads are called RT-2 
and have been developed by Tata Institute of Fundamental Research (TIFR), 
Indian Centre for Space Physics (ICSP) in association with Vikram Sarabhai 
Space Centre (VSSC). RT-2 onboard Russian CORONAS-PHOTON Mission Satellite 
was successfully launched on $30^{th}$ January, 2009. These payloads observe the
Sun mainly in the X-ray band of 15-150 keV.
I have worked on the development of the RT-2 system from its very beginning 
stage to its final Flight Model stage at ICSP (Kolkata) and at the same time
had to visit several national laboratories, namely, VSSC, SAC, PRL, TIFR (Mumbai).
My thesis contains the details of the development of RT-2 payloads along 
with some interesting scientific results what we got so far from these payloads.

RT-2/CZT is the only imaging payload of the RT-2 system with very high spacial and 
spectral resolutions. For the first time in the history of space-borne instruments, 
Fresnel Zone Plates (FZPs) were used as an imaging coder in RT-2/CZT. In RT-2/CZT, 
coded aperture masks (CAMs) were also used as coders for Cadmium Zinc Telluride (CZT) 
detectors. Before using these shadow cast imaging techniques, we made theoretical 
simulations (using Monte Carlo method) and experimental set-ups. These results 
has been discussed in Chapter 2, briefly. 
 
Due to completely different reason, astrophysical black holes (BHs), which are 
not supposed to emit any radiation, are also sources of very high energy X-rays. 
It turns out that the matter which falls into the black holes emits the 
radiation. The process of capturing matter by gravitating objects is called 
{\it accretion}. In steady states, a disk-like structure is formed around black 
holes. This is called the {\it accretion disk}. These disks primarily contain 
two components: one having a Keplerian angular momentum distribution, emits a 
multi-colour black body radiation due to its high optical thickness. The other 
component is sub-Keplerian, which is hot and lesser efficient in emitting 
radiation. Even though there is no boundary layer of a black hole in the 
usual sense, the matter between the centrifugal barrier and the horizon 
behaves as the boundary layer. This is called CENtrifugal pressure supported 
BOundary Layer or {\bf CENBOL}. The hot electrons of CENBOL intercept soft 
photons from the Keplerian disk and the synchrotron radiation from gyrating 
electrons and re-emit them as hard X-rays or $\gamma$-rays in the form of a 
power-law component. In my Thesis, I am going to present the results on the 
stellar mass black holes with its mass ranging from 3 to 20 times the mass 
of the sun. Out of a dozen of confirmed black hole candidates in our own 
galaxy, our aim is to study outburst and variable sources such as GRO J1655-40 
and GX 339-4, which show flare like events in a much shorter time scale. 

In Chapter 1 of my Thesis, we  gave an introduction of the subject ``Astronomy and Astrophysics”. 
Also we gave introductions on the X-ray properties, physical processes, theoretical models and 
missions involved for studying the Sun, the GRBs and the black hole X-ray binaries.

In Chapter 2, we discuss the details of the space instruments (RT-2 for solar science and GRB 
study \&  RXTE for black hole study), their data acquisition methods and analysis procedures, 
whose data we have used during the Ph.D. period. 

In Chapter 3, we present the observational results (only solar science) obtained so far using 
the RT-2 instruments.

Chapter 4 is devoted for the black hole study. In this Chapter, we have discussed the results 
obtained from our detailed timing and spectral study of 2005 outburst of the black hole candidate 
GRO J 1655-40.

In Chapter 5, we discuss the results obtained from timing and spectral study of the transient 
outburst source GX 339-4, during the initial phase of the on-going 2010 outburst.

Finally, in Chapter 6, we make concluding remarks and a brief plan of my future works.

\newpage
\vskip 2cm
\centerline{\Large\bf ACKNOWLEDGMENTS}
\vskip 1cm
%\centerline{\bf \large I want to dedicate this thesis to parents.}

It is great pleasure for me to express my heartiest gratitude to my
thesis supervisor Prof. Sandip K. Chakrabarti for his inspiring guidance,
constant support and immense patience throughout the years of my Ph.D. 
period. I am very grateful to him for setting up ambitious goals for me 
and helping me find out my own way to achieve them.
 
It is also my pleasure to thank all the academic and non-academic staff
of Indian Centre for Space Physics (ICSP) for their enthusiastic support 
during my studies here. I also thank Prof. J. N. Chakrabarty (Vice President, ICSP).
 
My sincerest greetings goes to Prof. A. R. Rao of Tata Institute
of Fundamental Research (TIFR), Mumbai, PI and Prof. S.K. Chakrabarti (SNBNCBS and ICSP)
for giving me opportunity to work as a team member of the RT-2 Developmental team. 
It is a rare opportunity to develop space instruments which worked 
well and sent valuable data.

I am thankful to Prof. A. Vacchi of INFN, Trieste, Italy and Director of ICTP 
for giving me opportunities to work at mlab, the Abdus Salam International Centre 
for Theoretical Physics (ICTP), Trieste, Italy on Silicon Drift Detector (SDD) 
during my repeated visits. This also helped me very much for 
enriching my knowledge for developing large area, high sensitive and high resolution 
of space borne instrument. I also want to thank Dr. N. Zampa for his kind 
collaboration at mlab.

I am also thankful to Dr. Anuj Nandi, Dr. Samir Mandal, Mr. Broja Gopal Dutta,
Mr. Ritabrata Sarkar and Mr. Partha Sarathi Pal of ICSP for their kind 
collaboration during my Ph.D. activities at ICSP.
My heartiest thanks go to my friends and colleagues at ICSP, with whom I 
have spend beautiful moments during my Ph.D. period. In this regard, I do not
like to mention anyone name or events as they are truly countless.
I have shared many `unforgettable' moments and I would like to keep them in my memory forever.
But I should mention the name of Dr. Ankan Das whose company enlivened me very often
and Mr. Debashis Bhowmick who has shared beautiful moments with me at ICSP and ICTP.

I should thank all my colleagues, past and present, in the astrophysics
group of ICSP. Specially I would like to mention the name of Dr. Anuj Nandi 
whose friendly guidance helped me to carry out my thesis work smoothly.
Among the others, I must mention the name of Dr. Samir Mandal, 
Dr. Sabyasachi Pal and Mr. Tilak Ch. Kotoch of ICSP
with whom I had fruitful discussions on various aspects of 
astronomy and astrophysics. 

Finally, I should acknowledge ``Council for Scientific and Industrial Research" 
(CSIR) for giving me full financial support and ICTP, Italy for giving a partial 
financial support for carry out my thesis work.

I thank my loving parents, Late Nanda Dulal Debnath and Dulali Debnath, for their continuous 
support from my childhood and making me as a good human. They help me to become an independent 
person to take all the decisions in my life. But it is a great sorrow to me that I lost my 
father few months before the submission of this Thesis, who motivated me a lot in doing research.
Although he was a good student, did not able to continue his study due to poverty. So, I always
tried to reduce his unhappiness with my success. I also thank to my elder brother Hari Har Debnath 
and my sister-in-law, for their continuous support to my study. Also I want to thank to my niece 
Hritika Debnath, who gave me a lot of love and pleasure to my life. 
A special thank goes to my wife Mrs. Moumita Debnath for her constant inspiration.
Among the other family members, I want to thank to my grant mother Brojabala Debnath, father-in-law 
Parimal Debnath, mother-in-law ITI Debnath and brother-in-law Partha Debnath.

Last but not the least I would like to acknowledge `myself', without whose
unfailing inner motivation this work could not have been completed.

\newpage
\vskip 3cm
\centerline{\Large\bf PUBLICATIONS IN REFEREED JOURNALS}
\vskip 1cm
\begin{enumerate}

\item {\textbf {\textit {Propagating oscillatory shock model for QPOs in GRO J1655-40
during the March 2005 outburst}}} by S. K. Chakrabarti, A. Nandi, {\bf D. Debnath},
R. Sarkar and B. G. Datta in {\bf Indian J. Phys., 79(8), 841-845 (2005)} 
(arXiv: astro-ph/0508024).

\item {\textbf {\textit {Evolution of the quasi-periodic oscillation frequency in
GRO J1655-40 - - Implications for accretion disk dynamics}}} by S. K. Chakrabarti,
{\bf D. Debnath}, A. Nandi and P. S. Pal in {\bf A$\&$A, 489, L41-L44 (2008)} 
(arXiv: astro-ph/0809.0876).

\item {\textbf {\textit {Timing and Spectral evolution of GRO J1655-40 during recent
2005 outburst}}} by {\bf D. Debnath}, S. K. Chakrabarti, A. Nandi and S. Mandal in
{\bf BASI, 36, 151 (2008)} (arXiv: astro-ph/0902.3791).

\item {\textbf {\textit {Fresnel Zone Plate Telescopes for X-ray Imaging I: Experiments
with a quasi-parallel beam}}} by S. K. Chakrabarti, S. Palit, {\bf D. Debnath}, A. Nandi
and V. Yadav in {\bf Exp. Astron., 24, 109 (DOI 10.1007/s10696-009-9144-y) (2009)} 
(arXiv: astro-ph/0910.1987).

\item {\textbf {\textit {Fresnel Zone Plate Telescopes for X-ray Imaging II: Results
of numerical simulations}}} by S. Palit, S. K. Chakrabarti, {\bf D. Debnath}, A. Nandi,
V. Yadav, V. Girish and A. R. Rao in {\bf Exp. Astron., 27, 77 (DOI 10.1007/s10686-009-9176-3) 
(2009)} (arXiv: astro-ph/0910.2353).

\item {\textbf {\textit {RT-2 Detection of Quasi-Periodic Pulsations in the 2009 July 5
Solar Hard X-ray Flare}}} by A. R. Rao, J. P. Malkar, M. K. Hingar, V. K. Agrawal, S. K. Chakrabarti,
A. Nandi, {\bf D. Debnath}, T. B. Kotoch, T. R. Chidambaram, P. Vinod, S. Sreekumar, Y. D. Kotov, 
A. S. Buslov, V. N. Yurov, V. G. Tyshkevich , A. I. Arkhangelskij, R. A. Zyatkov, S. S. Begum, 
P. K. Manoharan in {\bf ApJ, 714, 1142 (2010)} (arXiv: astro-ph/1003.3992).

\item {\textbf {\textit {Properties of the Propagating Shock wave in the accretion flow around 
GX 339-4 in 2010 outburst}}} by {\bf D. Debnath}, S. K. Chakrabarti and A. Nandi in {\bf A$\&$A, 
520, 98 (DOI 10.1051/0004-6361/201014990) (2010)} (arXiv: astro-ph/1009.3351).

\item {\textbf {\textit {Instruments of RT-2 Experiment onboard CORONAS-PHOTON and their 
test and evaluation I: RT-2/S and RT-2/G Payloads}}} by {\bf D. Debnath}, A. Nandi,
A. R. Rao, J. P. Malkar, M. K. Hingar, T. B. Kotoch, S. Sreekumar, V. P. Madhav and  
S. K. Chakrabarti in {\bf Exp. Astron. (in press) (DOI 10.1007/s10686-010-9205-2) (2010)}
(arXiv: astro-ph/1011.3326).

\item {\textbf {\textit {Instruments of RT-2 Experiment onboard CORONAS-PHOTON and their test 
and evaluation II: RT-2/CZT Payload}}} by T. B. Kotoch, Anuj Nandi, {\bf D. Debnath},
J. P. Malkar, A. R. Rao, M. K. Hingar, V. P. Madhav, S. Sreekumar and S. K. Chakrabarti, 
in {\bf Exp. Astron. (in press) (DOI 10.1007/s10686-010-9189-y) (2010)}
(arXiv: astro-ph/1011.3331).

\item {\textbf {\textit {Instruments of RT-2 Experiment onboard CORONAS-PHOTON and their 
test and evaluation III: Coded Aperture Mask and Fresnel Zone Plates in RT-2/CZT Payload}}}
by A. Nandi, S. Palit, {\bf D. Debnath}, S. K. Chakrabarti, T. B. Kotoch, R. Sarkar, 
V. Yadav, V. Girish, A. R. Rao and D. Bhattacherya in {\bf Exp. Astron. (in press) 
(DOI 10.1007/s10686-010-9184-3) (2010)} (arXiv: astro-ph/1011.3338).

\item {\textbf {\textit {Instruments of RT-2 Experiment onboard CORONAS-PHOTON and their 
test and evaluation IV: Background Simulations using GEANT-4 Toolkit}}} by R. Sarkar,
S. Mandal, {\bf D. Debnath}, T. B. Kotoch, A. Nandi, A. R. Rao, S. K. Chakrabarti, in 
{\bf Exp. Astron. (in press) (DOI 10.1007/s10686-010-9208-z) (2010)} (arXiv: astro-ph/1011.3340).

\item {\textbf {\textit {Instruments of RT-2 Experiment onboard CORONAS-PHOTON and their test 
and evaluation V: Onboard software, Data Structure, Telemetry and Telecommand}}} by S. Sreekumar, 
P. Vinod, E. Samuel, J. P. Malkar, A. R. Rao, M. K. Hingar, V. P. Madhav, {\bf D. Debnath}, 
T. B. Kotoch, A. Nandi, S. S. Begum and S. K. Chakrabarti, in {\bf Exp. Astron. (in press) 
(DOI 10.1007/s10686-010-9185-2) (2010)} (arXiv: astro-ph/1011.3344).

\item {\textbf {\textit {Detection of GRB~090618 with RT-2 Experiment Onboard the
Coronas - Photon Satellite}}} by A. R. Rao, J. P. Malkar, M. K. Hingar, V. K. Agrawal, 
S. K. Chakrabarti, A. Nandi, {\bf D. Debnath}, T. B. Kotoch, R. Sarkar, T. R. Chidambaram, 
P. Vinod, S. Sreekumar, Y. D. Kotov, A. S. Buslov, V. N. Yurov, V. G. Tyshkevich, 
A. I. Arkhangelskij, R. A. Zyatkov, S. Naik {\bf ApJ (in press) (2010)} (arXiv: astro-ph/1012.0461).

\item {\textbf {\textit {Onboard performance of the RT-2 detectors}}} by A. R. Rao, J. P. Malkar,
M. K. Hingar, V. K. Agrawal, S. K. Chakrabarti, A. Nandi, {\bf D. Debnath}, T. B. Kotoch, R. Sarkar, 
T. R. Chidambaram, P. Vinod, S. Sreekumar, Y. D. Kotov, A. S. Buslov, V. N. Yurov, V. G. Tyshkevich, 
A. I. Arkhangelskij, R. A. Zyatkov {\it (2010, Submitted to Solar System Research)}.

\item {\textbf {\textit {Spectral and Timing evolution of GX 339-4 during its 2010 outburst}}} by
{\bf D. Debnath}, S.K. Chakrabarti and A. Nandi {\it (2010, in preparation)}.

\item {\textbf {\textit {Oscillations of the Compton Cloud During Quasi-Periodic
Oscillations in black hole candidates}}} by {\bf D. Debnath}, S. K. Chakrabarti and P. S. Pal
{\it (2010, in preparation)}.

\end{enumerate}

%\newpage
\vskip 2cm
\centerline{\Large\bf PUBLICATIONS IN JOURNAL PROCEEDINGS}
\vskip 1cm
\begin{enumerate}

\item {\textbf {\textit {Spectral and QPO Properties of GRO J1655-40 in the 2005
Outburst}}} by S. K. Chakrabarti, A. Nandi, {\bf D. Debnath}, R. Sarkar and B. G. Datta
in {\it VI Microquasar Workshop (Microquasars and Beyond) from September 18-22, 2006
at Societ\`{a} Casino, Como, Italy [POS 103 (2006)].}

\item {\textbf {\textit {Quasi Periodic Oscillations due to Axisymmetric and
non-axisymmetric shock oscillations in black hole accretion}}} by S. K. Chakrabarti,
{\bf D. Debnath}, P. S. Pal, A. Nandi, R. Sarkar, M. M. Samanta, P. J. Witta, H. Ghosh
and D. Som in {\it $11^{th}$ Marcel Grossman Meeting on General Relativity from July 
23-29, 2006 at Freie Universitaet, Berlin, Germany [World Scientific, 569 (2008)].}

\item {\textbf {\textit {Solar Science using RT-2 payloads abroad Coronas-photon satellite}}} 
by {\bf D. Debnath}, A. Nandi, S. K. Chakrabarti, A. R. Rao and P. K. Manoharan, in 
{\it Proc. $25^{th}$ meeting of ASI [BASI, 25S, 82 (2008)]}.

\item {\textbf {\textit {Fresnel zone plates: their suitability for X-ray imaging}}}
by P. S. Pal, {\bf D. Debnath}, A. Nandi, V. Yadav, S. K. Chakrabarti, A. R. Rao and
V. Girish in {\it Proc. $25^{th}$ meeting of ASI [BASI, 25S, 83 (2008)]}.

\item {\textbf {\textit {Background simulation of the X-ray detectors using Geant4
toolkit}}} by R. Sarkar, S. Mandal, A. Nandi, {\bf D. Debnath}, S. K. Chakrabarti and
A. R. Rao in {\it Proc. $25^{th}$ meeting of ASI [BASI, 25S, 83 (2008)]}.

\item {\textbf {\textit {QPO Evolution in 2005 Outburst of the Galactic Nano Quasar
GRO J1655-40}}} by {\bf D. Debnath}, A. Nandi, P. S. Pal and S. K. Chakrabarti in
proceeding of {\it the Second Kolkata Conference on ``Observational Evidence for
Black Holes in the Universe"} from February 10-15, 2008 at Vedic village and Radisson
fFort, Kolkata, India [AIP Conf. Proc. 1053, 171 (2008)].

\item {\textbf {\textit {Fresnel Zone Plates for Achromatic Imaging Survey of X-ray
sources}}} by S. Palit, S. K. Chakrabarti, {\bf D. Debnath}, V. Yadav and A. Nandi
in proceeding of {\it the Second Kolkata Conference on ``Observational Evidence for
Black Holes in the Universe"} from February 10-15, 2008 at Vedic village and Radisson
fFort, Kolkata, India [AIP Conf. Proc. 1053, 391 (2008)].

\item {\textbf {\textit {CSPOB - Continuous Spectrophotometry of Black Holes}}} by
S. K. Chakrabarti, D. Bhowmick, {\bf D. Debnath}, R. Sarkar, A. Nandi, V. Yadav and
A. R. Rao in proceeding of {\it the Second Kolkata Conference on ``Observational
Evidence for Black Holes in the Universe"} from February 10-15, 2008 at Vedic village
and Radisson fFort, Kolkata, India [AIP Conf. Proc. 1053, 409 (2008)].

\item {\textbf {\textit {Indian Payloads (RT-2 Experiment) Onboard CORONAS-PHOTON
Mission}}} by A. Nandi, A. R. Rao, S. K. Chakrabarti, J. P. Malker, S. Sreekumar,
{\bf D. Debnath}, T. B. Kotoch, Y. Kotov and A. Arkhangelsky, in proceeding of 
{\it 1st International Conference of Space Technology} from August 24-26, 2009 
at Electra Palace Hotel, Thessaloniki, Greece (IEEE) (2009) (arXiv: astro-ph/0912.4126).

\item {\textbf {\textit {Fresnel Zone Plate Telescopes as high resolution imaging
devices}}} by S.K. Chakrabarti, S. Palit, A. Nandi, V. Yadav and {\bf D. Debnath},
in proceeding of {\it 1st International Conference of Space Technology} from August
24-26, 2009 at Electra Palace Hotel, Thessaloniki, Greece (IEEE) (2009) (arXiv: astro-ph/0912.4127).

\item {\textbf {\textit {RT-2 Observations of gamma-ray bursts}}} by 
S.K. Chakrabarti, A.R. Rao, V.K. Agrawal, A. Nandi, {\bf D. Debnath}, T. B. Kotoch,
S. Sreekumar, Y. Kotov and A.S. Buslov, in proceeding of {\it 38th COSPAR Scientific 
Assembly} from 18-25 July, 2010 at Bremen, Germany (2010).

\item {\textbf {\textit {RT-2 Observations of solar flares}}} by 
S.K. Chakrabarti, A.R. Rao, V.K. Agrawal, A. Nandi, {\bf D. Debnath}, T. B. Kotoch,
S. Sreekumar, Y. Kotov, A. Arkhangelsky, A. S. Buslov, E. M. Oreshnikov, V. Yurov, 
V. Tyshkevich, P. K. Manoharan and S. S. Begum, in proceeding of {\it 38th COSPAR 
Scientific Assembly} from 18-25 July, 2010 at Bremen, Germany (2010).

\item {\textbf {\textit {Simultaneous observation of Solar Events by Indian Payload (RT-2)
and by ICSP-VLF receiver}}} by A. Nandi, S. K. Chakrabarti, {\bf D. Debnath}, T. B. Kotoch, 
A. R. Rao, S. K. Mondal, S. Maji and S. Sasmal in proceeding of {\it Very Low Frequency Radio 
Waves: Theory $\&$ Observations (VELFRATO-10)} from March 13-18, 2010 at S.N. Bose National 
Centre for Basic Sciences, Kolkata, India [AIP Conf. Proc. 1286, 200 (2010)].

\item {\textbf {\textit {Gamma-Ray Bursts from RT-2 payloads and VLF signals}}} by T. B. Kotoch, 
S. K. Chakrabarti, A. Nandi, {\bf D. Debnath} in proceeding of {\it Very Low Frequency Radio Waves: 
Theory $\&$ Observations (VELFRATO-10)} from March 13-18, 2010 at S.N. Bose National Centre for 
Basic Sciences, Kolkata, India [AIP Conf. Proc. 1286, 339 (2010)].

\item {\bf The Use of Reconfigurable Virtual Instruments for Low Noise, High Resolution
Charge Sensitive Amplification } by A. Olufemi, D. Bhowmick, {\bf D. Debnath}, M. L. Crespo,
A. Cicuttin and A. Sen in proceeding of {\it Programme FPGAworld'2010 Stockholm}
on 2010 September 8 at Electrum Kista, Stockholm, Sweden.                                                                    

\end{enumerate}
               
%\newpage
\vskip 1cm
\centerline{\Large\bf PUBLICATIONS IN GCN CIRCULARS ARCHIVE}
\vskip 1cm
\begin{enumerate}

\item {\textbf {\textit {Detection of GRB 090618 by RT-2 Experiment onboard the CORONAS PHOTON
Satellite}}} by A. R. Rao, J. P. Malkar, M. K. Hingar, V. K. Agrawal, S. K. Chakrabarti,
A. Nandi, {\bf D. Debnath}, T. B. Kotoch, T. R. Chidambaram, P. Vinod, S. Sreekumar, Y. D. Kotov, 
A. S. Buslov, V. N. Yurov, V. G. Tyshkevich, A. I. Arkhangelskij, R. A. Zyatkov in 
{\it GCN circulars archive, (GCN no. 9665, 2009)}.\\

\item {\textbf {\textit {GRB 090820: detection of a strong burst by RT-2 on board CORONAS
PHOTON}}} is reported by S. K. Chakrabarti, A. Nandi, {\bf D. Debnath}, T. B. Kotoch,
A. R. Rao, J. P. Malkar, M. K. Hingar, V. K. Agrawal, T. R. Chidambaram, P. Vinod, S. Sreekumar, 
Y. D. Kotov, A. S. Buslov, V. N. Yurov, V. G. Tyshkevich, A. I. Arkhangelskij, R. A. Zyatkov 
in {\it GCN circulars archive, (GCN no. 9833, 2009)}.\\

\item {\textbf {\textit {RT-2 observation of the bright GRB 090926A}}} is reported by
S. K. Chakrabarti, A. Nandi, {\bf D. Debnath}, T. B. Kotoch, A. R. Rao, J. P. Malkar, M. K. Hingar, 
V. K. Agrawal, T. R. Chidambaram, P. Vinod, S. Sreekumar, Y. D. Kotov, A. S. Buslov, V. N. Yurov, 
V. G. Tyshkevich, A. I. Arkhangelskij, R. A. Zyatkov in 
{\it GCN circulars archive, (GCN no. 10009, 2009)}.\\

\item {\textbf {\textit {Detection of a short GRB 090929A by RT-2 Experiment}}} is
reported by S. K. Chakrabarti, A. Nandi, {\bf D. Debnath}, T. B. Kotoch, A. R. Rao, J. P. Malkar, 
M. K. Hingar, V. K. Agrawal, T. R. Chidambaram, P. Vinod, S. Sreekumar, Y. D. Kotov, A. S. Buslov, 
V. N. Yurov, V. G. Tyshkevich, A. I. Arkhangelskij, R. A. Zyatkov
in {\it GCN circulars archive, (GCN no. 10010, 2009)}.

\end{enumerate}

\newpage
\tableofcontents
\listoffigures
\listoftables

\newpage
\pagestyle{myheadings}
\pagenumbering{arabic}
\alpheqn
\resec
\refig
\retab
\def\k{{\bf k}}
\def\aug{{\tilde{\cal H}}}
\def\ahat{{\mathaccent "7E {\cal A}}}
\def\bhat{{\mathaccent "7E {\cal B}}}
\def\chat{{\mathaccent "7E {\cal F}}}
\def\dhat{{\mathaccent "7E {\cal D}}}
\def\shat{{\mathaccent "7E {\cal S}}}
\def\proj{{\cal P}}
\def\ket{\vert \vert  \{ \emptyset \} \rangle}
\def\ket2{\vert \vert \otimes \{ R \} \rangle}
\def\sqr{$^{2}$}
\def\T{$T$}
\def\P{$P$}
\def\Q{$Q$}
\def\G{$G$}
\def\M{{\bf M}}
\def\H{{\bf H}}
\def\I{{\bf I}}
\def\Pr{{\bf P}}
\def\Tr{{\bf T}}
\def\eq{\enskip =\enskip}
\def\pls{\enskip+\enskip}
\def\mns{\enskip -\enskip}
\def\oper{O^{\bf k}_{R}}
\def\aug{\tilde{\cal H}}
\def\envfn{\chi^{\alpha}_{RL}}
\def\struc{S^{\alpha}_{RL,R^{\prime}L^{\prime}}}
\def\rR{r_{R}}

\newpage
\markboth{\it Introduction}
{\it Introduction }
\chapter{Introduction}

%**********************extra*******************
\hskip 4.0cm {\it ``It is difficult to say what is impossible, for the dream of

\hskip 4.0cm yesterday is the hope of today and the reality of tomorrow".}

\hskip 9.0cm 
- - Robert Hutchings Goddard
\vskip 1.0cm
%% Taken from the preface of book "SPACECRAFT FOR ASTRONOMY" by Joseph A. Angelo, JR.
%**********************extra*******************

\section{Astronomy and Astrophysics: A brief introduction}
\index{Astronomy}{\bf Astronomy} 
is the scientific study of celestial objects 
(such as stars, planets, comets, and galaxies) and phenomena that originate 
outside the Earth's atmosphere (such as the cosmic background radiation). 
It also includes the observation of strange and exotic objects and events, 
\index{Solar Flare}such as pulsating stars, flares in stars (e.g. Solar Flares) supernovae, 
compact objects (e.g. white dwarfs, neutron stars, black holes), 
Gamma Ray Bursts (GRBs), Active Galactic Nuclei (AGNs), Quasars and 
the universe as a whole. It concerns the evolution, physics, 
chemistry, and motion of celestial objects, as well as the 
formation and evolution of the universe.

There is a considerable difference between the science of astrophysics and the other 
sciences, such as biology, chemistry and physics. While most of the scientists 
can perform experiments in laboratory, where they can change the experimental 
parameters of the system to see what the effect is, astronomers cannot make
such a change, they only can observe what is happening to the objects they are 
studying/observing. In a way, it can be said that the astronomers can treat 
the whole universe as a laboratory.

Astronomy is one of the oldest sciences. However, the invention of the 
telescope was required before astronomy was able to develop into a modern 
science. Historically, astronomy has included disciplines as diverse as 
astrometry, celestial navigation and observational astronomy.
Since the $20^{th}$ century, the field of professional astronomy split into 
{\bf observational} and {\bf theoretical} branches. 
{\bf Astrophysics} is the branch of astronomy that deals with the physics 
of the universe, including the physical properties (luminosity, density, 
temperature, and chemical composition) of galaxies, 
stars, planets, exoplanets, and the interstellar medium, as well as their 
interactions. Rapid progress in Astronomy and Astrophysics over the past several decades have been made 
possible because of advances in our understanding of fundamental physics 
and improvement in the equipments like telescopes (ground based as well 
as space-borne), remote sensing systems, computers etc. With the advent 
of modern technology and space-age, the space-borne telescopes i.e., 
satellites are capable of seeing the objects in different energy bands: 
$\gamma$-rays, X-rays, Ultraviolet (UV), Infrared (IR) and also in Optical, 
and thus our focus has been intensified to understand the physical processes, 
which are happening on around or inside the objects. Also, sophisticated 
ground-based telescopes are used as an effective tool for capturing 
optical and radio signals which are coming from the same source.

\subsection{Observational astrophysics}

The majority of astrophysical observations are made using the electromagnetic 
spectrum.

\subsubsection{\bf $\bullet$ Radio astronomy} Radio astronomy studies radiation with 
a wavelength greater than a few millimeters. Radio waves are usually emitted by 
cooler objects, including interstellar gas and dust clouds. The cosmic microwave 
background radiation is the red-shifted light from the Big-Bang. Pulsars were 
first detected at microwave frequencies. The study of these waves requires very 
large radio telescopes.

\subsubsection{\bf $\bullet$ Infrared astronomy} Infrared astronomy studies radiation 
with a wavelength that is too long to be visible but shorter than radio waves. 
Infrared observations are usually made with telescopes similar to the usual 
optical telescopes. Star forming regions, planets, etc. are normally 
studied at infrared frequencies.

\subsubsection{\bf $\bullet$ Optical astronomy} Optical astronomy is the oldest kind 
of astronomy. Telescopes paired with a charge-coupled device or spectroscopes are the 
most common instruments used. The Earths atmosphere interferes somewhat with optical 
observations, so adaptive optics and space telescopes are used to obtain the 
highest possible image quality. In this range, stars are highly visible, and 
many chemical spectra can be observed to study the chemical composition of stars, 
galaxies and nebulae.

\subsubsection{\bf $\bullet$ Ultraviolet, X-ray and Gamma-ray astronomy} Ultraviolet, 
X-ray and Gamma-ray astronomy study very energetic processes such as binary pulsars, 
black holes, magnetars, and many others. These kinds of radiations do not penetrate 
the Earth's atmosphere well. There are two possibilities to observe this part of the 
electromagnetic spectra: either by using space-based telescopes or by using 
ground-based imaging air Cherenkov telescopes (IACT). Observatories 
of the first type are the Rossi X-ray Timing Explorer (RXTE), the Chandra X-ray Observatory 
and the Compton Gamma-Ray Observatory etc. The High Energy Stereoscopic System (H.E.S.S.) 
and the MAGIC telescope etc. are examples of IACTs.

Other than electromagnetic radiations, few things may be observed from the Earth 
that originate from great distances. A few gravitational wave observatories have 
been constructed, but gravitational waves are extremely difficult to detect. 
Neutrino observatories have also been built, primarily to study our Sun and supernovae 
explosions. Cosmic rays consisting of very high energy particles, can be observed when 
they interact with the Earth's atmosphere and produce cosmic-ray showers.

Observations can also vary in their time scales. Most optical observations take 
minutes to hours, due to integration time constraints. Hence, phenomena that change 
faster than this cannot readily be observed. However, historical data on some objects 
are available spanning centuries or millennia. On the other hand, radio observations may 
look at events on a millisecond timescale (millisecond pulsars) or combine years of data 
(pulsar deceleration studies). The information obtained from these different 
timescales is very different.

The topic of the stellar evolution, is often modeled by placing 
the varieties of star types in their respective positions on the Hertzsprung-Russell 
diagram, which can be viewed as representing the state of a stellar object, 
from birth to destruction. 

The study of our nearest star, namely, the Sun has a special place in observational 
astrophysics. Due to the tremendous distance of all other stars, the Sun can be observed 
in a kind of detail unparalleled by any other star. Our understanding of the
Sun serves as a guide to our understanding of other stars. So, we choose the study of the
Sun to be a part of my Thesis work.

Similarly, studying black hole X-ray binaries are equally important and interesting.
It is not possible to detect any direct radiations from the black holes, although 
there have an indirect observational methods. At the time of mass accretion 
from its companion objects, black hole binaries form an 
accretion disk and emit electromagnetic radiations in a wide energy bands from radio to 
$\gamma$-rays. These emitted radiations we can observe and analyze using ground based data
and satellite data. From the detailed 
analysis of these radiation data, we can get an idea about the properties of the emitting 
object. So, in my Thesis, we have studied a few very interesting and fascinating black hole 
candidates (BHCs), such as, GRO J1655-40, GX 339-4 \& GRS 1915+105.

\subsection{Theoretical astrophysics}

Theoretical astrophysicists use a wide variety of tools which include analytical 
models (for example, polytropes to approximate the behaviors of a star) and 
computational numerical simulations. Each has some advantages. Analytical models 
of a process are generally better for giving insight into the heart of what is 
going on. Numerical models can reveal the existence of phenomena and effects 
that would otherwise not be seen. Theorists in astrophysics create theoretical 
models and figure out 
the observational consequences of those models. This helps allow observers to 
look for data that can refute a model or help in choosing between several 
alternate or conflicting models. Theorists also try to generate or modify models 
to take into account new data. 
In case of an inconsistency, the general tendency is to try to make minimal 
modifications to the models to fit the data. In some cases, a large amount of 
inconsistent data over time may lead to total abandonment of a model.

The goal of my thesis is to study X-ray properties (spectral and timing 
properties) of Sun and compact objects (mainly black hole candidates GRO J1655-40, GX 339-4
and gamma-ray burst (GRB) objects). This study was done by analyzing the observational 
data from the space-borne X-ray telescopes. For studying X-ray properties of the Sun 
and gamma-ray bursts, we used our Solar X-ray payloads (RT-2/S, RT-2/G 
and RT-2/CZT) data and for the black hole study, NASA's {\it Rossi X-ray Timing Explorer} 
({\it RXTE}) satellite data was used. Apart from the X-ray study, I participated in
the development, test and evaluation and calibration of RT-2 payloads from its very 
beginning stage to its final flight model stage. For the development of RT-2 
systems, I took part in some theoretical works such as: background simulations 
for three RT-2 detector payloads and theoretical (mainly simulation) and experimental 
characterizations of Fresnel Zone Plate (FZP) and Coded Aperture Mask (CAM) imaging 
techniques which was used in RT-2/CZT imaging. Both of these simulations were done by 
Monte-Carlo methods.

\paragraph{\rm In \S 1.2, we briefly discuss the life cycle of 
a star (from its birth to death stages). In \S 1.3, we present a brief discussion 
on the Sun and its properties. We also presented a list of X-ray 
astronomy missions dedicated to the observations of the X-ray properties of the Sun. 
In \S 1.4, we discuss X-ray properties of the compact objects (mainly black 
hole binaries). We also discuss radiative processes involved with it and 
the accretion processes and the models involved with it. We give a list of
the X-ray missions since 1949 to the present era and how these missions developed and 
strengthen our views about the X-ray observation of compact objects.
\S 1.5, is dedicated to the introduction of the study of gamma-ray bursts, their
classifications and theoretical models mostly accepted by the scientific people for
describing the origin of the GRBs. In \S 1.6, we present the method of analysis which
was followed in this Thesis for analyzing the data from the Sun, GRBs and black holes.}

\section{Life Cycle of a Star}

At Big Bang, all matters and energies of the observable universe were concentrated 
in one point of virtually infinite density. After the Big Bang, the universe started 
to expand and reached its present form. Some heavier isotopes 
of hydrogen were produced. No heavier elements, known as ``metals", were formed 
since the universe expanded rapidly and became too cold. The heavier elements 
were produced through various stages of the stellar evolutions.

Stars are mainly formed in the relatively dense part of the interstellar cloud. 
These regions are extremely cold (temperatures are about $10$ to $20$ $^\circ$~K). At these 
temperatures and densities, gases are mainly in the molecular form. Central region of 
a collapsing cloud fragment, which is in the process of formation of a star is called a
{\it protostar}. This has not yet become hot enough and does not have enough mass in 
the core to initiate the process of nuclear fusion (temperature needed to be $\sim~10^7~^\circ$~K) 
in order to halt its gravitational collapse. When the density reaches above a critical value, 
stars are formed. As the protostar continues to condense and the rise in temperature continue 
until the temperature of the star reaches about $10^7$ degrees Celsius ($1.8 \times 10^7$ 
degrees Fahrenheit). At this point, the nuclear fusion occurs in a process called the 
{\it proton-proton} reaction and the star stops collapsing because the outward 
force of heat balances the gravity. This stage is known as the {\it main sequence phase}.
Stars like to spend most of their life in this stable phase but the life span is highly 
dependent on the size and weight of the star. Massive stars burn their fuel much faster than 
the lighter stars. In {\it massive stars}, the great amount of weight puts a large amount of 
pressure on their core, raises up the temperature and speeds up the fusion process. These 
massive stars are very bright, but only live for a short time. Their main sequence phase
may last only a few hundred thousand years. {\it Lighter stars} will live on for billions of 
years because they burn their fuel much more slowly. Eventually, the stars fuel will begin 
to run out. After finishing most of its fuel, lighter star will expand into what is 
known as a {\it red giant} and a massive star will become {\it red supergiant}. 
This phase will last until the star exhausts its remaining fuel. At this point, 
the pressure of the nuclear reaction is not strong enough to equalize the force of 
gravity and the star will collapse. Most average stars will blow away their outer 
atmospheres to form a {\it planetary nebula}. Their cores will remain behind and radiate
as a {\it white dwarf} until they cool down. The left over is a dark ball of matter 
known as a {\it black dwarf}. 
If the star is massive enough, the collapse will trigger a violent explosion known as 
a {\it supernova}. If the remaining mass of the star is about $\sim$1.44 times that 
of our Sun ($M_\odot$) (the Chandrasekhar limit), the core is unable to support itself 
and it will collapse further to become a {\it neutron star}. The matter inside the star 
will be compressed so tightly that its atoms are compacted into a dense shell of 
neutrons. If the remaining mass of the star is more than about $2 -3$ times that of the Sun 
(the Tolman-Oppenheimer-Volkoff limit), it will collapse so completely that it will 
literally disappear from the universe. What is left behind is an intense region of 
gravity called a {\it black hole}.
 
\begin{figure}[h]
\begin{center}
\vspace{-0.0cm}
\epsfxsize=5.4in% \epsfysize=3.5in
\rotatebox{-0.2}{\epsfbox{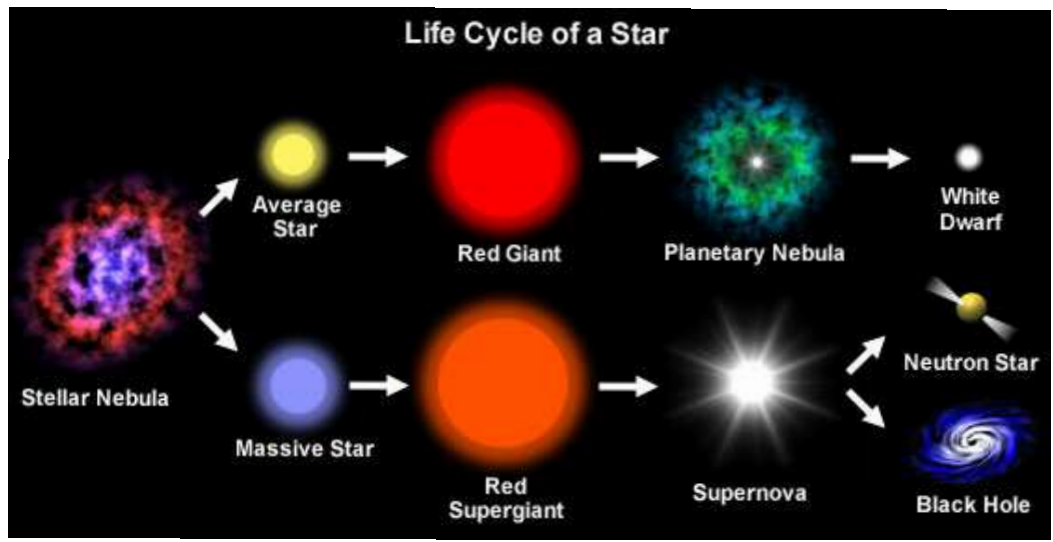}}
\vspace{-0.1cm}
\caption{Artists conception of a star life cycle. It shows the life cycle of the star birth from 
Stellar Nebula to its death at White Dwaft/Neutron Star/Black Hole (courtesy: NASA).}
\label{kn : fig1.1}
\end{center}
\end{figure}

Figure 1.1 gives an artistic concept of the life cycle of a star. 
The nebula that was expelled from the star may continue to expand for millions of years. 
Eventually, the gravity of a passing star or the shock wave from a nearby 
supernova may cause it to contract, starting the entire process all over again. 
This process repeats itself throughout the universe in an endless cycle of 
birth, death, and rebirth. It is this cycle of stellar evolution that produces 
all of the heavy elements required for life. Our Solar System was formed from such 
a second or third generation nebula, leaving an abundance of heavy elements 
here on Earth and throughout the Solar System. This means that we are all 
made of stellar material.

\section{Sun and its properties}

My thesis works are based on the two main sections of a star life cycle: {\bf (a)} Ordinary star 
(e.g. Sun) and {\bf (b)} Massive star end product (Black Holes). In this current section we 
introduce our nearest star the Sun.

Sun is a Population I G2 star. It is located at the center of our Solar System. The 
Earth and other objects (including other planets, asteroids, meteoroids, comets, and 
dust) orbit the Sun, which by itself accounts for about $99.86\%$ of the solar 
systems mass. Its mean distance from the Earth is $\sim 1.496 \times 10^{11}$~m. The Sun 
consists of hydrogen (about $74\%$ of its mass, or $92\%$ of its volume), helium 
(about $24\%$ of mass, $7\%$ of volume), and trace quantities of other elements, 
including iron, nickel, oxygen, silicon, sulfur, magnesium, carbon, neon, calcium, 
and chromium. Its mass is $1.9891 \times 10^{30}$~kg ($332,900$ times the mass of 
Earth). Solar radius is equal to $6.96 \times 10^{8}$~m (109 times Earth's radius) and its 
luminosity (power output) is $\sim 3.8 \times 10^{26}$ watts ($10$ trillion times the 
power consumption of all Earth’s nations combined). The Sun has a spectral class of G2V. 
G2 means that it has a surface temperature of approximately $5,780$~K giving it a white 
color, which often appears as yellow when seen from the surface of the Earth 
because of atmospheric scattering.

\subsection{Different layers of the Sun and their brief introductions}

In Figure 1.2, different parts of the Sun are shown, which mainly consists of {\it core, raiative 
zone, convective zone, photosphere, chromosphere, corona and coronal loops etc.}. 

\begin{figure}[h]
\begin{center}
\vspace{-0.3cm}
\epsfxsize=2.2in \epsfysize=2.2in
\rotatebox{-0.2}{\epsfbox{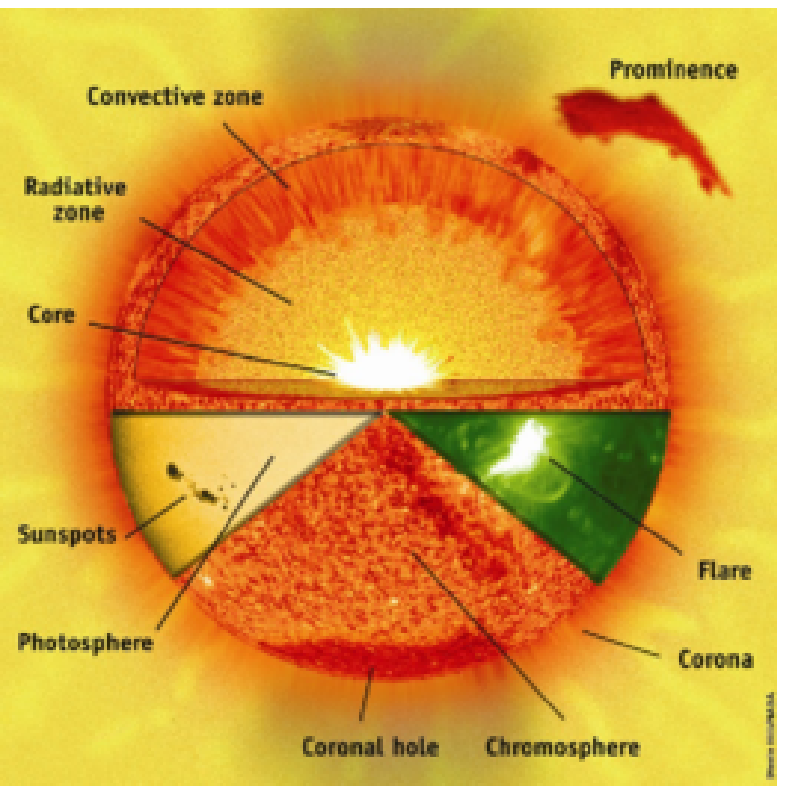}}
\hspace{0.1cm}
\epsfxsize=3.3in \epsfysize=2.2in
\rotatebox{-0.2}{\epsfbox{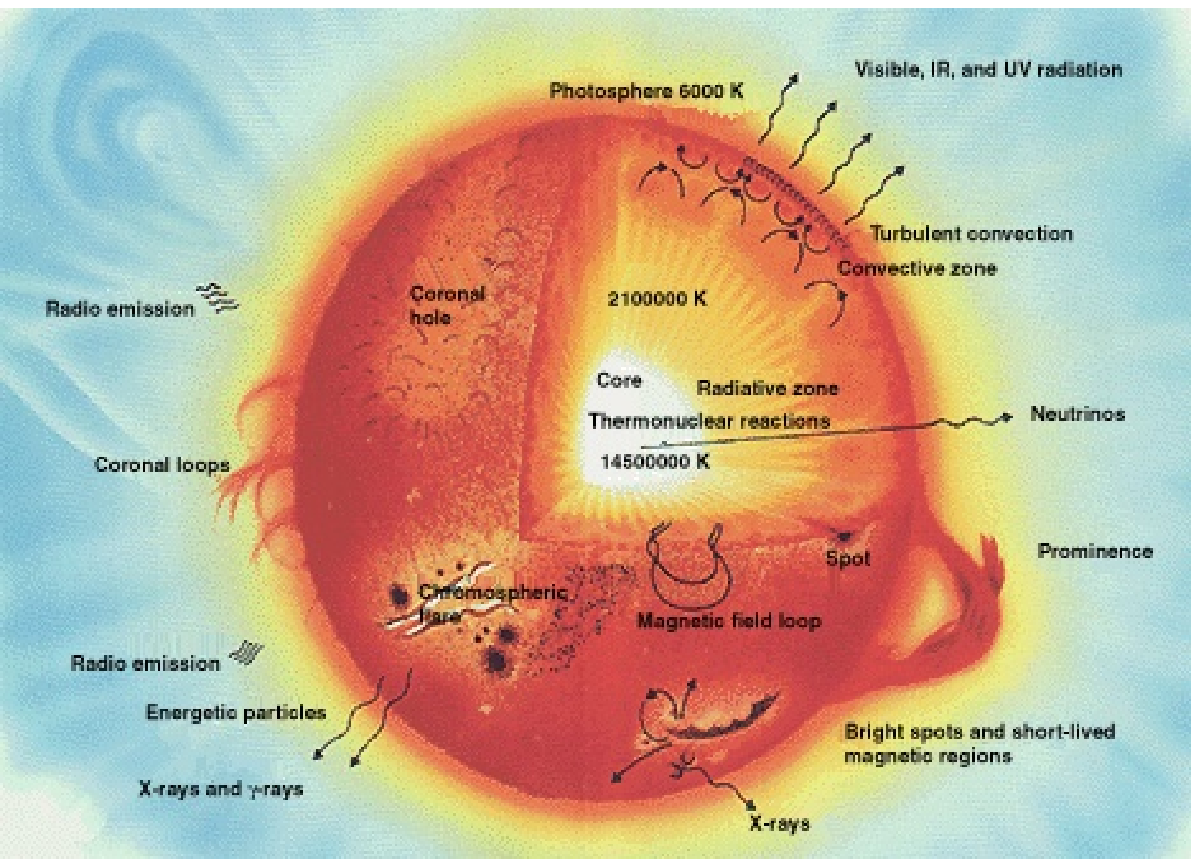}}
\vspace{-0.1cm}
\caption{(a) Left panel shows the gross view and the (b) Right panel shows the detailed view 
of the Solar Cross-Section (courtesy: NASA).}
\label{kn : fig1.2}
\end{center}
\end{figure}

\subsubsection{Core :}
The core of the Sun is extended from the center to about $0.2$ to $0.25$ 
solar radius ($R_\odot$). It has a density of up to $150$ g/cm$^3$ 
and a temperature of close to $1.5 \times 10^7$ K. At the core, about $9.2 \times 10^{37}$ 
protons (hydrogen nuclei) are converted into helium ashes in every second 
(out of $\sim 8.9 \times 10^{56}$ total amount of free protons in the Sun) via the p–p 
(proton-proton) chain reactions, which release total amount of energy $3.83 \times 10^{33}$ 
ergs per second. 

\subsubsection{Radiative Zone :}
The radiative zone extends about $0.25$ to about $0.7$ $R_\odot$. 
In this zone, there is no thermal convection, while the material grows cooler as altitude
increases (from $7\times 10^6~^\circ$C to about $2\times 10^6~^\circ$C). 
This temperature gradient is less than the value of adiabatic loss rate and hence cannot 
drive convection. Heat is transferred by radiation-ions of hydrogen and helium, emit photons.

Between the radiative zone and the convection zone, there is a transition layer called 
the {\it tachocline}. This is the region, where the sharp regime changes between the uniform 
rotation of the radiative zone and the differential rotation of the convection zone, 
results in a large sheared layer - - a condition where successive layers slide past one 
another. The flow motion is found in the convection zone above, slowly disappear from the 
top of this layer to its bottom, matching the calm characteristics of the radiative zone 
on the bottom. Presently, it is believed that a magnetic dynamo within this layer 
generates the solar magnetic field.

\subsubsection{Convective Zone :}

The convection zone is the layer at solar surface, where the solar plasma 
is not dense enough or hot enough to transfer the heat energy of the interior 
outward via radiation. As a result, the thermal convection 
occurs as thermal columns carry hot material to the surface (photosphere) of the 
Sun. Once the material cools off at the surface, it plunges back downward to 
the base of the convection zone, to receive more heat from the top of 
the radiative zone. At the visible surface of the Sun, the 
temperature has dropped to $\sim 5700$ K and the density to only $0.2~g/m^3$.

\subsubsection{Photosphere :}

The visible surface of the Sun, the photosphere, is the layer below which the Sun 
becomes opaque to visible light. Above the photosphere visible sunlight is free to 
propagate into space, and its energy escapes the Sun entirely. The change in opacity 
is due to the decreasing amount of H$^-$ ions, which absorb visible light easily. 
The photosphere is actually ten to hundreds of kilometers 
thick, being slightly less opaque than air on Earth. Because the upper part of the 
photosphere is cooler than the lower part, an image of the Sun appears brighter in 
the center than on the edge or limb of the solar disk, in a phenomenon known as limb 
darkening. Sunlight has approximately a black-body spectrum that indicates its 
temperature is about $6000$~K, interspersed with atomic absorption lines from the 
tenuous layers above the photosphere. The photosphere has a particle density of 
$\sim10^{23} m^{-3}$. 

\subsubsection{Chromosphere :}

The chromosphere layer extends nearly 
$2 \times 10^4$~km above the photosphere with gas density gradually thinning out but the 
temperature rising rapidly upwards. The temperature in the chromosphere increases 
gradually with altitude, ranging up to around $2 \times 10^4$~K near the top. In the 
upper part of chromosphere helium becomes partially ionized. Faculae and solar flares 
are observable in the chromosphere regime. {\it Faculae} are bright luminous hydrogen clouds which 
form above regions where sunspots are about to form. {\it Flares} are bright filaments 
of hot gas emerging from sunspot regions. {\it Sunspots} are dark depressions on the 
photosphere with a typical temperature of $\sim 4000$~K.      

\subsubsection{Corona :}

The corona is the extended outer atmosphere of the Sun, which is much larger in 
volume than the Sun itself. The corona continuously expands into the space forming the 
solar wind, which fills all the Solar System. The low corona, which is very near the 
surface of the Sun, has a particle density around $10^{15} - 10^{16}~m^{-3}$. The 
average temperature of the corona and solar wind is about $1 - 2$ million Kelvins, 
however in the hottest regions, it is around $8 - 20$ million Kelvins. While no complete theory 
yet exists to account for the temperature of the corona, at least some of its heat is 
known to be from magnetic reconnection.

\subsubsection{Coronal loop :}

Coronal loops form the basic structure of the lower corona and the transition region of the Sun. 
These highly structured and elegant loops are a direct consequence of the twisted solar 
magnetic flux within the solar body. The population of coronal loops can be directly linked 
with the solar cycle, it is for this reason coronal loops are often found with sunspots at 
their footpoints. The upwelling magnetic flux pushes through the photosphere, exposing the 
cooler plasma below. The contrast between the photosphere and solar interior gives the 
impression of dark spots, or sunspots.

Coronal loops are a rarity on the solar surface as the majority of closed flux structures 
are empty i.e. the mechanism that heats the corona and injects chromospheric plasma into 
the closed magnetic flux, is highly localized. The mechanism behind plasma filling, dynamic 
flows and coronal heating remains a mystery. The mechanism(s) must be stable enough to 
continue to feed the corona with chromospheric plasma and powerful enough to accelerate 
and therefore heat the plasma from $6000$~K to well over $1$ MK over the short distance from 
chromosphere, transition region to the corona. So, coronal loops are 
targeted for intense study. They are anchored to the photosphere, fed by chromospheric 
plasma, protruded into the transition region and they exist at coronal temperatures after 
undergoing intensive heating.

\subsection{Solar Magnetic Fields}

The existence of large magnetic fields in the sunspots probably provides the most 
important characteristics of an active Sun. It supports a strong, changing magnetic 
field that varies year-to-year and reverses direction about every eleven years around 
solar maximum. The solar magnetic field gives rise to many effects that are 
collectively called the solar activity, including sunspots on the surface of the Sun, 
solar flares, and variations in solar wind that carry material through the solar 
system. Effects of solar activity on Earth include auroras at moderate to high 
latitudes, and the disruption of radio communications and electric power. Solar 
activity is could have played a large role in the formation and evolution of 
the Solar System. This activity also changes the structure of Earth's outer atmosphere.

All the matter in the Sun is in the form of gas and plasma because of its high 
temperatures. This makes it possible for the Sun to rotate faster at its equator 
(about 25 days) than it does at higher latitudes (about 35 days near its poles). 
The differential rotation of the Sun's latitudes causes its magnetic field lines to 
become twisted over a length of time, causing the magnetic field loops to erupt from the 
Sun's surface and trigger the formation of the Sun's dramatic sunspots and solar 
prominences. This twisting action gives rise to the 
solar dynamo and an 11-year solar cycle of magnetic activity as the Sun's magnetic 
field reverses itself about every 11 years.

\subsection{Solar Flare}

A solar flare is a explosion in the Sun's outer atmosphere that can release as much as 
$6 \times 10^{32}$ ergs of energy. Most flares occur in active regions around 
sunspots, where intense magnetic fields penetrate the photosphere to link the corona 
to the solar interior. Flares are powered by the sudden (timescales of minutes to tens 
of minutes) release of magnetic energy stored in the corona. If a solar flare is 
exceptionally powerful, it can cause coronal mass ejections.
Flares produce a burst of continuous radiations across the electromagnetic spectra, from radio 
waves to X-rays and $\gamma$-rays. Flares are believed to originate in chromosphere 
but rise far up into the lower corona. The theory about the origin of solar flares and 
some observational results are discussed in Chapter 3.

Solar flares are classified as A, B, C, M or X according to the peak flux of $100$ to 
$800$ picometer X-rays near Earth, as measured on the GOES spacecraft (see, Fig. 1.3). 
Each class has a peak flux ten times greater than the preceding one, with X class flares 
having a peak flux of order $10^{−4}~W~m^{-2}$. Within a class there is a linear scale 
from $1$ to $9$, so an X2 flare is twice as powerful as an X1 flare, and is four times 
more powerful than an M5 flare. The more powerful M and X class flares are often associated 
with a variety of effects on the near-Earth space environment. Although the GOES 
classification is commonly used to indicate the size of a flare, it is only one 
measure. 

\begin{figure}[h]
\begin{center}
\vspace{-0.3cm}
\epsfxsize=3.0in %\epsfysize=2.5in
\rotatebox{-0.2}{\epsfbox{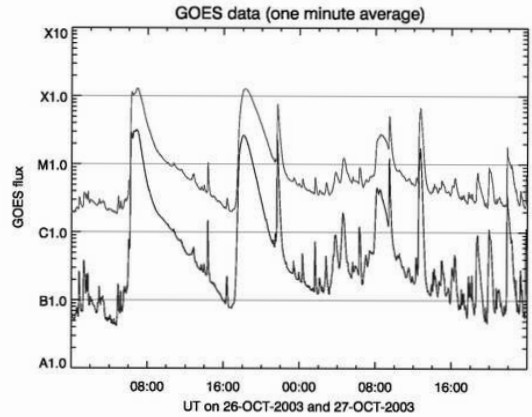}}
\vspace{-0.1cm}
\caption{Classification of Solar Flares using GOES observation. Upper line shows 
$1.55 - 12.5$~keV (Soft X-ray) and lower line shows $3.5 - 24$~keV (Hard X-ray) light
curves respectively (courtesy: www.spaceweather.com).}
\label{kn : fig1.3}
\end{center}
\end{figure}

\subsection{Solar Wind and CME}

The solar wind is a stream of charged particles, plasma - ejected from the upper 
atmosphere of the Sun. It consists mostly of electrons and protons with energies 
of about $1$~keV. The stream of particles varies in temperature and speed with the 
passage of time. These particles are able to escape the Sun's gravity, in part 
because of the high temperature of the corona.

Solar {\it coronal mass ejections} (CMEs), which are caused by release of magnetic 
energy at the Sun. CMEs are often called ``solar storms" or ``space storms" in the 
popular media. They are sometimes, but not always, associated with solar flares, 
which are another manifestation of magnetic energy release at the Sun. 

The Sun appears to have been active for $4.57$ billion years and has enough fuel to 
live on for another $\sim 5$ billion years or so. At the end of its life, the Sun will start 
to fuse helium into heavier elements and begin to swell up, ultimately growing so 
large that it will swallow the Earth. After a billion years as a red giant, it will 
suddenly collapse into a white dwarf -- the final end product of a star like ours. 
It may take a trillion years to cool off completely. 

\subsection{X-ray Astronomy Missions for Solar study}

The satellites named Pioneer 6, 7, 8, and 9 were created to make the first detailed, comprehensive 
measurements of the solar wind, solar magnetic field and cosmic rays. These were 
NASA's mission. Pioneer 6 was launched on December 16, 1965 from Cape Canaveral 
to a circular solar orbit with a mean distance of 0.8 AU. Pioneer 7 was launched on 
August 17, 1966 from Cape Canaveral into solar orbit with a mean distance 
of 1.1 AU. Pioneer 8 was launched on December 13, 1967 from Cape Canaveral into 
solar orbit with a mean distance of 1.1 AU from the Sun. 
Pioneer 9 was launched on November 8, 1968 from Cape Canaveral into solar 
orbit with a mean distance of 0.8 AU. Pioneers 6, 7 \& 8 were partially communicating 
with the ground station till end of the last century, but Pioneer 9 has lost its 
functionality from 1983.

The Solar Maximum Mission satellite (or SolarMax) was designed to investigate solar 
phenomenon, particularly solar flares. It was launched on February 14, 1980. The 
SolarMax mission was ended on December 2, 1989. 

The Global Geospace Science (GGS) WIND satellite is a NASA science spacecraft 
launched on November 1, 1994 from launch pad 17B at Cape Canaveral Air Force Station 
(CCAFS) in Merritt Island, Florida. It was deployed to study radio and plasma that 
occur in the solar wind and in the Earth's magnetosphere before the solar wind 
reaches the Earth. As on April 2008, the mission is still operating.

The Solar and Heliospheric Observatory (SOHO) is a spacecraft that was launched on 
a Lockheed Martin Atlas IIAS launch vehicle on December 2, 1995 to study the Sun. 
It is a joint project of international cooperation between the European Space Agency 
(ESA) and NASA. SOHO currently continues to operate after over ten years in space. 
In addition to its scientific mission, it is currently the main source of near-real 
time solar data for space weather prediction. Originally it was planned SOHO as a 
two-year mission, but currently it continues to operate after over ten years in space. 
In October 2009, a mission extension lasting until December 2012 was approved.

Advanced Composition Explorer (ACE) is a space exploration mission being conducted 
as part of the Explorer program to study matter comprising energetic 
particles from the solar wind, the interplanetary medium, and other sources {\it in situ}.
Real-time data from ACE is used by the Space Weather Prediction Center to improve 
forecasts and warnings of solar storms. The spacecraft is still in generally good condition, 
and has enough fuel to maintain its orbit until 2024.

TRACE (Transition Region and Coronal Explorer) is a NASA space telescope designed 
to investigate the connections between fine-scale magnetic fields and the associated 
plasma structures on the Sun by providing high resolution images and observation of 
the solar photosphere and transition region to the corona. Currently, the mission is 
in working phase with good conditions of detectors.

Reuven Ramaty High Energy Solar Spectroscopic Imager (RHESSI) is the sixth mission 
in the line of NASA Small Explorer missions (also known as SMEX). Launched on 5 
February 2002, its primary mission is to explore the basic physics of particle 
acceleration and explosive energy release in solar flares. RHESSI is designed to 
image solar flares in energetic photons from soft X rays ($\sim3$ keV) to $\gamma$-rays 
(up to $\sim20$ MeV) and to provide high resolution spectroscopy up to $\gamma$-ray 
energies of $\sim20$ MeV. Furthermore, it has the capability to perform spatially 
resolved spectroscopy with high spectral resolution. Currently, this mission is also
in good working phase.

Hinode formerly known as ``Solar-B", is a Japan Aerospace Exploration Agency Solar 
mission with United States and United Kingdom collaboration. It is the follow-up to 
the Yohkoh (``Solar-A") mission. It was launched on the final flight of the M-V 
rocket from Uchinoura Space Center, Japan on September 22, 2006. Hinode was planned 
as a three-year mission to explore the magnetic fields of the Sun. It consists of a 
coordinated set of optical, extreme ultraviolet (EUV), and X-ray instruments to 
investigate the interaction between the Sun's magnetic field and its corona. 

STEREO (Solar TErrestrial RElations Observatory) is a solar observation mission was 
launched on 26 October 2006. This will enable stereoscopic imaging of the Sun and 
solar phenomena, such as coronal mass ejections. It has already passed more than 
three and half years in orbit successfully.

Koronas-Foton, also known as CORONAS-Photon (Complex ORbital Observations Near-earth 
of Activity of the Sun), is a Russian Solar research satellite. It is the third 
satellite in the Russian Coronas programme, and part of the international living 
with a star programme. It was launched on $30^{th}$ January 2009, 
from the Plesetsk Cosmodrome, aboard the final flight of the Tsyklon-3 rocket. 
The satellite's scientific payload includes an array of 12 instruments. Eight 
instruments were designed for registering electromagnetic radiation from the Sun in 
a wide range of spectrum from near electromagnetic waves to $\gamma$-radiation, as 
well as solar neutrons. Two instruments were designed to detect charged particles 
such as protons and electrons. Coronas-Photon also carries three Indian payloads, namely,
R\"{o}ntgen Telescope (RT-2) instruments: RT-2/S, RT-2/G, and RT-2/CZT. These will be used to 
conduct photometric and spectrometric research into the Sun, and for low-energy 
$\gamma$-ray imagery. Unfortunately the mission has lost its functionality after December, 
2009, due to power circuit failure.

\section{Compact objects}

The term Compact Star or Compact object is used to refer collectively to white dwarfs, 
neutron stars, other exotic dense stars, Gamma-Ray Bursts (GRBs), and black holes. 
These objects are very dense and massive, although their radius is small.

Compact stars form at the endpoint of stellar evolution. A star shines and thus 
loses energy. The loss from the radiating surface is compensated by the production 
of energy from nuclear fusion in the interior of the star. When a star has exhausted 
all its energy and undergoes stellar death, the gas pressure of the hot interior 
can no longer support the weight of the star and the star collapses to a denser 
state: a compact star. 

\subsection {X-ray binary}

X-ray binaries are a class of binary stars that are luminous in X-rays. The 
X-rays are produced by matter falling from one component, called the donor (usually 
a relatively normal star) to the other component, called the accretor, which is 
compact: a white dwarf, a neutron star, or a black hole. The infalling matter releases 
gravitational potential energy, up to several tenths of its rest mass as X-rays. 
There are three types of X-ray binaries: High-Mass X-ray Binary (HMXB), 
Intermediate-Mass X-ray Binary (IMXB) and Low-Mass X-ray Binary (LMXB).

\subsubsection{HMXB:} A HMXB is a binary star system that is strong in X rays
in which the normal stellar component is a massive star: usually a O or B star, 
a Be star, or a blue supergiant. The compact, X-ray emitting, component is generally 
a neutron star, black hole, or possibly a white dwarf. A fraction of the stellar 
wind of the massive normal star is captured by the compact object, and produces 
X-rays as it falls onto the compact object.

In a high-mass X-ray binary, the massive star dominates the emission of optical light, 
while the compact object is the dominant source of X-rays. The massive stars are very 
luminous and therefore easily detected. One of the most famous high-mass X-ray binaries 
is Cygnus X-1, which was the first identified black hole. Other HMXBs include Vela X-1 
(not to be confused with Vela X), and 4U 1700-37.

\subsubsection{IMXB:} An IMXB is a binary star system where one of the components is 
a neutron star or a black hole. The other component is an intermediate mass star.

\subsubsection{LMXB:} A LMXB is a binary star where one of the components is either 
a black hole or neutron star. The other, donor, component usually fills its Roche 
lobe and therefore transfers mass to the compact star. The donor is less massive than 
the compact object, and can be on the main sequence, a degenerate dwarf (white dwarf), 
or an evolved star (red giant). Approximately $100$ LMXBs have been detected in 
the Milky Way, and of these, $13$ LMXBs have been discovered in globular clusters. 
Recently, the Chandra X-ray Observatory has revealed LMXBs in many distant galaxies.

A typical low-mass X-ray binary emits almost all of its radiation in X-rays, and 
typically less than one percent in visible light, so they are among the brightest 
objects in the X-ray sky, but relatively faint in visible light. The apparent 
magnitude is typically around 15 to 20. The brightest part of the system is the 
accretion disk around the compact object. The orbital periods of LMXBs range from 
ten minutes to hundreds of days. GRO J1655-40 (primary mass is $\sim7.02 M_\odot$ \& 
companion star mass is $\sim2.3 M_\odot$) and GRS 1915+105 (primary mass is 
$\sim14 M_\odot$ \& companion star mass is $\sim1.2 M_\odot$) are the two well 
known and well studied LMXBs.

\subsubsection{White dwarf}
A white dwarf, also called a degenerate dwarf, is a small star composed mostly of 
electron-degenerate matter. Because a white dwarf's mass is comparable to that of 
the Sun and its volume is comparable to that of the Earth, it is very dense. Its 
faint luminosity comes from the emission of stored thermal energy.

White dwarfs are thought to be the final evolutionary state of all stars whose mass
is less than $1.44$ solar mass ($M_\odot$) (the Chandrasekhar limit). 
A white dwarf is very hot when it is formed but since it has no source of energy, 
it will gradually radiate away its energy and cool down. This means that its radiation, 
which initially has a high color temperature, will lessen and redden with time. 

\subsubsection{Neutron Star}
A neutron star is a type of remnant that can result from the gravitational collapse 
of a massive star during a Type II, Type Ib or Type Ic supernova event. Such stars 
are composed almost entirely of neutrons, which are subatomic particles without 
electrical charge and roughly the same mass as protons. Neutron stars are very hot 
and are supported against further collapse because of the Pauli exclusion principle. 

A typical neutron star has a mass between $1.35 - 2.1$ solar masses, with a 
corresponding radius of about 12 km. In general, compact stars of less than $1.44$ 
solar masses, the Chandrasekhar limit, are white dwarfs, above $2 - 3$ solar masses 
(the Tolman-Oppenheimer-Volkoff limit), produces black hole.
%(the Landau-Oppenheimer-Volkov limit), produces black hole.

\subsubsection{Black Hole}

Black hole (BH) is a region of space which is so dense and has a strong gravitational 
field that not even light or any other kind of radiation can escape: its escape velocity 
exceeds the speed of light. The black hole has a one-way surface, called an {\it event 
horizon}, into which objects can fall but cannot come out. Black holes are predicted 
by Einstein’s theory of general relativity, which shows that if a quantity of matter 
is compressed within a critical radius, no signal can ever escape from it. 

Thus, by definition,
it is impossible to directly observe a black hole. However, it is possible to infer 
its presence by its gravitational action on the surrounding environment, particularly 
with microquasars and active galactic nuclei, where matter falling into a nearby 
black hole is significantly heated and emits a large amount of X-ray radiation. This 
observation method allows astronomers to detect their existence. The only objects 
that agree with these observations and are consistent within the framework of general 
relativity are black holes. Figure 1.4 shows the binary system of BHC GRO J1655-40.

\begin{table}[h]
\small
%\scriptsize
\centering
\caption{\label{table 1.1} Classes of the blackhole}
%\centerline {Table 1.1}
%\centerline {Fitted data with two component flow model}
\vskip 0.1cm
\begin{tabular}{|l|c|c|}
\hline
{\bf ~~~~~~~Class} & {\bf Mass} & {\bf Size}\\
\hline
Supermassive BH & $10^5 - 10^9 M_{Sun}$ & 0.001 - 10 AU \\
\hline
Intermediate-mass BH & $10^3 M_{Sun}$ & $10^3 km = R_{Earth}$ \\
\hline
Stallar-mass BH & $10 M_{Sun}$ & 30 km \\
\hline
Primordial BH & up to $M_{Moon}$ & 0.001 - 10 AU \\
\hline
\end{tabular}
\end{table}

Depending on their physical masses, there are four classes of black holes: 
stellar, intermediate, supermassive and primordial (or mini).
A {\it stellar black hole} is a region of space into which a star (or collection of 
stars or other bodies) has collapsed. This can happen after a star massive enough to 
have a remnant core of more than $2-3$ solar masses %(the Landau-Oppenheimer-Volkov limit 
(the Tolman-Oppenheimer-Volkoff limit for neutron stars) reaches the end of its 
thermonuclear life. It collapses to a critical size, overcoming both electron and 
neutron degeneracy pressure, whereupon gravity overwhelms all other forces. 
Black holes found at the center of galaxies have a mass up to $100$ million solar 
masses and are called {\it supermassive black holes}.
Between these two scales, there are believed to be {\it intermediate black holes} 
with a mass of several thousand solar masses. 
{\it Primordial black holes}, proposed by Stephen Hawking, could have been created 
at the time of the BIG-BANG, when some regions might have got so compressed that
they underwent gravitational collapse. With original masses comparable to that of 
earth or less, these mini black holes could be of the order of $1$~cm (about half an 
inch) or smaller. Their existence is, at yet, not confirmed. 

\begin{figure}[h]
\begin{center}
\vspace{-0.2cm}
\epsfxsize=3.4in %\epsfysize=5.5in
\rotatebox{-0.2}{\epsfbox{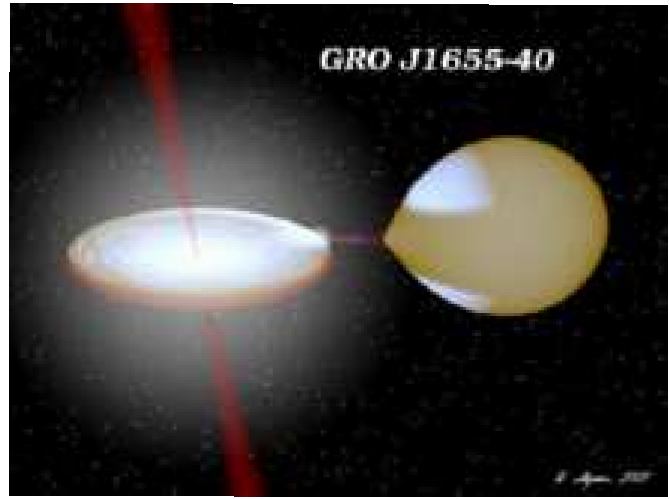}}
%\epsfxsize=3in \epsfysize=4in
%\rotatebox{270}{\epsfbox{fig3.1b.ps}}
\vspace{-0.1cm}
\caption{Artists' conception of X-ray binary system ``GRO J1655-40" where matters 
are stripped off and transferred via Roche lobe overflow into an accretion disk 
around GRO J1655-40 black hole from its companion star. A torus of material is 
shown spiraling onto the black hole. This torus is the hot inner region of the
accretion disk, wherefrom most of the X-rays are produced. It also shows the jet
outflow from the binary system (courtesy: NASA).}
\label{kn : fig1.4}
\end{center}
\end{figure}

We can also classify black holes according to their physical properties. The simplest 
massive black hole has neither charge nor angular momentum. These non-rotating black 
holes are often referred to as {\it Schwarzschild black holes} after the physicist 
Karl Schwarzschild who discovered this solution in 1915. 
There are also another two types of black hole which are rotating. One type of this 
class don't have any charge, called {\it Kerr black holes}. Another type of rotating 
black holes consist of changes are called {\it Kerr-Newman black holes}. These rotating 
black holes obey exact black hole solutions of Einstein's equations of General 
Relativity. These rotating black holes are formed in the gravitational collapse of 
a massive spinning star or from the collapse of a collection of stars or gas with 
an average non-zero angular momentum. As most stars rotate it is expected that most 
black holes in nature are rotating black holes. Well known galactic black hole 
candidate GRS 1915+105, may be rotating $1150$ times per second.

\subsection{Radiative Processes associated with black hole}

All celestial objects in our Universe emit radiations in the entire electromagnetic 
spectrum from radio to $\gamma$-rays. Although the production mechanism of the 
radiations of different wave bands, depends on the nature of the surrounding medium 
and physical processes associated with the object. Most the observed radiations in 
black holes are in the X-rays and this emission mechanism in this wave band are of 
mainly two types: Thermal emission and Non-thermal emission (Rybicki \& Lightman, 1979).

\subsubsection{\textbf A. Thermal Emission}

Thermal emission is electromagnetic radiation emitted from the surface of an object 
which is due to the object's temperature. From observation point of view it is 
confirmed that are in general four type of thermal radiations emit from black 
hole, are: blackbody radiation, thermal bremsstrahlung, thermal Comptonization, 
and line emission.

\noindent{\bf a$)$ Blackbody Radiation:}

In this case, the radiation is emitted by an idealized (i.e. system with 
thermodynamic equilibrium) perfect medium (object). It has a continuous spectrum
that depends only on the 
temperature of the source. The emission from a black body, known as the black body
radiation. Different regions of many astronomical objects emit 
this type of radiation. Black body radiation follows the Planck distribution law. 
The peak of the distribution shifts to shorter wavelengths as the temperature 
increases (Wien's law). This leads to the common experience that at moderate 
temperatures objects glow a dull red, then change colour successively through 
bright red, yellow, white to blue-white as the temperature is increased. 
The total emitted energy increases rapidly with temperature, leading to 
the Stefan-Boltzmann law. According to the Planck's law blackbody spectrum 
can be described as
$$
%\centerline {$I(\nu,T) = (\frac{2h \nu^3}{c^2}) (\frac{1}{e^{\frac{h\nu}{kT}} - 1})$} 
I(\nu,T) = (\frac{2h \nu^3}{c^2}) (\frac{1}{e^{\frac{h\nu}{kT}} - 1}) ,
\eqno{(1.1)}
$$
where, $I(\nu,T)$ is the amount of energy emitted by the black body per unit surface 
area per unit time per unit solid angle, $h$ is the Planck constant, $c$ is the speed of
light in vacuum, $k$ is the Boltzmann constant, $T$ is the temperature in Kelvin and 
$\nu$ is the frequency of the emitted electromagnetic radiation.

It has been observed that most of galactic black hole candidates emit blackbody 
radiation as a continuous energy spectrum. This is actually emitted from their hot 
($\sim kT$), optically thick moving plasmas from their accretion disks. But observed
black body spectrum highly depends on the accretion disk temperature. So, it is the 
modified version of the black body spectrum. This composite photon spectrum (called 
{\it disk black body} (diskbb) spectrum) emitted from the BH disk at an inclination angle of 
$\theta$ and inner, outermost disk radii as $r_{in}$ and $r_{out}$ respectively
can be defined as (Makishima et al. 1986):
$$
%\centerline {$f(E) = \int_{r_{in}}^{r_{out}} 2\pi r~cos \theta~I[E,T(r)]dr =	
%	\frac{8\pi}{3}r_{in}^2 cos \theta\int_{T_{out}}^{T_{in}}(T/T_{in})^{-11/3}I(E,T)dT/T_{in}$}
f(E)=\int_{r_{in}}^{r_{out}} 2\pi r~cos \theta~I[E,T(r)]dr= \frac{8\pi}{3}r_{in}^2 cos \theta
\int_{T_{out}}^{T_{in}}(\frac{T}{T_{in}})^{-11/3}I(E,T)\frac{dT}{T_{in}} ,
\eqno{(1.2)}
$$
where, $T_{in}$ = $T(r_{in}$) and $T_{out}$ = $T(r_{out}$) are the innermost and outermost disk 
temperatures, respectively, $E$ is the emitted photon energy, and $I(E,T)$ is the black body photon
flux per unit photon energy from a unit surface area of temperature $T$. $I(E,T)$ can be derived
from the above mentioned black body equation as 
{$I(E,T) = \frac{2E^3}{(hc)^2} (\frac{1}{e^{\frac{E}{kT}} - 1})$} (using energy equation $E = h\nu$).

\noindent{\bf b$)$ Thermal Bremsstrahlung:}

When an electron is accelerated in the Coulomb field of an ion or other charge particle, it
emits a radiation, which is called the bremsstrahlung (or free-free) emission.
In astrophysics, thermal bremsstrahlung radiation occurs when the particles are 
populating the optically thin emitting plasma and are at a uniform temperature. 
They follow the Maxwell-Boltzmann distribution.

The power emitted per cubic centimeter per second can be written as
(Rybicki \& Lightman, 1979),
$$
%\centerline {$\epsilon_{ff} = 1.4\times 10^{-27} T^{1/2} n_{e} n_{i} Z^{2} g_B$}
\epsilon_{ff} = 1.4\times 10^{-27} T^{1/2} n_{e} n_{i} Z^{2} g_B ,
\eqno{(1.3)}
$$
where, `ff' stands for free-free, $1.4\times 10^{-27}$ is the condensed form 
of the physical constants and geometrical constants associated with integrating 
over the power per unit area per unit frequency, n$_e$ and n$_i$ are the 
electron and ion densities respectively, $Z$ is the number of protons of 
the bending charge, $g_B$ is the frequency averaged {\it Gaunt factor} 
and is of order unity and $T$ is the global X-ray temperature determined from the 
spectral cut-off frequency $\hbar\nu = kT$, above which exponentially small amount 
of photons are created because the energy required for creation of such a photon 
is available only for electrons belonging to the tail of the Maxwell-Boltzmann distribution.

This process is also known as Bremsstrahlung cooling since the plasma is usually optically 
thin to photons at these energies and the energy radiated is emitted freely into 
the Universe.

%\vskip 1.5cm
\noindent{\bf c$)$ Thermal Comptonization:}

Comptonization is a simple radiative process, occurs when a X-ray or $\gamma$-ray 
photon undergo in a matter and encounters with an electron. The inelastic scattering 
of photon with electron results decrease of photon energy, called {\it Compton scattering} 
where a part of the photon energy is transferred to the scattered electron. 
When the photon gains energy, it is called the {\it inverse Compton scattering}.

If $E (= h\nu = hc/\lambda)$ and $E' (= h\nu' = hc/\lambda')$ are the incident 
and scattered photon energies respectively, then $\Delta E = E' - E$ can be defined as
$$
%\Delta E = E' - E \simeq -\frac{E^2}{m_e c^2}(1-cos\theta)
\Delta E = E' - E \simeq \frac{E^2}{m_e c^2}(1-cos\theta) ,
\eqno{(1.4)}
$$
where, $m_e$ is the rest mass of electron, $c$ is the velocity of light and $\theta$ is 
the scattering angle of the incident photon, i.e., angle between its initial final directions.
When $\Delta E < 0 $, i.e., the scattered photon loses its energy, Compton scattering occurs 
and when $\Delta E > 0 $, inverse Compton scattering occurs.
Inverse Compton scattering is very important in astrophysics. In X-ray astronomy, the accretion
disk surrounding a black hole is believed to produce a thermal spectrum. The lower energy
photons produced from this spectrum are scattered to higher energies by relativistic 
electrons in the surrounding matter. This is believed to cause the power-law component in 
the black hole X-ray spectra.

{\it Thermal Comptonization} is the a method of Compton scattering, when it occurs on a 
thermal plasma full of electrons characterized by temperature $T$ and optical depth $\tau$. 
The mean relativistic energy gain per collision can be expressed as
$$\frac{\Delta E}{E} \simeq (\frac{4kT}{m_e c^2}) + 16 (\frac{kT}{m_e c^2})^2
~~~~~~~~~~~~~~~~~for~~E\ll kT ,$$
$$~~~~~~\leq 0~~~~~~~~~~~~~~~~~~~~~~~~~~~~~~~~~~~~~~~~~for~~E\geq kT .
\eqno{(1.5)}
$$

Number of scatterings depend on the optical depth of the medium. The relation of the number
of scatterings ($N$) with optical depth ($\tau$) is $N \simeq (\tau + \tau^2)$. In the 
Compton scattering process, there is another important parameter, called Compton `$y$' 
parameter, which signifies if during the time of traversing through a medium, a photon 
will be able to change its energy significantly or not. It can be defined as: 
$$y \equiv (av.~energy~change~per~scattering)\times(number~of~scatterings) ,$$
$$\hspace{-7.25 cm}= \frac{\Delta E}{E} N \simeq (\frac{4kT}{m_e c^2})\tau .
\eqno{(1.6)}
$$

\noindent{\bf d$)$ Line Emission:}

The {\it emission spectrum} of a chemical element or compound is the relative intensity of 
the frequency of radiation emitted by the element's atom or the compound's molecules when 
they return back to their ground state. Each element's emission spectrum is unique. So, from 
the spectral analysis one can easily identify the emitted chemical element or compound.

In X-ray astronomy, the {\it line emission} is also an important source of radiation. 
In a hot gas ($\sim10^7$ K), the elements heavier than hydrogen are not completely 
ionized except at high temperatures. When a fast electron strikes an ion with bound
electrons, it often transfers energy to that ion, causing a transition to a
higher energy level. After a short while, the excited ion decays rapidly to
the ground state by radiating photons of energy characteristics of the spacing
of energy levels through which the excited electron passes. This radiation
appears as spectral lines with energies determined by the radiating ion
species (material present in the hot gas).

So, for example, in a `He-like' ions, the so-called resonance line is produced
when an electron jumps from $1s2p~^1P_1$ level to the $1s^2~^1S_0$ level. This
emission mechanism is somehow complex whereas the emission of H$_\alpha$
line is rather simple, which requires an electron to jump from $n=3$ to $n=2$
(n is the principle quantum number).

Most of the cases, it is found that the radiation from the thermal gas is a blend
of the thermal bremsstrahlung and the line radiation (from different ion 
species). Line emission appears predominantly in plasmas that have temperature 
less than $5\times 10^7$ K. Above this temperature, almost all the ions are 
stripped off their bound electrons that causes them to radiate the energy as 
an X-ray continuum. Thus observing the X-ray spectra, the shape of the continuum 
and the presence of lines can identify the origin as a hot gas of plasma. The
temperature of the gas can be calculated from the particular lines present and
from the shape of the high energy end of the bremsstrahlung continuum. The
strength and energies of the lines also reveal the elemental composition of
the hot gas.

\subsubsection{\textbf B. Non-thermal Emission}

The term `non-thermal' emission generally refers to the radiation from
particles whose distribution do not follow the Maxwell-Boltzmann distribution. 
The non-thermal emissions are very important in high energy astrophysics.

Impulsive particle acceleration and the consequent sudden release of energy through 
electromagnetic radiation is an important observational aspect in present day astrophysics. 
The impulsive acceleration takes place in diverse settings like planetary atmospheres, 
solar active regions, accretion disc surrounding compact objects like neutron stars and 
black holes, magnetic quakes in magnetars and possibly coalescing compact objects 
causing the release of large amount of energy at cosmological distances in gamma-ray 
bursts. One of the common features of such a phenomenon is the emission of hard X-rays 
and soft gamma-rays from mildly relativistic electrons and gamma-rays from relativistic 
protons and nucleons. It is the dream of every observational high energy astrophysicist 
to measure the shape and time evolution of the spectrum of such radiation and get to the 
basic physics governing the particle acceleration, say, for example, extract the very 
fundamental Physics going on near the event horizon of a black hole. 

In recent times, in the astrophysical context, Compton Gamma-ray Observatory (CGRO), 
BeppoSAX, Rossi X-ray Timing Explorer (RXTE) have made very important observational 
studies in the field of accretion on to black holes (stellar mass and super massive), 
gamma-ray bursts etc. These observations have gone a long way in identifying the sources 
of emission and associating and correlating emissions in other wave-bands, but, 
the extraction of the hard X-ray or gamma-ray continuum spectra and deconvolving 
the source physical processes has made some modest progress. 

\noindent{\bf a$)$ Cyclotron Radiation:}

Cyclotron radiation is the electromagnetic radiation emitted by a charged particle 
circling in a magnetic field substantially below the speed of light (non-relativistic). 
The Lorentz force on the particles acts perpendicular to both the magnetic field lines 
and the particle's motion through them, creating an acceleration of charged particles 
that causes them to emit radiation (and to spiral around the magnetic field lines). 
The radiation is circularly polarized and appears at a single frequency, the 
{\it gyro-frequency}, which is independent of the velocity of the particle, but depends 
on the strength of the magnetic field ($B$), and is given by $\omega_g = eB/m_e$. 
The cyclotron makes use of the circular orbits that charged particles exhibit in a 
uniform magnetic field. Furthermore, the period of the orbit is independent of the 
energy of the particles, allowing the cyclotron to operate at a set frequency, and 
not worry about the energy of the particles at a given time. Cyclotron radiation is 
emitted by all charged particles traveling through magnetic fields, however, not 
just those in cyclotrons. Cyclotron radiation from plasma in interstellar space or 
around black holes and other astronomical phenomena are an important source of 
information about distant magnetic fields.

Cyclotron radiation has a spectrum with its main spike at the same fundamental
frequency as that of the particle's orbit, and harmonics at higher integral factors.
Harmonics are the result of imperfections in the actual emission environment, which
also create a broadening of the spectral lines. The most obvious source of line
broadening is non-uniformities in the magnetic field, as an electron passes from one
area of the field to another, its emission frequency will change with the strength
of the field. Other sources of broadening include collisional broadening from the
electron failing to follow a perfect orbit, distortions of the emission caused
interactions with the surrounding plasma, and relativistic effects if the charged
particles are sufficiently energetic. In some X-ray binaries, such as Her X-1,
$B$ can be order of $10^{12}$ Gauss, so that $\nu_g$ corresponds to hard X-rays at $10 - 30$ keV.

\noindent{\bf b$)$ Synchrotron Radiation:}

Synchrotron radiation is the electromagnetic radiation, similar to cyclotron radiation, 
but generated by the acceleration of relativistic (i.e. moving near the speed of light) 
charge particles through magnetic fields. The radiated energy is proportional to the 
fourth power of the particle speed and is inversely proportional to the square of the 
radius of the circulating path. The radiation produced in the process may range over 
the entire electromagnetic spectrum, from radio waves to infrared, visible, ultraviolet, 
X-rays and high energy $\gamma$-rays. It is distinguished by its two characteristics 
nature, polarization and non-thermal power-law spectra. It was first detected 
in the jet of M87 in 1956 by G. R. Burbidge. Environs of supermassive black holes 
produce synchrotron radiation, especially by relativistic beaming of jets produced 
by accelerating ions through magnetic fields.

The classical formula for the radiated power from an accelerated electron is 
$$ P = \frac{2ke^2}{3c^2} a^2 .
\eqno{(1.7)}$$ 
For a non-relativistic circular orbit, the acceleration is just the centripetal acceleration, 
$v^2/r$. However, for relativistic case, the acceleration can be written as 
$a = \gamma^2\frac{v^2}{r}$, where $\gamma = \frac{1}{\sqrt{1-\frac{v^2}{c^2}}}$.  So, for the 
relativistic limit, the radiated power for the synchrotron process becomes 
$$P = \frac{2ke^2}{3c^2}[\gamma^2\frac{v^2}{r}]^2 = \frac{2ke^2\gamma^4v^4}{3c^3r^2} . \eqno{(1.8)}$$

\noindent{\bf c$)$ Non-thermal Comptonization:}

In the previous section we have discussed thermal Comptonization processes, which occurs 
in the presence of thermal electrons (obeying Maxell-Boltzmann's distribution). However,
with the presence of non-thermal electrons in the plasma, the process of Comptonization 
will be modified. This is called non-thermal Comptonization.

In case of thermal distribution, there is a upper limit of the spectral energy at $\sim 400$ keV.
However, the effect of non-thermal electrons on Comptonization is to produce a high-energy 
tail. This is quite above the thermal cut-off. The high-energy tail is simply the characteristics
of the superposition of the individual electron spectra of non-thermal electrons which have 
optical depth $\tau_{es}$ $\ll$ 1. Therefore, the  spectral shape depends on the the energy
index $p$ of the power-law distribution of the electrons, and the resultant spectrum is of 
power-law like nature, with an spectral index $s = (p-1)/2$.

Also in this case, the seed-photon flux (compared to the thermal case) will be much higher, 
so the luminosity in the non-thermal Comptonization spectrum will be more. The high energy 
tails (beyond 400 keV) in the X-ray spectrum of black holes are modeled as power-law 
distribution of non-thermal electrons that are present in the hot plasmas of the accretion disks.

\subsection{Accretion processes around a black hole}

In an accretion process in a binary system, a black hole attracts matter from its 
companion. This matter releases energy in different wavelengths (from radio to
gamma rays). So, it is very important to know what fraction of 
gravitational potential energy is released via accretion process and is
converted into energetic radiation.
On the other hand, studying the energetic radiation one can 
measure physical properties of the black hole and also its companion object. 
Here, we will discuss all these effects.

In the process of accretion, matter falls from companion object to black hole 
due to the gravitational force. In the process, elements of infalling matter gains kinetic 
energy with the loss of its potential energy. The rate of the radiated energy
i.e. {\it luminosity} in the accretion process is given by,
$$
L = \frac {1}{2} \dot{m} v_{ff}^2 = \frac {GM\dot{m}}{R}
= \frac {1}{2} \frac {(2GM/c^2)}{R} \dot{m}c^2
= \frac {1}{2} (\frac {r_g}{R}) \dot{m}c^2 = \eta \dot{m}c^2,
\eqno (1.9)
$$
where $r_g = 2GM/c^2$ is the Schwarzschild radius, $\dot{m}$ is the 
{\it accretion rate} and $\eta$ is a parameter known as {\it efficiency}, which 
is the measure of the fractional change of gravitational energy into radiation. 
Again, $\eta$ can be written as,
$$
\eta = \frac {1}{2} (\frac {r_g}{R}) = \frac {G}{c^2} (\frac {M}{R}),
\eqno (1.10)
$$
where, $M/R$ is the measure of the compactness of the star.
The calculated values of the efficiency factor ($\eta$) for a white dwarf 
($M \sim M_\odot$, $R \sim 5\times 10^3$ km), a neutron star ($M \sim M_\odot$, 
$R \sim 14$ km) and a black hole ($R = r_g = 2GM/c^2$) are approximately 
equal to $0.0003$, $0.1$ and $0.5$ respectively.

The characteristic luminosity of any compact object is known as 
{\it Eddington luminosity} ($L_{Edd}$) can be expressed as,
$$
L_{Edd} = \frac {4\pi GMm_p c}{\sigma_T} \simeq 1.3\times10^{38}
(\frac {M_{BH}}{M_\odot})~{\rm ergs~sec^{-1}} 
\simeq 3.3\times 10^4(\frac {M_{BH}}{M_\odot})L_\odot ,
\eqno (1.11)
$$
where, $m_p$ is the proton mass, $\sigma_T$ is the Thomson scattering
cross-section, $M$ is the mass of the gravitating object, $M_{BH}$ is the mass
of a black hole and $L_\odot (=4\times 10^{33}~{\rm ergs~sec^{-1}}$), is the solar 
luminosity. The typical Eddington luminosity for the Galactic black hole candidate 
GRO J1655-40 of mass $7.02M_\odot$, is $9.386\times 10^{38}~{\rm ergs~sec^{-1}}$.
The corresponding mass accretion rate is called the {\it Eddington accretion rate} 
and is given by,
$$
\dot{M}_{Edd} \simeq 1.44\times10^{17} (\frac {M_{BH}}{M_\odot}) ~{\rm gm/sec}.
\eqno (1.12)
$$
In observational astronomy, the {\it Eddington luminosity} and the corresponding 
accretion rate are treated as yardsticks to measure many physical properties of 
the stars as well as the compact X-ray binary system.

In the following sub-sections, we will briefly discuss the real accretion processes 
along with the development of accretion disk models, from Bondi flow to Two Component
Advective Flow (TCAF) paradigm.

\subsubsection{$\bullet$ Spherically symmetric accretion flow: Bondi flow}

This is simplest model for the accretion flow dynamics, where the flow is spherically 
symmetric, and adiabatic in nature. Here, the matter flows sub-sonically 
at a large distance, and becomes supersonic near a black hole. Under some 
conditions, it can become supersonic to subsonic. This model was introduced by {\it Sir Hermann 
Bondi} in his 1952 published classic paper (Bondi, 1952). A detailed description of 
this particular type of flow has been given in the book of {\it Theory of Transonic 
Astrophysical Flows} (1990). The mass accretion rate relationship for the Bondi Flow is given by,
$$
\dot{M} = 4\pi r^2 \rho u,
\eqno (1.13)
$$
which is constant throughout the flow. During the mass accretion process from companion 
stars towards compact objects, the flow matter velocity ($u$) and density ($\rho$) increases.
In the process of Bondi flow on a Schwarzchild black hole, the accreting matter velocity 
reaches the velocity of light ($c$) at the horizon with density $\sim$~0 (as most of the 
matters are sucked in by the black hole at horizon). However, the maximum attainable 
sound speed for the flow is $c/\sqrt3$. At infinity (where matter is at almost rest), 
the flow velocity and density are characterized by $v_\infty \sim 0$ and $\rho_\infty$
respectively. The flow thus is essentially transonic in nature. There exists a point between
infinity and the horizon, where the flow
velocity becomes equal to the sound speed ($a=\sqrt{\gamma P/\rho}$, where $\gamma$, $P$ 
and $\rho$ are the adiabatic index, pressure and density respectively) of the medium. 
This is known as the sonic point ($r_s$).
After integration on the Euler's equation for adiabatic flows
and after using above boundary conditions, $a_s$ can be derived as,
$$
a_s(r_s) = a_s(\infty) (\frac {2}{5-3\gamma})^{1/2},
\eqno (1.14)
$$
where, $\gamma$ is the adiabatic index. The corresponding mass accretion rate,
in terms of $\rho(\infty)$ and $a_s(\infty)$, is given by,
$$
\dot{M} \approx 1.4\times10^{44} (\frac {M}{M_\odot})^2 [\frac {\rho(\infty)}
{10^{-24}}] [\frac {a_s(\infty)}{10 {\rm km/s}}]^{-3}~{\rm gm}~{\rm sec^{-1}}.
\eqno (1.15)
$$

\subsubsection{$\bullet$ Disk accretion: Keplerian/sub-Keplerian accretion flows}

There are mainly three well known and generally accepted accretion disk models.
{\it (a)} Shakura-Sunyav disk or Standard Keplerian disk model, {\it (b)} 
Advection Dominated Accretion Flow (ADAF) model and {\it (b)} Two Component Advective 
Flow (TCAF) model. 
In the following sub-sections, we will briefly discuss about these models.

\subsubsection{(a) Shakura-Sunyav disk or standard Keplerian disk model} 

N.I. Shakura \& R.A. Sunyaev proposed the so-called {\it standard 
Keplerian disk} model in their 1973 published classic paper (Shakura \& Sunyaev, 1973). 
This model is also known as the Shakura-Sunyav or SS disk model. In this model, it is 
assumed that the accreted matter forms a geometrically thin accretion disk around the
compact object and the inner edge of the disk truncates at $6GM/c^2$, i.e., $\sim$ 3 
Schwarzchild radii ($r_g$) for a non-rotating Schwarzchild black hole. In the accretion 
process under this model, the infall velocity of the matter is much smaller than the 
free-fall velocity. Here, the matter flows towards the black hole in viscous time scale. 
In this model, it is assumed that the excess angular momentum is transported
outward via viscous processes. The viscous stress 
($w_{r \phi}$) is proportional to the pressure $P_{tot}$ :
$$
w_{r \phi} = -\alpha P_{tot} = -\alpha \rho a_s^2,
\eqno (1.16)
$$
where, $\alpha$ is the viscosity parameter (always $>$ 1), $\rho$ is the density and 
$a_s$ is the iso-thermal sound speed. 

Novikov \& Throne (1973) improved this model by including general relativity. However,
the basic assumption of a thin and subsonic disk having a cutoff at 3 Schwarzschild 
radii remained the same.

\subsubsection{(b) Advection Dominated Accretion Flow (ADAF) model}

The standard Keplerian disk model certainly explains much of the observations. But it is not 
a global solution for black hole accretion. A self-similar solution was presented by Narayan 
\& Yi (1994) which shows that very low radiative efficiency is possible as very low accretion 
rates. This is known as the advection dominated accretion flow or ADAF. Since a complete and 
more accurate transonic flow solution achieves much of the same thing and more (due to the 
shock formation), the ADAF model will not be discussed here in detail.

As the accretion rate increases, the optical depth ($\tau$) 
to absorption also increases. Consequently, matter travel time also increases  as it is 
roughly proportional to $\tau^2$. The photon escape time exceeds the advection 
time (Jaroszy$\acute{\rm n}$ski et al. 1980) and so a fraction of radiation is advected 
through the black hole event horizon. In 1988, Abramowicz et al. in their 
so-called `slim accretion disk' model tried to show that as a result of advection a new 
stationary equilibrium solution is possible at the super-Eddington accretion rates.

\subsubsection{(c) Two Component Advective Flow (TCAF) model}

About three decades ago, Paczy\'nski \& Wiita (1980), Jaroszy$\acute{\rm n}$ski et al. in (1980)
and others tried to introduce a general transonic flow solution as they realized
that the flow from a Keplerian disk must become supersonic close to a black hole, 
However, they succeeded only partially as only the inner sonic point was studied.
In order to achieve global and complete black hole matter accretion solution,
Chakrabarti and his collaborators (Chakrabarti, 1990, Chakrabarti \& Molteni, 1993, 
Chakrabarti \& Titarchuk, 1995, Chakrabarti, 1996a, 1996b, Molteni, Sponholz \& Chakrabarti, 
1996, Ryu, Chakrabarti \& Molteni, 1997) in the nineties 
introduced a globally complete solution which is known as the {\it two component 
advective flow} or TCAF model. They included all the physical processes while solving 
the most general and basic flow equations. It has been observed that the flow remains 
stable under TCAF model in most of the parameter space even when nuclear 
reactions is turned on. Two Component Advective Flow (TCAF) is the combination 
of two types of flows: Keplerian and sub-Keplerian. In the Keplerian component,
the matter flows in Keplerian orbit and it radially flows toward the black hole 
slowly in the viscous time scale. However, in the sub-Keplerian flow,
matter does not follow the Keplerian orbit and the matter flow velocity is faster.

In an advective disk, the flow advects (or carries) some physical quantities e.g., mass, 
energy, angular momentum, entropy etc. It is assumed that matter flow velocity at a large 
distance from the black hole near about zero (subsonic), but before entering into the black 
hole, its velocity would have to be supersonic (i.e., Mach Number $M = u/a > 1$, where $u$ 
is the average radial flow velocity and $a$ is the sound speed of the medium). So, the matter 
has to cross in its flow path at least one sonic point (where $M = 1$). 
From the previous discussion, we know that the sub-Keplerian flow moves in almost free-fall time scale
which carries less angular momentum than the Keplerian flow (carries more angular momentum and 
moves in viscous time scale). It is known that the centrifugal force ($\sim 1/r^2$) 
grows much faster than the gravitational force ($\sim 1/r$), so for a stable flow solution 
a standing shock wave is required, which indeed forms a `boundary layer'. This shock may be
present and may be oscillating in nature or absent altogether, depending on the physical conditions. This
shock oscillation creates the temporal variability in the form of `quasi periodic oscillations' 
(QPOs) which have been observed for many black hole candidates. In this boundary layer, 
the kinetic energy of the flow is converted into the thermal energy. This forms a 
hot Compton cloud region above the disk which can inverse-Comptonize  
soft photons into hard photons and produce outflows and winds (Chakrabarti, 1999). 
This boundary layer is the CENtrifugal pressure supported BOundary Layer 
(CENBOL) and is an essential component of the accretion process.

The propagating and oscillating shock (POS) model (derived from this general 
TCAF model) is able to explain QPO evolutions in the transient black hole candidates (Chakrabarti 
et al. 2005, 2006a, 2008a, 2009a, Debnath et al. 2008a, 2010a) in many outburst sources. 
These results will be discussed in subsequent Chapters.

\subsection{X-ray Astronomy Missions for Black Hole Study}

The Astronomical Netherlands Satellite (ANS) was a space-based X-ray and Ultraviolet 
telescope. It was launched into Earth orbit on 30th August $1974$ from Vandenberg Air 
Force Base, United States. The mission was on for 20 months until June 1976, and was 
jointly funded by the Netherlands Institute for Space Research and NASA.

Cos-B was the first European Space Research Organization mission to study $\gamma$-ray 
sources. The mission consisted of a satellite containing $\gamma$-ray detectors, which 
was launched by NASA on behalf of the ESRO on August 9, 1975. The mission was 
completed on April 25, 1982.

Einstein Observatory (HEAO-2) was the first fully imaging X-ray telescope put into 
space and the second of NASA's three High Energy Astrophysical Observatories. 
HEAO-2, was launched on November 13, 1978, from Cape Canaveral, Florida.
The mission was completed on March 25, 1982. 

The Exosat satellite was operational from May 1983 until April 1986 and in that 
time made 1780 observations in the X-ray band of most classes of astronomical object 
including active galactic nuclei, stellar coronae, cataclysmic variables, white dwarfs, 
X-ray binaries, clusters of galaxies, and supernova remnants.

ASTRO-C, renamed Ginga (Japanese for `galaxy'), was an X-ray astronomy satellite 
launched from the Kagoshima Space Center on 5 February 1987. Ginga reentered the Earth's 
atmosphere on 1 November 1991.

ROSAT was a German X-ray satellite telescope. It was named in honour of Wilhelm 
R\"{o}ntgen. It was launched on June 1, 1990 with a Delta II rocket from Cape 
Canaveral. 

ASCA (formerly named ASTRO-D) was Japan's fourth cosmic X-ray astronomy mission, 
and the second for which the United States provided part of the scientific payloads. 
The satellite was successfully launched February 20, 1993. The mission has lost its 
life in 2001. 

The Array of Low Energy X-ray Imaging Sensors (ALEXIS) USA X-ray telescopes feature 
curved mirrors whose multi-layer coatings reflect and focus low-energy X-rays or 
Proportional Counter Array (PCA), the High-Energy X-ray Timing Experiment (HEXTE), 
and one instrument called the All Sky Monitor (ASM). The RXTE observes X-rays from 
black holes, neutron stars, X-ray pulsars and X-ray bursts. RXTE was launched from 
Cape Canaveral on 30 December 1995 on a Delta rocket. RXTE is a great successful 
mission. The discovery list of the Satellite is huge. On the basis of RXTE results, 
as of late 2007, more than 1700 scientific papers have been published.

BeppoSAX was an Italian-Dutch satellite for X-ray astronomy. This was launched in 
1996, the expected operating life of two years was extended to April 30, 2002. 
One of BeppoSAX's main achievements was the identification of numerous 
gamma-ray-bursts with extra-galactic objects.

The Chandra X-ray Observatory is a satellite launched on STS-93 by NASA on July 23, 
1999. It was named in honor of Indian-American physicist Subrahmanyan Chandrasekhar 
who is known for determining the mass limit for white dwarf stars.
Chandra Observatory is the third of NASA's four Great Observatories. Chandra 
is sensitive to X-ray sources 100 times fainter than any previous X-ray telescope, 
due primarily to the high angular resolution of the Chandra mirrors. It is a highly 
successful mission and the list of its discoveries is already very big.

The XMM-Newton (X-ray Multi-Mirror Mission - Newton) is an orbiting X-ray observatory, 
named in honor of Sir Isaac Newton. It was launched by the European Space Agency from 
the Guiana Space Centre at Kourou on 10 December 1999.

The European Space Agency's INTErnational Gamma-Ray Astrophysics Laboratory (INTEGRAL) 
is detecting some of the most energetic radiation that comes from space. It is the 
most sensitive $\gamma$-ray observatory ever launched. INTEGRAL was launched from 
Baikonur spaceport, Kazakhstan in 2002.

Astro-E2 (presently known as SUZAKU) is Japanese X-ray astronomy satellite was 
launched on July 10, 2005. It has a high spectroscopic resolution together with a 
very wide energy band, ranging from soft X-rays up to $\gamma$-rays (0.3 - 600 keV). 
High resolution spectroscopy and wide-band are essential factors to physically 
investigate high energy astronomical phenomena, such as black holes and supernovae. 
One such feature, the broad Iron K line, may be the key to more direct imaging of black 
holes.

AGILE is an X-ray and $\gamma$-ray astronomical satellite of the Italian Space Agency 
(ASI). It was launched successfully into orbit on April 23, 2007 by the Indian 
PSLV-C8 launch vehicle, developed by the ISRO. AGILE is capable of imaging distant 
celestial objects in the X-ray and $\gamma$-ray regions of the electromagnetic 
spectrum.

\section{Gamma-Ray Bursts (GRBs)}

{\it Gamma-Ray Bursts (GRBs)} are short-lived bursts of $\gamma$-ray photons, associated
with extremely energetic explosions in distant galaxies. They are the most luminous
electromagnetic events occurring in the universe. GRBs can last from milliseconds to nearly
an hour, although typical burst lasts a few seconds. The initial burst is usually followed
by a longer-lived afterglow emitting at all types of electromagnetic wavelengths (X-ray,
ultraviolet, optical, infrared and radio etc.). GRBs were first detected by the U.S. military
Vela satellites in 1967. Klebesadel et al. (1973) published first scientific paper on the 
detection of short GRBs using Vela 5 \& 6, which were observed in July, 1969 and in July, 1972 
in a wide hard X-ray energy band (0.2 - 1.5 MeV). The detection of GRBs has opened up a new 
interesting and challenging domain of astrophysical field of research in a wide energy bands of 
electromagnetic spectra from high energy gamma-rays to radio wavelengths (Gehrels et al. 2009).
Also isospheric disturbances caused by the high-energy GRB sources can be measured with 
very low frequency (VLF) receivers. GRB830801 (Fishman \& Inan, 1988) is the first  
GRB observed by this technique.

\subsection{Classification of GRBs}

It is observed that the lightcurves of any two GRBs are not identical, with large variation in
almost every property: the duration of the observable emission can vary from milliseconds to
tens of minutes, there can be a single pulse or several individual sub-pulses, and individual
peaks can be symmetric or with fast brightening and very slow fading. Some bursts are preceded
by a ``precursor" event, a weak burst much before the actual burst. Many classification
schemes have been proposed, but these are often based solely on differences in the appearance
of the lightcurves and may not always reflect the reality.

From the duration of the observed GRBs, it is clear that there are mainly two types of GRBs:
1. ``long GRBs", and 2. ``short GRBs". The average duration of the short lived GRBs is about
$0.3$ seconds, where as the average duration of the long lived GRBs is about $30$ seconds.

\begin{enumerate}

\item Most of the observed GRBs which have a duration of greater than $2$ seconds are termed as 
{\bf long gamma-ray bursts}. Long GRBs are well studied because of their frequent occurrence 
and of their brightest long duration afterglows than their counter parts. Almost every well 
studied long GRB has been associated with a star forming galaxy and in many cases a 
core-collapse supernova as well, unambiguously linking long GRBs with the death of massive stars.

\item The GRB events having duration less that $2$ seconds are classified as {\bf short gamma-ray bursts}.
The true nature of these events remained unknown so far, although the leading hypothesis is that
they originate from the merger of binary neutron stars. A small fraction of short GRBs are
probably associated with giant flares from soft gamma-ray repeaters in nearby galaxies.

\end{enumerate}

\subsection{Origin of GRBs and theoretical models}

After around four decades of the discovery, so far exact origin of GRBs are not very much clear. 
At present, many theoretical models are available in literature, but people generally 
accept `Fireball' (FB) and `Cannonball' (CB) models (see Dar, 2006 and references therein). Although
these two models sound similar, they are completely different. FB model is the more accepted
model by the GRB scientific community.  

\begin{figure}[h]
%\vskip -0.2cm
\centering
\includegraphics[height=2.2in,width=3.0in,angle=0]{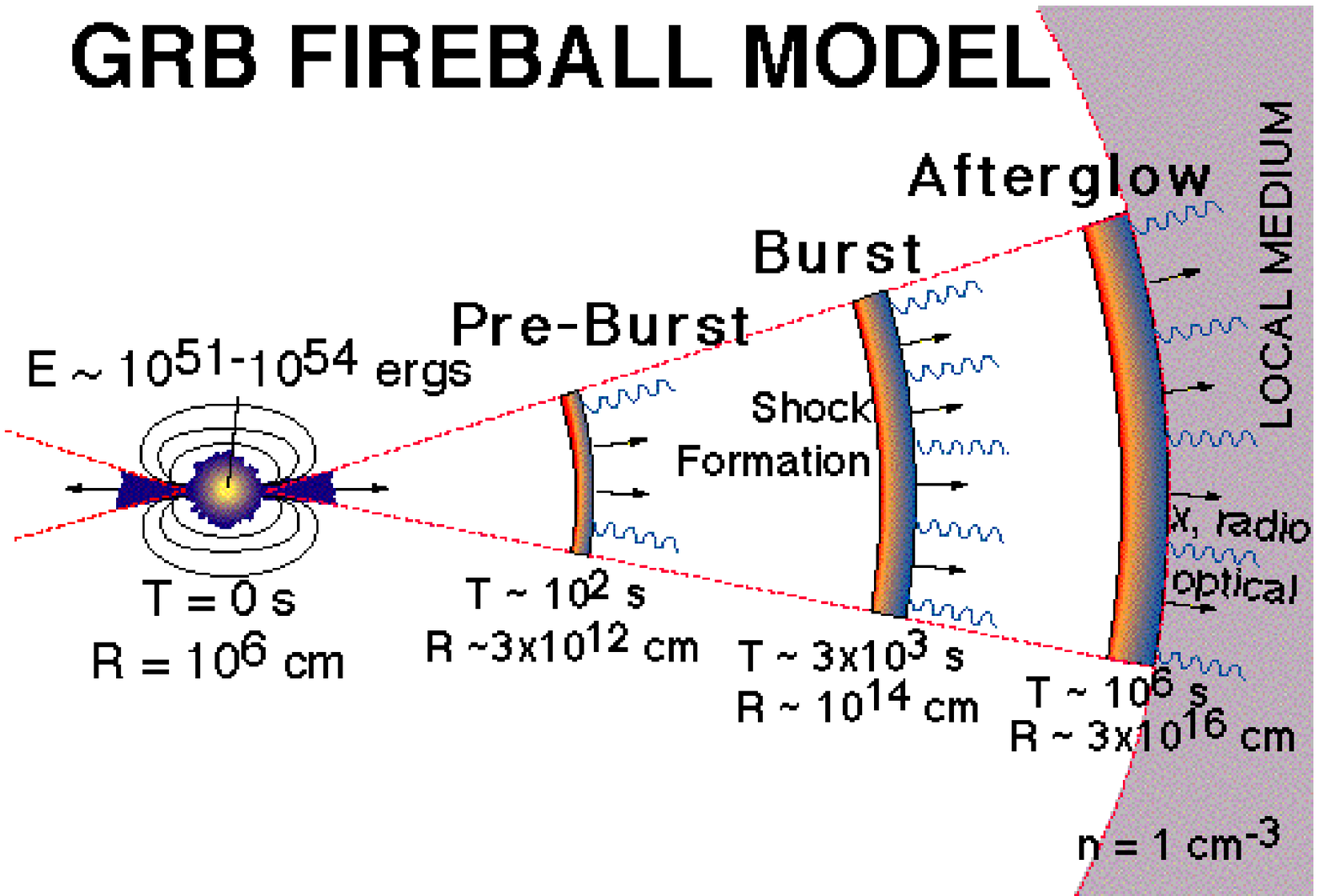}\hspace{0.1 cm}
\includegraphics[height=2.2in,width=2.0in,angle=0]{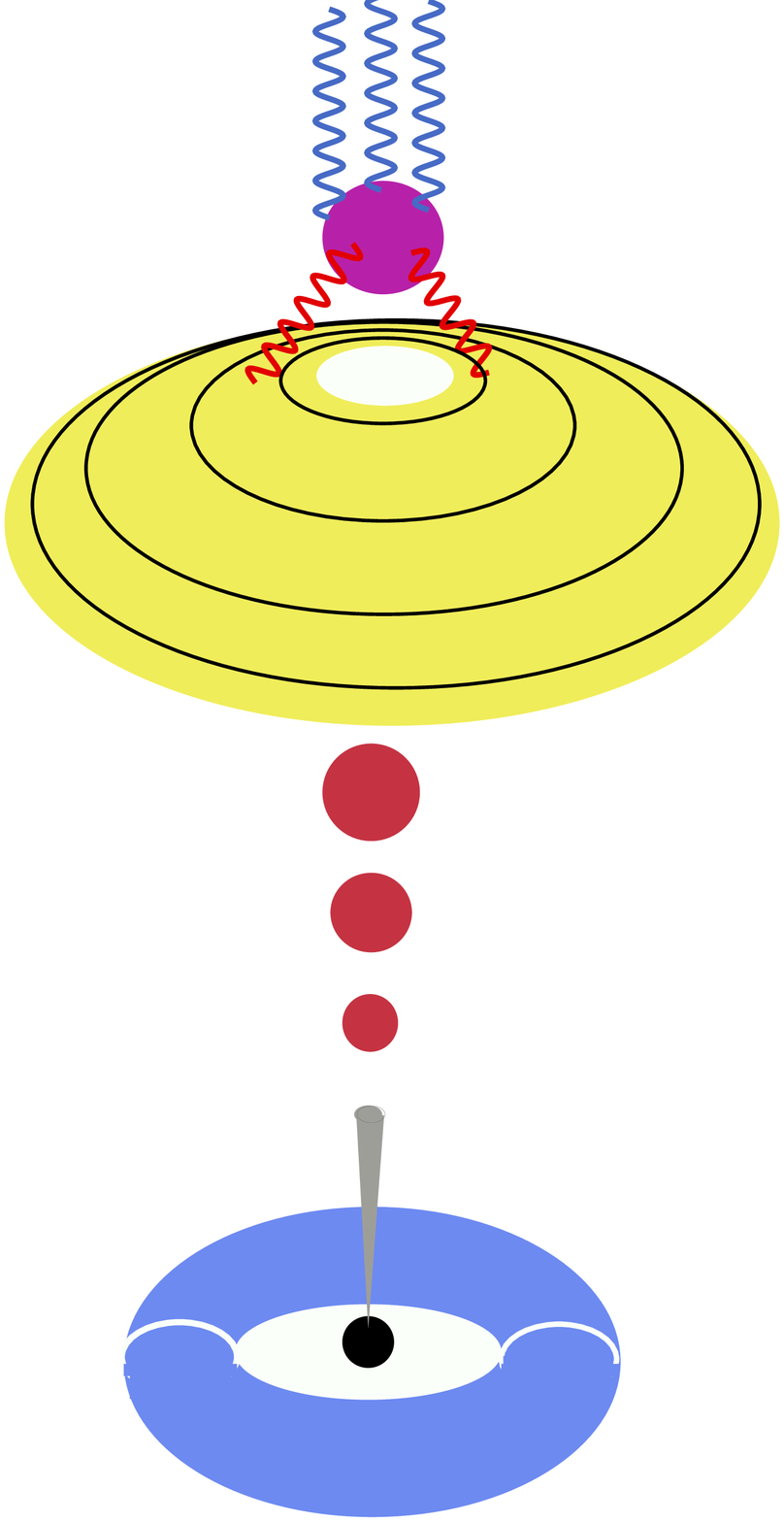}
\caption{(a-b) Left panel (a) shows artists concept of the FB model with its internal/external 
shock formation and afterglow radiations in different energy bands and their time scales. 
Right panel (b) also shows artists concept of the CB model with its rotating accretion disk 
around the newly formed compact object in a result of core-collapse via supernova explosion and 
fall back matter (Dar 2006).}
\label{kn : fig1.5}
\end{figure}

\subsubsection{$\bullet$ Fireball (FB) model}

FB model (Fig. 1.5a) was first proposed by Paczy\'nski (1986) and Goodman (1986). But the 
present FB model is the vastly modified version of the old FB model (see, review articles 
of Zhang \& M\'{e}sz\'{a}ros, 2004 and Piran, 2005). Most of scientific community believe 
that GRBs emit from the highly relativistic conical fireballs (M\'{e}sz\'{a}ros \& Rees, 1992, 
Levinson \& Eichler, 1993). These fireballs are produced by a rare class of massive energetic 
supernovae (SNe type Ic) explosions due to direct collapse of massive stars to black holes 
(Paczy\'nski, 1988). In this case, observed GRB pulses are supposed to be produced by 
synchrotron emission due to the collisions between highly relativistic fireballs (conical 
shells), while GRB afterglow is also due to synchrotron radiation, but emits when the merged 
shells collide with the interstellar medium (ISM).

\subsubsection{$\bullet$ Cannonball (CB) model}

Present CB model (Dar \& De R\'{u}jula, 2000, 2004) is modified version of the theorem 
introduced by De R\'{u}jula (1987) and his collaborator. In the CB model (Fig. 1.5b), it is assumed
that long GRBs and their afterglows are produced by bipolar jets of cannonballs, which 
are believed to be produced due to the supernova explosions caused by the core collapse 
of ordinary stars. It is also believed that in the CB model, an accretion disk is formed
around the newly borne compact object (like BH) by stellar materials near the surface of 
the collapsing core or by distant stellar mass fall into its passage (De R\'{u}jula, 1987).

\section{Observational Data Analysis} 

The data analysis is essential to carry out work in observational astronomy. The observational 
study of astrophysical objects in any electromagnetic energy bands can be classified
mainly in two types: $i)$ time series study/timing analysis and $ii)$ spectroscopic 
study/spectral analysis.

\subsection{Timing Analysis}

Timing analysis is the statistical study of measurements (equally) spaced over time. 
There can be various aims which include understanding the phenomenon represented in 
a single series, forecasting the future, and explaining the relationships between 
variables that change over time. The major feature of time-series data is that 
observations are not independent of previous observations. The pattern of serial 
dependence, or auto-correlation, must be accounted for in statistical analysis for 
inferences regarding the relationships between variables to be valid.
In the astrophysical study timing analysis means the study of the astrophysical objects 
in time domain, i.e., how physical properties (luminosity, density, temperature, and 
chemical composition etc.) of the objects change with time. In our study of the X-ray 
properties of Sun and black holes, we concentrated on lightcurves, power density spectra, 
quasi-periodic oscillations, hardness softness diagrams for the timing analysis.

\subsubsection{Light curve}

Light curves (Fig 1.6a) are the graphs of the intensity or photon counts of astronomical objects 
as they change with time. The information contained in the 
lightcurve includes the timing of events, such as eclipses or pulses, and the amplitude 
of changes in the radiation received at Earth. For black holes the intensity of the light 
curve depends on the mass accretion rate (i.e. flow of matter from its companion objects). 

\begin{figure}[h]
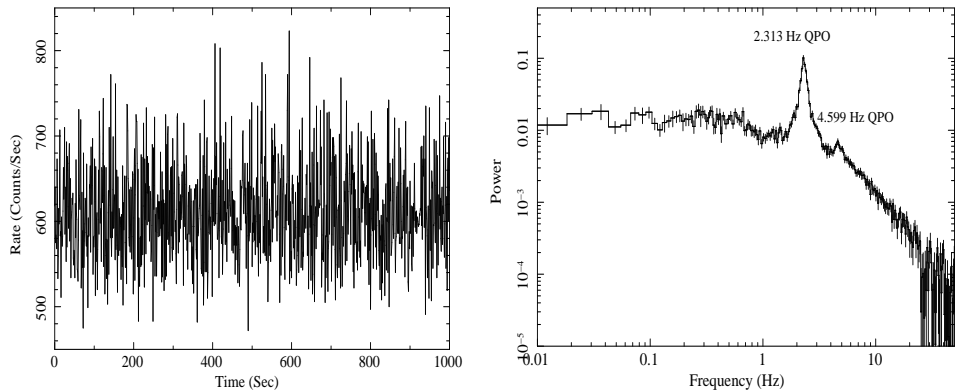

%\vskip -0.2cm
\centering
\includegraphics[height=2.4in,width=2.0in,angle=270]{fig1.6a.ps}\hspace{0.1 cm}
\includegraphics[height=2.4in,width=2.0in,angle=270]{fig1.6b.ps}
\caption{(a-b) Left panel (a) shows GRO J1655-40 lightcurve (LC) and right panel (b) shows 
its Fourier transformed power density spectrum (PDS), observed on March 10, 2005. The LC shows 
periodicity, but oscillation frequency is not clear. From the PDS it is clear that LC has 
oscillation at 2.313 Hz. It also has a harmonic at frequency 4.599 Hz.}
\label{kn : fig1.6}
\end{figure}

Each data point in a lightcurve is a photometric measurement, recorded at a particular time. 
These points represent measurements of the amount of radiation from the source received at 
the detector per second per area in a particular energy band.

\subsubsection{Power Density Spectra}

Power spectra (Fig 1.6b) tell us how the power is distributed at
different frequencies. In the frequency domain, this is the square of the 
magnitude obtained by Fourier Transform. The power spectrum 
is computed either by a Fast Fourier Transform (FFT) algorithm or a direct
Fourier algorithm on a lightcurve.

\subsubsection{Quasi-Periodic Oscillations/Pulsations}

In X-ray astronomy, quasi-periodic oscillations/pulsations (QPOs/QPPs) refer to the way 
the X-ray light from the vicinity of an astronomical object 
(white dwarf, neutron star or black hole) oscillates.
The QPO phenomenon helps us to understand the innermost structure of accretion disks, nature of the
accretion flow. Also from thorough study of this timing feature one can get an idea about
the masses, radii and spin periods of the compact objects. A QPO is identified by performing 
a power spectrum on the lightcurve. A periodic pulsation appears in the power spectrum 
at exactly one frequency (a Dirac delta function), but on the other hand, a QPO appears as 
a broader peak of Lorentzian shape (see Fig. 16b).

There are many theories of the origin of QPOs, but it is generally accepted that 
the X-ray pulsation occurs due to an oscillating shock wave, generated by the infalling matter at 
the accretion disk of the compact object (Chakrabarti et. al. 2005, 2008a, Debnath et al. 2010a). 
According to the theory, the oscillation frequency is inversely proportional 
to the infall time scale of the matter from region of oscillation (i.e. shock location) 
to black hole. For the black hole candidates, observed QPO frequencies vary from mHZ to 
kHZ. The highest QPO frequency observed for any black hole candidate is 450 Hz (for 
GRO J1655-40). It showed QPO frequencies in the range of $0.01$ Hz - $450$ HZ 
(Remillard et. al. 1999, Strohmayer 2001, Debnath et. al. 2008a). For the Sun, people 
also have observed low frequency QPPs at the time of solar flares whose periods may vary 
from $10$s to $150$s (Young et al. 1961, Aschwanden 1987, Rao et. al. 2010a etc.).

As an alternative explanation for the origin of QPOs, some scientific people believe that 
the QPOs are caused by hot spots orbiting around a compact object.
Since such a physical process is not sustainable for days after days,
it is difficult to conceive that this is what is going on.

\subsection{Spectral Analysis}

Spectral analysis means the study of the astrophysical objects in energy domain. X-ray
spectroscopy offers an important channel of information about our Universe. It is receiving
increasing importance as X-ray astronomy has matured and technology  is well developed.
Now, we have the ability to make measurements not just of the X-ray continua, 
but of the discrete line features as well.  
From the analysis of the X-ray line spectra of an astrophysical object we can get an idea 
about the elemental composition and abundances, the temperature and electron temperature 
and density, magnetic field etc. Spectral analysis also tells us the physical processes 
occurring around (e.g. accretion disk around a compact object) or inside a celestial object.

\begin{figure}[h]
%\vskip -0.2cm
\centering
\includegraphics[height=2.5in,width=2.70in,angle=270]{fig1.7a.ps}\hspace{0.1 cm}
\includegraphics[height=2.35in,width=2.72in,angle=270]{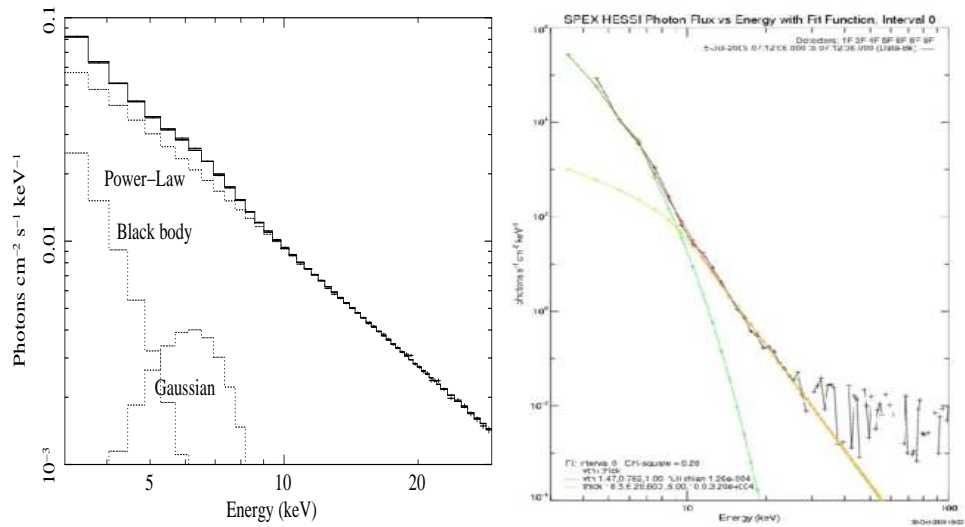}
\caption{(a-b) Left panel (a) shows RXTE PCA 3 - 10 keV 2005 GRO J1655-40 outburst spectrum, 
observed on March 10, 2005 and the right panel (b) shows RHESSI observed spectrum of 5th July, 
2009, C2.7 class solar flare.}
\label{kn : fig1.7}
\end{figure}

A black hole spectrum contains mainly two components: thermal (black body nature) and 
non-thermal (power-law nature) (see, Fig. 1.7a). According to two component 
advective flow model (TCAF), thermal component is dominated by Keplerian photons 
(below, $\sim$ 10 keV), whereas the non-thermal component is dominated by sub-Keplerian 
photons (above $\sim 10$ keV). Based on the degree of importance of the black body 
and the power-law components black hole spectra can be divided into few classes. 
such as, hard, soft, intermediate etc.

Similarly, a solar spectrum also consists of two components: thermal and non-thermal. The
thermal component consists of black body photons, where as the non-thermal component contains 
power-law photons due to bremsstrahlung or synchrotron emission (see Fig. 1.7b).

	\reseteqn
	\resetsec
	\resetfig
	\resettab
\alpheqn
\resec
\refig
\retab
\def\k{{\bf k}}
\def\aug{{\tilde{\cal H}}}
\def\ahat{{\mathaccent "7E {\cal A}}}
\def\bhat{{\mathaccent "7E {\cal B}}}
\def\chat{{\mathaccent "7E {\cal F}}}
\def\dhat{{\mathaccent "7E {\cal D}}}
\def\shat{{\mathaccent "7E {\cal S}}}
\def\proj{{\cal P}}
\def\ket{\vert \vert  \{ \emptyset \} \rangle}
\def\ket2{\vert \vert \otimes \{ R \} \rangle}
\def\sqr{$^{2}$}
\def\T{$T$}
\def\P{$P$}
\def\Q{$Q$}
\def\G{$G$}
\def\M{{\bf M}}
\def\H{{\bf H}}
\def\I{{\bf I}}
\def\Pr{{\bf P}}
\def\Tr{{\bf T}}
\def\eq{\enskip =\enskip}
\def\pls{\enskip+\enskip}
\def\mns{\enskip -\enskip}
\def\oper{O^{\bf k}_{R}}
\def\aug{\tilde{\cal H}}
\def\envfn{\chi^{\alpha}_{RL}}
\def\struc{S^{\alpha}_{RL,R^{\prime}L^{\prime}}}
\def\rR{r_{R}}

\newpage
\markboth{\it Space Instruments, Data Acquisition and Analysis Procedure}
{\it Space Instruments, Data Acquisition and Analysis Procedure}
\chapter{Space Instruments, Data Acquisition and Analysis Procedure}

The goal of my thesis is to study X-ray properties (spectral and timing properties) of Sun 
and some compact objects (mainly black hole candidates GRO J1655-40, GX 339-4 and gamma-ray bursts (GRBs))
of our Galaxy. These studies were done by analyzing the observational data from the space-borne 
X-ray telescopes. For studying X-ray properties of the Sun and gamma-ray bursts, we used X-ray data 
of the Indo-Russian collaborative {\it RT-2 payloads (RT-2/S, RT-2/G and RT-2/CZT)} 
and for the black hole study, NASA's {\it Rossi X-ray Timing Explorer} ({\it RXTE}) satellite 
data were used. Apart from the X-ray data analysis, I participated in the development, test and evaluation
and calibration of RT-2 payloads from its very initial stage to its final flight model stage. 
For the development of RT-2 systems, I also took part in some theoretical works such as: 
characterization of the detectors of RT-2/CZT imaging detectors using Fresnel Zone Plate (FZP) and 
Coded Aperture Mask (CAM) as coders for the hard X-ray imaging. This characterization was done by 
using Monte-Carlo simulation methods. In this Chapter, we will discuss about the descriptions of 
the RT-2 and the RXTE space instruments and their data acquisition, analysis procedures.

\section{RT-2 Experiment: A mission for solar science}

Sun occupies a special place in astrophysical studies as it can be more closely
studied than any other star. Several dedicated satellites (RHESSI, SOHO, GOES etc.)
have been launched to understand its behavior in a wide energy band of electromagnetic
radiations. Even though its surface temperature is only $\sim$6000 K, it emits
X-rays and $\gamma$-rays up to a few MeV. This is primarily because of rapid magnetic
reconnection which produces energetic solar activities.
Apart from the thermal electrons which obey Maxwell-Boltzmann distribution,
the charged particles, especially electrons, are accelerated by shocks
and acquire a non-thermal (power-law) distribution. These non-thermal electrons
emit energetic synchrotron emissions. With time, the energy shifts from one wavelength
to another. One of our goals is to understand the
energy transport processes on the solar surface. It is due to plasma oscillations,
or pinching or sausage instabilities in the magnetic field.

RT-2 (named after R\"{o}ntgen) payload (Debnath et. al. 2008b, 2010b, Nandi et. al. 2009, 2010, 
Kotoch et. al. 2010, Sarkar et. al. 2010, Sreekumar et. al. 2010, Rao et al. 2010c) 
is a part of Russian Solar mission satellite named Coronas-Photon (Kotov et al. 2008, Nandi et 
al. 2009.). It was launched into $\sim$550~Km polar LEO (Low Earth Orbit) on 
30$^{th}$ January, 2009 from Plesetsk Cosmodrome, Russia. The mission is dedicated to 
study mainly the solar flares in a wide energy band of electromagnetic spectrum ranging 
from UV to high-energy $\gamma$-rays ($\sim$2000 MeV). 

The main objective of the RT-2 Experiment is to study (i) time resolved hard X-ray spectra 
of solar flares, (ii) galactic and extra-galactic sources near the ecliptic plane, (iii) 
gamma ray bursts (GRBs) and (iv) diffused cosmic X-ray background, in a wide energy band 
of 15 keV to $\sim$ 1 MeV.

\begin{figure}[h]
\begin{center}
\vspace{-0.3cm}
\epsfxsize=4.5in % \epsfysize=1.5in
\rotatebox{-0.2}{\epsfbox{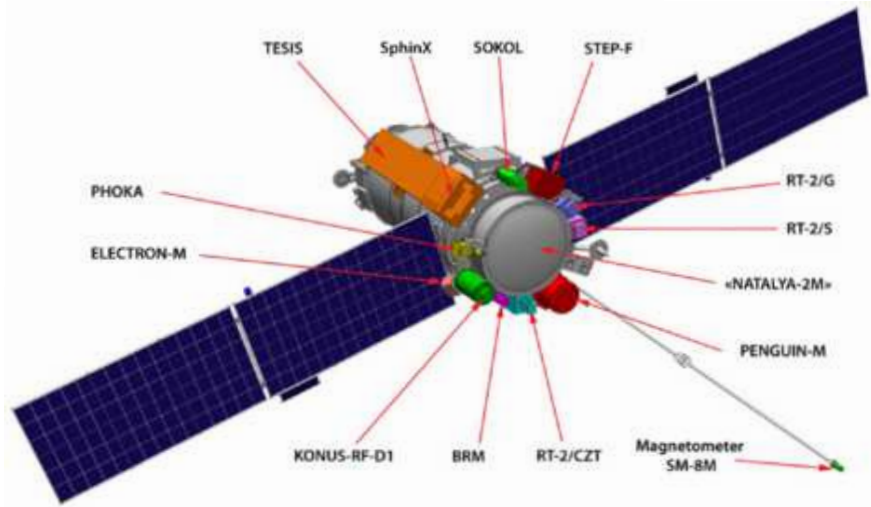}}
\vspace{-0.1cm}
\caption{ A realistic diagram of the CORONAS-PHOTON satellite.}
\label{kn : fig2.1}
\end{center}
\end{figure}

The RT-2 payloads have been developed by Tata Institute of Fundamental Research (TIFR),
Indian Centre for Space Physics (ICSP) in association with Vikram Sarabhai Space Centre (VSSC).
I worked on the development of the RT-2 system as a team member, in laboratories of different 
Indian institutes, namely, VSSC (Thiruvananthapuram), SAC (Ahmedabad), 
PRL (Ahmedabad), TIFR (Mumbai) and ICSP (Kolkata).
%The details of these developments along with scientific results related to the spectral
%and temporal properties of the Solar X-rays from the payloads, are discussed in this Chapter.

\section{RT-2 scientific payloads}

RT-2 is an Indian Space Research Organization (ISRO) sponsored X-ray and low-energy $\gamma$-ray 
(up to 1 MeV) solar experiment in collaboration with MPhei, Russia. It consists of three 
scientific payloads (RT-2/S, RT-2/G \& RT-2/CZT) and one master control electronic payload 
(RT-2/E). Both RT-2/S and RT-2/G payloads consist of identical NaI(Tl) and CsI(Na) 
scintillation crystals in a phoswich combination, having the same diameter ($116$ mm) 
but different thicknesses (Debnath et. al. 2010b). RT-2/CZT consists of solid state imaging detectors CZT
(Cadmium Zinc Telluride) and CMOS (Complimentary Metal Oxide Semiconductor) (Kotoch et al. 2010). Main working
energy range of RT-2/S \& RT-2/G is of 15 - 100 keV (higher energy limit is extendable up
to $\sim 1$ MeV) and for RT-CZT is of 20 - 100 keV.

All the three payloads are placed outside the hermetically sealed vessel of the satellite
and co-aligned with the Sun pointing axis. The instruments have different viewing angles
with Field of View (FOV) $4^\circ \times 4^\circ$ for RT-2/S, $6^\circ \times 6^\circ$
for RT-2/G and ranging from $6' \times 6'$ to $6^\circ \times 6^\circ$ for RT-2/CZT-CMOS.

\begin{figure}[h]
\begin{center}
\vspace{-0.3cm}
\epsfxsize=4.5in % \epsfysize=1.5in
\rotatebox{-0.2}{\epsfbox{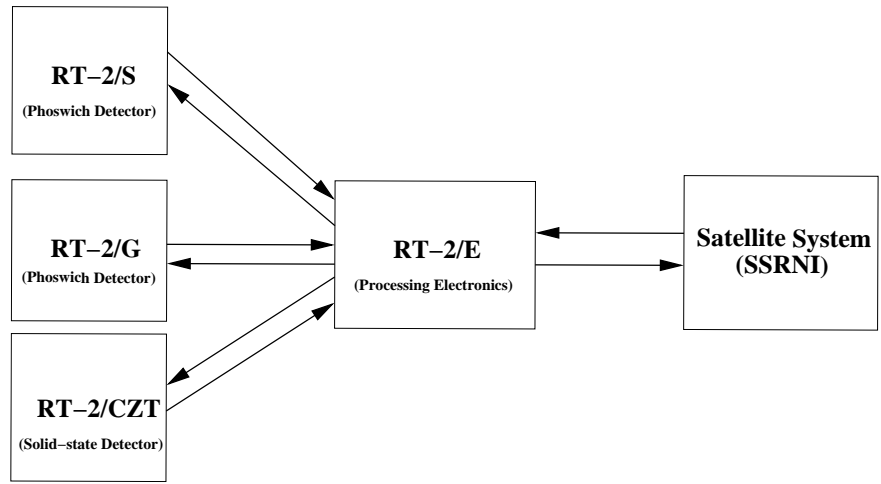}}
\vspace{-0.1cm}
\caption{Schematic diagram of RT-2 system (Sreekumar et al. 2010).}
\label{kn : fig2.2}
\end{center}
\end{figure}

\subsection {RT-2/S (RT-2/G) payload}

RT-2/S and RT-2/G payloads consist of NaI(Tl) and CsI(Na) crystals mounted in a
scintillator-phoswich assembly viewed by a photo-multiplier tube (PMT). The
NaI crystal is of $3$~mm thickness and $116$~mm diameter is coupled with the
CsI crystal of $25$~mm thick and of same diameter, to provide mechanical stability of
the phoswich detector. These two crystals are optically coupled and hermetically sealed
with an entrance window on NaI(Tl) side and a viewing glass window coupled to the PMT
through an appropriate light guide on the CsI(Na) side. The NaI(Tl) crystal is sensitive
to X-ray photons of $15 - 100$~keV, whereas CsI(Na) is effective for high energy
X-rays/$\gamma$-rays of $30 - 1000$~keV as well as to charge particle background.

Both the detector assemblies sit behind a mechanical slat collimator surrounded by
a uniform shield of Tantalum material and having different viewing angles of 4$^\circ$ x
4$^\circ$ (RT-2/S) and 6$^\circ$ x 6$^\circ$ (RT-2/G). As per design and scientific
requirement, RT-2/S will work in the energy range of 15 keV to 100 keV, extendable
up to 1 MeV, whereas aluminum filter is used to cut-off low energy photons
($\leq$ 20 keV) for RT-2/G payload.

Another important use of both the crystals in the Phoswich mode is for the background rejection.
The light signal from the CsI(Na) crystal has a different scintillation decay time (650 ns)
than that from the NaI(Tl) crystal (250 ns) and this distinction is used to eliminate
non X-ray background in the NaI crystal. The charge particle background will be removed by
the upper energy threshold. The Phoswich assembly is viewed by a $76.2$~mm diameter
photomultiplier through a $10$~mm thick light guide.

\begin{figure}[h]
\vspace{-0.3cm}
\centering
\includegraphics[height=3.0in,width=2.4in,angle=0]{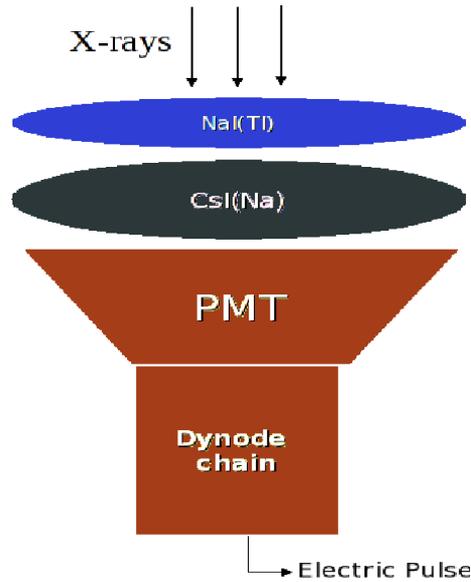}
\caption{A Schematic drawing of Phoswich assembly (NaI and CsI crystals) along with the PMT 
(Debnath et al. 2010b).}
\label{kn : fig2.3}
\end{figure}

The interaction of X-ray photons of energy up to 100 keV with NaI and CsI crystals is fully
dominated by the photo-electric process and thus absorbed radiation (secondary electron-hole
pair absorbed by the impurities) converts into light photons (due to the decay of the excited
impurities). These light photons eventually strike the photo-cathode of the PMT (Knoll 1999)
(gain $\sim$10$^6$) and converted into narrow electrical pulse whose magnitude (pulse height)
is proportional to the energy of the incident radiation. Light outputs from NaI and CsI crystals
have decay time scales of 250 ns and 650 ns respectively. Pulses from the detector (crystal)
are amplified in a pre-amplifier and two post amplifiers (G1 and G2).
Simultaneously, the pulse shape discriminator (PSD) is also measured as the width of the pulse.
Output from one of the amplifier is selected based on the outputs from a Lower Level Discriminator
(LLD) and an Upper Level Discriminator (ULD). If ULD is triggered, G2 is selected, otherwise
G1 is selected. The output from the selected amplifier is digitized using an analog to digital
converter (ADC). The ADC is 12 bits wide and has a typical conversion time of 10 $\mu$s.

The energy resolution (FWHM) of the scintillator Phoswich is expected to be $18\%$ at $60$~keV
and the pulse height variation across the crystal will be less than $3\%$. The radioactive
isotopes $^{241}$Am (13.95, 17.74, 26.35 \& 59.54 keV), $^{57}$Co ($122$~keV),
isotopes $^{241}$Am ($13.95$, $17.74$, $20.8$, $26.35$ \& $59.54$ keV), $^{57}$Co ($122$~keV),
$^{109}Cd$ ($22$ \& $88$~keV) were used for laboratory calibration. The entire system

Electronics of both the instruments are identical, the only difference is that, RT-2/G has one
Al sheet (thickness $\sim 2~mm$) above its collimator to increase its low energy cutoff.
Both RT-2/S \& RT-2/G detectors have 8 timing counters. Out of these 8 counters first 4 counters
consist of G1-NaI(Tl) spectral information, next 2 counters consist of G1-CsI(Na) spectral
information and the remaining 2 counters consist of G2 spectral information. The default spectral
channel numbers of the counters and their calibrated energies are shown in Tab. 2.1.

\begin{table}[h]
\small
%\scriptsize
\centering
%centerline {Table 2.1}
%\centerline {RT-2/S (RT-2/G) channel boundary (default) and energy ranges}
\caption{\label{table2.1} RT-2/S (RT-2/G) channel boundary (default) and energy ranges}
\vskip 0.2cm
\begin{tabular}{|l|c|c|c|c|}
\hline
Amplifiers & Counters & Channel ranges &\multicolumn{2}{|c|}{Energy Ranges (keV)}\\
\cline{4-5}
 & &(default)&~~~~~RT-2/S~~~~~&RT-2/G\\
\hline
        & C1 & 0 - 140   & $<$ 13  & $<$ 13 \\
        & C2 & 141 - 280 & 13 - 27 & 13 - 27\\
G1-NaI  & C3 & 281 - 560 & 27 - 55 & 27 - 56\\
        & C4 & 561 - 1023& 55 - 102& 56 - 104\\
\hline
        & C5 & 0 - 256   & $<$ 33   & $<$ 34\\
G1-CsI  & C6 & 257 - 1023& 33 - 209 & 34 - 209\\
\hline
        & C7 & 0 - 64    & $<$ 48 & $<$ 246\\
G2      & C8 & 65 - 255  & 65 - 570& 246 - 1000\\
\hline
\end{tabular}
\end{table}

\subsection {RT-2/CZT payload}

The RT-2/CZT payload consists of three CZT detector modules (OMS40G256, procured from
Orbotech Medical Solutions Ltd., Israel) and one CMOS detector (RadEye-1, Rad-icon Imaging
Corp., USA) arranged in a $2 \times 2$ array. Each module of CZT detector consists of 256
individual pixels (detectors) of $2.5~mm \times 2.5~mm$, which are controlled by 2 ASICs and
one CMOS detector consists of $512 \times 512$ pixels of individual pixel dimension of
$50 \mu m$. The entire CZT-CMOS detector assembly sits behind a collimator with two different
types of coding devices, namely Coded Aperture Mask (CAM) and Fresnel Zone Plate (FZP),
surrounded by a uniform shield of Tantalum and has varying viewing angle of
6$'$ - 6$^\circ$. For the four detectors of the RT-2/CZT payload, four different
combinations of imaging techniques has been used (Nandi et al. 2010).

RT-2/CZT payload is the only imaging device in the CORONAS-PHOTON mission to image the
solar flares in hard X-rays of energy range from $20$ to $100$ keV. The effective area of 3 CZT
modules is 48~cm$^2$ with an average energy resolution of 8$\%$~@60~keV (at 10$^\circ$C).
All three CZT detectors have the spectral information along with high resolution imaging
capabilities. On the other hand, the CMOS detector has an effective area of 4.5~cm$^2$ with
high resolution imaging capability only.

\begin{figure}[h]
\vspace{-0.3cm}
\centering
\includegraphics[height=2.2in,width=2.2in,angle=270]{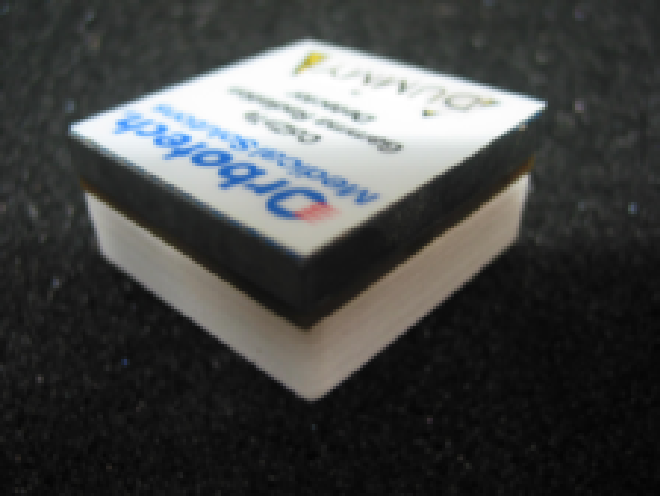}
\includegraphics[height=2.2in,width=2.2in,angle=270]{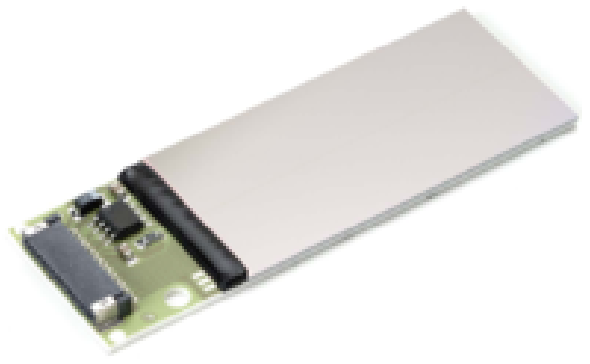}
\caption{(a-b): Left panel (a) shows that the isometric view of a CZT Module used in RT-2/CZT of 
dimension 3.96 cm $\times$ 3.96 cm $\times$ 5 mm and right panel (b) shows the RedEye1 CMOS Detector, 
used in RT-2/CZT (Kotoch et al. 2010).}
\label{kn : fig2.4}
\end{figure}

\begin{itemize}
\item {} Configuration 1 (CONFIG-1) contains CAM1 as coder and CZT1 as a detector. Spacing
between CAM1 and CZT1 is $40~cm$, the angular resolution is $21.5~arcmin$ and FOV is 5.72$^\circ$.
\item {} Configuration 2 (CONFIG-2) contains CAM2 as coder and CZT2 as detector. The
detector-coder spacing, angular resolution and FOV for this configuration is same as Configuration 1.
\item {} Configuration 3 (CONFIG-3) contains dual FZP (FZP1) with diameter $3.0~cm$ as coder
and CZT3 as detector. The spacing between the two zone plates is $32~cm$ and the angular
resolution is $64~arcsec$. The FOV for this configuration is found to be $409~arcsec$ wide.
\item {} Configuration 4 (CONFIG-4) is designed with dual FZP (FZP2) as shadow-caster of
diameter $2.4~cm$ and a high spatial resolution pixilated CMOS as a detector. The spacing
between two FZPs is $32~cm$, the angular resolution is around $54~arcsec$ and FOV is 4.29$^\circ$.
\end{itemize}

In the following sub-section, we will discuss the physibility studies made for RT-2/CZT payload.
In Nandi et al. (2010), we showed these results in details. Here I will discuss the Monte-Carlo
simulation results made for only two combinations (CONFIG-1 \& CONFIG-4) imaging detectors and 
coders out of four.

\subsubsection{$\diamond$ Monte Carlo simulations for CAM and FZP}

All the RT-2/CZT detectors along with coders will receive parallel rays of radiation as the
sources are effectively at infinite distance when satellite is at its orbit. 
In general, it is difficult to make an experimental arrangement to keep source at infinite
distance so that the shadow caster can receive parallel beam of X-rays.
The only way we can verify that all the characteristics of the coders explained in above are 
true and examine their efficiency in reproducing exact replica of what they observe in their 
field of view is to simulate extensively all the required cases for all four configurations 
keeping all the parameters and environment intact (Nandi et al. 2010, also see, Chakrabarti et 
al. 2009b, 2009c, Palit et al. 2008, 2009).

Simulations using Monte Carlo method are done for every configuration and for all cases
involving varying number of sources, with large off-axisness of source for the verification
concerning FOVs and for closely placed sources to verify the mathematically obtained
angular resolutions. The infalling photon number on the front coders (CAM1, CAM2, FZP1 and
FZP2) for all four configurations is chosen to be 5$\times$$10^{5}$, while considering a
single source. For double or multiple sources with varying intensity, photon numbers for 
the brightest source remain the same. The photon numbers for relatively lower intensity sources
are mentioned in respective sections. This number is sufficient for hundred second
(onboard accumulation time for each frame) data accumulation by the imager (4 different
configuration for imaging in RT-2/CZT payload) from a $C$ class flare (and above) that
occurred in the surface of the Sun.

Simulation results are interpreted based on the two-dimensional (2D) and three-dimensional (3D) 
representations of the reconstructed source positions and relative strength of the peaks. Source 
intensity variation in reconstructed image plane is plotted in arbitrary units. Detailed 
simulation results for CONFIG-1 \& CONFIG-4 are presented in the following sub-sections.

{$\bullet$ RT-2/CZT CONFIG-1 (CAM1 + CZT1):} 

CONFIG-1 consists of a single coder CAM (CAM1) and a CZT (CZT1) module and both are placed
at $40~cm$ apart in the first quadrant of the collimator. CAM pattern for this configuration
is shown in Fig. 2.5(a). A source position is generated at a position of
$\theta$ = 42$^\circ$ and $\phi$ = $1^\circ 47'$. The shadow of the CAM due to this source
on detector plane is shown in Fig. 2.5(b).
The shift of shadow pattern of the CAM from the central position codes the information on
the position of source with respect to the central point in FOV. Reconstruction of the image
(source position) from the CAM pattern is done in accordance with the method discussed above.
In Fig. 2.5(c) and Fig. 2.5(d), we show the 2D and 3D views of the reconstructed
source (image). In both Figs. of the reconstructed sky
plane, the FOV is 5.72$^\circ$ wide along in each sides. The reconstructed source position
as we evaluated from Fig. 2.5(c), exactly matches with the actual source position assigned
during simulation.

\begin{figure}[h]
\centering
\includegraphics[height=1.2in,width=1.2in]{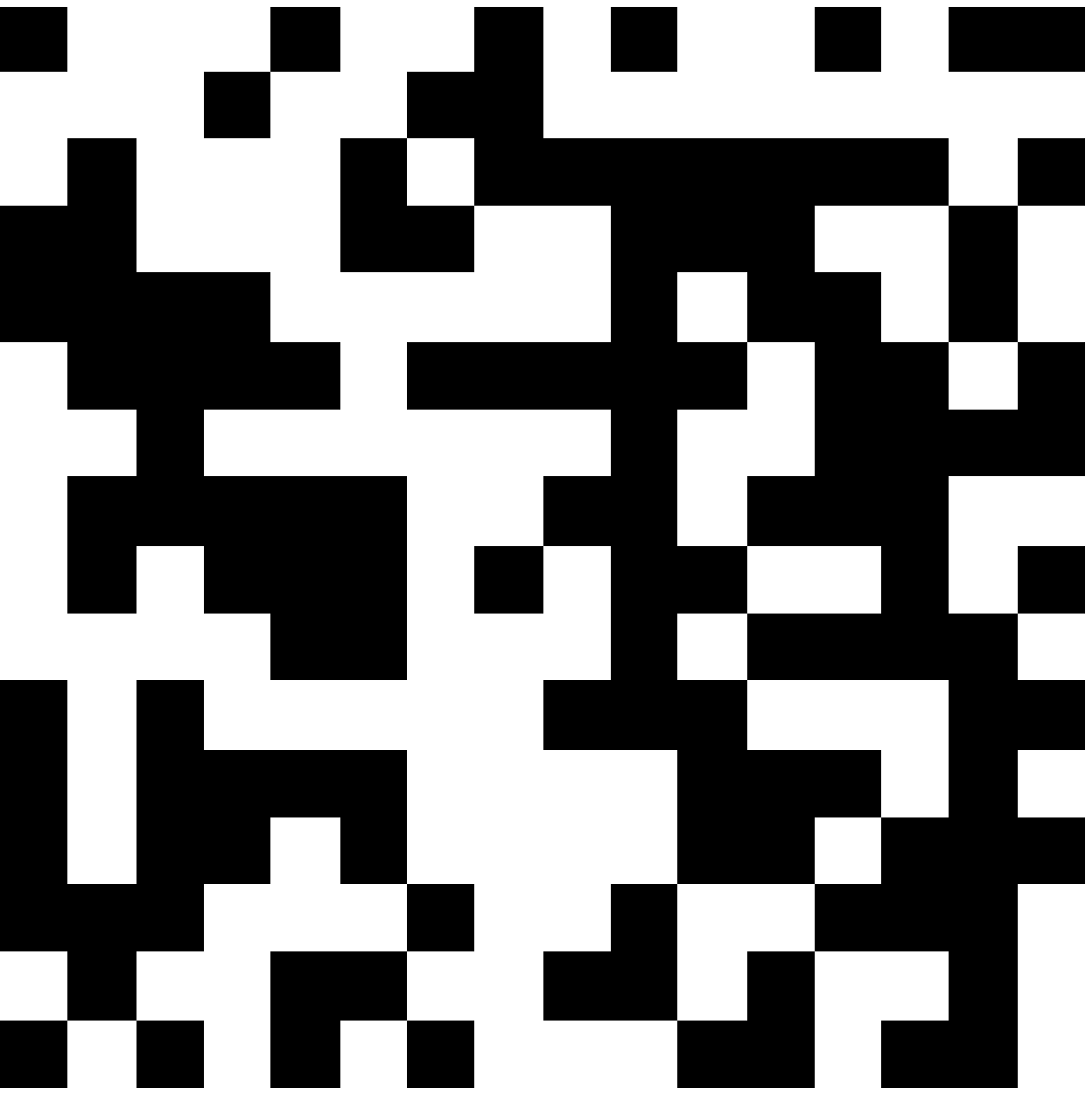}\hspace{0.2 cm} %1
\includegraphics[height=1.2in,width=1.2in]{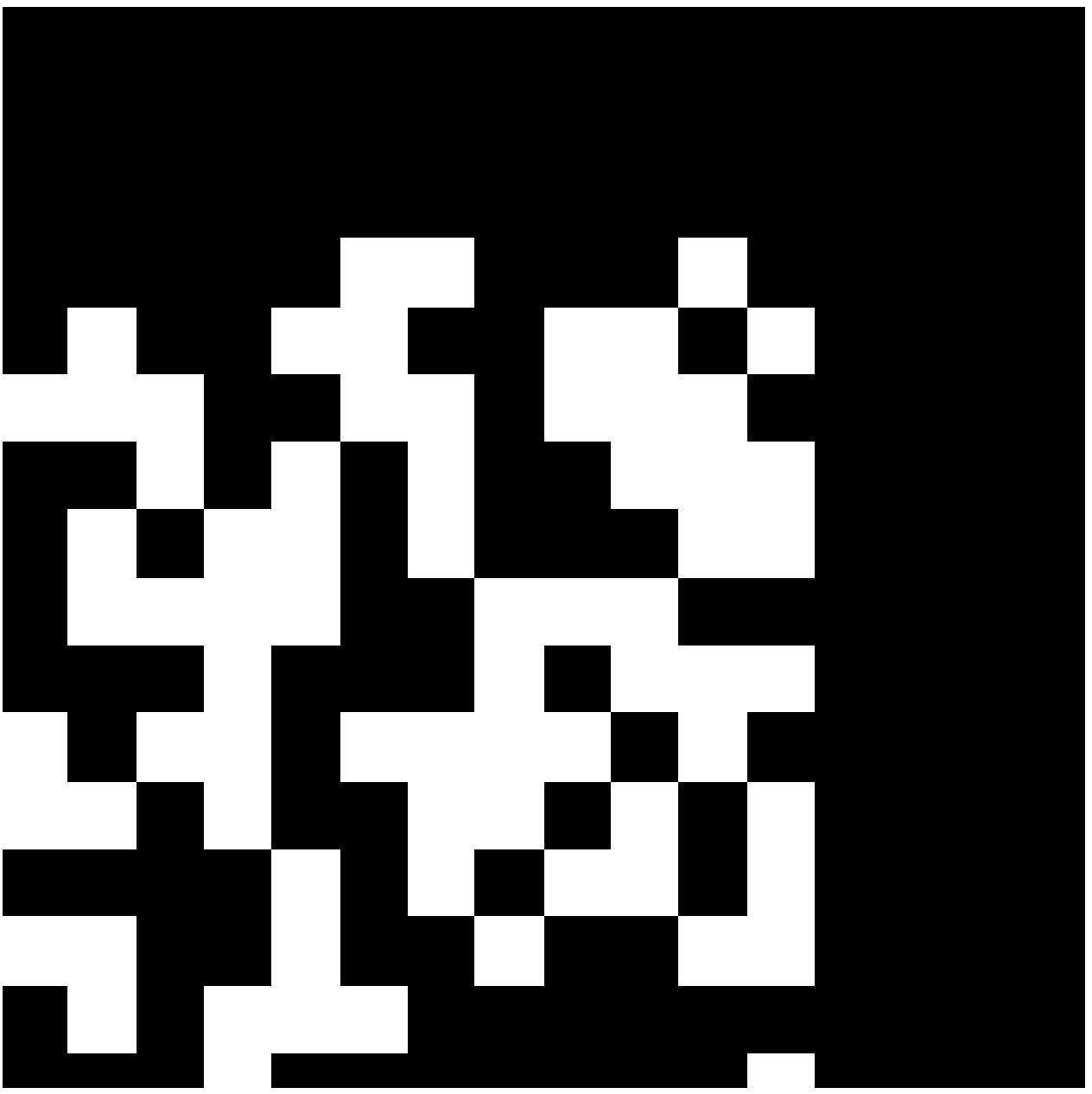}\hspace{0.2 cm}
\includegraphics[height=1.2in,width=1.2in]{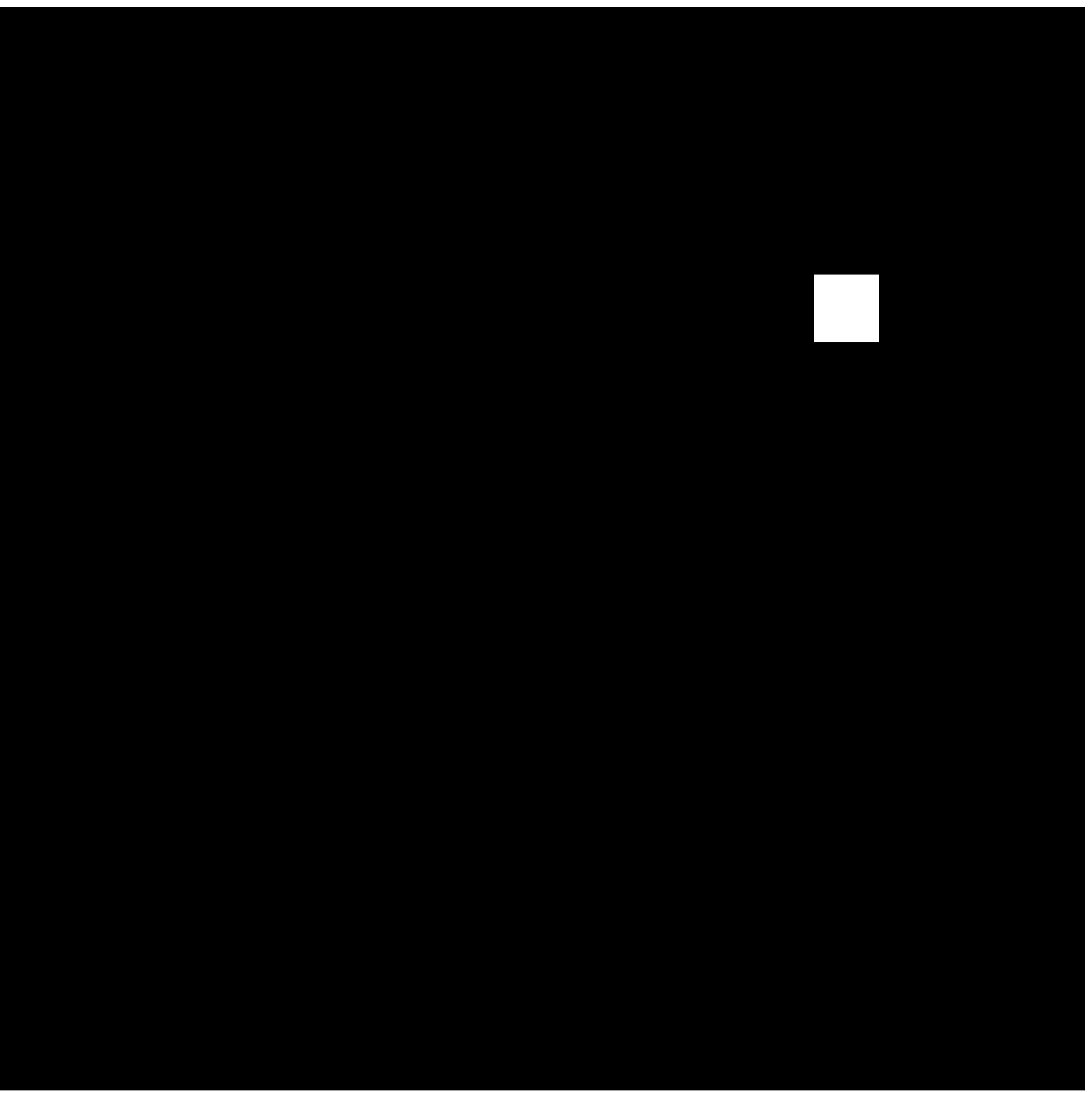}
\includegraphics[height=1.4in,width=1.6in]{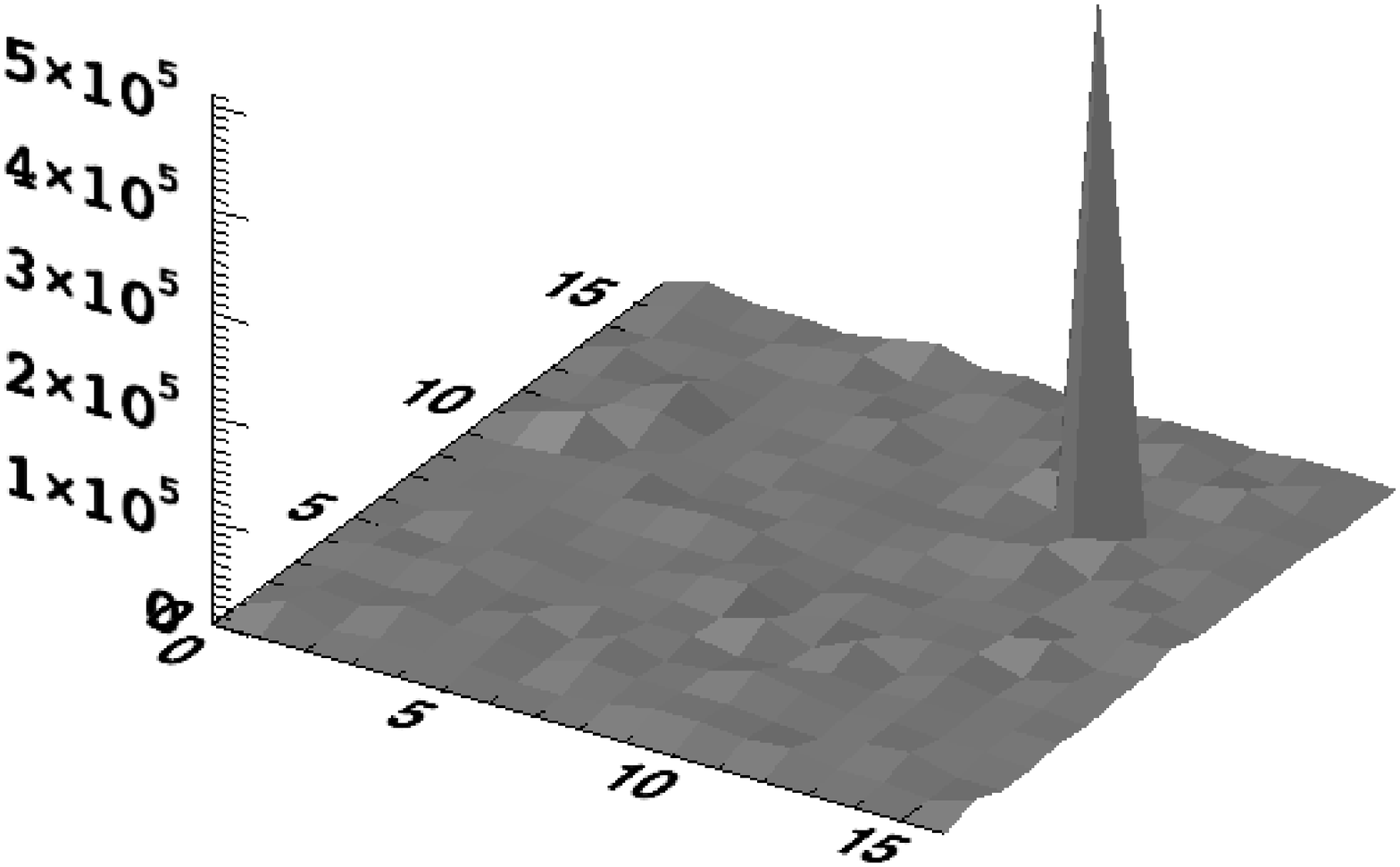}
\caption{(a-d) shows, from left: (a) simulated picture of CAM1 pattern, (b) shadow pattern 
obtained in the CZT detector for CONFIG-1 for a single source, (c) 2D view of reconstructed
sky plane, and (d) 3D picture of the source plane obtained by reconstruction (Nandi et al. 2010).}
\label{kn : fig2.5}
\end{figure}

From Fig. 2.5(d), we get an estimation of the relation between the height of the peak of the
reconstructed source and the actual intensity of the source.

In Fig. 2.6(a-c), we present the double source simulation results to verify the accuracy of the
mathematically obtained angular resolution of the configuration containing CAM1 and CZT1. For
this, we placed two sources at an angular separation of 21.5 arc minute from each other. The
number of photons falling on the CAM from the brighter source is 5$\times$$10^{5}$ and that from the
lighter one is 2.5$\times$$10^{5}$. In Fig. 2.6(a), we showed the shadow pattern obtained for
two sources which are placed very close to each other. The reconstructed source (image) in 2D
and 3D view of both the sources are shown in Fig. 2.6(b) and 2.6(c) . It can be seen from the
Figs. that two sources are separated by one pixel between them. So the sources can be said to
be just resolved.

\begin{figure}[h]
%\vskip -1.0cm
\centering
\includegraphics[height=1.2in,width=1.2in]{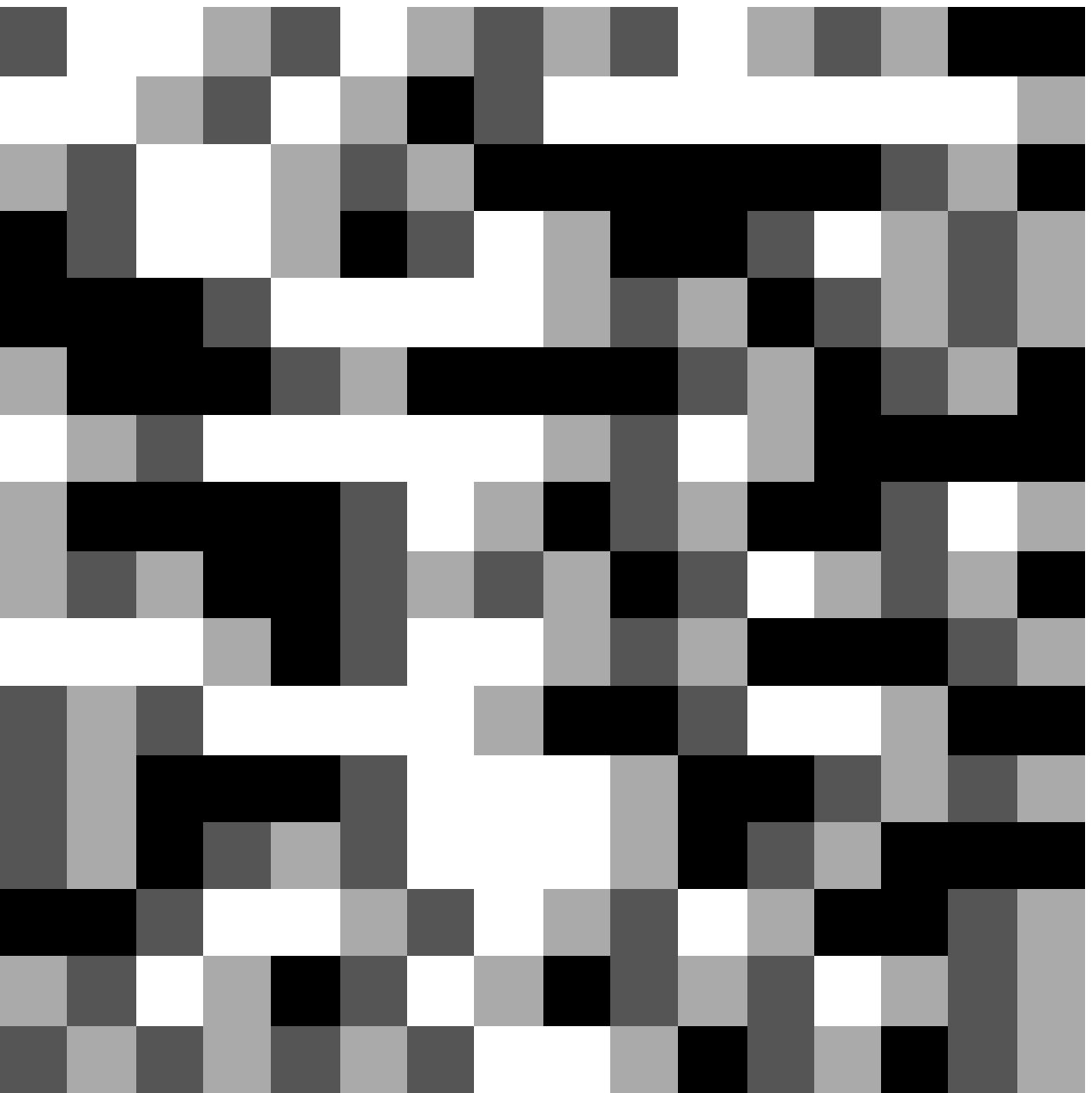}\hspace{0.2 cm} %2
\includegraphics[height=1.2in,width=1.2in]{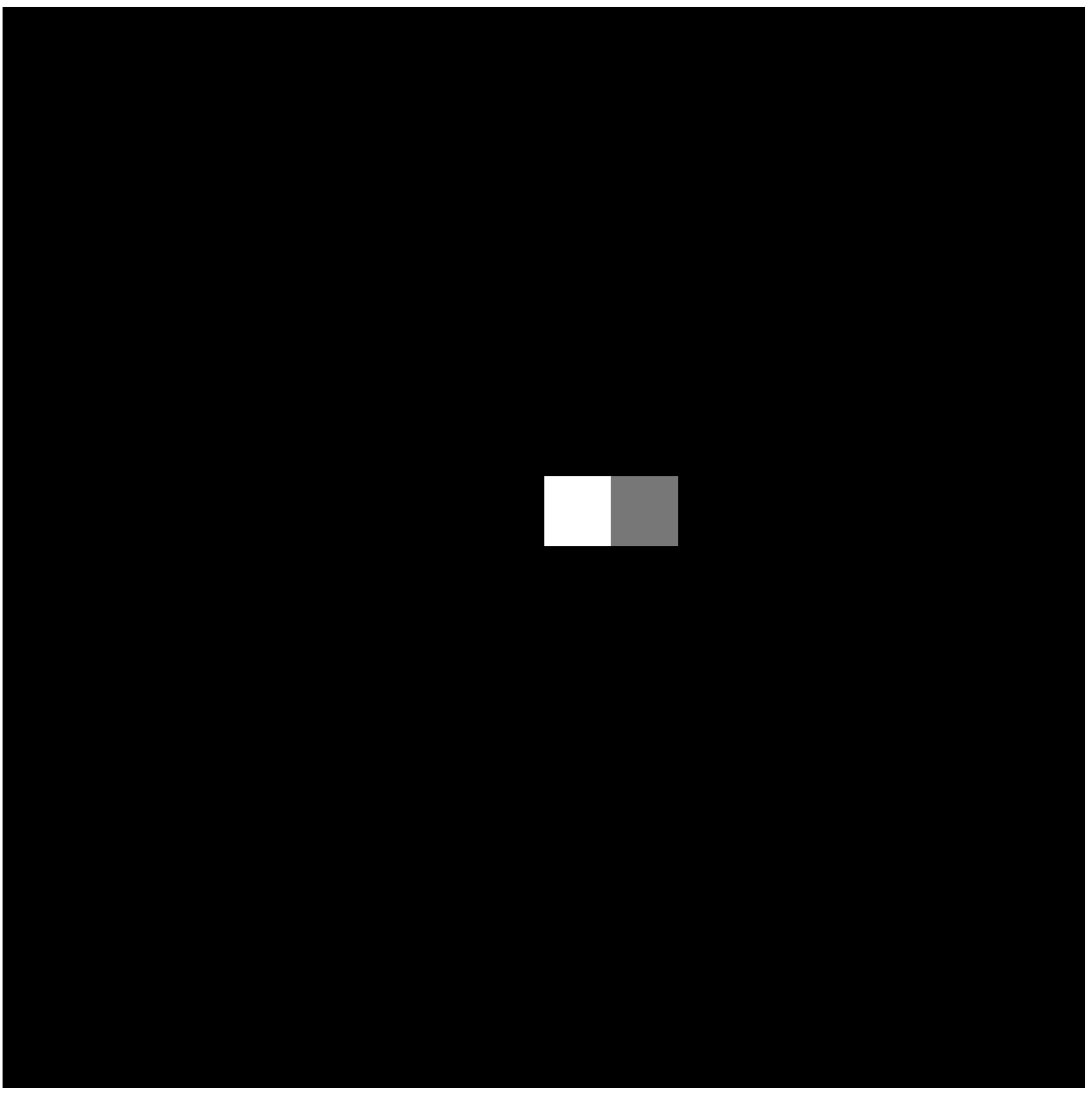}
\includegraphics[height=1.4in,width=1.6in]{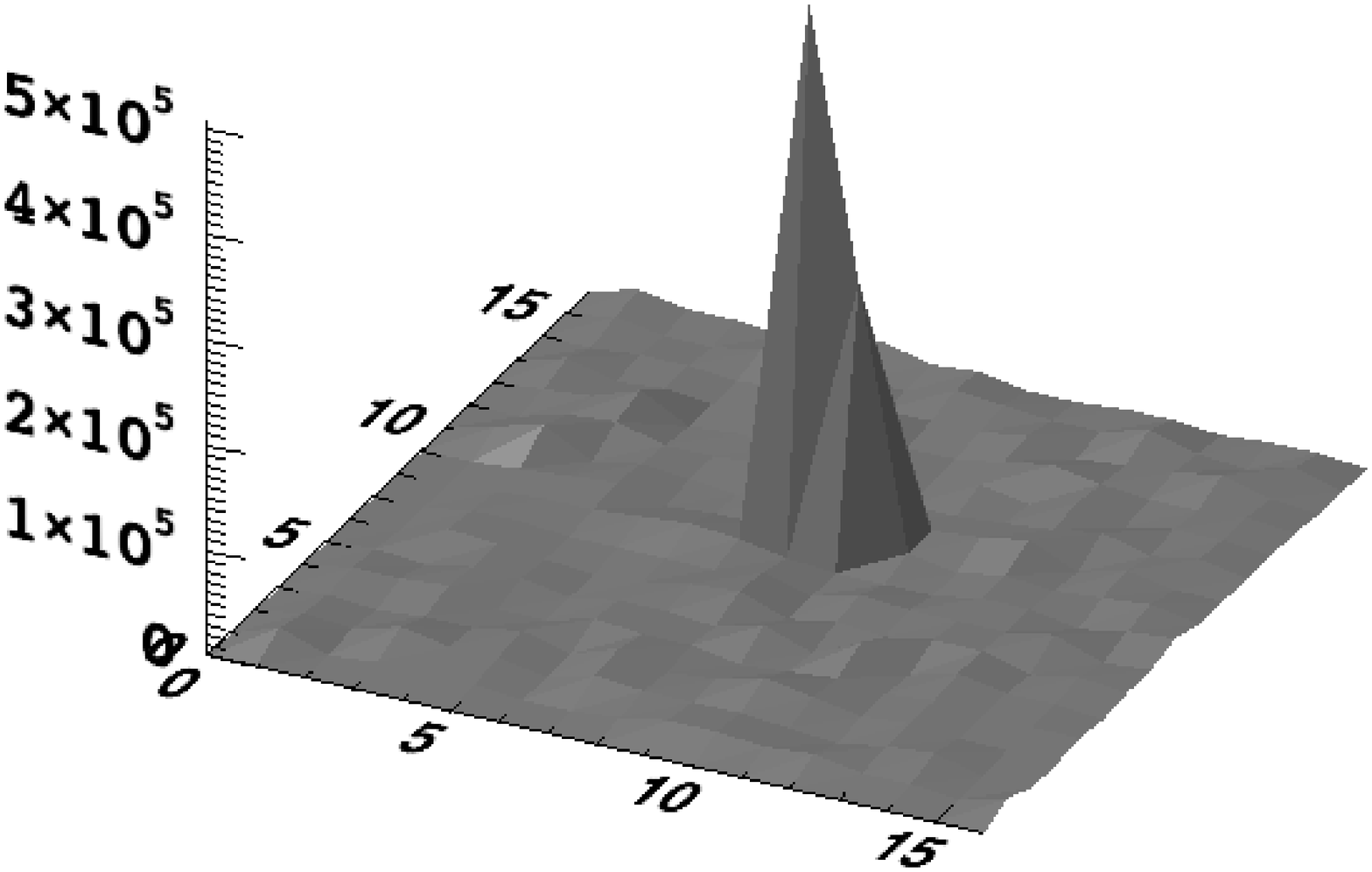}
\caption{(a) Shadow pattern obtained for two sources placed very closed to each other with CAM1
and CZT1 (left panel) configuration. (b) 2D view of reconstructed sky plane (middle panel). (c) 
3D view of the source intensities obtained by reconstruction (right panel) (Nandi et al. 2010).}
\label{kn : fig2.6}
\end{figure}

The CONFIG-1 and CONFIG-2 are identical as both configurations use CAM and CZT except that the CAM
patterns are different in CAM1 and CAM2. Therefore, the FOV and angular resolution in both
configuration are mathematically the same.

{$\bullet$ RT-2/CZT CONFIG-4 (FZP2 + CMOS):} 

CONFIG-4 is the best possible configuration for imaging a hard X-ray source in terms of
the achievable angular resolution. This configuration consists of dual FZP (FZP2) coder with a
finest zone width of $0.0041$ cm and high position sensitive CMOS detector with the smallest
pixel size of $0.005$ cm. The FZP coder is of negative cosine type. Inner zone radius of
each zones is $0.1$ cm and the number of zones is $144$.

\begin{figure}[h]
\vskip -0.1cm
\centering
\includegraphics[height=1.2in,width=1.2in]{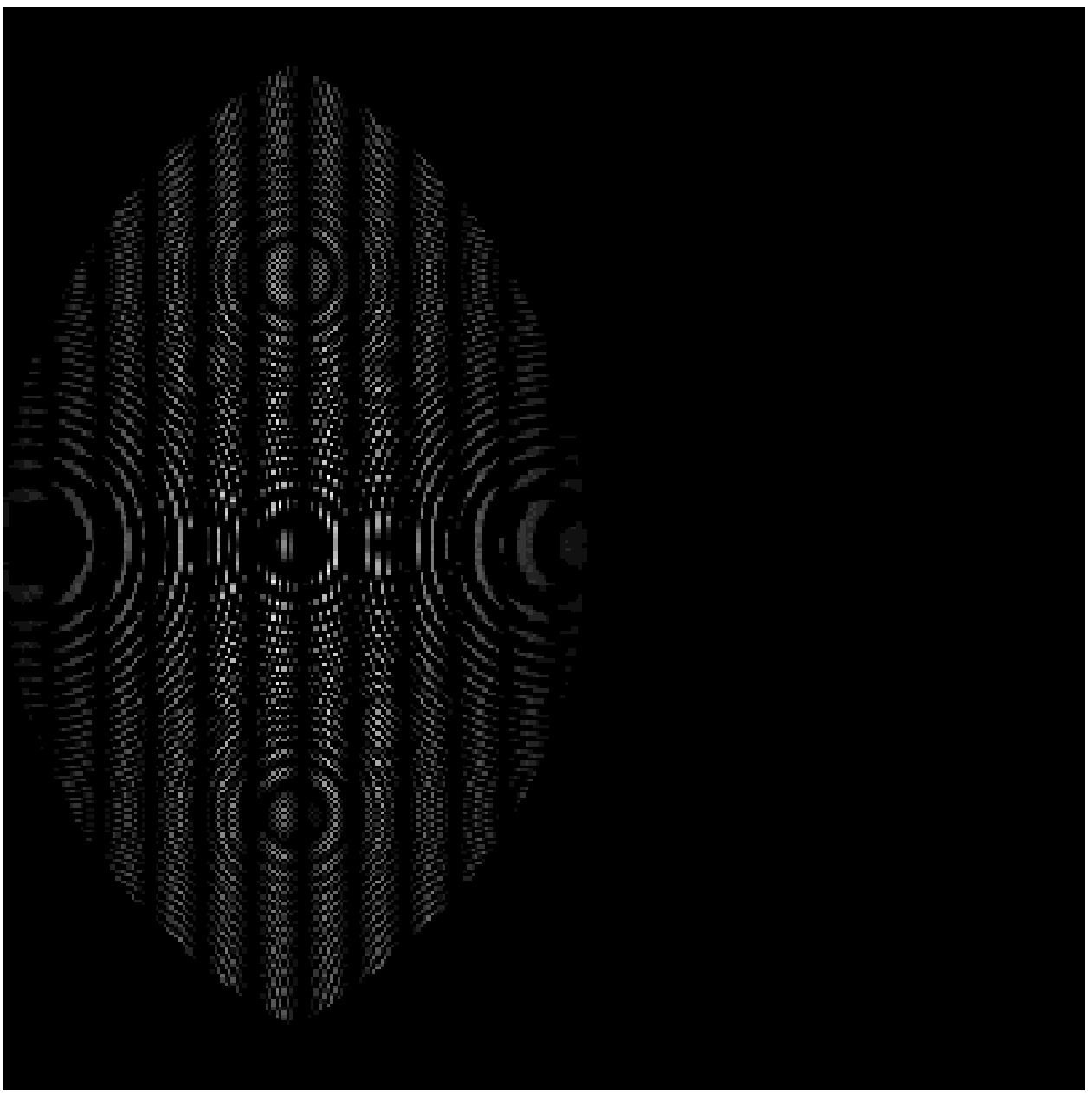}\hspace{0.2 cm}
\includegraphics[height=1.4in,width=1.8in]{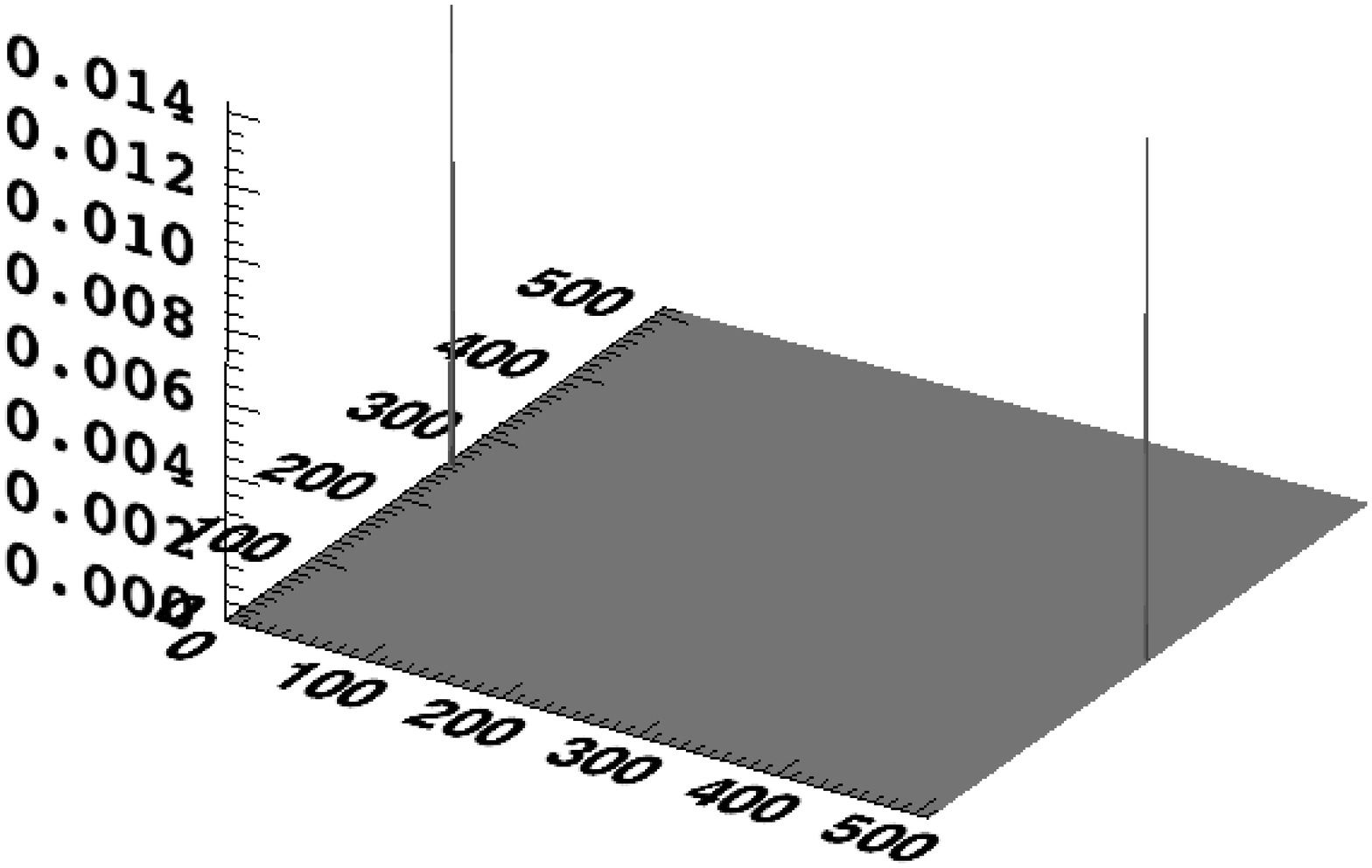}
\caption{(a) Fringes obtained on a CMOS detector with a pair of zone plates as coded aperture
(left). (b) 3D picture of the reconstructed source along with pseudo source which also appears
in the reconstructed source plane. The central DC offset is chopped out (Nandi et al. 2010).}
\label{kn : fig2.7}
\end{figure}

The Moire fringe pattern for the source with offset $\phi$ = 2.145$^\circ$ is shown in
Fig. 2.7(a). In Fig. 2.7(b), 3D view of the reconstructed source plane is shown along
with the pseudo source (ghost source). Reconstructed source plane shows that the source is
at extreme end of the FOV of the collimator, which confirms that the FOV is actually
4.29$^\circ$ (twice the $\phi$ value).

\begin{figure}[h]
\vskip -0.5cm
\centering
\includegraphics[height=1.2in,width=1.2in]{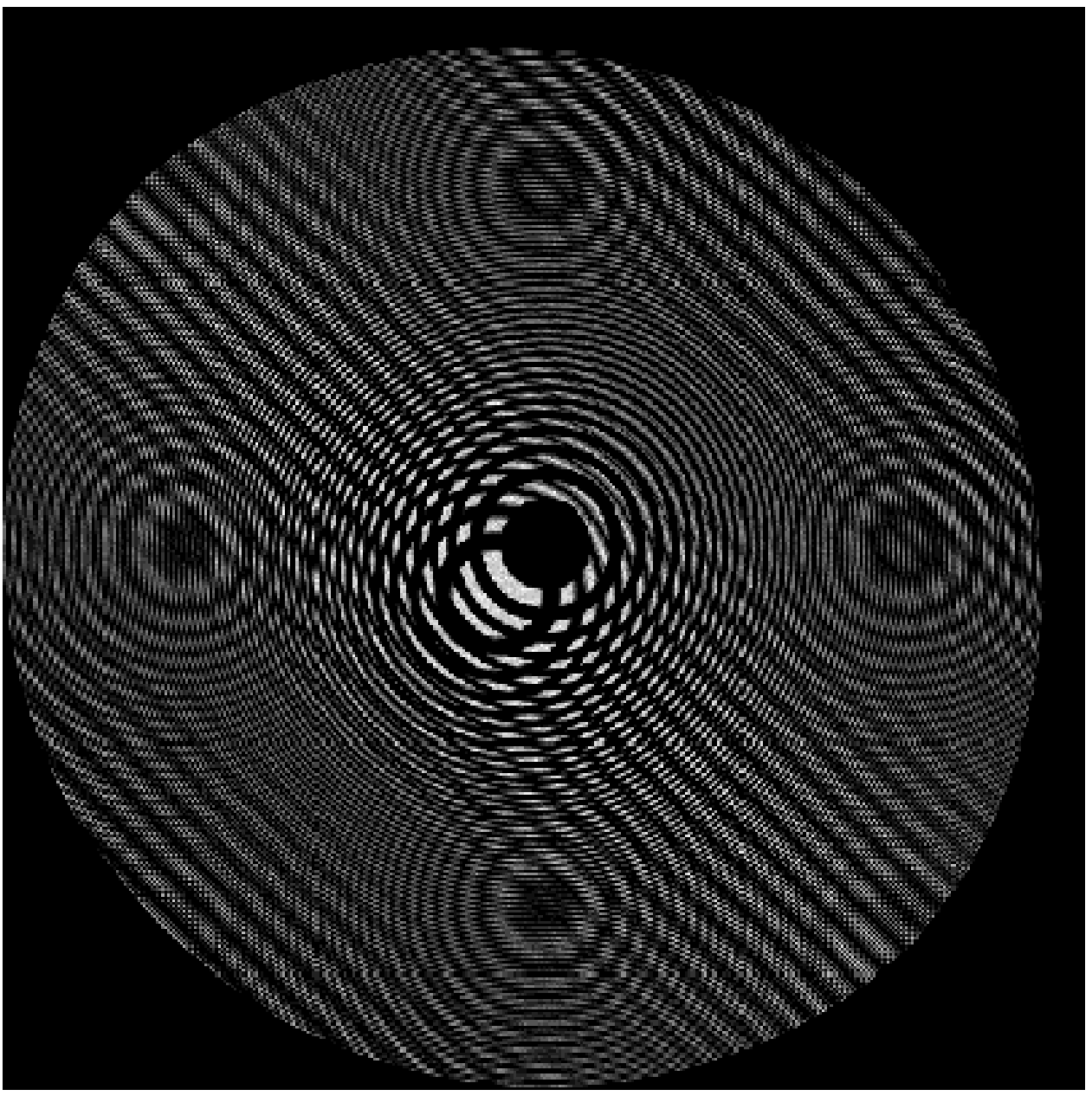}\hspace{0.2 cm}
\includegraphics[height=1.6in,width=1.6in]{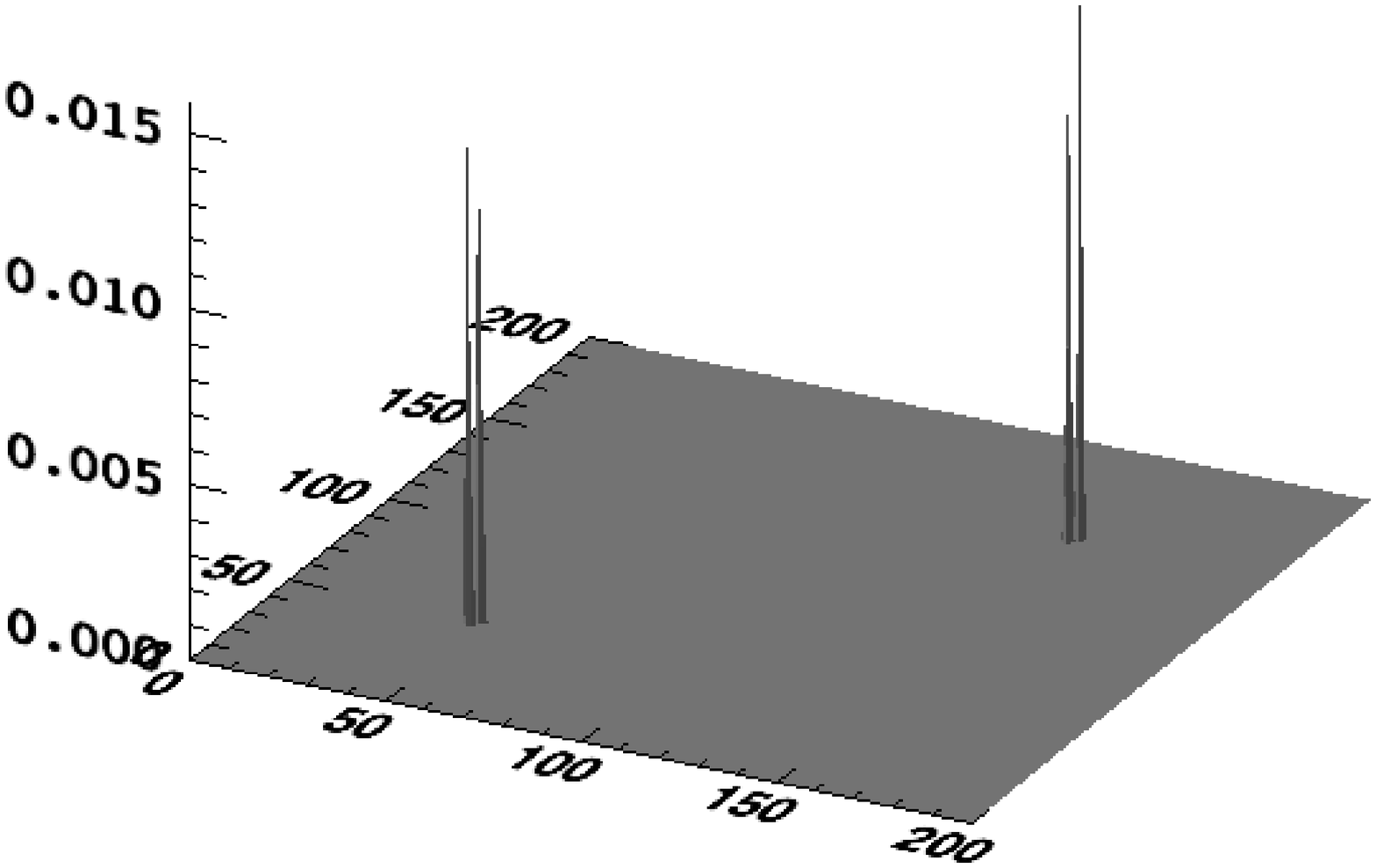}
\caption{(a) Fringes obtained (left) with a pair of sources at an angular distance from each other
equal to the calculated angular resolution (54'') of the FZP2-CMOS combination. (b) 3D view of
the reconstructed sources. To get a closer view, the part of the reconstructed plane containing
the sources is zoomed, so that out of 600 pixels along each sides only 200 pixels
are observed (Nandi et al. 2010).}
\label{kn : fig2.8}
\end{figure}

The most important aspect of this configuration is the best possible angular resolution which could
be around 54''. To verify the mathematically calculated angular resolution value, we simulate
with two sources placed 54'' apart and less than that. From simulation it is found that the
sources which are placed less than 54'' apart are not resolvable at all. In Fig. 2.8(a,b),
we showed the fringe pattern and reconstructed the sky plane of two sources which are separated
by 54''. The double pseudo source (ghost image) is also seen in the 3D view. Separation
between the two closely placed sources are found to be equal to one detector pixel dimension.

\begin{figure}[h]
\vskip -0.5cm
\centering
\includegraphics[height=1.2in,width=1.2in]{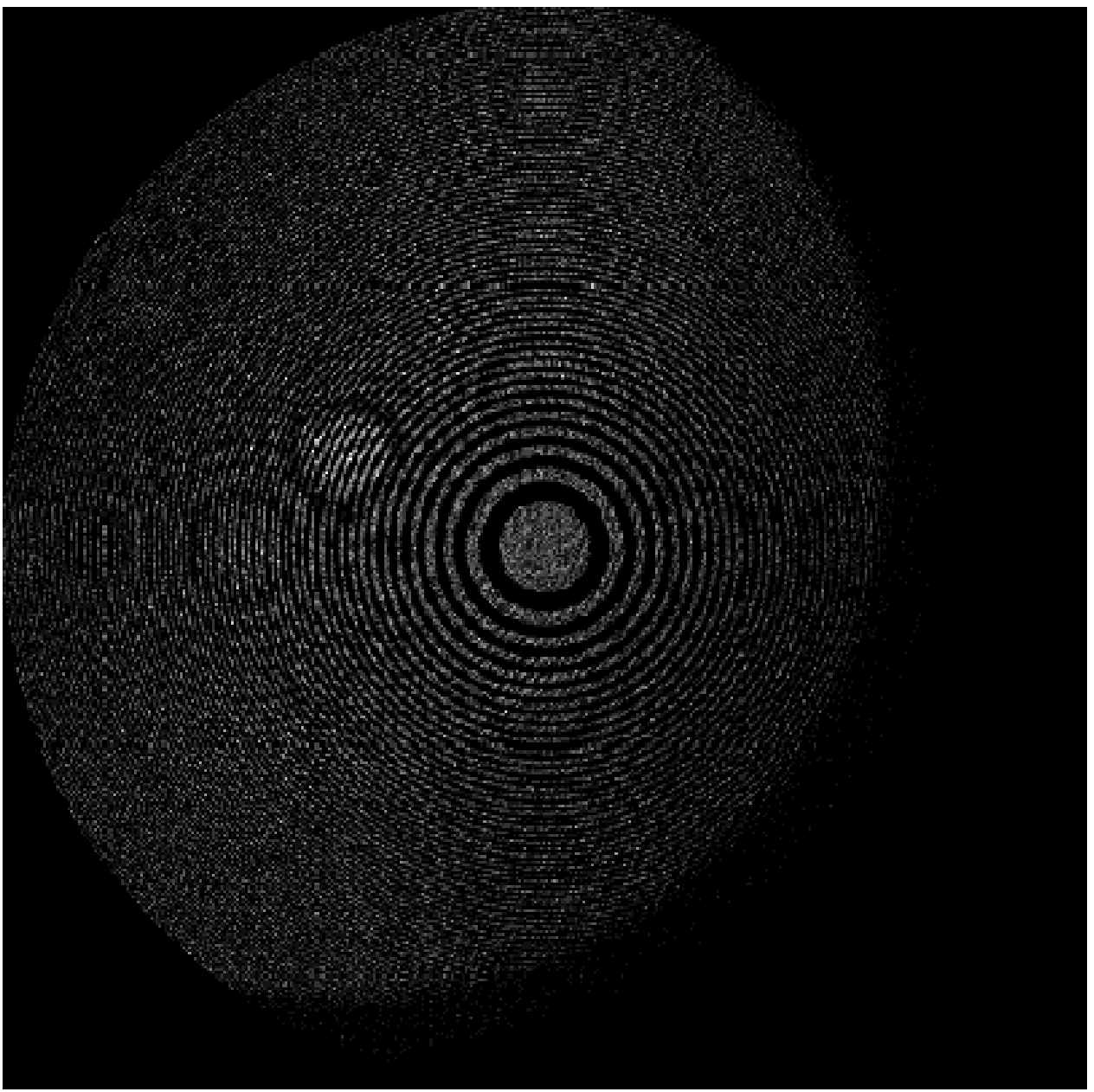}\hspace{0.2 cm}
\includegraphics[height=1.2in,width=1.2in]{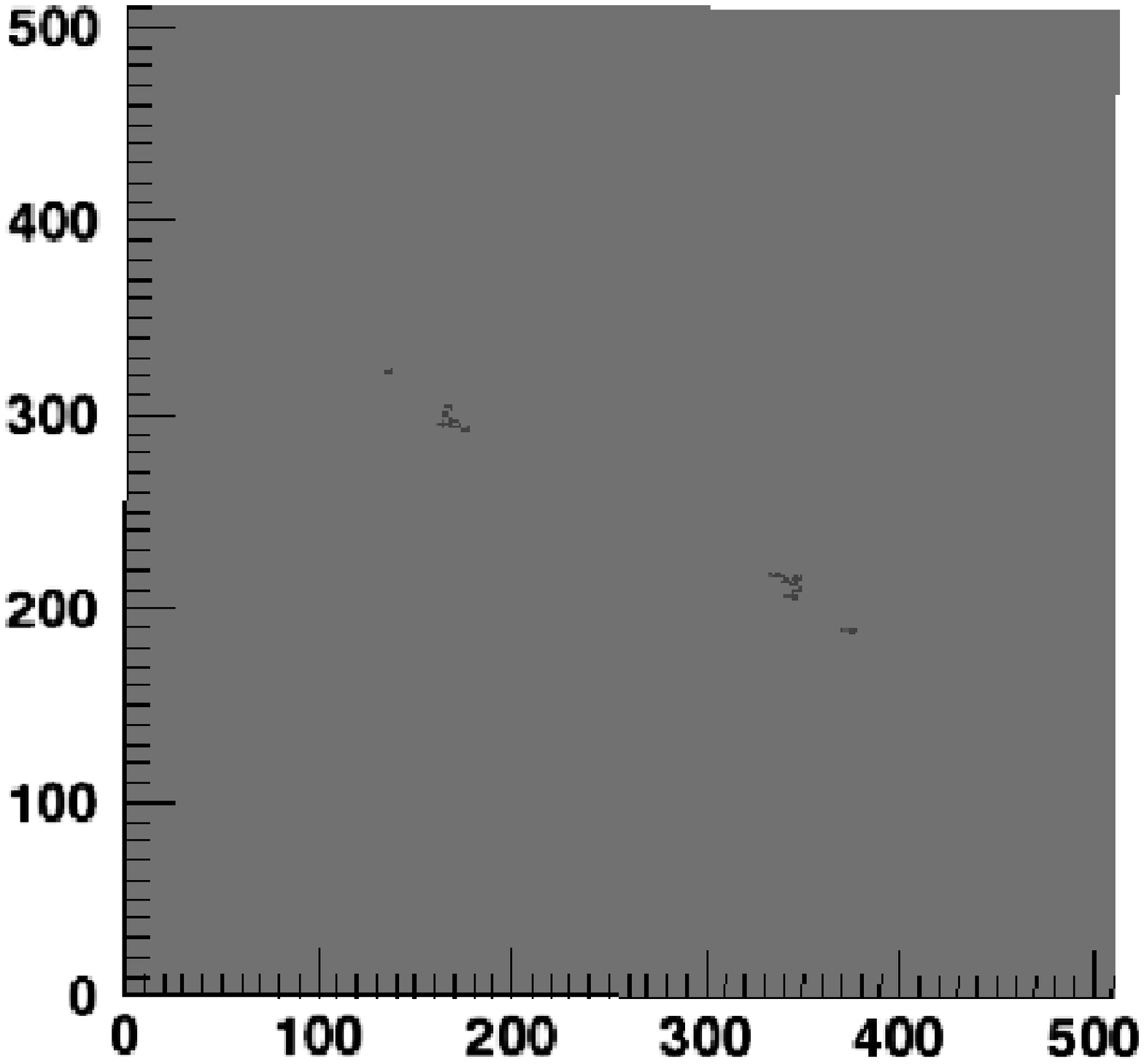}
\includegraphics[height=1.6in,width=1.8in]{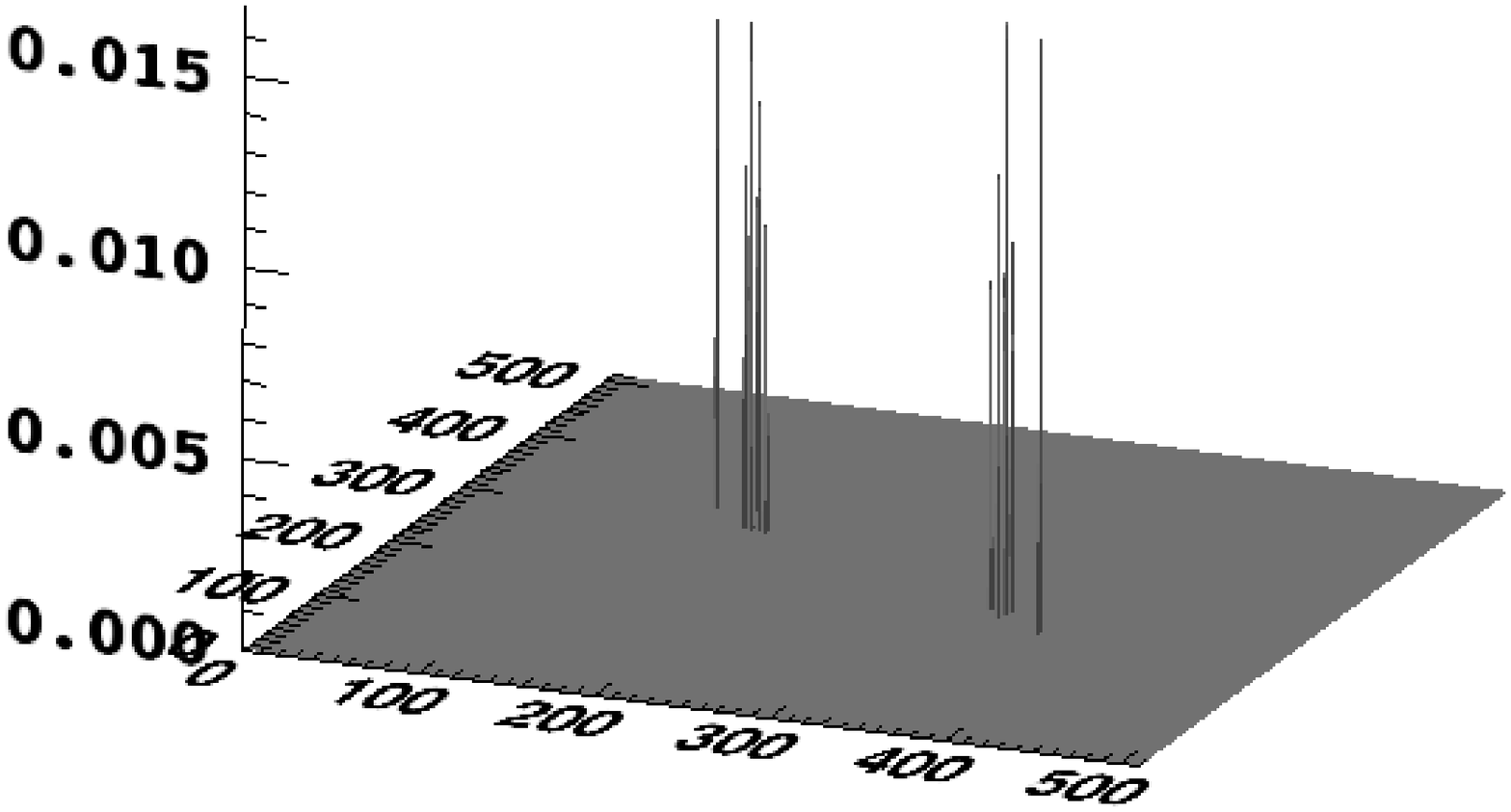}
\caption{(a) Fringes obtained with a pairs of zone plates on CMOS detector for multiple sources
(extended source) (left panel). (b) 2D view of the reconstructed sources (middle panel).
(c) 3D view of reconstructed sources of the sky plane (right panel) (Nandi et al. 2010).}
\label{kn : fig2.9}
\end{figure}

So far, we considered a point source to do the simulation. In principle, it is also possible
to do simulation for extended sources. Simulation is done for the same configuration (CONFIG-4)
with large number of point sources, which can be conveniently taken as an extended source. The
fringe pattern along with 2D and 3D views of reconstructed sources are given in Fig. 2.9(a-c).
The amplitude of individual reconstructed sources gives their relative intensities in the extended
source distribution.

\section{RT-2 data acquisition}

RT-2/E contains master control electronics for the RT-2 system. It receives various operational 
commands from satellite ground communication system (SSRNI) and sends these commands to the RT-2 
scientific payloads. It also sends `sec' data send command to the detectors and receives data in 
every second. It also packetizes and compresses the RT-2 data and sends to the SSRNI. Finally 
satellite data can be downloaded from SSRNI, normally twice in a day. In the following sub-sections
we will discuss about the RT-2 data structures and the modes of operations. 

\subsection{Data structure}

Because the on-board memory size is 10MB, RT-2 data are sub-divided into two telemetry 
channels: 18 \& 19, of each memory size 5MB. Channel 18 contains RT-2/S and RT-2/G spectral and timing
data along with the health parameters of the payloads. Similarly channel 19 consists of RT-2/CZT 
spectral, timing and imaging data along with its health parameters.

{$\bullet$ Data specifications of RT-2/S (RT-2/G):}

Data specification for the two phoswich detector payloads RT-2/S \& G are the same. Each
photon event at RT-2/S (RT-2/G) detector generates 32 bit data: 12 bit ADC data, 1 bit
for selecting G1/G2, 7 bit Pulse Shape (PS) and 12 bit Timing. These 32 bit event data
along with the health parameters encoded in VCO, are packaged by FPGA and kept in a memory
page. They contain a header block containing health parameters and counts, a spectrum block,
a timing block with high time resolution counting and an event block. In detector Event mode 
maximum events that can be stored in memory per sec are 7360 events. The detailed data 
specifications for the RT-2/S and RT-2/G in Normal and Event mode (Sreekumar et. al, 2010, 
Debnath et al. 2010b) are given in the following Tab. 2.2. 

\begin{table}[h]
\small
%\scriptsize
\centering
\caption{\label{table2.2} RT-2/S (RT-2/G) data specifications}
\vskip 0.2cm
\begin{tabular}{|l|c|c|c|}
\hline
\multicolumn{4}{|l|}{\bf VCO data: (2 bytes):}\\
\hline
D15 &\multicolumn{2}{|c|}{D14-D12} & D11-D0 \\
\hline
Mode Id (0/1) &\multicolumn{2}{|c|}{VCO Channel Numbers (0-7)} & VCO Counts (values) \\
\hline
\hline
\multicolumn{4}{|l|}{\bf Scientific Data in NORMAL Mode:}\\
\hline
         & G1-NaI & 2048 bytes & 1024 spectral ch. $\times$ 1 word\\
\cline{2-4}
Spectrum & G1-CsI & 2048 bytes & 1024 spectral ch. $\times$ 1 word\\
\cline{2-4}
         & G2     & 512 bytes & 256 spectral ch. $\times$ 1 word\\
\cline{2-4}
         & PSD    & 256 bytes & 128 spectral ch. $\times$ 1 word\\
\hline
TIMING & 800 words & 1600 bytes & 8 ch. $\times$ 100 blocks $\times$ 1 word\\
\hline
COUNTERS & 16 words & 16 bytes & 8 counters $\times$ 1 word\\
\hline
 & & {\bf Total 6480 bytes} & \\
\hline
\hline
\multicolumn{4}{|l|}{\bf Scientific Data in EVENT Mode: (4 bytes/event):}\\
\hline
D20-D31 & D13-D19 & D12 & D0-D11 \\
\hline
Time & PSD & G1/G2 sel. & ADC \\ 
\hline
\end{tabular}
\end{table}

Detector data are send to RT-2/E for processing as `sec' command is received. Header, spectral 
and timing blocks are sent in the Normal mode and header, event block are sent in the Event mode. 
These modes could be decided by ground command (Sreekumar et al. 2010).

{$\bullet$ Data specifications of RT-2/CZT:}

RT-2/CZT can be operated in two main detector modes: Event Mode and Normal Mode. Each
photon event generates two words (32 bit) information. From these 32 bit data, we got
header, spectral, image and timing data blocks. Data stored in the detector FPGA memory, 
are send to the RT-2/E at every second in response of `sec' data send command. A CZT
detector can accumulate maximum 4032 events. In Tab. 2.3, RT-2/CZT data specification
(Kotoch et al. 2010) are given.

\begin{table}[h]
\small
%\scriptsize
\centering
\caption{\label{table2.3} RT-2/CZT data specifications}
\vskip 0.2cm
\begin{tabular}{|l|c|c|c|}
\hline
\multicolumn{4}{|l|}{\bf Event Mode: (4 bytes/event):}\\
\hline
D31-D20 & D19-D10 & D9-D2 & D1-D0 \\
\hline
Time & ADC value & Pixel Id & Detector Id (0-2) \\
\hline
\hline
\multicolumn{4}{|l|}{\bf Normal Mode: CZT}\\
\hline
Image  & 3072 words & 1K words per CZT & 4 ch. $\times$ 256 pixels $\times$ 1 word\\
\hline
Spectrum & 1536 words & 512 words per CZT & \\
\hline
Timing & 1200 words & 400 words per CZT & 4 ch. $\times$ 100 timing words\\
\hline
Counter & 24 words & 8 words per CZT & 4 counters $\times$ 2 words\\
\hline
VCO  & 1 word &\multicolumn{2}{|c|}{D15 (Mode Id: 0/1), D14-D12 (ch. No.), D11-D0 (counts)}\\
\hline

Special & 8 words &\multicolumn{2}{|l|}{Telemetry, Temperature, Command sent, Data read against}\\
                     & &\multicolumn{2}{|l|}{command, event number, CMOS line number, Calibration } \\
                     & &\multicolumn{2}{|l|}{result identification word and Calibration status}\\
\hline
\multicolumn{4}{|l|}{\bf Normal Mode: CMOS}\\
\hline
Image  & 4096 words &\multicolumn{2}{|l|}{(256 $\times$ 256) pixels $\times$ 1 bit} \\
\hline
Sum  & 512 words &\multicolumn{2}{|l|}{Vertical sum and horizontal sum (256 words each)}\\
\hline
\end{tabular}
\end{table}

\subsection{RT-2 operational modes} 

Depending upon different processor commands and detector commands, RT-2 system can work in 
different solar modes. Detectors can work mainly in two modes: NORMAL and EVENT (or Test). 
The processor can work in five different solar modes (which involve different data packets): 
Bad Mode, Test Mode, Debug Mode, Solar Quiet Mode (SQM), Solar Flare Mode (SFM).

RT-2 system mainly works on SQM and SFM. In both these operational modes, RT-2 detectors work 
on NORMAL mode. Only difference is that in SFM data comes from RT-2 scientific payloads 
(only RT-2/S \& RT-2/G) to RT-2/E in 10 times faster rate.

\subsubsection {Solar Quiet Mode - SQM (100 sec/frame)}

This is the primary accumulation mode since the Sun is quiet in hard X-rays most of the time.
In this mode, spectrum is obtained for every $100$ sec and count rates for every second in
RT-2/S \& RT-2/G. Similarly, from CZT detectors, the spectrum and the image are obtained in 
every $100$ sec and count rates are obtained in every second. From CMOS detectors, only image is 
obtained in every $100$ sec.

\subsubsection{Solar Flare Mode - SFM (10 sec/frame)}

The major science requirement for this experiment is the availability of high temporal and
spectral resolution data during solar flares. Since such flares occur randomly, the onboard
software has a built-in mechanism for checking the current count rate against the present
thresholds to detect the flares. The flare search is carried out at every second.
In this mode, data frame structure is identical to the solar quiet mode except that both
the time resolutions are reduced by a factor of 10, i.e. in this mode count rates are every
0.1 sec and spectra are stored at every 10 sec. So, in this mode RT-2/S \& RT-2/G data comes
after every 10 sec, but RT-2/CZT data comes in every 100 sec since it does not have this 
operational mode.

Other than these two main operational modes, RT-2 system can also work in Test Mode (detectors 
work on EVENT mode, data comes in every second), Debug Mode (detectors work on NORMAL mode, 
but data comes in every one second). Apart from all these scientific functional modes, there 
have one safe mode (Bad Mode) and night mode (Shadow Mode). When the satellite enters into 
high flux region (SAA, North and South polar regions), satellite generates a BAD signal and 
depending on that signal, the whole RT-2 system operates into the BAD Mode. In this mode, data 
comes in every $100$ sec. Shadow Mode is activated when the Sun is out of satellite's field of 
view, i.e., during the instruments night time and also at solar occultation time. In this 
mode data comes mainly in every $100$ sec.

\section{RT-2 data analysis procedure}

The scientific data from the three RT-2 detector payloads is stored temporarily in the
memory of RT-2/E. After proper packetization, data is compressed with the onboard
software (Sreekumar et al., 2010). The compressed data is transferred to the satellite
system (SSRNI) for down-link to the ground station as soon as the telecommand is received.

RT-2 scientific and satellite telemetry data normally downloaded to the ground station
twice a day. Scientific telemetry data contains scientific timing, spectral and imaging
(only for RT-2/CZT) data for all the three detectors. Similarly satellite telemetry data
contain every $4$ sec monitored satellite health status, i.e., physical parameters (temperature,
memory, Flare, Corona, HV status etc). RT-2 data comes from the satellite via two telemetry
channels: 18 \& 19. The allotted memory size for each channel is $5 MB$. The Channel 18 contains
RT-2/S \& G data, where as 19 channel contains RT-2/CZT data.

RT-2 scientific data which comes from the satellite are compressed. So, before analyzing 
scientific data we need to run the task {\it `decomp'} to decompress the downloaded compressed 
data, which could be 10 to 12 times more in size than the compressed data depending on the mode 
of operation. This decompressed data is analyzed using a software written in LabVIEW.
These LabVIEW programs can analyse and save health parameters, timing, spectral and
image data of the three detectors in ASCII formatted files.

\begin{figure}[h]
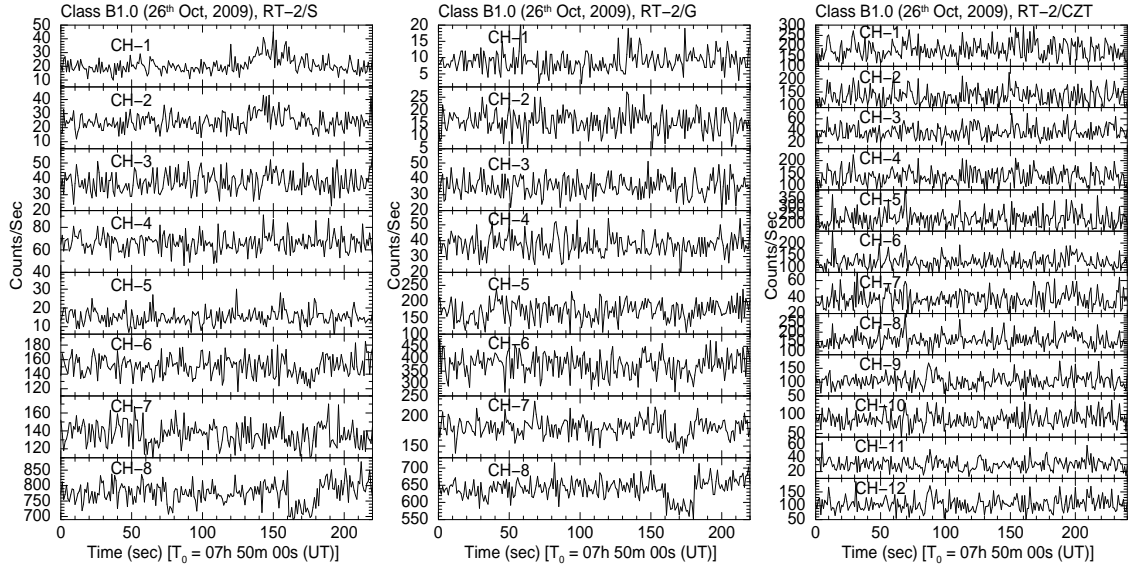

\centering
\includegraphics[height=1.9in,width=3.0in,angle=270]{fig2.10a.ps}
\includegraphics[height=1.9in,width=3.0in,angle=270]{fig2.10b.ps}
\includegraphics[height=1.9in,width=3.0in,angle=270]{fig2.10c.ps}
\caption{(a-c) Left, middle and right panels show (a) RT-2/S, (b) RT-2/G and RT-2/CZT
lightcurves for the 4048 CORONAS-PHOTON orbit data, observed on $26^{th}$ October, 2009.}
\label{kn : fig2.10}
\end{figure}

\subsection {Timing Data Analysis}

RT-2/S \& RT-2/G has 8 timing counters, bear different energy bands of spectral channel/energy
informations. Default pre-launch counter energy ranges for RT-2/S \& RT-2/G have been given in
Tab. 2.1. For SQM and SFM we get respectively 1 sec and 0.1 sec time resolution for
RT-2/S \& RT-2/G detectors. Figure 2.10(a-b) shows RT-2/S \& RT-2/G lightcurves of the flight 
data of orbit 4048. The observation date of the data is $26^{th}$ October, 2009. RT-2/S \& RT-2/G 
channels 1-5 have detected a B1.0 class solar flare on this data at 07:52:00 UT, are shown in 
Figure.

RT-2/CZT has 12 timing counters (4 for each CZT detectors). Counters 1-4 are allocated
for CZT module 1, 5-8 are for CZT 2 and rest 9-12 are for CZT 3. Each counter corresponds to
some specific channels of CZT1, CZT2 \& CZT3 spectra. Default channel ranges for
counters 1, 5 \& 9 are: 0 - 64 channels; for counters 2, 6 \& 10 are: 65 - 128 channels;
for counters 3, 7 \& 11 are: 129 - 192 channels; and for counters 4, 8 \& 12 are: 193 - 511
channels. Fig. 2.5 (c) shows RT-2/CZT lightcurve of the satellite orbit 4048
(observed on $26^{th}$ October, 2009). Due to the noise at the CZT pixels, we have not
been able to detect the solar flare observed by RT-2/S \& RT-2/G
in RT-2/CZT lightcurve data.

\begin{figure}[h]
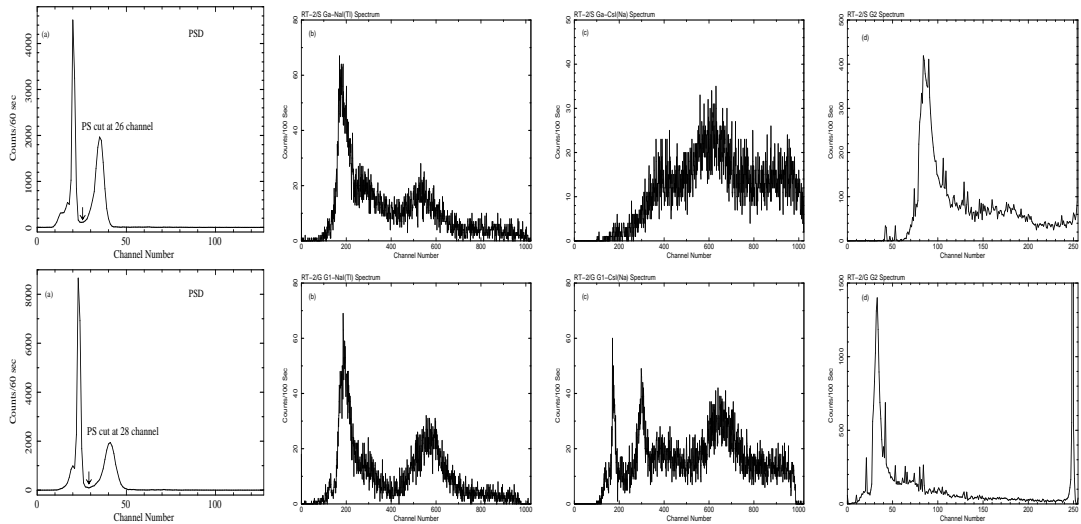

\centering
\includegraphics[height=1.32in,width=1.32in,angle=270]{fig2.11Aa.ps}\hspace{0.1 cm}
\includegraphics[height=1.32in,width=1.32in,angle=270]{fig2.11Ab.ps}\hspace{0.1 cm}
\includegraphics[height=1.32in,width=1.32in,angle=270]{fig2.11Ac.ps}\hspace{0.1 cm}
\includegraphics[height=1.32in,width=1.32in,angle=270]{fig2.11Ad.ps}\vspace{0.1 cm}
\includegraphics[height=1.32in,width=1.32in,angle=270]{fig2.11Ba.ps}\hspace{0.1 cm}
\includegraphics[height=1.32in,width=1.32in,angle=270]{fig2.11Bb.ps}\hspace{0.1 cm}
\includegraphics[height=1.32in,width=1.32in,angle=270]{fig2.11Bc.ps}\hspace{0.1 cm}
\includegraphics[height=1.32in,width=1.32in,angle=270]{fig2.11Bd.ps}
\caption{(A-B) Top panel four spectra are of RT-2/S detector: (a) PSD spectrum, (b) G1-NaI(Tl)
spectrum, (c) G1-CsI(Na) spectrum and (d) G2 spectrum and lower panel four spectra are of RT-2/G.
Onboard calibration source peak of $^{57}$Co (122 keV) is detected at G1-CsI \& G2 spectra and
$I^{121}$ decay peak (58 keV) is detected at G1-NaI spectra.}
\label{kn : fig2.11}
\end{figure}

\subsection {Spectral Data Analysis}

Both RT-2/S and RT-2/G have 3 energy spectra (G1-NaI, G1-CsI \& G2) and one PSD
spectrum. PSD spectrum shows the pulse shape of the events from NaI and CsI crystals and
the channel value at the `valley' of the PSD spectrum is applied to separate the NaI \&
CsI photon counts, which are registered in G1 amplifier as NaI and CsI spectra. PSD
spectrum is of total 128 channels. Pulse Shape (PS) {\it cut value} for RT-2/S is 26
channel and for RT-2/G is 28 channel. The spectra of G1-NaI \& G1-CsI has 1024 channel
informations, whereas G2 has spectral information in 256 Channel. Energy calibrated G1-NaI
spectrum is capable of detecting photons up to $\sim$ 100 keV. G1-CsI \& G2 spectra are
capable of detecting high energy photons up to 1 MeV. The strong peak in the G1-NaI
spectrum is of emission from $^{241}$Am radio-active source (59.5 keV). Other strong peaks
in the G1-CsI and G2 spectra are of emission from $^{57}$Co radio-active source (122 keV),
which is used as onboard calibration source. The energy resolution of both the instruments
(RT-2/S \& RT-2/G) are found to be around $18.72\%$ (RT-2/S) and $18.36\%$ (RT-2/G) at
60 keV.

\begin{figure}[h]
\vskip -0.3cm
\centering
\vbox{
\includegraphics[height=1.32in,width=1.32in,angle=270]{fig2.12a.ps}}\vspace{0.5 cm}
\includegraphics[height=1.32in,width=1.32in,angle=0]{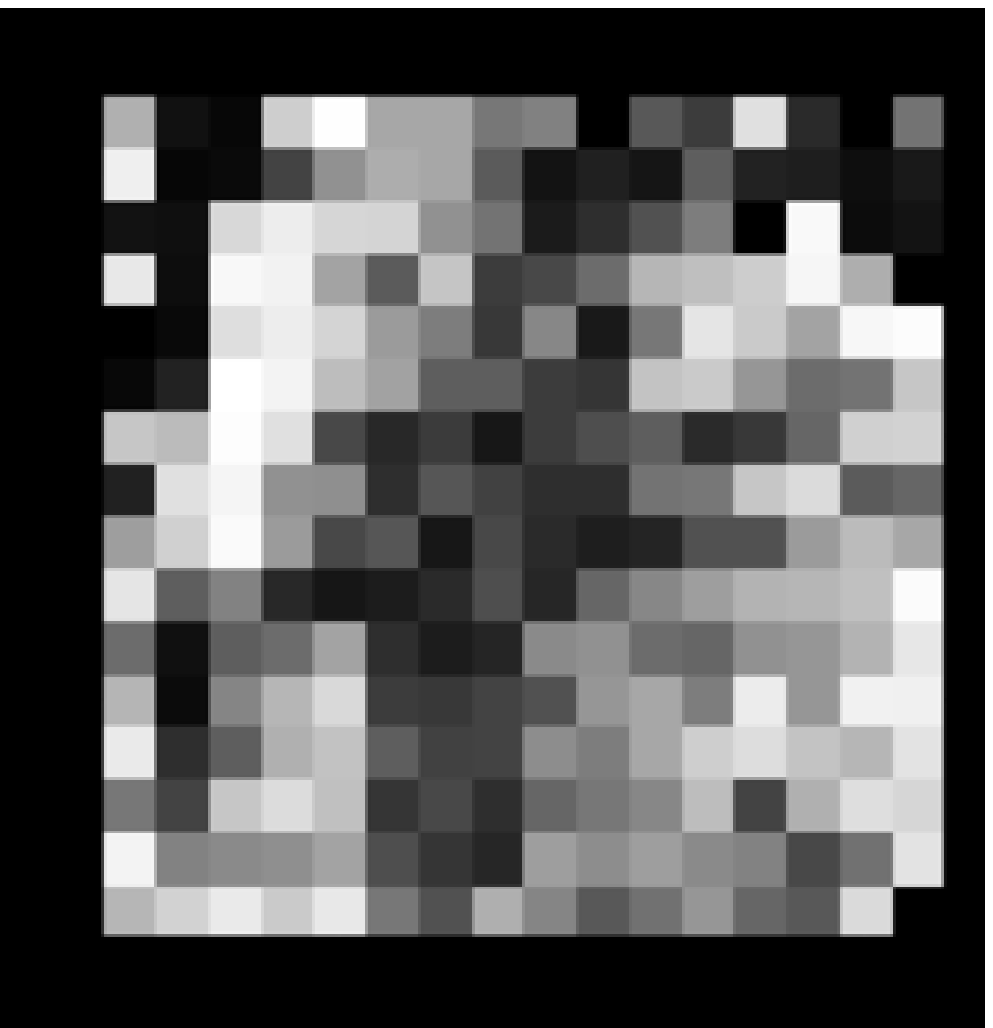}\hspace{0.1 cm}
\includegraphics[height=1.32in,width=1.32in,angle=0]{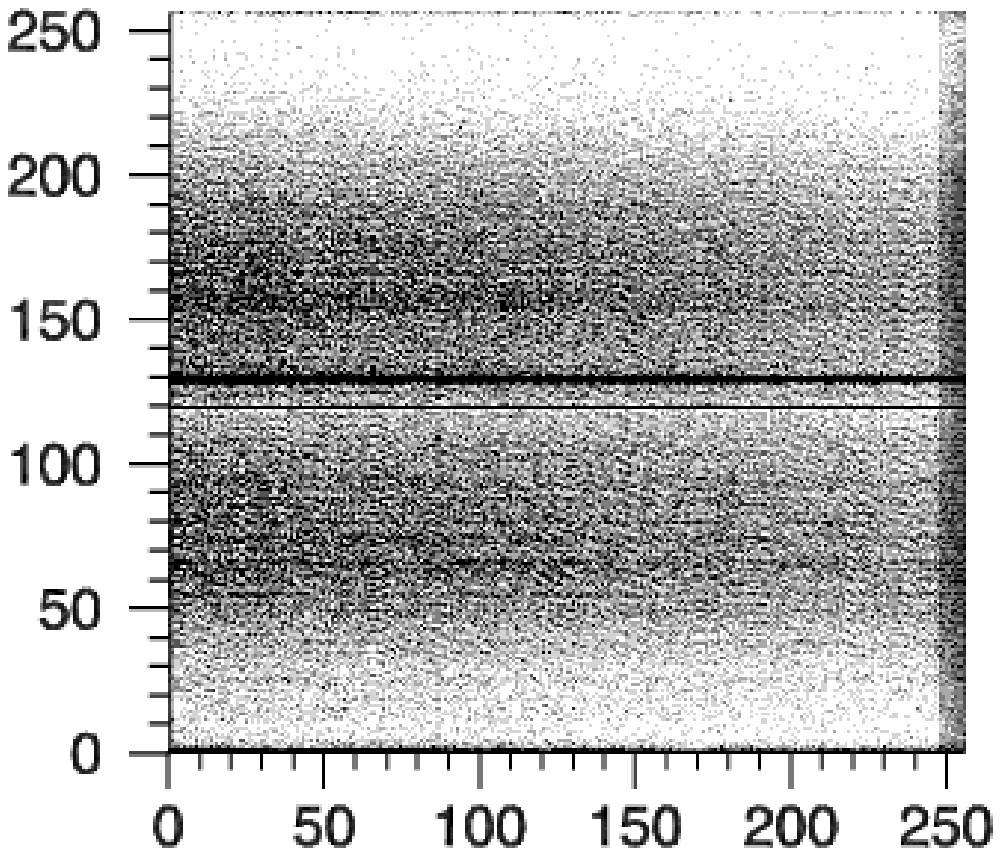}
\caption{(a-c) Top panel (a) shows CZT module 2 spectrum of the satellite orbit 441, observed on 
$28^{th}$ February, 2009. Bottom left panel (b) shows the CZT 2 image and right panel (c) shows CMOS 
background image for the same satellite orbit. In the images, Y \& X axes indicate pixel numbers 
of the detector modules.} 
%CZT image contains $16 \times 16$ pixels informations, where as CMOS has 
%the information of $256 \times 256$ pixels data.}
\label{kn : fig2.12}
\end{figure}

In Fig. 2.11(a-b), the spectra of PSD, G1-NaI, G1-CsI \& G2 for RT-2/S and RT-2/G detectors
of the satellite orbit 2051 (observed on $15^{th}$ June, 2009) are plotted. From the PSD
spectra it is clear that PS cutoff value of RT-2/G1-NaI(Tl) \& G1-CsI(Na) spectra for RT-2/S 
\& RT-2/G detectors are at 26 \& 28 channels respectively.

RT-2/CZT payload consists of three solid state imaging CZT detectors and one CMOS
detector. All the three CZT detectors (CZT1, CZT2 \& CZT3) can acquire both solar spectral
and image informations but the CMOS has only imaging capability. The energy range for the
CZT detectors are $20 - 100$ keV. RT-2/CZT spectrum for CZT 2 module is shown in Fig. 2.12(a), 
for the in-flight RT-2 data of the satellite orbit 441 (first RT-2/CZT high voltage switch ON 
orbit, was observed on $28^{th}$ February, 2009).

\subsection {Image Data Analysis}

As mentioned before, all the three CZT detectors and the CMOS detector of the RT-2/CZT payload
receive imaging information. In Fig. 2.12(b), CZT module 2 image of $28^{th}$ February, 2009,
orbit 441 is plotted. In Fig. 2.12(c), CMOS background image has been plotted. Image of the 
CZT detector contains $16 \times 16$ pixel matrix (total 256 pixels) of each pixel size $2.5~mm$ 
and CMOS detector contains $256 \times 256$ pixel matrix of pixel size $50~\mu m$.

%%%%%%%%%%%%%%% RXTE %%%%%%%%%%%%%%%%%%%%%%%%%%
\newpage
\section{RXTE: A mission for compact object study}

My thesis work for the observational black holes study is based on {\it Rossi X-ray Timing
Explorer} (RXTE) data. RXTE is a NASA's astronomy mission satellite, was launched from Cape
Canaveral on 30 December, 1995, on a Delta rocket. This mission gives a huge bust in our
understanding about the astrophysical objects. This satellite was launched to study the most
luminous sources of our Galaxy (i.e. X-ray binaries), Active Galactic Nucleies (AGNs),
quasars, pulsars etc. Major advantage of this satellite over previous X-ray instruments is due
to its unparalleled detector characteristics and wide energy range (2 - 200 keV). RXTE detectors
have unique capability of acquiring X-ray photons with $\mu$s time resolution over a wide energy
band, which is quite able to probe temperature, magnetic fields and other fundamental physical
properties close to the observed sources (mostly compact objects) as well as the geometry of
the systems, nature and characteristics of the compact object itself (e.g., masses, spin
rotation periods and size of the compact objects etc.). RXTE consists of three main scientific
instruments: the All-Sky Monitor (ASM) (Levine et al. 1996), the Proportional Counter Array
(PCA) (Jahoda et al. 1996), the High Energy X-Ray Timing Experiment (HEXTE) (Rothschild et al. 1998).

RXTE is a polar satellite which orbits the earth at an altitude of $\sim$580 km with an inclination
angle 23$^\circ$ and a period of $\sim$100 min. Therefore it passes over the South Atlantic Anomaly (SAA)
region in 6 of its daily 14 - 15 orbits for each time duration of 10-20 min. During SAA region
time, all instruments are put into the safe mode to protect them against the highly charge
particle SAA flux.

The satellite already has worked more than 13 years and still it is working. More than
1700 scientific papers have published on the basis of RXTE data, near about 100 Ph.D
thesis have been submitted on the basis of the study of RXTE data. Also it has discovered
more than 50 astrophysical X-ray sources.

\section{Brief discussion about RXTE scientific payloads}

NASA astronomy satellite RXTE consists of three scientific payloads: ASM, PCA and HEXTE.
In this Section we will discuss briefly about the configurations of these three instruments.

\subsection{All-Sky Monitor (ASM)}

The ASM (Fig. 2.13) of RXTE has been designed to monitor the sky in 1.5 - 2.0 keV energy band. 
The ASM consists of three Scanning Shadow Cameras (SSCs). A motorized drive assembly is used 
to rotate these SSCs for viewing different parts of the sky. Each SSC contains a Position
Sensitive Proportional Counter (PSPC) system which views the sky through coded aperture mask.
The mask is of a thin Al sheet of 15 transparent (open) and 16 opaque (closed) matrix elements
of equal dimensions. Each PSPC contains 8 resistive carbon-coated quartz fiber anodes, each
end of which is connected to a dedicated electronic measurement chain.

\begin{figure}[h]
\vskip 0.6cm
\centering
\includegraphics[height=2.4in,width=2.4in,angle=0]{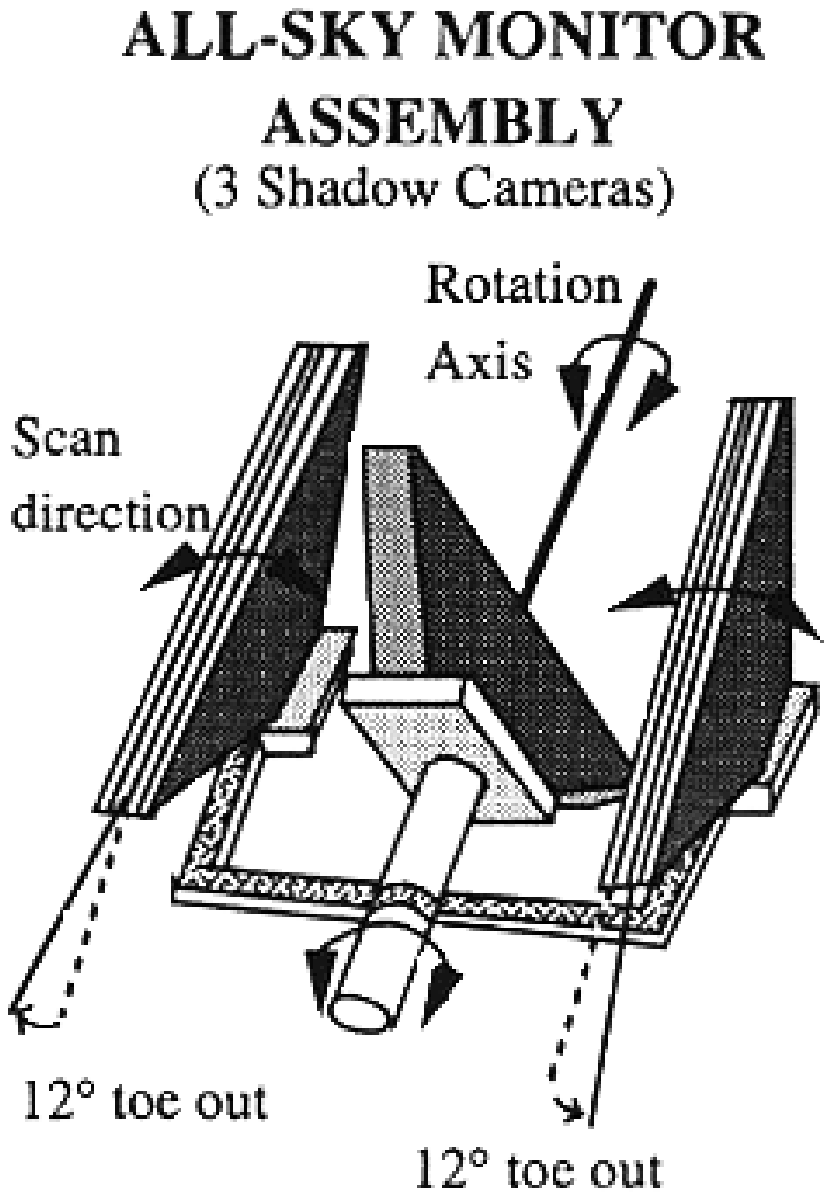}\hspace{0.2cm}
\includegraphics[height=2.4in,width=2.4in,angle=0]{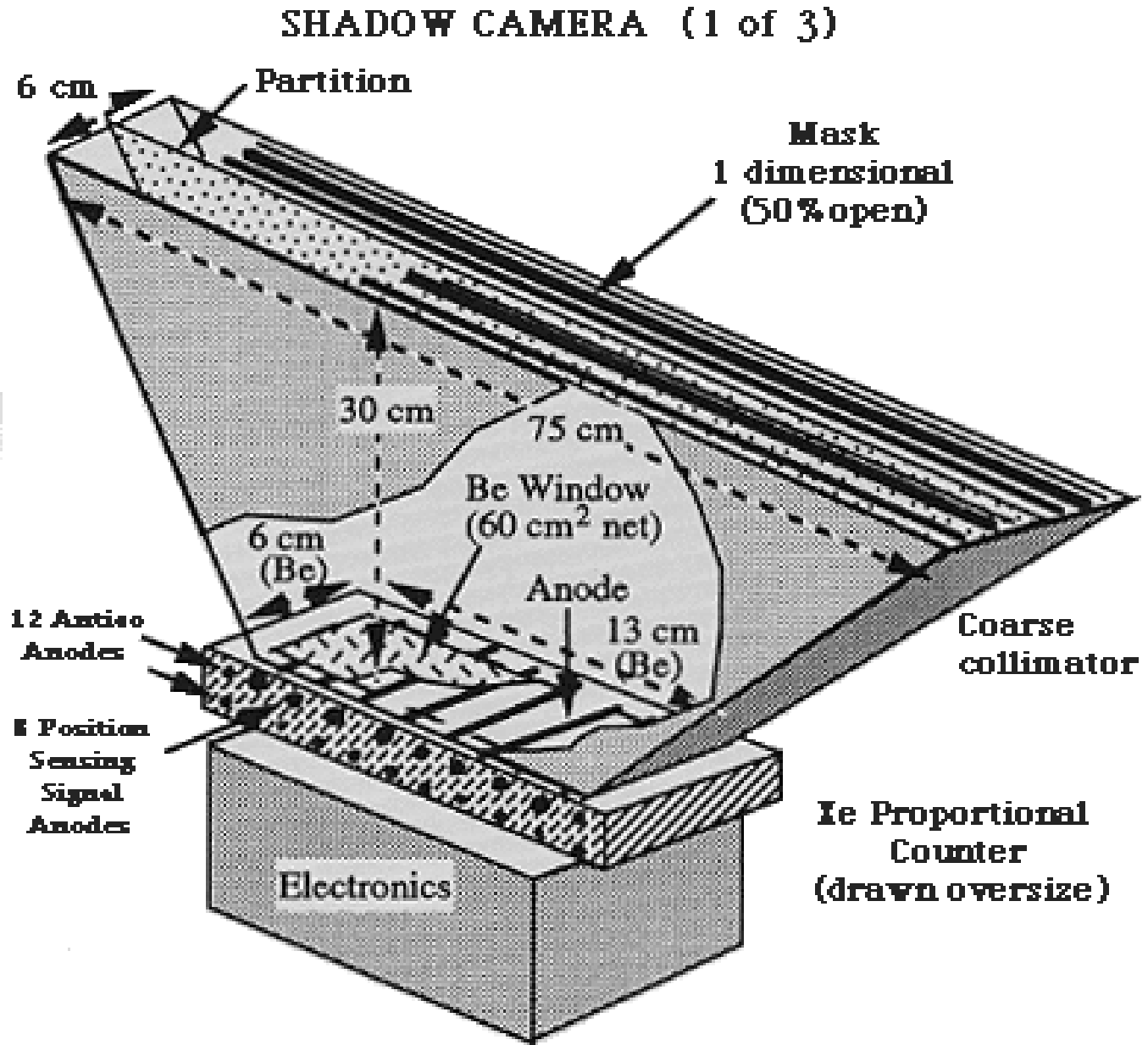}
\caption{(a-b): Left panel (a) shows the ASM assembly with 3 shadow cameras and right panel 
(b) shows the detailed cross-sectional view of one shadow camera (courtesy: NASA).}
\label{kn : fig2.13}
\end{figure}

Observing X-ray source intensity and direction, measurements are derived from the displacements
and strengths of the shadow patterns cast by the X-ray sources with the camera's field of view
(FOV). Each camera has FOV of $6^\circ \times 90^\circ$ FWHM and detection sensitivity up to
30 mCrab (1 Crab = 75.5 SSC cts/sec). The X-ray events detected by ASM are normally processed
on-board by two ASM Event Analyzers (EAs) in the Experiment Data System (EDS).
Typically, a source is observed $5-10$ times a day, and the full coverage is obtained in
a series of 90 sec stationary exposures known as `dwells'. The ASM also provides the unequally
sampled data in one-day average from a number (typically $5-10$) of individual ASM dwells.
ASM has proved to be a unique instrument on-board {\it RXTE} due to its fascinating
first-results or quick look capability, such as detection of a new transient in the sky,
state change of known sources and quasi-periodic or highly flux variations in a longer
time scale etc.

The ASM archival data for the observed sources are available at the NASA site
(http://xte.mit.edu/ASM\_lc.html). Data are available in three different energy bands:
a-band (1.5 - 3.0 keV), b-band (3 - 5 keV) and c-band (5 - 12 keV).

\subsection{Proportional Counter Array (PCA)}

\begin{figure}[h]
\vskip 0.6cm
\centering
\includegraphics[height=2.4in,width=2.4in,angle=0]{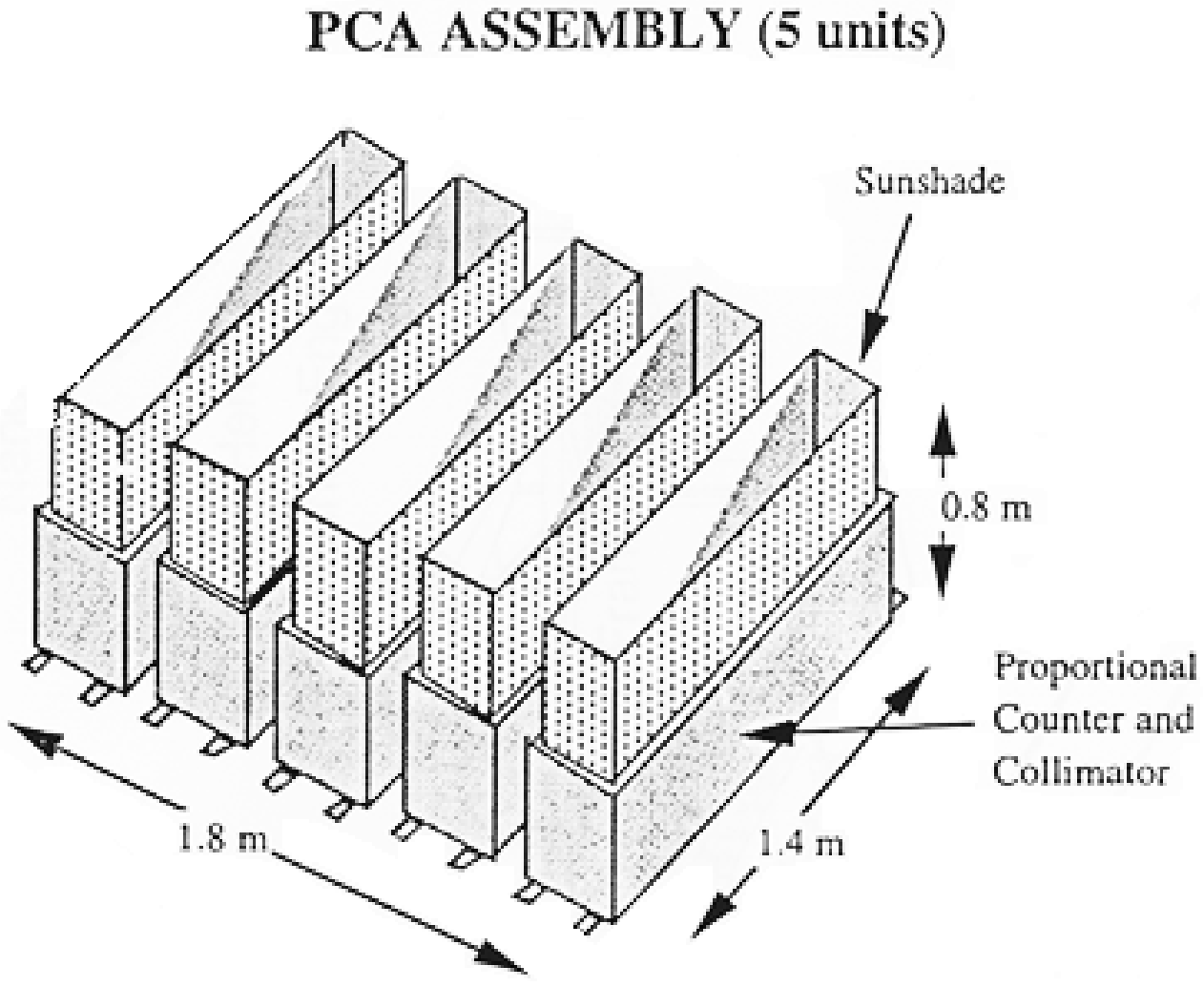}\hspace{0.2cm}
\includegraphics[height=2.4in,width=2.4in,angle=0]{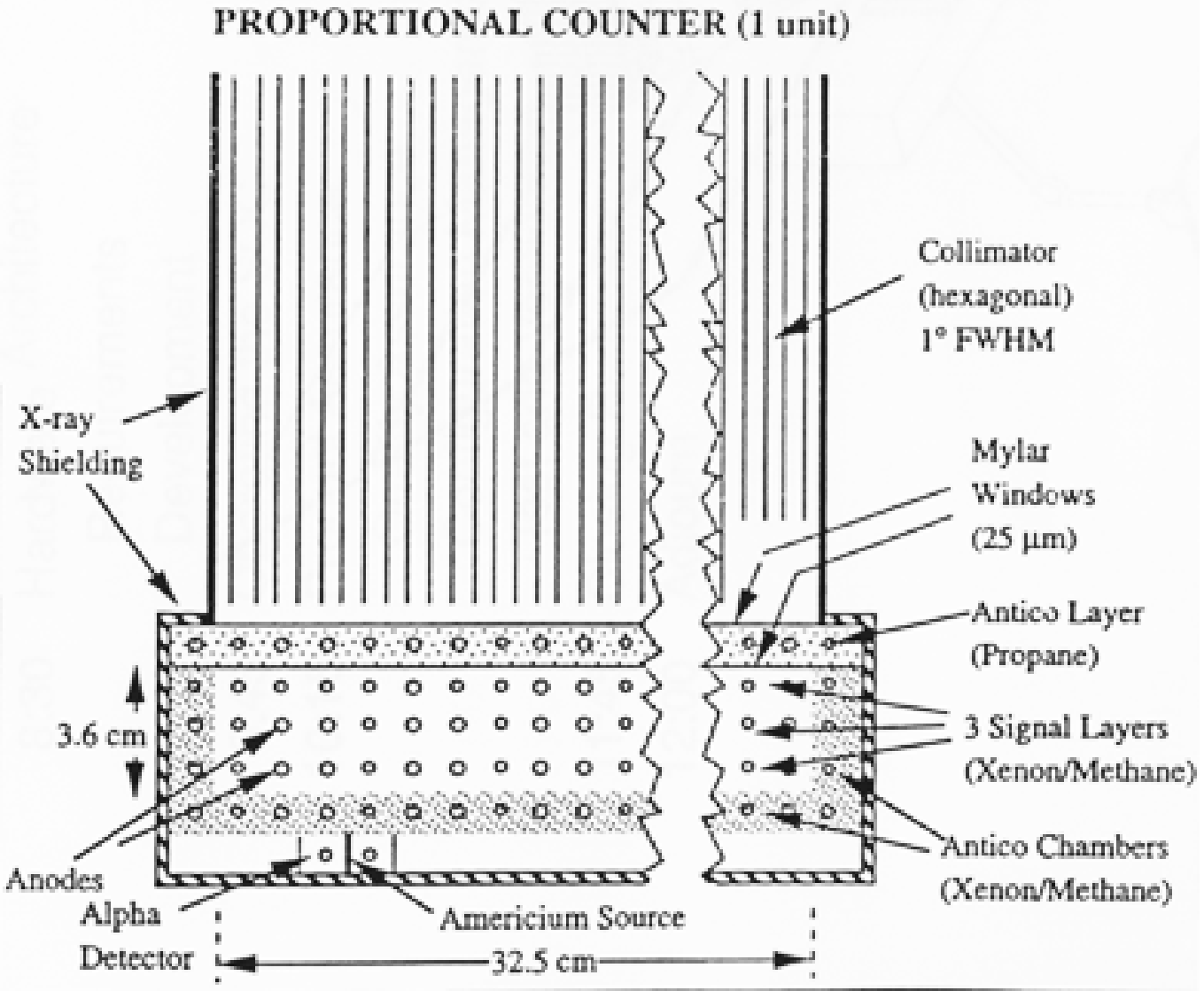}
\caption{(a-b): Left panel (a) shows the PCA assembly with five PCUs and right panel 
(b) shows the detailed cross-sectional view of one PCU (courtesy: NASA).}
\label{kn : fig2.14}
\end{figure}

The PCA (Fig. 2.14) of RXTE has been designed mainly to study timing properties of bright 
galactic sources and also for faint extra-galactic sources in 2 - 60 keV X-ray energy band 
with unique time resolution of 1 $\mu$sec. PCA consists of 5 Proportional Counter array Units 
(PCUs). Each detector has an effective area of $\sim~1300~cm^2$. PCA has a good energy 
resolution of $18\%$ @ 6 keV.

Each PCU consists of a hexagonal beryllium copper made collimator with $1^\circ$ FWHM FOV
and two gas-filled chambers mounted on top of each other. The first chamber is filled with
propane (veto layer) and it protects the other from background radiations.
The second chamber is filled with a mixture of Xenon/Methane (90/10) gas and it is
the main X-ray detector for each PCU. This chamber contains 4 layers of anode grids
separated by cathode wires (per grid 20 anodes are present). The upper 3 layers, each
splitted into two (left and right), are used for X-ray detection. The X-rays impacting on
a Xenon atom generate a cascade of electrons and they are collected by the high voltage anode.
The higher the X-ray photon energy, the more electrons are generated,
and thus the higher the electron pulse-amplitude. The anodes in the grid of the last layer serve 
as a discriminator i.e., veto layer for charged particles entering through the detector walls.

My thesis work for the observational study of black hole candidates are mainly on 
the basis of this PCA data analysis results. The PCA archival public data for 
the observed sources are available at the NASA site 
(http://heasarc.gsfc.nasa.gov/cgi-bin/xte/all\_too.pl).

\subsection{High Energy X-Ray Timing Experiment (HEXTE)}

\begin{figure}[h]
\vskip 0.6cm
\centering
\includegraphics[height=2.4in,width=2.4in,angle=0]{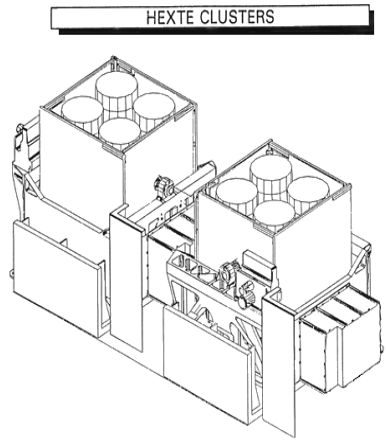}\hspace{0.2cm}
\caption{HEXTE cluster assembly with eight ($2\times 4$) phoswich detectors, each cluster contains 
4 phoswich detectors are shown (courtesy: NASA).}
\label{kn : fig2.15}
\end{figure}

The HEXTE (Fig. 2.15) of RXTE has been designed to study the timing and the spectral properties 
of bright sources in hard X-ray regime of 15 - 250 keV. The HEXTE consists of 2 independent 
clusters (A or 0 and B or 1) of detectors, whereas each cluster contain 4 NaI(Tl)/CsI(Na) 
phoswich scintillation counters sharing a common 1$^\circ$ FWHM FOV. Total detection area 
of the 8 detectors is 1600~$cm^2$. Each detector covers the whole energy range of 15 - 250 keV
with an average energy resolution of 15.4\% FWHM @ 60 keV. The cluster orientation of the
HEXTE instrument permits effective immunity to systematic background variations through the
use of continuous gain control, chopping of the source signal and anti-coincident shielding
for the charge particle events.

The HEXTE has provided important spectral information in the high energy band, but the
nominal count rates (e.g., 289 counts/sec for the Crab) and low sensitivity
(1 Crab = 360 count/sec per HEXTE cluster) limit its use for timing studies.
HEXTE archival data can be downloaded from the same NASA site of PCA data.

\section{RXTE data analysis procedure}

RXTE archival public data for all observed astrophysical sources are available at
NASA RXTE TOO page (http://heasarc.gsfc.nasa.gov/cgi-bin/xte/all\_too.pl). All the
data are in fits format. Data analysis FTOOLS (HEASoft) software package is also
available at NASA site (http://heasarc.gsfc.nasa.gov/docs/software/lheasoft).

Satellite on-board processor which processes ASM, PCA and HEXTE event data is called
Experiment Data System (EDS). EDS consists of 8 Event Analyzers (EAs): six for
PCA \& HEXTE data system and two for ASM, winch can process incoming event data into
8 different modes. Two preset EAs of PCA data are:

\begin{enumerate}
\item
the Standard-1 data mode provides data in the energy range of $2-60$ keV with
a time resolution of 0.125 sec and no energy resolution since all 256
channels are combined into one channel,
\item
the Standard-2 data mode provides data in the same energy range of $2-60$ keV
with a time resolution of 16 sec and high spectral resolution (the pulse
height is recorded in 129 channels covering the $2-60$ keV energy range,
although maximum energy limit changes from one Epoch to another Epoch).
\end{enumerate}

Other four EAs for PCA data can be set by the principal investigator (PI) of
the observation or by the RXTE ground operation staffs. In general most used EAs'
are `Event',`Binned' and `Single-Bit'. In Event mode, data is transferred to the
ground station as a  time series of unevenly spaced events described by arrival
time, pulse height, PCU ID etc. and this mode has different configurations like:
Generic event ($E\_62us\_32M\_36\_1s$ or $E\_125us\_64M\_0\_1s$), Good Xenon
(Good\_Xenon1\_2s), Transparent.
In a Binned mode, the actual photon arrival time is lost. The events are binned
with respect to time, photon energy and detector number. The Generic binned
($B\_8ms\_16A\_0\_35\_H$), Standard-1 and Standard-2 are different
configurations of Binned data mode. Data of Single-Bit mode consists of a
stream of ones and zeros representing detector
events and clock ticks (set the time resolution). The generic Single-Bit
configurations are designated with a character string beginning with SB
($SB\_500us\_50\_249\_2s$) and it has the same format as that of the Binned
mode with higher time resolution between the two events.

The data analysis is to be done for data obtained 
in 'good time' of the detector.  We need to make a good time
intervals (GTI) file. The GTI file is created with ftool task {\it maketime}.
The ftool task {\it xtefilt} is needed to create the filter file,
automatically consulting the appropriate housekeeping files. This task requires
the Observation Id no., FITS Master Index (FMI) file and a file listing the Application
IDs (appids) as input. For normal science binned data ftools task {\it saextrct}
is used to create lightcurve (.lc) and for the same purpose task {\it seextrct}
is needed for event mode data. To run these commands pca SAA information file is
needed to pass. Also for the spectral study creating spectral file (.pha), it is
needed to run task {\it saextrct} on Standard-2 data. For
the spectral study, the background file in very important. For creating the background file
from raw data task {\it runpcabackest} is required to run. For this purpose pca
background calibration file of data epoch is needed to pass.

Similarly for HEXTE data, the ftools task {\it hxtback} is needed to create the background
file. Since HEXTE instrument is designed to study the high energy data, the dead time
correction is essential. For that purpose, ftools task {\it hxtdead} is required to
run. Other commands are the same as for PCA, such as making GTI files (using {\it maketime}),
creating lightcurves (.lc) and spectral (.pha) files (using {\it saextrct}) etc.

\section{Models used in fittings of RXTE timing and spectral data}

Data reduction and analysis for the black hole candidates were carried out with the
FTOOLS version of HEADAS-6.1.1 software and XSPEC version 12.3.0. For the detailed
timing and spectral analysis we used FTOOLS inbuilt package models. In the following
sub-sections we will discuss about nature of the fitting models, which we used
for the temporal and spectral studies of black hole X-ray binaries.

\subsection{Timing analysis models}

After making the lightcurves for the source data using {\it saextrct} or {\it seextrct}
commands we load (or plot) the lightcurves into the {\it ``PLT"} (pgplot) mode.
Then we can fit them using standard FTOOLS XRONOS package models or user defined local
models. In general for the timing analysis, we fit power density spectra using {\it CONS}
(frequency independent constant factor), {\it POWR} (due to non-thermal power-law emissions),
{\it LORE} (Lorentzian line profile) and {\it GAUS} (Gaussian line profile). The best fit
results are obtained by using the least square fit technique. To find $\chi^2$ fit value,
we need to run command {\it `stat'} and to find model parameters' uncertainties, we need
to run {\it `uncer'} command (for e.g., `uncer 1' to see uncertainty for fitted
parameter 1). Also to calculate $\pm$ error values for the fitted model parameters, we use
{\it ``fit err"} task, which gives us the $90\%$ confidence range of any fitted parameter.
Now, we will discuss about the above fitted models in details.

\subsubsection{$\bullet$ CONS}

The model $``CONS"$ represents nothing but an energy independent multiplicative constant
factor, needs for improving fit statistics. It can also be defined as power-law model
(POWR) with `zero' slope value.

\subsubsection{$\bullet$ POWR}

This model represents the power-law nature of the radiating photon distribution. This
power-law nature of the emitted photons occurs mainly due to the non-thermal radiative
processes. The $``POWR"$ model uses two parameters: power-law slope index ($\alpha$) and
normalization (PN). The power-law distribution can be defined by the following simple
relation:
$$
A(\nu) = K~\nu^{-\alpha},
\eqno{(2.1)}
$$
where $\alpha$ is the dimensionless power-law slope index and $K$ is the power-law normalization
factor PN.

\subsubsection{$\bullet$ LORE}

This model indicates the Lorentzian line profile. In our analysis, we used this model to find
QPO centroid frequencies. It uses three parameters: line frequency (LC), line full width at 
half maximum (FWHM) and line normalization (LN). The $``LORE"$ model profile with frequency ($\nu$) 
can be defined as:
$$
%A(E) = K\frac{\sigma/2\pi}{[(E-E_L)^2 + (\sigma/2)^2]},
A(\nu) = \frac{LN}{(1 + [\frac{2(\nu-LC)}{LW}]^2)}.
\eqno{(2.2)}
$$
%where $E_L$ is the line frequency, $\sigma$ is the FWHM line width and $K$ is the normalization.

The sharpness of the QPO peak is calculated by the quality (Q) factor, which is defined as
$$
Q = \frac{\nu_0}{FWHM} = \frac{\nu_0}{2\Delta} = \frac{LC}{LW},
\eqno{(2.3)}
$$
where $\nu_0$ is the QPO centroid frequency and $\Delta$ is the half width at half maximum.

Also {\it rms amplitude} of the Lorentzian fit profile can be calculated by the relation:
$$
rms~amplitude = 100(\sqrt{I/{\rm mean~count~rate}}),
\eqno{(2.4)}
$$
where $I$ (=$\pi*LN*LW/2$) is the integral of the Lorentzian.

\subsubsection{$\bullet$ GAUS}

The model $``GAUS"$ represents Gaussian line profile. It has also three model parameters:
line frequency (GC), line full width (GW) and line normalization (GN). This model
profile with frequency ($\nu$) can also be defined as
$$
A(\nu) = GN~exp[-0.5(\frac{\nu-GC}{GW})^2] = GN~exp[-Z^2 /2],
\eqno{(2.5)}
$$
where $Z = (\nu-GC)/GW$. Also we can calculate the integral value of this Gaussian model
profile by the factor $\sqrt{2\pi}*GN*GW$.

\subsection{Spectral analysis models}

In the same manner we extract spectral .pha files from the source and corresponding
background ``Standard-2" mode data files using FTOOLS task {\it saextrct}. These pha
files are first rebinned using command {\it rbnpha} and then loaded in the XSPEC along
with energy response file. Now this loaded spectrum can be fitted by using default XSPEC
models or by user defined local models. In general for the spectral data fittings, we
fit energy spectra for the black hole candidates by using {\it wabs} (photo-electric
absorption due to intermediate media between source and observer), {\it diskbb}
(black body model modified for the black hole accretion disk), {\it powerlaw}
(non-thermal power-law distribution), {\it gauss} (Gaussian peak for line emissions)
and {\it compST} (Sunyaev-Titarchuk Comptonization model).
The best-fit can be obtained by using appropriate combination of models and by observing
the value of the reduced $\chi^2$, which can be defined as
$$
\chi^2 = \sum_{i=1}^N (\frac{\Delta_i}{\sigma_i})^2 = (\frac{\Delta_1}{\sigma_1})^2
+ (\frac{\Delta_2}{\sigma_2})^2 + \ldots + (\frac{\Delta_N}{\sigma_N})^2,
\eqno{(2.6)}
$$
where $\Delta$s are the difference between observed and theoretical values, $\sigma$s
are the size of the error bars. The reduced $\chi^2$ value is calculated by the relation
$\chi^2_{red}$~=~$\chi^2$/$DOF$, where Degrees of Freedom ($DOF$) = $N-P$; $N$ is the
number of data points and $P$ is the number of free model parameters required to fit the
data. For the best-fit, this $\chi^2_{red}$ value should be $\sim 1$. Sometimes to obtain
the best-fit results we need to add extra model component and that addition of extra model can
be verified via F-test by using {\it ftest} command, which requires to pass $\chi^2_{red}$
values. In the following sub-sections we will discuss about the nature of the above mentioned 
fitting models.

\subsubsection{$\bullet$ wabs}

The $``wabs"$ model, uses photo-electric absorptions using Wisconsin cross-sections (Morrison \&
McCammon 1983). Energetic radiations from the astrophysical sources get absorbed by the
intermediate media between observers and sources. This absorption mainly occurs due to
the photo-electric process. This model uses only one fitting parameter and can be defined by
the following simple relation:
$$
M(E) = exp[{-{n_H}~\sigma(E)}],
\eqno{(2.7)}
$$
where $\sigma(E)$ is the photo-electric Wisconsin cross-section (not including Thomson
Scattering) and fitted parameter ($n_H$) is equivalent to the hydrogen column density in
units of $10^{22}~atoms~cm^{-2}$.

\subsubsection{$\bullet$ diskbb}

The $``diskbb"$ model represents the multi-colour black body spectrum comes from the accretion
disk. This model is known as the disk black body model. It basically represents the modified version of
the black body model spectrum which is radiated from the black hole accretion disks. The Eqn. 1.2
of the introductory Chapter 1, is the governing equation of this model and there we also
briefly discussed about this model. This model uses two parameters: disk black body temperature
($T_{in}$) at inner disk radius in keV and disk black body normalization. This normalization
factor is the value of $(r_{in}~in~km / D~in~10kpc)^2 cos\theta$, where $r_{in}$ is the
apparent inner disk radius, $D$ is the distance of the source, and $\theta$ is the disk
inclination angle (see, Mitsuda et al. 1984, Makishima et al. 1986 and Kubota et al. 1998).
So, from the model fitting we can easily calculate the inner disk radius and temperature.

\subsubsection{$\bullet$ powerlaw}

This model represents the power-law nature of the non-thermal radiating nature of the accretion
processes. This may occur by the processes non-thermal inverse Comptonization, synchrotron or
cyclotron emissions (discussed in the introductory Chapter 1). This model uses two parameters:
power-law photon index ($\Gamma$) and power-law normalization ($K$). The power-law distribution
spectrum can be defined as:
$$
A(E) = K~E^{-\Gamma},
\eqno{(2.8)}
$$
where $\Gamma$ is the dimensionless power-law photon index
%(which is equivalent to power-law slope $\alpha$ plus 1)
and $K$ is the power-law normalization factor. This $\Gamma$ value increases as the spectrum
becomes softer. For the hard state spectra this $\Gamma$ value is less than $\sim 2$ and for
the soft state this value is above $\sim 3$.

\subsubsection{$\bullet$ gauss}

This model indicates the simple Gaussian line profile. We use this model to fit Iron emission
line $\sim 6.5~keV$. This model uses three parameters: line energy in keV, line width in keV
and normalization in unit of $photons~cm^{-2}~s^{-1}$. The $gauss$ model spectral distribution
can be defined as:
$$
A(E) = K \frac{1}{\sigma \sqrt{2\pi}} exp[-(E-E_l)^2 /2\sigma^2],
\eqno{(2.9)}
$$
where $E_l$ is the parameter 1 (line energy) in keV, $\sigma$ is the parameter 2 (line width)
in keV and $K$ is the parameter 3, normalization factor. This $K$ represents the value of the total
number of photons in $cm^{-2}~s^{-1}$ at the line.

\subsubsection{$\bullet$ compST}

This $compST$ model is the Sunyaev-Titarchuk (1980) Comptonization model. This model
indicates the presence of the Comptonized cool photon cloud on the disk. This model also
uses three parameters: Compton cloud temperature ($kT$) in keV, optical depth ($\tau$), and
normalization. The normalization represents the value of the function $Nf/4\pi d^2$, where
$N$ is the total number of photons from the source, $d$ is the distance to the source, and
$f$ is the factor $z(z+3)y^z / \Gamma (2z+4)\Gamma(z)$. In the factor $f$; $z$ is the
spectral index, $y$ is the injected photon energy in units of temperature, and $\Gamma$
is the incomplete gamma function.

	\reseteqn
	\resetsec
	\resetfig
	\resettab
\alpheqn
\resec
\refig
\retab
\def\k{{\bf k}}
\def\aug{{\tilde{\cal H}}}
\def\ahat{{\mathaccent "7E {\cal A}}}
\def\bhat{{\mathaccent "7E {\cal B}}}
\def\chat{{\mathaccent "7E {\cal F}}}
\def\dhat{{\mathaccent "7E {\cal D}}}
\def\shat{{\mathaccent "7E {\cal S}}}
\def\proj{{\cal P}}
\def\ket{\vert \vert  \{ \emptyset \} \rangle}
\def\ket2{\vert \vert \otimes \{ R \} \rangle}
\def\sqr{$^{2}$}
\def\T{$T$}
\def\P{$P$}
\def\Q{$Q$}
\def\G{$G$}
\def\M{{\bf M}}
\def\H{{\bf H}}
\def\I{{\bf I}}
\def\Pr{{\bf P}}
\def\Tr{{\bf T}}
\def\eq{\enskip =\enskip}
\def\pls{\enskip+\enskip}
\def\mns{\enskip -\enskip}
\def\oper{O^{\bf k}_{R}}
\def\aug{\tilde{\cal H}}
\def\envfn{\chi^{\alpha}_{RL}}
\def\struc{S^{\alpha}_{RL,R^{\prime}L^{\prime}}}
\def\rR{r_{R}}

\newpage
\markboth{\it Solar Science using RT-2}
{\it Solar Science using RT-2}
\chapter{Solar Science using RT-2}

RT-2 instruments onboard CORONAS-PHOTON mission satellite was launched on $30^{th}$ of 
January, 2010 into a $\sim$550~Km polar LEO orbit mainly for the detailed timing and spectral
study of the Sun in soft and hard X-ray regime. So far, RT-2 has detected a few hard X-ray solar 
flares and gamma-ray bursts. In this Chapter, we will discuss about the results obtained
due to the solar flares.

\section{Solar flares and physics behind its origin}

A flare is defined as a sudden, rapid, and intense variation of brightness in the timing
data of an astrophysical object. A solar flare occurs when the magnetic energy that has built up
in the solar atmosphere is suddenly released. Radiation is emitted across the entire 
electromagnetic spectrum, from radio waves to high energy X-rays, $\gamma$-rays.
The first solar flare recorded in astronomical literature was on September 1, 1859.
Two scientists, Richard C. Carrington and Richard Hodgson, were independently observing 
sunspots at the same time, when they were viewing a large flare in white light.
 
Solar flares are the most powerful emissions in the entire solar system and they
release energies of $10^{32}$ to $10^{33}$ ergs in $100$ to $1000$~seconds. It is
also found that during the flare, electrons are accelerated up to $10 - 100$~keV using
a significant fraction of this energy budget.

There are typically three stages of a solar flare. First is the {\it precursor} stage, where the 
release of magnetic energy is triggered. Soft X-ray emission is detected in this stage. In the 
second {\it impulsive} stage, protons and electrons are accelerated to energies exceeding 1 MeV. 
During the impulsive stage, radio waves, hard X-rays, and $\gamma$-rays are emitted. The gradual 
build up and decay of soft X-rays can be detected in the third, {\it decay} stage. The duration of 
these stages can be as short as few seconds or as long as an hour.

Solar flares extend out to the layer of the Sun called the {\it corona}. The corona is the outermost 
atmosphere of the Sun, consisting of highly rarefied gas. Inside a flare, the temperature typically 
reaches 10-20 million degrees Kelvin, and can be as high as 100 million degree Kelvin. The corona 
is visible in soft X-rays. In general corona is not uniformly bright, but concentrated around the 
solar equator in the loop-shaped features. These bright loops are located within and connect areas 
of strong magnetic filed called {\it active} regions. Sunspots are located within these active regions. 
Solar flares are also occur in these active regions.

The frequency of the flares coincides with the Sun's 11 year cycle. When the solar cycle is at a 
minimum, active regions are small and rare and few solar flares are detected. The increase in 
number occurs as the Sun approaches the maximum part of its cycle. The Sun is expected to reach maximum 
activity period in the year 2011.

\begin{figure}[h]
\centering
\includegraphics[height=2.8in,width=2.8in,angle=0]{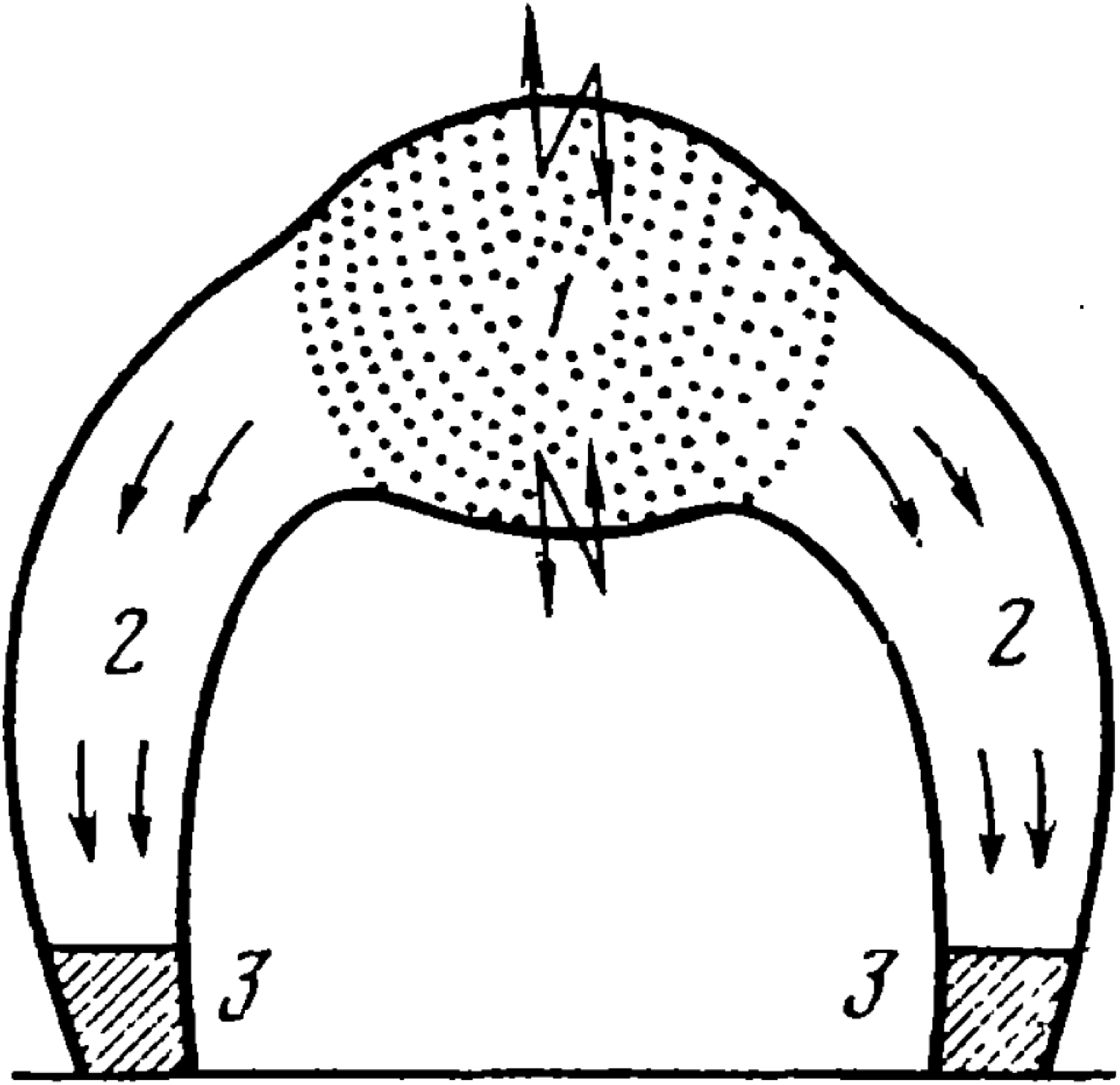}\hspace{0.1 cm}
\caption{A model for hard X-ray emissions and quasi-periodic pulsations during solar flare 
in the coronal loop. $1)$ is the region of the initial energy release, where fast 
magnetic-reconnection process occurs and a fraction of hard X-ray emission comes out, 
$2)$s are regions of modulated fast electron streams, $3)$s are the regions, from where 
maximum hard X-ray radiations are emitted (Zaitsev \& Stepanov 1982).}
\label{kn : fig3.1}
\end{figure}

In most observed solar flares, the hard X-ray radiation comes from the footpoints of the magnetic 
loops (regions marked as $3$ in Fig. 3.1) (Krucker et al. 2008, Duijveman et al. 1982), 
presumably as the the thick-target bremsstrahlung of fast electrons stopped in the cold and 
collisionally dense chromosphere (Brown 1971, Hudson 1972). Recent RHESSI observations have 
revealed a new class of events in which the hard X-ray emission comes predominantly from the 
coronal flare loop (region marked as $1$ in Fig. 3.1) itself, with little or no emission from 
the footpoints (Veroing \& Brown 2004, Sui et al. 2004, Goff et al. 2005).

According to Krucker et al. (2008), the coronal hard X-ray sources can be divided into two classes. 
The first class concentrates on properties of hard X-ray sources produced only by transport 
effects (e.g., binary collisions, magnetic mirroring etc.), leaves coronal acceleration as a 
``black box". This theory is known as the {\it classical thick-target theory} (Brown 1971, Hudson 1972). 
Alternatively, one can include self-consistent particle acceleration processes. 

X-ray spectra provide a way to distinguish the radiation emitted by hot, thermal electrons 
from that emitted by accelerated, non-thermal electrons. The shape of the X-ray spectrum emitted 
by thermal electrons is distinct from that emitted by the non-thermal electrons. The radiation 
observed at soft X-ray energies is typically from thermal plasma with a temperature on the order 
of 10 - 30 million degrees Kelvin.% ($\simeq$ 1-25 keV, since 1 eV = (1/11604) deg. Kelvin). 
Recent RHESSI observational results (Hudford et al. 2002, Battaglia \& Benz 2006) tell us that the 
spectrum consists of a combination of thermal bremsstrahlung and spectral lines (e.g., Fe, O, 
Mg etc.) from the elements in the hot plasma. The hard x-ray spectrum is dominated by bremsstrahlung 
from the accelerated electrons and, at the lower hard x-ray energies, thermal bremsstrahlung from 
plasma with a temperature above around 30 million degrees Kelvin. Also, one may observe soft-hard-soft 
spectral evolution at the impulsive phase of coronal footpoint sources (Krucker et al. 2008). 
This spectral evolution strongly suggests the presence of the acceleration mechanism rather than 
to the coronal transport.

In some rare occasions, quasi-periodic pulsations (QPPs) can be observed during the solar flare.
There are theoretical models in literature for the QPP, out of them two models are most popular:
$1)$ {\it plasma model}, in which the modulation of emitted radiation is related to periodic regimes 
of plasma instabilities, and $2)$ {\it magneto-hydrodynamic} (MHD) model, in which the emitted 
radiation is modulated by the source (magnetic tube) MHD oscillations (Zaitsev et al. 1984).

From RHESSI observations, it has been established that the particle acceleration and the
energy release process (Lin et al., 2003) are linked together. Therefore, it is a
unique opportunity to investigate the energetic solar flares with RT-2/S \& RT-2/G (Debnath
et al. 2010) mainly in the energy range $15$ to $150$~keV in Phoswich mode and $\sim 100$~keV to
$1$~MeV in spectroscopic mode during the next $24^{th}$ solar cycle.
At the same time, it is also possible to image the hard X-ray solar flares with the
RT-2/CZT payload, which has been designed to work in the energy range of $20 - 150$~keV
(Kotoch et al. 2010).  It has very good spectral as well as spatial resolutions.

\section{Solar flares observed with RT-2}

After its successful launch on January 30, 2009, RT-2 has detected several hard X-ray 
solar flares. Both RT-2/S \& RT-2/G instruments of the RT-2 Experiment, onboard 
CORONAS-PHOTON mission satellite detected the GOES C2.7 class solar flare on 5th July, 2009
(Rao et al., 2010). In this solar flare, we have observed hard X-ray {\it Quasi-Periodic
Pulsations} (QPPs) of time period $\sim 12$ seconds. We have also detected a series of
low energetic solar flares during the recent solar eruptions from $22^{nd}$ October, 2009
to $2^{nd}$ November, 2009. These flares were detected mostly in the low energy X-ray band
($<$ 25 keV). In this week, we observed two solar flares on $26^{th}$ October, 2009.
On $26^{th}$ October, at 07:52:00 UT, a B1.0 class solar flare was detected by RT-2/S
and at 22:48:00 UT, a C1.3 class solar flare was detected by the both RT-2/S \& RT-2/G detectors.
In the following sub-Sections we will discuss the results obtained from the first detected,
$5^{th}$ July, 2009 solar flare (in details) and some preliminary results of the later
(October 26, 2009) observed low energetic solar flares.

\subsection{Hard X-ray QPPs observed in $5^{th}$ July, 2009 solar flare}

The C2.7 class solar flare was detected by the both phoswich detectors (RT-2/S \&
RT-2/G) of the RT-2 experiment. The flare was observed mainly in the hard X-ray energy band
ranging from 15 - 59 keV (see Fig. 3.2a). During the flare time both of the RT-2/S
and RT-2/G detectors were in Solar Quiet Mode (frames per 100 sec (spectra per 100 sec and 
timing per sec) and detectors are in NORMAL Mode). Light curves of RT-2/S \& RT-2/G 
in the energy range below 35 keV shows multiple peaks, with some signature of periodic pulses.
These periodic nature in the lightcurves motivate us to make deeper analysis in timing as well
as spectral domain. The strongest peak count rate for RT-2/S is $\sim 325$ counts/sec and that 
of RT-2/G is $\sim 90$ counts/sec. 
We carry out the Fourier analysis of these lightcurves and found hard X-ray Quasi-Periodic Pulsations 
(QPPs) of period $12$ seconds. We augmented these results using the publicly available data 
from the $RHESSI$ satellite. Also, we made spectral analysis and measured the spectral parameters.

\begin{figure}[h]
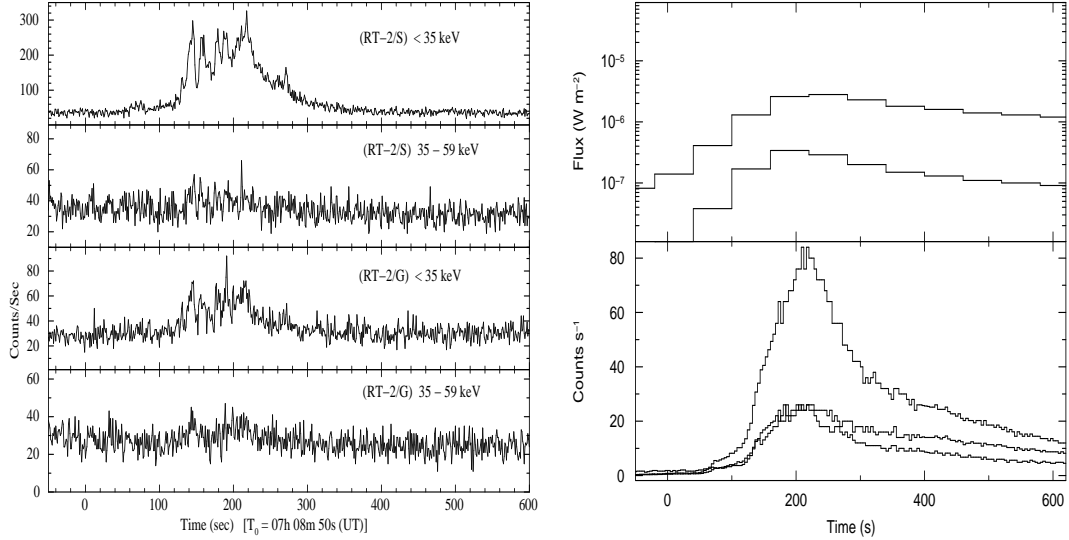

\centering
\includegraphics[height=2.8in,width=2.8in,angle=270]{fig3.2a.ps}\hspace{0.1 cm}
\includegraphics[height=2.6in,width=2.8in,angle=270]{fig3.2b.ps}
\caption{(a,b) Left panel (a) indicates RT-2/S and RT-2/G lightcurves in different
energy band (marked). In the energy range below 35 keV RT-2/S shows indication of a stronger flare
than what is observed in RT-2/G. This is because RT-2/G detector has a $2$ mm Aluminium filter above
its collimator and is effective only above $\sim$ 25 keV. The data bin size is 1 s
and T$_0$ is UT 07:08:50 on July 5, 2009.
Right panel (b) indicates GOES and RHESSI lightcurves of the flare.
Top panel: GOES lightcurves in the bands 1 - 8 $\AA$ (3.1 - 24.8 keV) and 0.5 - 4 $\AA$
(1.6 - 12.4 keV) (from top) with the time resolution of 1 minute. Bottom panel: RHESSI light 
curves in the energy bands: 6-12 keV, 3-6 keV  and 12-25 keV (from top). Bin size is 4 seconds 
and T$_0$ is UT 07:08:50 on 5th July, 2009 (Rao et al. 2010).}
\label{kn : fig3.2}
\end{figure}

The satellite was at high latitudes at the beginning of the observations and the
background rates slowly stabilized when the satellite approached towards low background
equatorial region. Since the detectors (RT-2/S \& RT-2/G) use the Phoswich technique,
the changing background has negligible impact on the $<$ 35 keV lightcurves.
For example, the background rate during the first $200$ second of observation
(in this energy band) is 36.6$\pm$0.4 s$^{-1}$ and it is 36.4$\pm$0.4 s$^{-1}$
towards the end of observation.

The RT-2/G detector has an Aluminium window to block X-rays below $\sim$25 keV (Chapter 2).
The lightcurves from RT-2/S and RT-2/G detectors in two channels are shown in
Fig. 3.2(a). The bin size is 1 second and T$_0$ is UT 07:08:50 on 2009 July 5.
Quasi-periodic pulsations are clearly seen in the lightcurve. We define the rising
phase of the flare as between 125 to 225 s (07:10:55 to 07:12:35) and the falling
phase as between 225 s to 325 s (07:12:35 to 07:14:15).

\subsubsection{GOES and RHESSI Observations }

The data from GOES and RHESSI satellites are shown in Fig. 3.2(b). According to RHESSI,
the flare is dominated mainly in the X-ray band of 12-25 keV range which correspond to channel
2 data of our RT-2/S, RT-2/G payloads. The flare started at 07:08:28 UT
and ended at 07:21:32 UT with peak at 07:12:22 UT on July 5, 2009. Total duration of the
flare is 784 sec, average counts is $\sim 84$ and sum counts is $113983$. This is the first 
major hard X-ray (above 12 keV range) observed by RHESSI in 2009 and also after the launch 
of our CORONAS-PHOTON Satellite. Fig. 3.2(b) clearly shows that both in RHESSI and GOES
hard X-ray photons dominate over soft X-ray photons. Also from the GOES lightcurves we
can classify the flare as of C2.7 class.

\subsubsection{Quasi-Periodic Pulsations}

Quasi-Periodic Pulsations (QPPs) during solar flares have been observed for many years
(Young et al. 1961) and their periodicities vary from milliseconds to several seconds.
The fast QPPs of millisecond durations are generally attributed to some wave-particle
interactions (Aschwanden 1987). On the other hand, the long period QPPs (periodicity
$>$ 10 s) observed in the microwave emission of solar flares are also seen in hard
X-rays (Nakariakov et al. 2003) and they could be resulting from some MHD oscillations
in the source region or due to modulation of electron acceleration and injection mechanisms.
Jakimiec and Tomczak (2009) have investigated QPPs in about 50 flares using $Yohkoh$ and
$BATSE$ hard X-ray data and have derived a correlation between the QPP periods
(ranging from 10 s to 150 s) and sizes of loop-top sources. They conclude that the
hard X-ray oscillations are confined to the loop-top sources and the observations
are described with a model of oscillating magnetic traps. Fleishman et al. (2008)
have made a detailed analysis of the 2003 June 15 solar flare (GOES X1.3 class)
and detected hard X-ray (based on $RHESSI$ data) and microwave oscillations
with periods ranging from 10s to 20s. They, however, conclude that QPPs are
associated with quasi-periodic acceleration and injection of electrons.

Since the basic cause of QPPs have implications for particle acceleration mechanism,
it is important to investigate QPPs at diverse source intensities. we observe QPP in 
the C2.7 solar flare, detected by RT-2 on 2009 July 5 (Rao et al. 2010). In the next 
sub-sections spectral and temporal characteristics of the flare will be discussed.

\subsubsection{Timing Analysis}

In the lightcurves of RT-2/S and RT-2/G (Fig. 3.2a), we observe multiple peaks, which
motivate us to make Fourier transformed power density spectra and find periodicities. 
To find the modulation power and the periodicities, we follow the method used in Fleishman 
et al. (2008). If $C(t)$ is the count rate at time $t$, the normalized modulation is
$$
S(t) = \frac {C(t) - <C(t)>}{<C(t)>}
% F_m(\frac {t}{t_m})^r[\frac {d}{d+r} + \frac {r}{d+r}(\frac {t}{t_m})^{(r+1)}]^{-(r+d)/(r+1)},
\eqno (2.1)
$$
where $<C(t)>$ is the running average taken over a number of bins, which is 20 s in our case. 
The modulation power over a period of time is equal to the average of S$^2$(t) and the square root 
of modulation power is the modulation amplitude (see, Fleishman et al. 2008).

\begin{figure}[h]
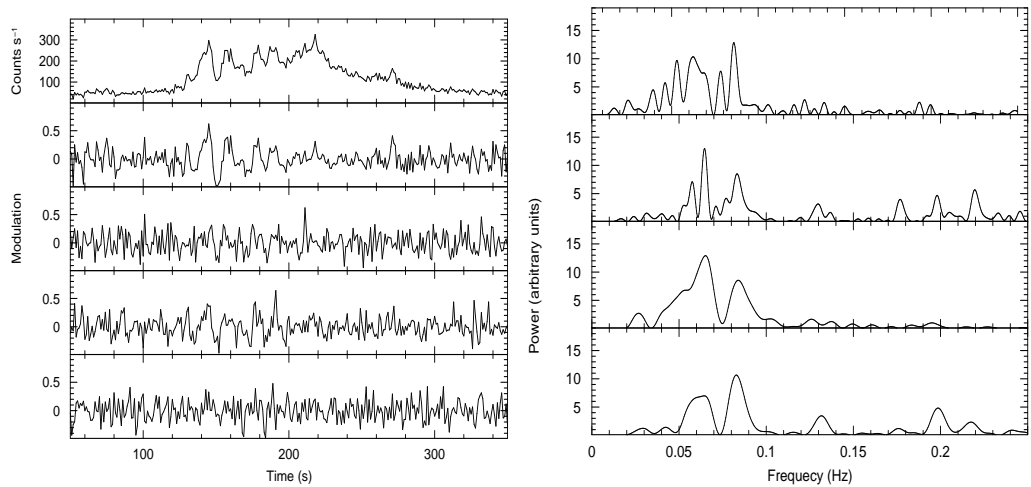

\centering
\includegraphics[height=2.6in,width=2.5in,angle=270]{fig3.3a.ps}\hspace{0.1 cm}
\includegraphics[height=2.6in,width=2.5in,angle=270]{fig3.3b.ps}
\caption{(a,b) Left panel (a) shows RT-2/S low energy lightcurves with normalized modulation
in the four energy bands, shown in Fig. 3.2(a) and right panel (b) shows the power spectra
of these lightcurves (Rao et al. 2010).}
\label{kn : fig3.3}
\end{figure}

In  Fig. 3.3(a), we plotted the normalized modulation along with the count rates. 
The top panel shows the count rates in the RT-2/S 20 -- 35 keV range and the successive panels 
downwards show the normalized modulation for RT-2/S 20 - 35 keV, 35 - 59 keV and similarly for 
RT-2/G, respectively. To estimate the errors in the modulation power, we calculated the
modulation power in the background regions for a period of $100$ sec, and the RMS deviation 
is deemed as the error in the measured values. The background subtracted modulation 
powers, for 4 second integration time, are given in Tab. 3.1, for the rising and falling 
phase of the flare, respectively.

In Fig. 3.3(b), we showed the Fourier transform of the normalized modulation,
for RT-2/S 20 - 35 keV total lightcurve (top panel) and rising phase (35 - 59 keV)
of the flare (second panel from top). The third and fourth panels show similar
power spectra for the RT-2/G lightcurves. The periods of the highest peaks in the power
spectra are also shown in Tab. 3.1.

\begin{table}[h]
\small
\centering
\caption{\label{table3.2} Modulation amplitudes and QPP Periods of the solar flare}
\vskip 0.2cm
\begin{tabular}{|l|c|c|c|c|}
\hline
Observation & \multicolumn{2}{|c|}{Rising Phase} & \multicolumn{2}{|c|}{Falling Phase} \\
\cline{2-5}
 & Amp. (\%)& Period (sec) & Amp. (\%)& Period (sec) \\
\hline
RT-2/S 20 -- 35 keV & 13.5$\pm$0.4 & 15.3$\pm$0.1 & 5.2$\pm$0.4 & 12.2$\pm$0.2  \\
 (2$^{nd}$ peak)    & - -          & 11.9$\pm$0.2 & - -         & - - \\
\hline
RT-2/S 35 -- 59 keV & 4.6$\pm$2.8  & - -          & $<$ 5.6     & - - \\ 
\hline
RT-2/G 25 -- 35 keV & 6.6$\pm$0.4  & 12.1$\pm$0.2 & $<$ 0.8     & 15.6$\pm$0.3 \\
(2$^{nd}$ peak)     & - -          & 15.5$\pm$0.3 & - -         & - - \\
\hline
\end{tabular}
\end{table}

\subsubsection{Spectral Analysis}

Solar hard X-ray spectrum is of isothermal and power-law type, can be fitted with the combination 
of isothermal and double-power-law (or broken power-law) models  (Lin et al. 2003, Krucker et al. 
2008). At the same time, the  black hole X-ray spectrum is a combination of black body and power-law 
types (This will be discussed in the following Chapters in detail.). In this Section, we will 
discuss the model fitted spectrum of the 5th July, 2009 solar flare.

Before going to fit RT-2/S \& G spectra, we generated appropriate response matrices for 
the detectors which were generated using FTOOLS `genrsp' task. For the channel - energy 
calibration we used the background line at 58 keV (due to $^{121}$I decay) of the on-board 
calibration source $^{57}$Co. The values of energy resolution function measured during the 
ground calibration and the effective areas from the known geometrical properties of the detectors 
are used for response matrix generation. Background spectrum obtained away from the solar flare
is used. The XSPEC tool of the $ftools$ package is used for spectral fitting. The
deconvolved 20 - 35 keV spectra are shown in Fig. 3.4 (a) for RT-2/S (filled circles) and
RT-2/G (open circles). The spectrum is very steep and it is best fitted by a simple
bremsstrahlung function of energy 3.43$\pm$0.30 keV. This model is shown as a dashed line 
in the Figure.

\begin{figure}[h]
\centering
\includegraphics[height=2.52in,width=2.55in,angle=000]{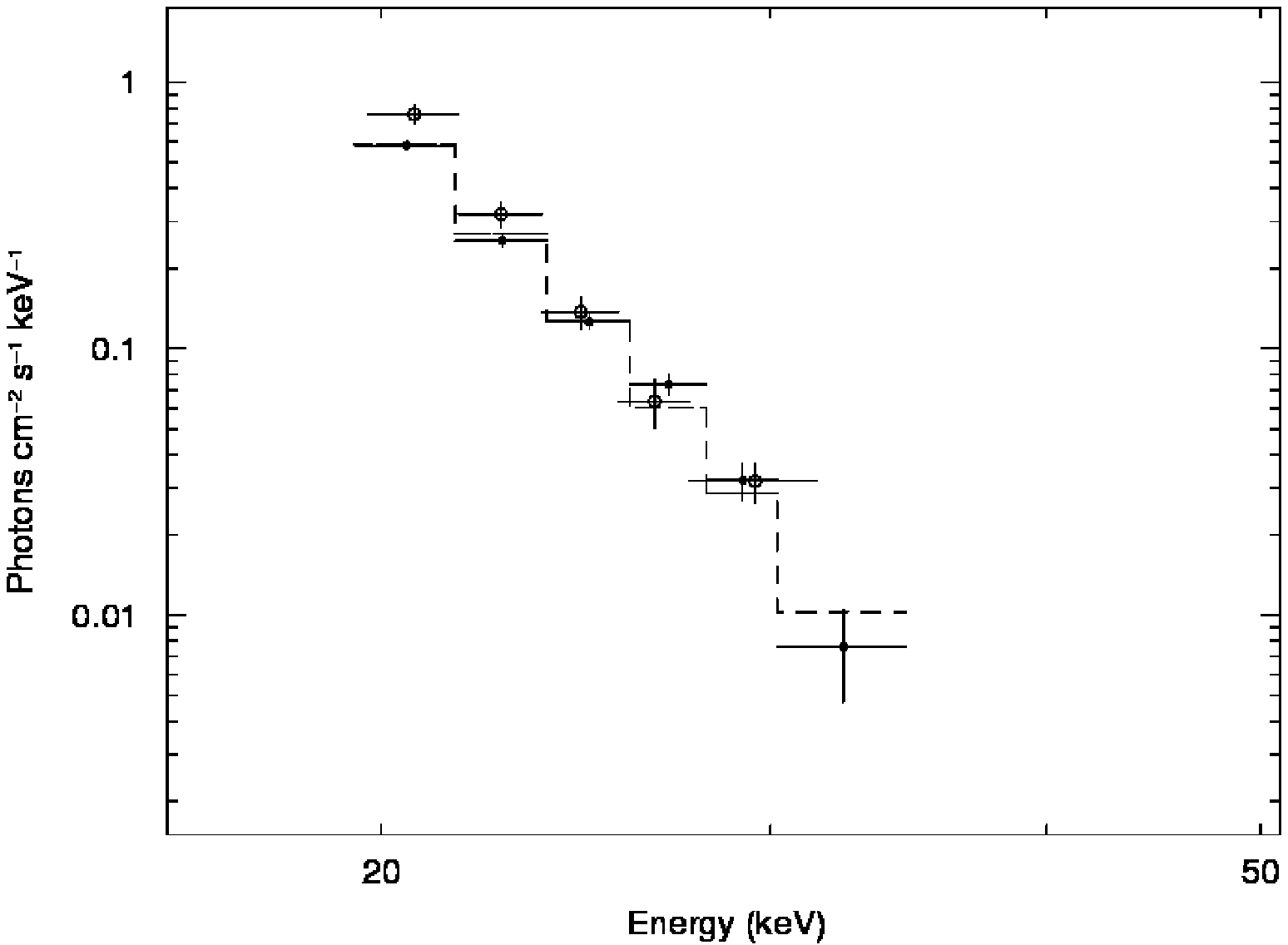}\hspace{0.1 cm}
\includegraphics[height=2.52in,width=2.45in,angle=000]{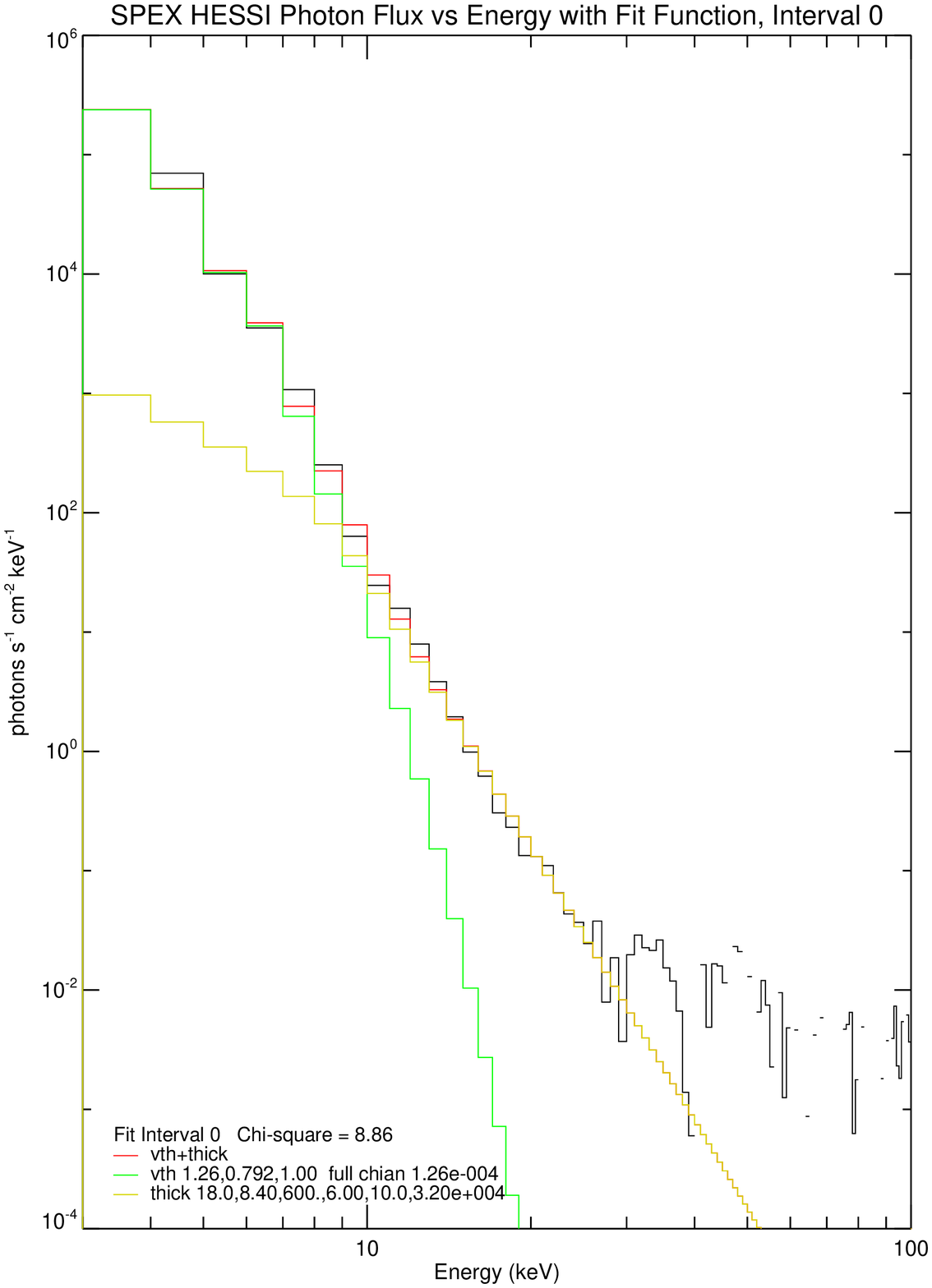}
\caption{(a-b) Left panel shows (a) the deconvolved spectra from RT-2/S (filled circles) and RT-2/G
(open circles) along with a simple bremsstrahlung spectrum (dashed line)
and right panel shows (b) the RHESSI photon energy spectra during the rising phase of the flare
(07:11:40 to 07:12:30). The two component thin and thick target bremsstrahlung model
is also shown in the Figure (Rao et al. 2010).}
\label{kn : fig3.4}
\end{figure}

We also made spectral study of the flare using RHESSI archival data. Fig. 3.4(b) shows
the background subtracted and spatially integrated 3 - 100 keV spectra from RHESSI
observations for the rising phase and the falling phase of the flare. We used
the standard RHESSI software of Solar SoftWare (SSW) for the creation of spectrum and
the RHESSI OSPEX package is used for the spectral fitting of the count spectra.

The RHESSI spectrum was fitted with a two component power-law model, consisting of 
an optically thin thermal bremsstrahlung radiation function, parametrized by the plasma
temperature $kT$ and the emission measure (EM) and a thick target bremsstrahlung, 
characterized by the electron flux and the power law index ($\Gamma$) of the electron
distribution function below the break energy.
It can be seen that the 20 - 30 keV RHESSI spectrum agrees quite closely with the
RT-2 data.

\subsection{Low energy solar flare detected on $26^{th}$ October, 2009}

A series of low energy solar flares was detected by the RT-2 instruments during the
recent solar eruptions from $22^{nd}$ October, 2009 to $2^{nd}$ November, 2009. These
observed flares are mostly in the low energy X-ray band ($<$ 25 keV). During this week,
RT-2 has detected two solar flares on $26^{th}$ October, 2009.

(a) A very weak flare of B1.0 class was detected by RT-2/S detector at 07:52:00 UT.
The flare was so weak that it was detected only by the channels 1 \& 2 of the RT-2/S
detector (see Fig. 3.5). RT-2/G signature is very weak (signal to noise ratio $\sim$ 1).
This is because RT-2/G detector has an extra $2~mm$ Aluminium sheet (for low-energy cutoff)
above its collimator (Chapter 2). During the flare time, both of our RT-2 phoswich instruments
(RT-2/S \& RT-2/G) were at Solar Quiet Mode. So, we got a highest time resolution of 1 sec.

\begin{figure}[h]
\centering
\includegraphics[height=2.5in,width=3.0in,angle=270]{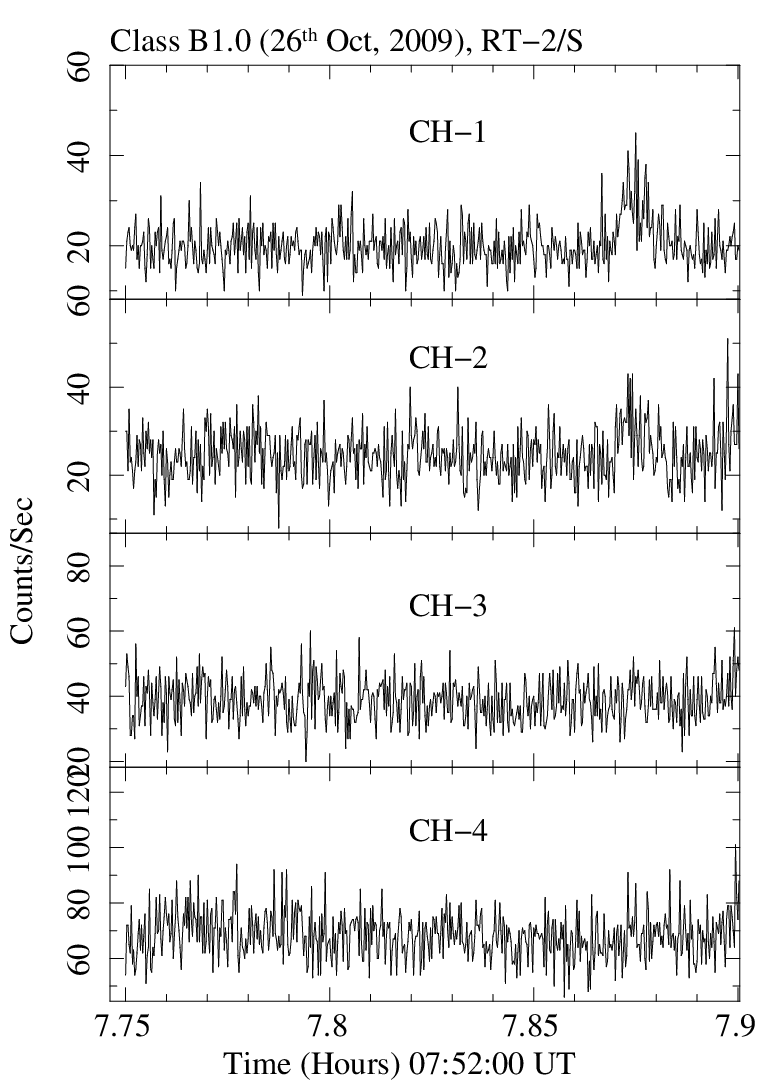}\hspace{0.1 cm}
\includegraphics[height=2.5in,width=3.0in,angle=270]{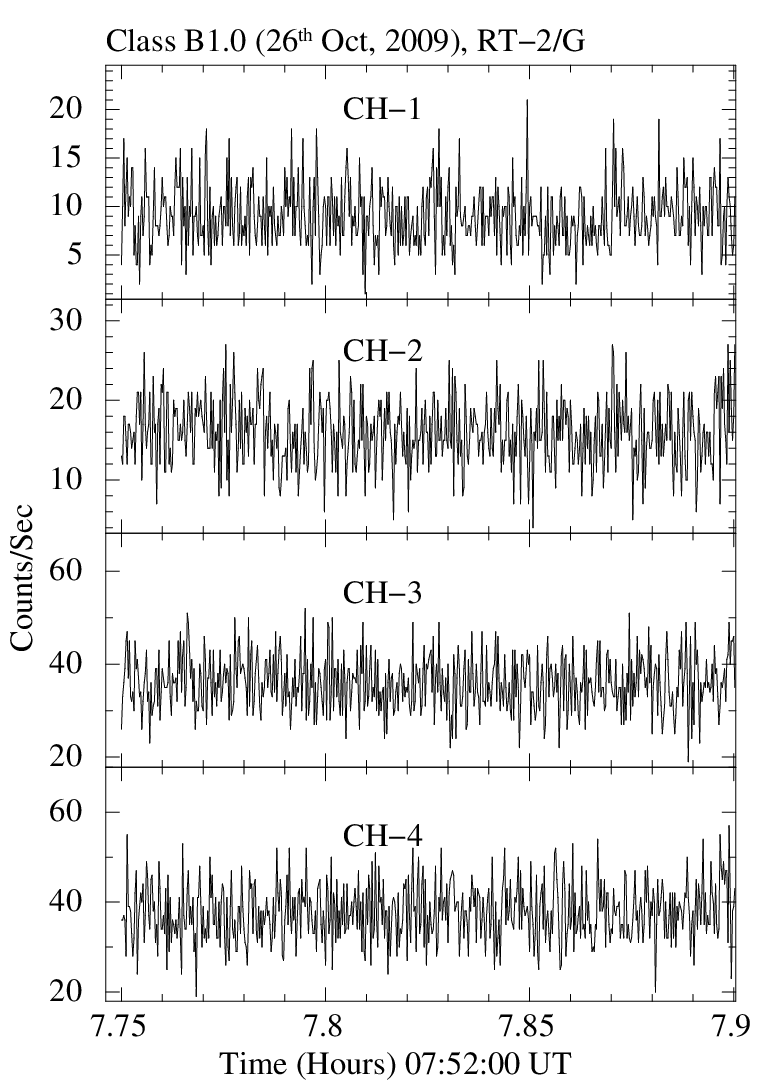}
\caption{(a-b) Left panel shows (a) the RT-2/S lightcurves for channels 1-4 and right panel
shows (b) the RT-2/G lightcurves for channels 1-4 of the B1.0 class solar flare, was detected
at 07:52:00 UT on $26^{th}$ October, 2009. Channels 1 \& 2 of RT-2/S show the clear detection
of the flare.}
\label{kn : fig3.5}
\end{figure}

(b) A flare of C1.3 class was also detected by the both phoswich detectors of the
RT-2 experiment (RT-2/S and RT-2/G) at 22:48:00 UT on $26^{th}$ October, 2009.
At the low energy bands (below 25 keV), flare was detected by the energy channels 1 \& 2
of the RT-2/S detector and by the channel 1 \& 5 of the RT-2/G detector (see Fig. 3.6).
Here also during the flare time both the instruments were at Solar Quiet Mode.

\begin{figure}[h]
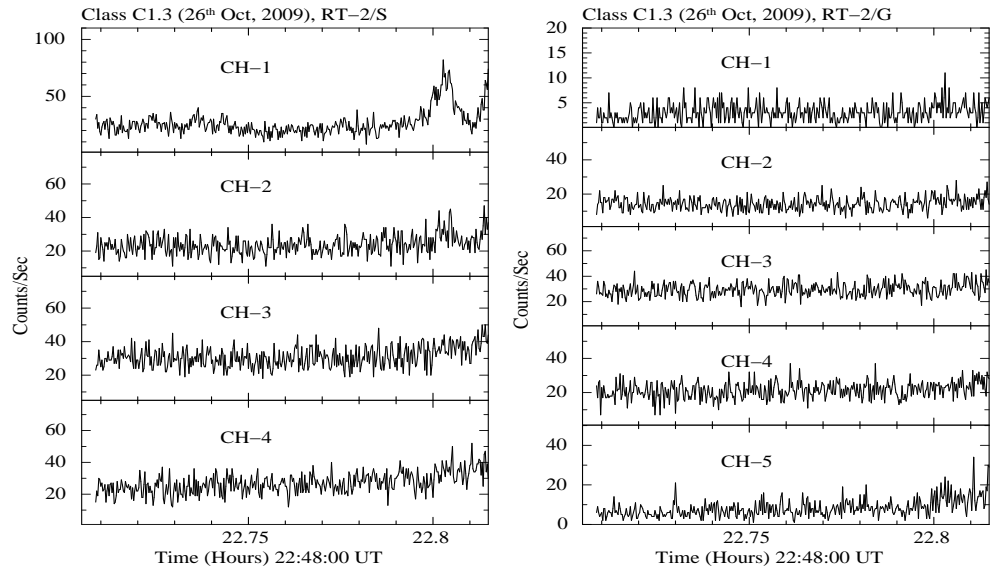

\centering
\includegraphics[height=2.5in,width=3.0in,angle=270]{fig3.6a.ps}\hspace{0.1 cm}
\includegraphics[height=2.5in,width=3.0in,angle=270]{fig3.6b.ps}
\caption{(a-b) Left panel shows (a) the RT-2/S lightcurves for channels 1-4 and right panel
shows (b) the RT-2/G lightcurves for channels 1-4 of the C1.3 class solar flare, was detected
at 22:48:00 UT on $26^{th}$ October, 2009. Channels 1 \& 2 of RT-2/S and Channel 1 \& 5 of RT-2/G
show the clear detection of the flare.}
\label{kn : fig3.6}
\end{figure}

Detailed analysis of the above mentioned solar flares and other flares are continuing.
The detailed timing and spectral analysis results will be published in near future 
(Debnath et al. 2010d).

	\reseteqn
	\resetsec
	\resetfig
	\resettab
\alpheqn
\resec
\refig
\retab
\def\k{{\bf k}}
\def\aug{{\tilde{\cal H}}}
\def\ahat{{\mathaccent "7E {\cal A}}}
\def\bhat{{\mathaccent "7E {\cal B}}}
\def\chat{{\mathaccent "7E {\cal F}}}
\def\dhat{{\mathaccent "7E {\cal D}}}
\def\shat{{\mathaccent "7E {\cal S}}}
\def\proj{{\cal P}}
\def\ket{\vert \vert  \{ \emptyset \} \rangle}
\def\ket2{\vert \vert \otimes \{ R \} \rangle}
\def\sqr{$^{2}$}
\def\T{$T$}
\def\P{$P$}
\def\Q{$Q$}
\def\G{$G$}
\def\M{{\bf M}}
\def\H{{\bf H}}
\def\I{{\bf I}}
\def\Pr{{\bf P}}
\def\Tr{{\bf T}}
\def\eq{\enskip =\enskip}
\def\pls{\enskip+\enskip}
\def\mns{\enskip -\enskip}
\def\oper{O^{\bf k}_{R}}
\def\aug{\tilde{\cal H}}
\def\envfn{\chi^{\alpha}_{RL}}
\def\struc{S^{\alpha}_{RL,R^{\prime}L^{\prime}}}
\def\rR{r_{R}}

\newpage
\markboth{\it GRO J1655-40: An Outbursting Galactic Black Hole Candidate}
{\it GRO J1655-40: An Outbursting Galactic Black Hole Candidate}
\chapter{GRO J1655-40: An Outbursting Galactic Black Hole Candidate}

The Galactic black hole candidates are the most fascinating objects to study 
in X-rays, as these sources undergo peculiar timing and spectral changes 
during their transient as well as the persistent phases. 
GRO J1655-40 is a well studied Galactic black hole candidate. 
The soft X-ray transient GRO J1655-40 was first observed by BATSE 
onboard CGRO on 27th July, 1994 (Zhang et al. 1994). This source was extensively observed 
with RXTE during its outburst of 1996-1997 and also in 2005 (Fig. 4.1). In 1996-97 outburst,  
it showed a very complex timing and spectral behaviour and was X-ray active at least 
for $16$ months. We made a detailed timing and spectral studies of the full 2005 
outburst using publicly available RXTE archival data (Debnath et al. 2008a, 
Chakrabarti et al. 2005, 2006a, 2008a). These analysis results will be discussed in this 
Chapter.

GRO J1655-40, an enigmatic Low Mass X-ray Binary (LMXB) system is located at $(l,b) = 
(344.98^\circ,2.45^\circ)$ (Bailyn et. al. 1995) with R.A.=$16^h54^m00^s$ and 
Dec.= $-39^\circ50^m 45^s$. Its mass (M = $7.02\pm0.22~M_\odot$, Orosz \& Bailyn 1997)
distance (D = $3.2\pm0.2$~kpc, Hjellming \& Rupen 1995), and inclination angle 
($\theta = 69.5^\circ\pm0.1^\circ$, Orosz \& Bailyn 1997) are well determined.
The mass of its companion star is = $2.3~M_\odot$ (Bailyn et. al. 1995). GRO J1655-40
may also showed signatures of the ejection of the superluminal radio jet 
(Tingay et. al. 1995, Hjellming \& Rupen 1995). The maximum speed of the jet was 
found to be $\sim 0.37 c$. Recent VLT-UVES spectroscopic observations suggest that 
the distance to the source is $\le 1.7$ kpc (Foellmi et. al., 2006) with a secondary 
star of spectral type of F6IV, making it one of the closest known black hole candidates.

\section{Major results obtained from the past outbursts}

The transient black hole candidate GRO J1655-40 showed two well studied X-ray outbursts in 1996-97 
and 2005. These two large outbursts were quiet extensively observed by RXTE satellite.
In the following sub-sections, we will briefly discuss the major results obtained so far 
from the above two outbursts. In particular, we studied the detailed timing and spectral properties of the 
2005 outburst.

\subsection{1996-97 GRO J1655-40 outburst}

The transient Galactic black hole GRO J1655-40 showed a major outburst in the years 1996-97 
for a total duration of $\sim$~16 months (see, Fig. 4.1A). The ASM lightcurve of the 
outburst shows a double peak profile and it is quite different from other black hole 
candidates. During the first peak in May, 1996, the source showed a strong flaring activity with 
non-thermal emission, whereas during the second peak in August, 1997, the source 
spectrum was softer and thermal, except near the end of the outburst
when its spectrum was hard (Sobczak et. al. 1999). At least three distinct spectral states,
namely, very high state, high/soft state and low/hard state (Sobczak et al. 1999)
have been reported. The luminosity variation of the outburst was of fast rise and exponential
decay (Chen et. al. 1997). Investigation of X-ray timing properties of GRO J1655-40 during
the 1996-97 outburst revealed QPOs varying from $0.1$ Hz to $300$ Hz (Remillard et. al. 1999).
Two very important discoveries  were found there: one is the superluminal radio jet
(Tingay et. al. 1995, Hjellming \& Rupen 1995) and the other is the existence of very high
QPO frequencies ($300$ \& $450$ Hz) (Remillard et. al. 1999, Strohmayer 2001).

\begin{figure}[h]
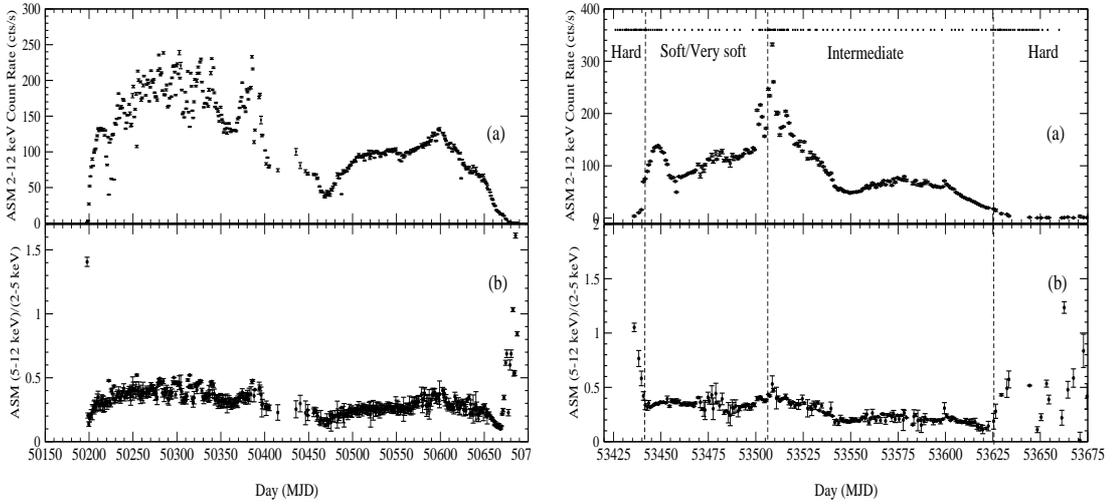

\vskip 0.7cm
\centering
\includegraphics[height=2.6in,width=2.8in,angle=000]{fig4.1a.eps}\hspace{0.1 cm}
\includegraphics[height=2.6in,width=2.8in,angle=000]{fig4.1b.eps}
\caption{(A,B) Left panel (A) shows the 1996-97 GRO J1655-40 outburst and right panel (B) shows 
the 2005 outburst. In both the Figures, (a) 2-12 keV ASM lightcurve and (b) hardness ratio 
(5-12 keV vs. 2-5 keV count ratio) as a function of the Modified Julian Day (MJD) of the event
are plotted. In the right panel, the vertical dashed lines indicate the spectral transition of 
states and dotted horizontal points are our analyse days (Fig.B: Debnath et al. 2008a).}
\label{kn : fig4.1}
\end{figure}

\subsection{2005 GRO J1655-40 outburst}

After remaining `dormant' for almost eight years, GRO J1655-40 showed a renewed X-ray 
activity in the late February 2005 (Markwardt \& Swank 2005, Chakrabarti, Nandi, Debnath
et al. 2005, Shaposhnikov et al. 2007, Chakrabarti, Debnath, Nandi \& Pal 2008a, Debnath,
Chakrabarti, Nandi \& Mandal 2008a, Debnath et al. 2008c).
The source remained active in X-rays for the next $260$ days and during this period
it was extensively observed with the RXTE satellite (see Fig. 4.1B).

In our communication (Debnath et al. 2008a), we presented a detailed analysis of the
RXTE (ASM and PCA) archival data of 122 days which spread over the full period of the
outburst and present the results for both the timing \& the spectral properties of
GRO J1655-40 during this outburst phase, which reveal several very important aspects of the 
nature of the transient accretion process around a black hole. 

From the lightcurves, hardness/softness diagrams, spectral slopes and most importantly 
the variation of the QPO frequency, one can come up with a very comprehensive picture of 
what might be happening when such an outburst takes place. In the entire outburst phase, 
we identified four spectral states characterized by the presence or absence of a soft 
multi-color disk black body component at low energy and the power-law component at 
higher energies above $\sim 10$ keV. The four identified states are termed as the 
{\it hard}, {\it soft}, {\it very-soft} and {\it intermediate states}. During the 
total outburst we observed the transitions in the sequence: {\it hard $\rightarrow$ 
soft/very soft $\rightarrow$ intermediate $\rightarrow$ hard}. In each of these 
spectral states, we carried out detailed timing and spectral analysis and find QPO frequencies. 

In the previous communications (Chakrabarti et al. 2005, Chakrabarti et al. 2008a), the evolution 
of the QPO frequencies with time was shown in the initial and final outburst stages. The rapid 
variation in QPO frequencies was explained by using an oscillating and propagating shock.

Shaposnikov et. al. (2007) also carried out a multi-wavelength study for the early stage 
(beginning with 21st of February, 2005) of the outburst of GRO J1655-40 for a total of 
25 days of data using instruments like RXTE \& INTEGRAL for X-rays, VLA for radio study 
and ROSTE \& SMARTS for optical region. On the basis of their multi-wavelength campaign 
they classified the spectral states of the observed period in four spectral states, namely, 
low-hard, hard intermediate, soft intermediate, high-soft. 
After correlating X-ray and radio fluxes they concluded that the physical origins of the radio
emission and the X-ray emission are not the same. The evidence of a closer coupling
between the power-law component and QPO as also observed by Vignarca et. al. (2003)
is totally consistent with the shock propagation model of Chakrabarti et al. (2005, 2008a)
as the shock does not propagate in the {\it disk} as they mentioned, but through the 
sub-Keplerian flow which surrounds the disk (e.g., Chakrabarti \& Titarchuk, 1995).

Our study, on the other hand, covers 122 days of the observational data spreading over 
the full period of the outburst. On the basis of the results of RXTE data, we classified 
the total outburst in a slightly different way with four distinct spectral states.
Furthermore, we thoroughly studied the QPO behaviour. We got QPOs in a total of $67$ 
observations out of a total of $150$ observations. We also studied the photon count
variation in different energy bands for different spectral states via hardness and
softness intensity diagrams. We identify the energy band in which QPOs are predominantly 
seen. We show spectral components and their flux variations. We claim that two components 
of the flow, namely, the Keplerian (disk) and the sub-Keplerian (halo) are necessary to 
explain the mass accretion dynamics. We theoretically estimate the disk and the halo rates 
from spectral fits of several observations.

In the following Sections, we will discuss the major results obtained from our study
in the both timing and spectral domains.

\section{Timing Analysis}

Timing analysis of 2005 GRO J16555-40 outburst depends on the data analysis results on the 
publicly available observational data from the RXTE instruments, the All Sky Monitor (ASM) 
and the Proportional Counter Array (PCA) covering the entire eight months of the outburst.
Our analysis covers from the 25$^{th}$ of February, 2005 (Modified Julian Day (MJD) = 53426) 
to 16$^{th}$ of October, 2005 (MJD = 53659). The ASM data has four energy bands corresponding 
to $2-3$ keV, $3-5$ keV, $5-12$ keV and $2-12$ keV. PCA contains five proportional counter 
units (PCUs 0-4). We used only PCU 2 data for both the timing and spectral analysis due to 
its reliability and it is on for $\sim 100\%$ of the good time. Data reduction and analysis were 
carried out with the FTOOLS version of HEADAS-6.1.1 software and XSPEC version 12.3.0.

We used the PCU2 data from the Event mode ($E\_125us\_64M\_0\_1s$) and Science Array mode
($B\_8ms\_16A\_0\_35\_H$) data for the timing analysis. Our timing analysis is mainly to
study the lightcurves with hardness and softness variations and the power density spectra 
(PDS) of each data. Out of these observations, we find QPOs in a total of $67$ observations 
made in $43$ days. Timing analysis results are discussed in the following sub-sections.

\subsection{ASM Light Curve}

We extracted and analyzed the ASM (Levine et al. 1998) data of different energy bands for
the entire observations. In Fig. 4.1(B), the total 2-12 keV ASM lightcurve (counts/sec) and 
the ASM hardness ratio (ratio of the photon count rates in 5-12 keV  and 2-5 keV bands) are
plotted. The origin of the time axis is MJD 53420 ($19^{th}$ February, 2005), which is six 
days before the initial rise of the X-ray intensity. The hardness ratio variation distinctly 
reflects the state transitions. The hard to soft transition takes place on the $13^{th}$ of 
March, 2005 (MJD = 53442), the soft to intermediate transition on the $16^{th}$ of May, 2005 
(MJD = 53506), and the intermediate to hard transition takes place on the $12^{th}$ of 
September, 2005 (MJD = 53625). These are marked on the plot, using vertical dashed lines. 
However, the local changes in the spectral features of different states are not evident 
from this plot. This leads us to conduct a robust spectral analysis using the PCA data and 
the results are discussed in the next sub-sections.

\subsection{Hardness-Intensity Diagram}

We extracted and analyzed the PCA (Jahoda et al., 1996) data of the full outburst from 
February 25, 2005 (MJD = 53426) to October 16, 2005 (MJD = 53659).
Figure 4.2(a-b), shows RXTE ASM and PCA hardness-intensity diagrams (HIDs) respectively. 
In Fig. 4.2(a), the ASM 2-12 keV count rates for the full period of 2005 GRO J1655-40 outburst 
are plotted against hardness ratio of 5-12 keV and 2-5 keV ASM photon counts. In the same 
way, in Fig. 4.2(b), the PCA 3-20 keV count rate of the 2005 outburst against X-ray color 
(PCA count ratio between 6-20 keV and 3-6 keV energy bands) are plotted. It is evident that 
the pre and post-outburst phases tend to appear and disappear from the low count region having 
harder spectrum. In this GRO J1655-40 outburst, the transition occurs from the spectral 
states {\it hard $\rightarrow$ soft (very soft) $\rightarrow$ intermediate $\rightarrow$ hard}. 
It is observed that the rapid changes in the hardness ratio occurs only in the hard states, 
whereas in the soft and intermediate states the hardness ratio changes very slowly. 
Both rising \& falling arms of the diagram corresponds to the hard state. In both the 
cases, we found the presence of QPOs.

\begin{figure}[h]
\centering
\includegraphics[height=2.5in,width=2.5in,angle=270]{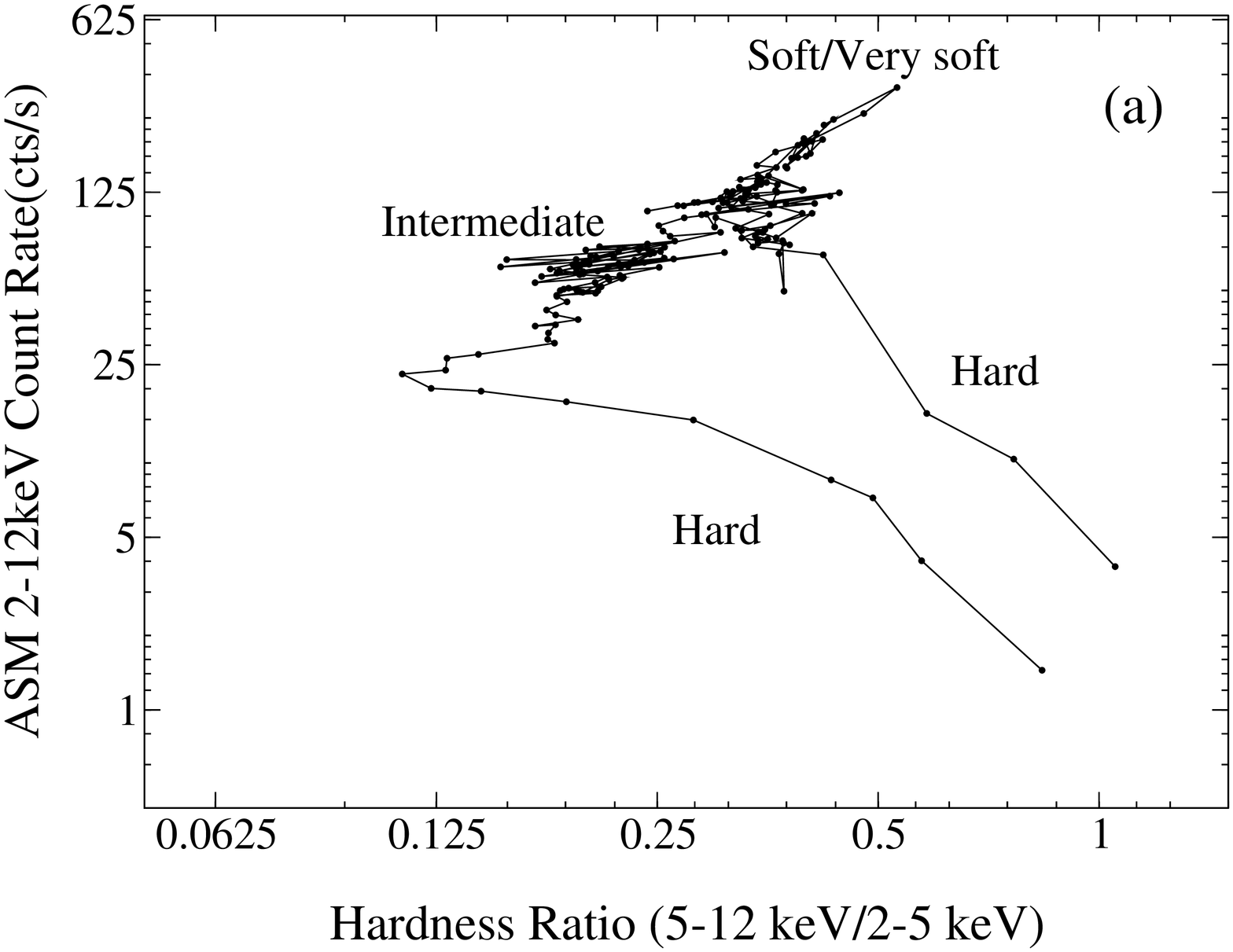}
\includegraphics[height=2.5in,width=2.5in,angle=270]{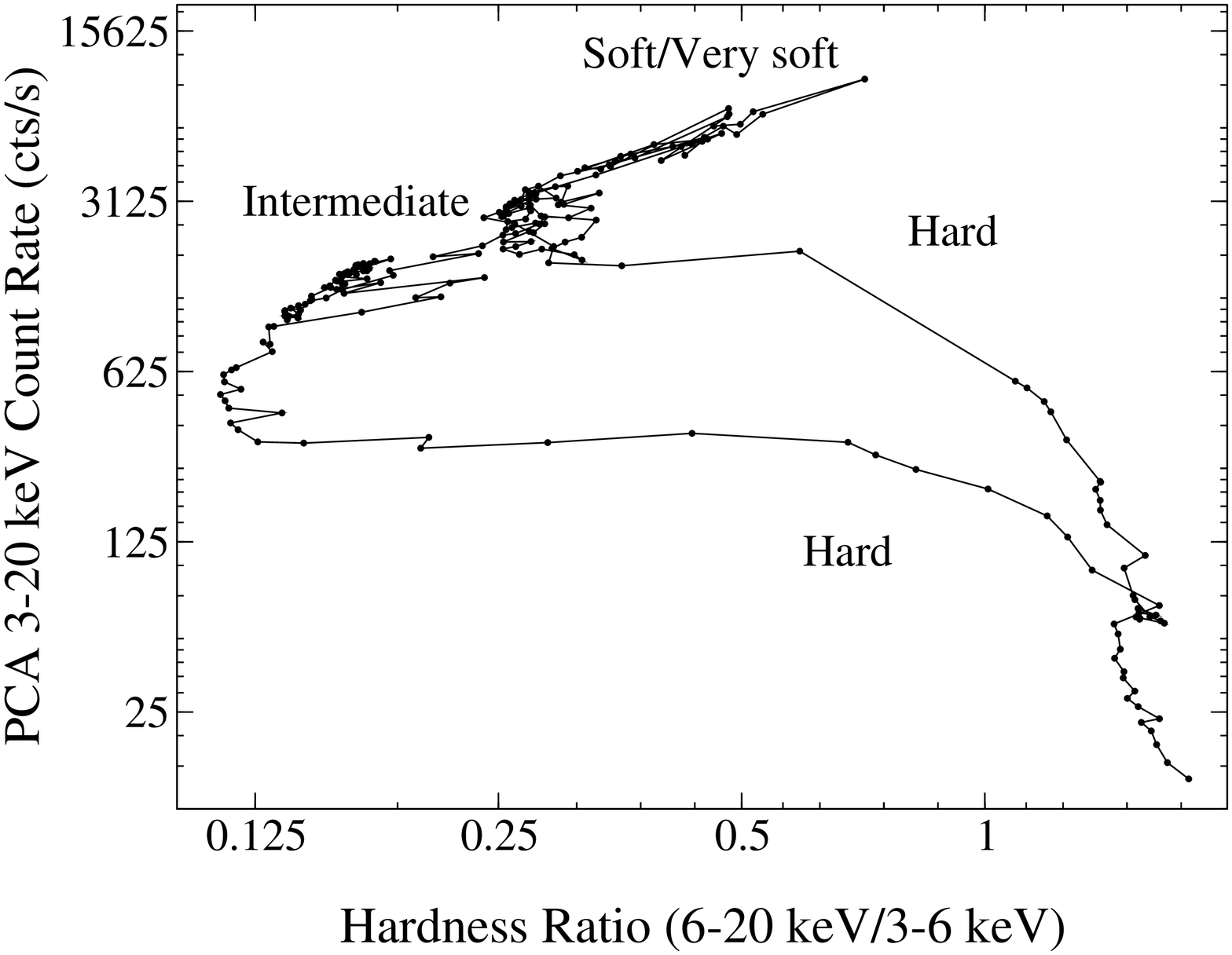}
\caption{(a-b) The Hardness Intensity Diagrams (HIDs) observed with RXTE ASM \& PCA. For ASM HID,
count rates are plotted (in y-axis) in the energy range of 2-12 keV and in the x-axis hardness 
ratio of the photon counts between 5-12 and 2-5 keV are plotted. Similarly in the PCA HID, the 
y-axis represents the PCA count rates in 3-20 keV energy band and the x-axis represents the 
hardness ratio between the count rates of the energy bands of 6-20 keV and 3-6 keV bands, are 
plotted (Fig.b: Debnath et al. 2008a).}
\label{kn : fig4.2}
\end{figure}

In the first phase of the hard state from the 25$^{th}$ of February, 2005 (MJD = 53426) to
the 12$^{th}$ of March, 2005 (MJD = 53441), we found QPOs from $34$ mHz to $17.78$ Hz.
The observed QPO frequencies were found to be increased monotonically with time
(day) from 0.082 Hz to 17.78 Hz (on the first day another QPO at $34$ mHz was also seen).
This QPO frequency rise was observed up to $\sim 15$ days from the initial observed day. 
This was because, after 12$^{th}$ of March 2005, the Keplerian component became dominant over 
the sub-Keplerian component and could been able to cool down the disk, which in fact 
stopped the shock oscillations and observed QPO signatures. So, we observed hard to soft 
state transition on 13$^{th}$ March, 2005 (MJD = 53442). The soft state started 
from the March 13, 2005 and continued till 15$^{th}$ of May, 2005 (MJD = 53505). 
In this region no QPO was observed. The intermediate state was seen from the 16$^{th}$ of 
May, 2005 (MJD = 53506) to 11$^{th}$ of September, 2005 (MJD = 53624). Interestingly, 
in this state, we observed QPOs only for $8$ days, from 16$^{th}$ of May, 2005 (MJD = 53506) 
to 20$^{th}$ of May, 2005 (MJD = 53510) and from 25$^{th}$ of May, 2005 (MJD = 53515) to 
27$^{th}$ of May, 2005 (MJD = 53517). In between, for four days we observed no signature 
of QPOs. The QPO frequencies varied were from $13.17$ Hz to $19.04$ Hz. In the PDS, we also 
found one broad QPO bump at frequencies near $7$ Hz. The final hard state observed was from 
the 12$^{th}$ September, 2005 (MJD = 53625) to October 16, 2005 (MJD = 53659). 
The QPOs of $0.023$ Hz to $20.20$ Hz were observed in this state. If we follow 
one of the QPO frequencies (of sharp peak and high ``Q" value), we found it to decrease 
monotonically from $13.14$ Hz to $0.034$ Hz within $20$ days.  

\subsection{Light curves and Hardness, Softness diagrams}

For the detailed study of the photon count variations over the outburst, we extracted 
2 - 15 keV (0-35 Channels) PCA lightcurves with a time bin of 1 sec also studied both 
the hardness and softness ratio variations. Also for the power density spectra (PDS), we 
extracted PCA lightcurves with 0.01 sec time bins. These lightcurves were extracted 
using FTOOLS task ``saextrct" on Science data and tasks ``sefilter" \& ``seextrct" 
on Event mode data. To plot the lightcurves ``fplot" task was used. The hardness 
and the softness ratio diagrams were made by extracting lightcurves for three energy 
bands: $A:0-8$ channels ($2-4$ keV), $B:9-35$ channels ($4-15$ keV) and $C:36-138$ 
channels ($15-60$ keV). A hardness diagram is the plot between $C/A$ vs. $B/A$ while 
the softness diagram is the plot between $B/C$ vs. $A/C$. Our motivation of splitting 
the energies in this way stems from the fact that the Keplerian disk primarily emits 
at a low energy ($\lsim 4$ Kev) for the mass of the black hole, we were interested in. 
Thus, $A$ would be emitted mostly from the Keplerian component. The component $B$ would 
be emitted from the region where the moderate thermal Comptonization of the Keplerian 
photons took place. The component $C$ would be emitted from the region which was definitely 
depleted or enhanced during state transitions as it was represented at the higher energy side 
of the pivotal energy [$\sim 15$ keV] in the spectrum. Thus, these diagrams were not directly 
connected to the spectral states - - rather, they were connected to the geometry, i.e. the 
number of soft photons produced by the Keplerian disk ($\sim A$) and the seed photons 
intercepted by the `Compton cloud' [$\sim (B + C)$] and the number of scatterings they 
undergo ($\sim B$ or $\sim C$).

\begin{figure}[h]
\vskip -0.2cm
\centering
\includegraphics[height=1.45in,width=1.44in,angle=0]{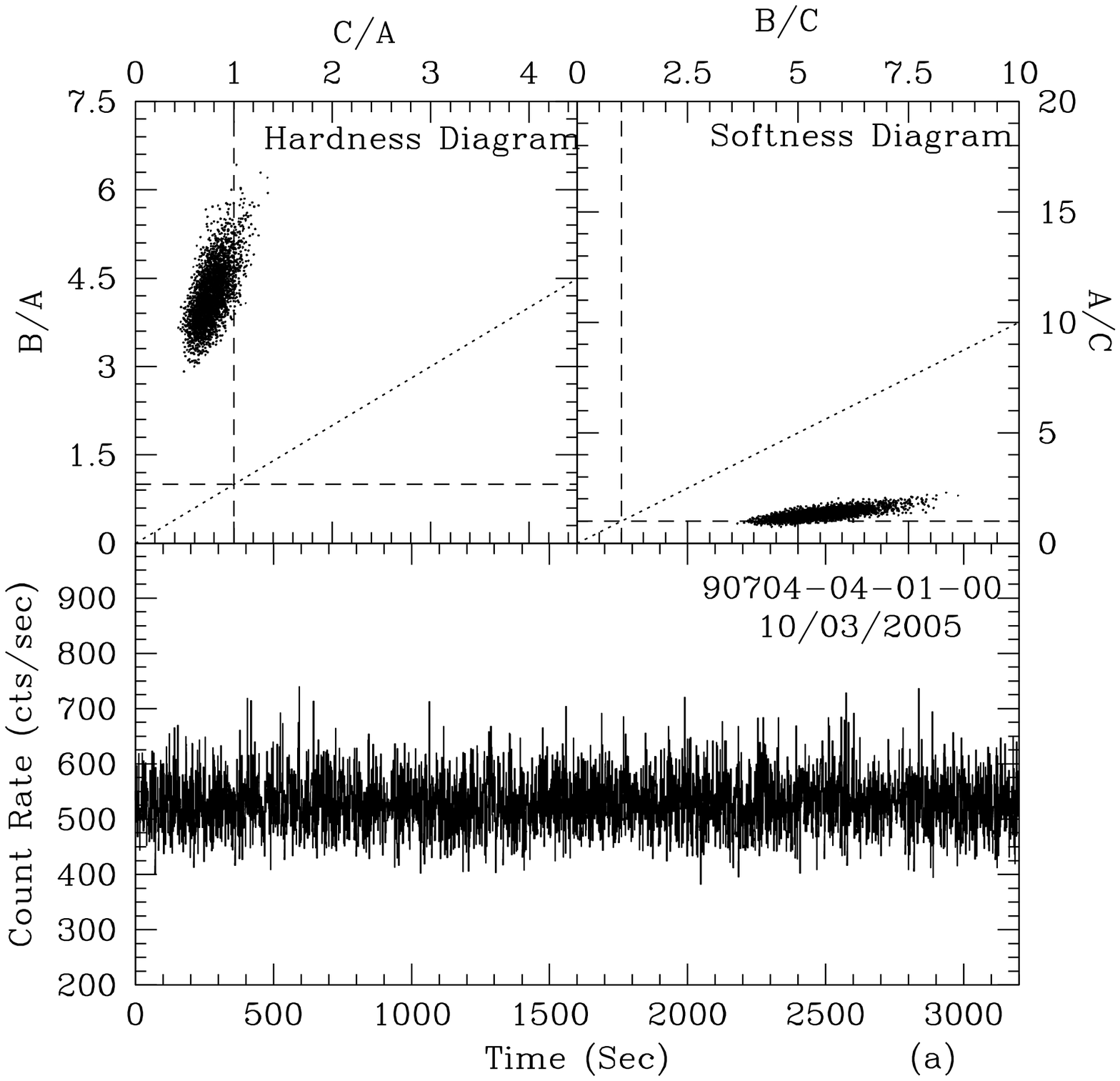}
\includegraphics[height=1.45in,width=1.44in,angle=0]{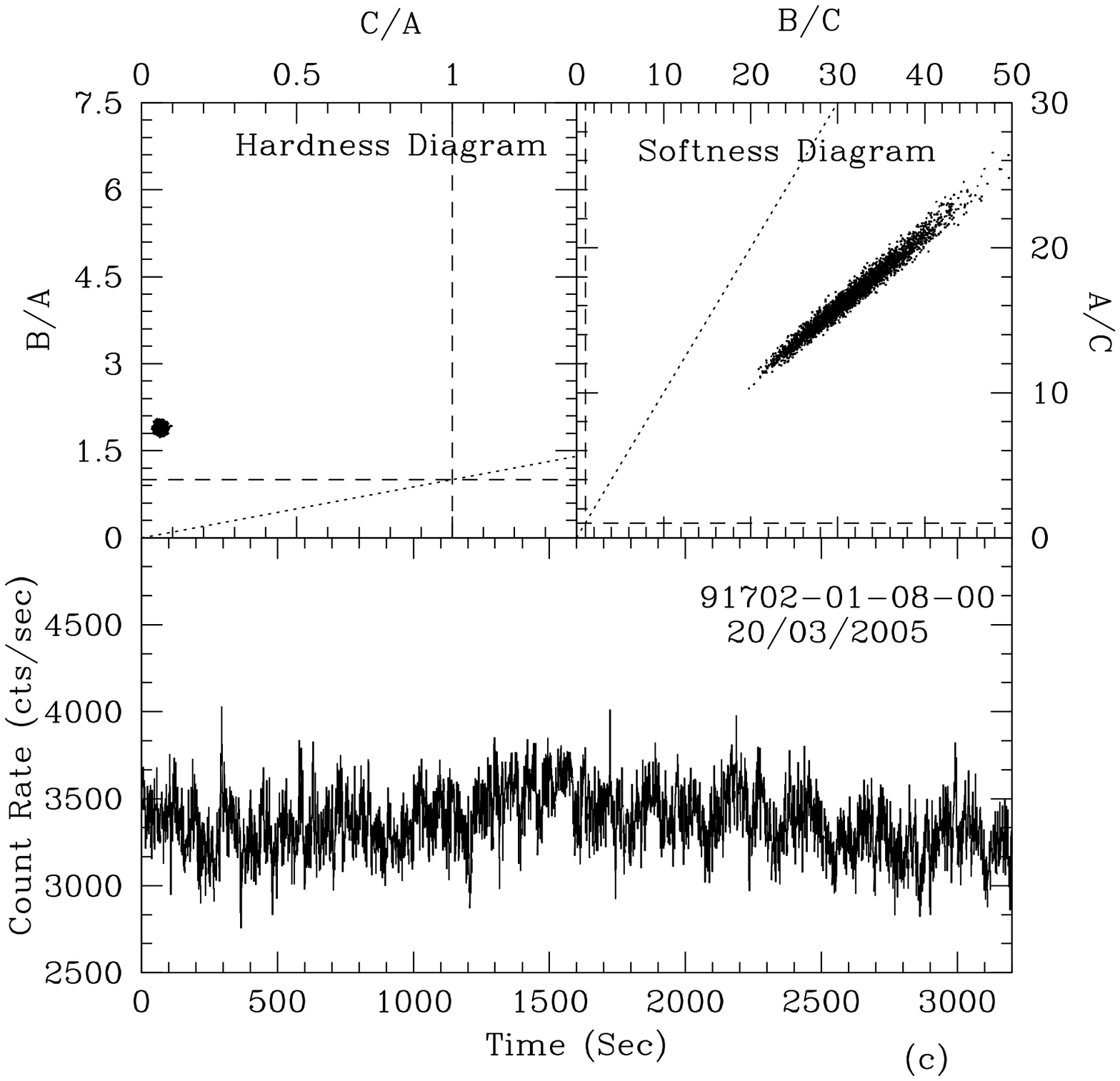}
\includegraphics[height=1.45in,width=1.44in,angle=0]{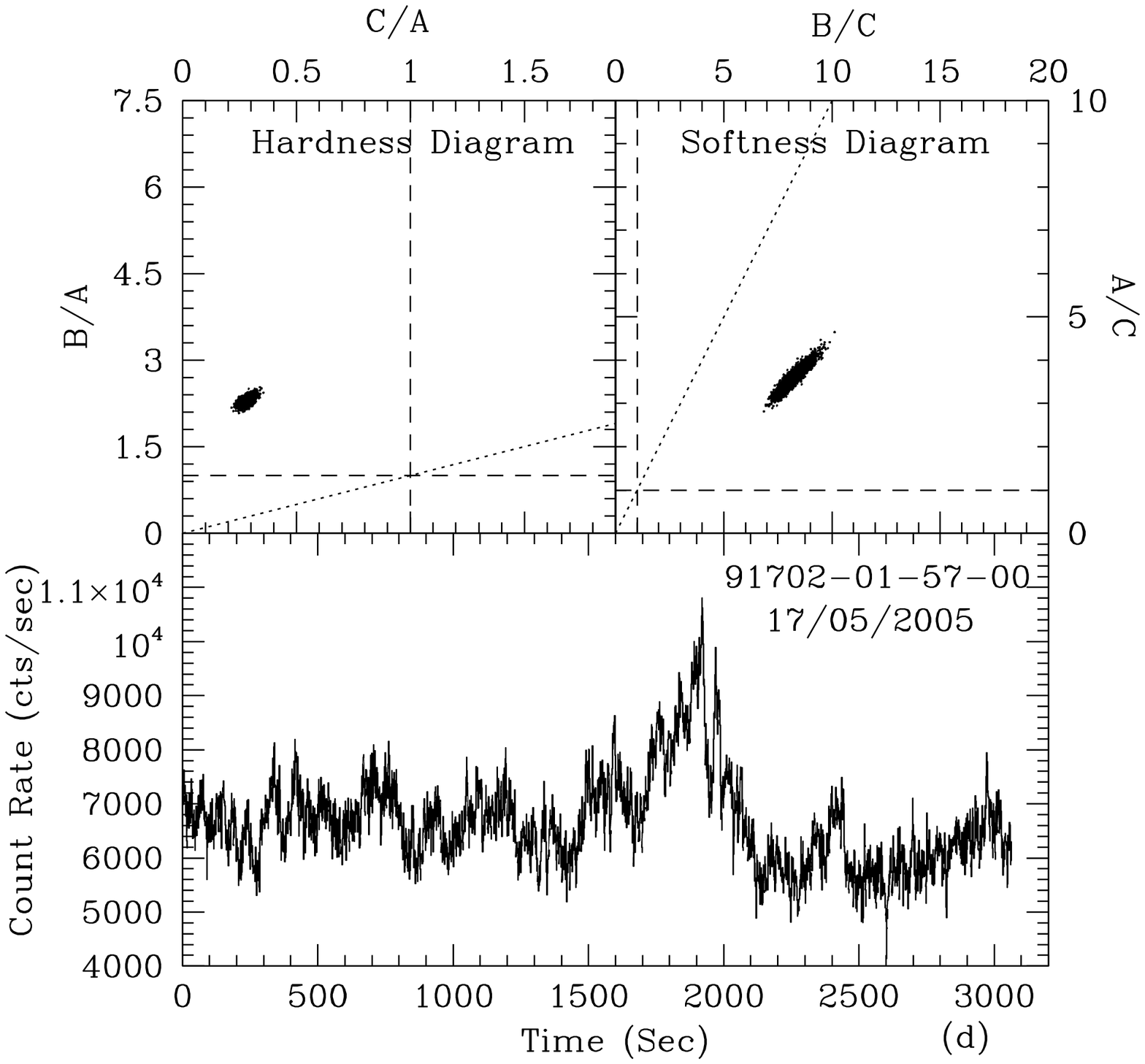}
\includegraphics[height=1.45in,width=1.44in,angle=0]{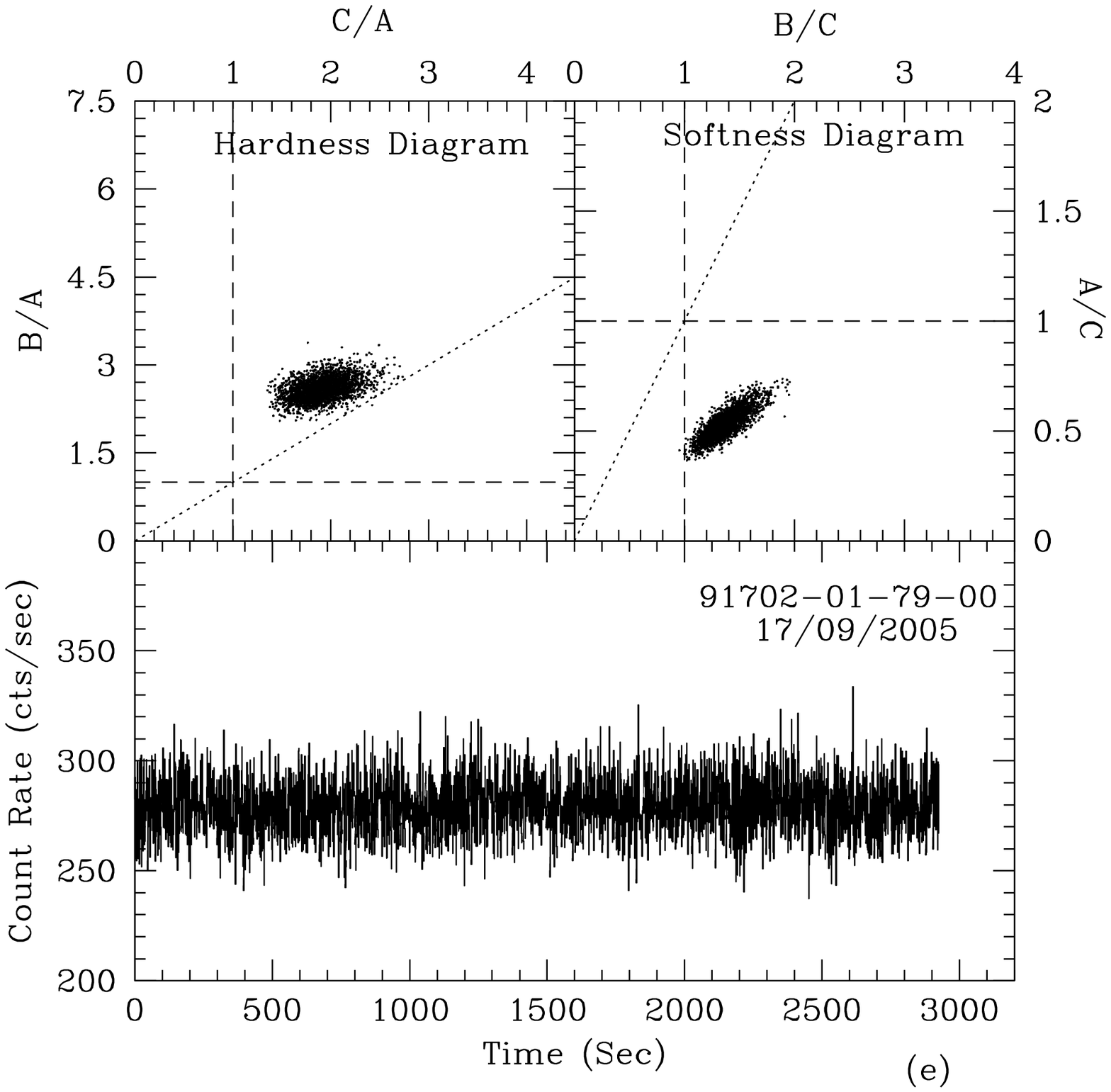}
\caption{(a-d):In the lower panel, 2 - 15 keV (0-35 Channels) PCA lightcurve and in the 
upper panel the hardness and softness diagrams are plotted. In hardness diagrams,
the dashed horizontal ($B=A$), vertical ($C=A$) and the dotted line ($B=C$) are for
reference purpose. In softness diagram they represent $C=A$, $B=C$ and $A=B$
respectively. Observation dates are: (left to right) (a) on 10th of March, 
2005 (Obs ID:90704040100) at the initial rising hard state phase, (b) on 
20th March, 2005 (Obs ID:91702-01-08-00) at the Soft/Very soft state,  (c) 
on 17th of May, 2005 (Obs ID:91702015700) at the intermediate state, and 
(d) on 17th of September, 2005 (Obs ID:91702017900) at falling hard state phase 
of the outburst (Debnath et al. 2008a).}
\label{kn : fig4.3}
\end{figure}

From Figs. 4.3(a-d) and 5.4 (a), we obtain an approximate idea on the evolution of high 
energy and low energy seed photons i.e. about the supply the Keplerian (disk) and the 
Sub-Keplerian (halo) matter/photon rates.

\subsubsection{(a) Hard State in the rising phase:}
Figure 4.3(a) shows initial hard state lightcurve and hardness-softness intensity diagrams 
of 10th March, 2005. From the Figure, it is clear that $B> A > C$. So, here the intensity of 
the intermediate energy photons is stronger than low and hard X-ray photons. Here we observed 
QPOs of frequencies 2.313 \& 4.59 Hz. During the full initial hard state observations, we 
observed QPOs with monotonically increasing frequencies. But just after three days (on 13th of
March) of the present observation, QPO vanished and the spectrum became softer and moved towards 
the soft/very soft state.

\subsubsection{(b) Soft/Very soft State:}
Figure 4.3(b) shows soft/very soft state lightcurve and hardness-softness intensity diagrams 
of 20th March, 2005, when the source was in the soft state. From the Figure, we can interpret 
that $B>A >>C$. The $C$ component is further reduced while $A$ and $B$ continue to go up 
with $A$ approaching $B$. So, as the day progresses, the supply of the Keplerian component 
$B \& A$ increased and the spectrum became softer. No QPO signature was observed in this state.

\subsubsection{(c) Intermediate State:}
Figure 4.3(c) represents the lightcurve and the hardness/softness diagram in the intermediate
state as observed on 17th May, 2005. Here the $C$ component is increased very rapidly while 
the others increasing very slowly. This state shows some evidence of QPOs on certain days.
Here also $B>A>C$.

\subsubsection{(d) Hard State in the decline phase:}
Figure 4.3(d) shows the lightcurve and the hardness-softness intensity diagrams when the source was 
in the hard state again on a typical day, 17$^{th}$ of September, 2005. Here, both the slopes 
of the hardness and the softness diagrams are flatter as compared to those in the intermediate 
state and have a similar characteristics as that of the hard state. In this case, the components 
$B$ and $C$ became dominant and the more than the photon counts in $A$. Here, $B > C > A$. The 
tendencies of the ratios $C/A$ and $B/A$ in relation to the count rate are consistent with that
shown in Fig. 4.3(a-b). Again, QPO has started appearing in this state and the frequency went 
down as the days progressed.  

\begin{figure}[h]
\vskip -0.2cm
\centering
\includegraphics[height=2.92in,width=2.92in,angle=0]{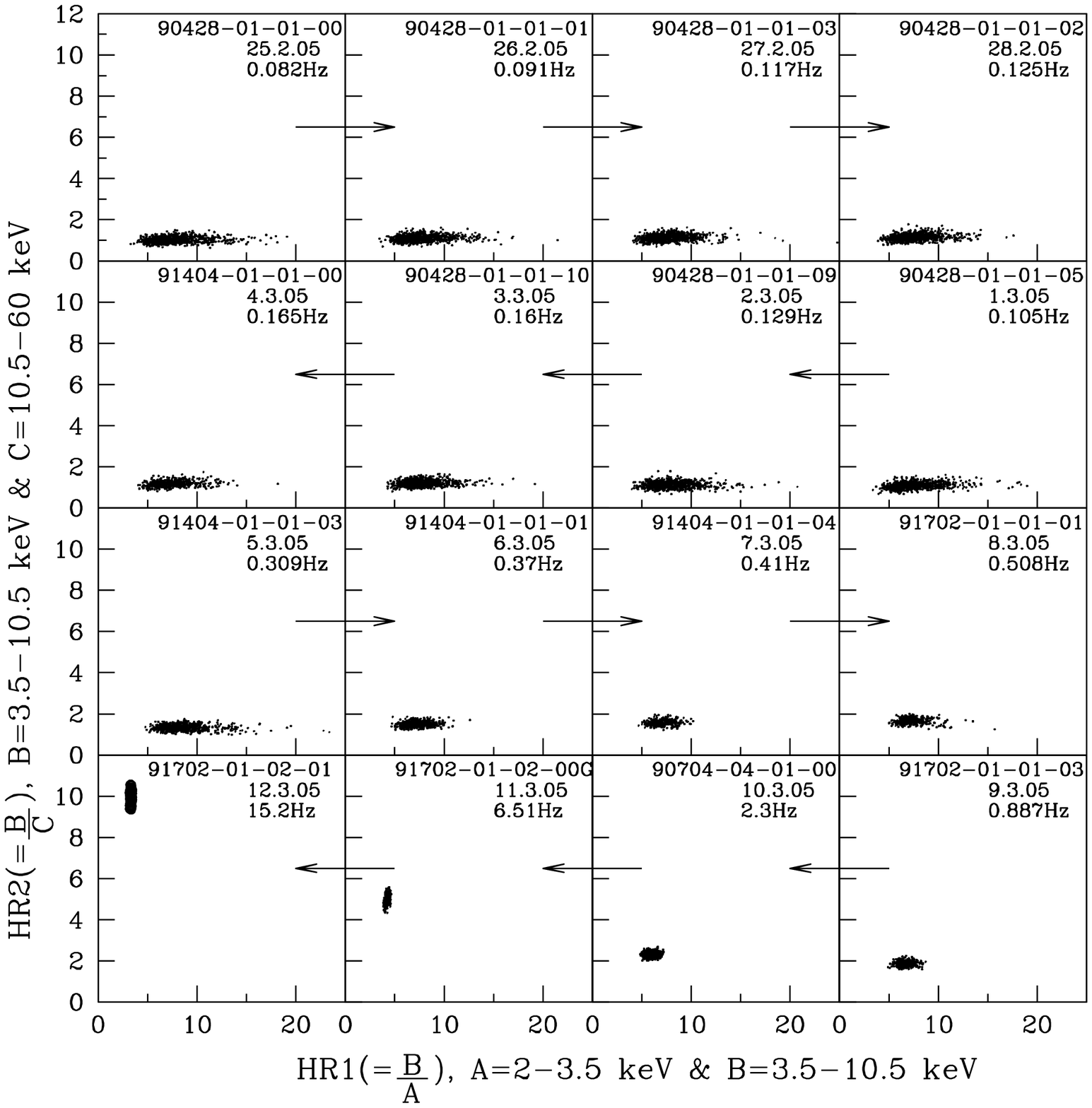}
\includegraphics[height=2.92in,width=2.92in,angle=0]{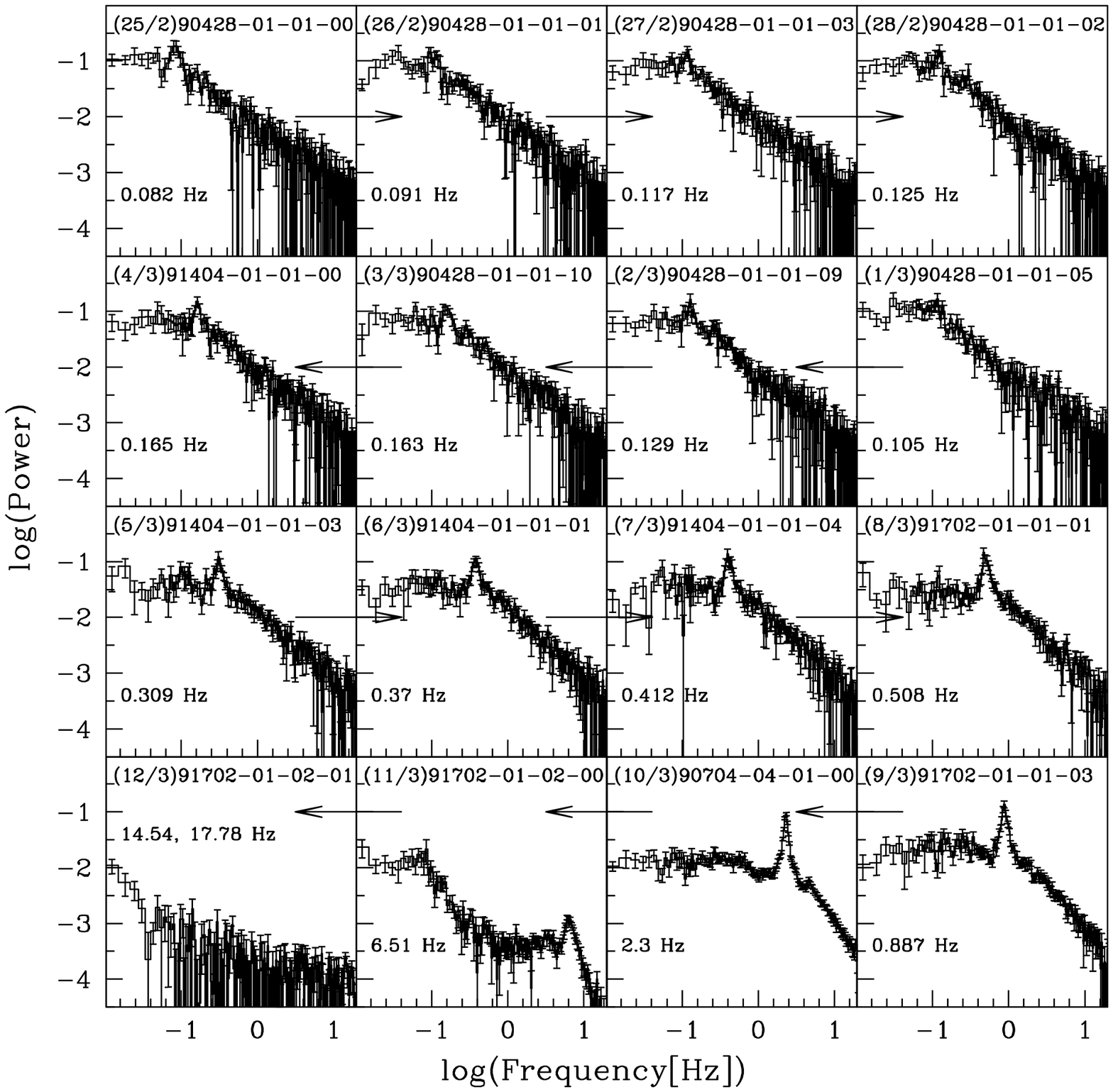}
\caption{(a-b) Left panel (a) shows the variation of the hardness ratios during first two weeks 
(i.e. rising hard state phase) of the outburst from $25^{th}$ of February, 2005 to $11^{th}$ 
of March, 2005. It clearly shows the increasing dominance of the soft photons (the Keplerian 
flows) over hard photons (the sub-Keplerian flows) as time passes by. The observation Ids, date 
and the QPO frequencies are also been written inside the individual plots.
Right panel (b) shows the variation of the PDS with QPO frequencies marked in the initial 
hard state of the outburst. The dates (dd/mm), the observation ID and the frequency of the 
QPOs are in the inset. Arrows indicate the direction in which the dates are increasing
(Chakrabarti et al. 2006a, Debnath et al. 2008a).}
\label{kn : fig4.4}
\end{figure}

The day-wise variations of the hardness ratios and power density spectra (discussed in details 
in the next sub-section) for the initial rising hard state of the outburst are shown in Fig. 4.4(a-b). 
For studying the rising hard state hardness ratio evolution, we divided 0-60 keV PCA 
data into 3 new intervals: $A:0-7$ channels ($2-3.5$ keV), $B:8-24$ channels ($3.5-10.5$ keV) 
and $C:25-138$ channels ($10.5-60$ keV). We extracted lightcurves for 16 days PCU2 PCA data 
in the above intervals in between February 25, 2005 to March 11, 2005. Then in Fig. 4.4(a) we 
plotted HR1 (B \& A band photon count ratio, B/A) vs. HR2 (B \& C band photon count ratio, B/C)
for all the 16 days data. From the hardness ratio evolution diagram it can be concluded that as 
initial few days ratio is stable (i.e. change of all the components are around same), but in 
the simultaneous days, as the day progressed $B$ \& $A$ components (i.e. low \& intermediate 
energy photon numbers) raised rapidly over $C$ component (very high energy photon numbers). 
This means that as as the day progressed Keplerian (soft) flow component became stronger over 
the hard component sub-Keplerian part.

\subsection{Power Density Spectra}

To generate the Power Density Spectrum (PDS), we used ``powspec" task of XRONOS package 
with a normalization factor of `-2' to have the `white' noise subtracted rms fractional
variability. The power obtained has the unit of rms$^2$/Hz. The lightcurve of X-ray 
variability from which PDS was obtained were binned at $0.01$ sec time resolution so that 
the Nyquist frequency is $50$ Hz. QPOs are generally Lorentzian type (Nowak 2000, van der 
Klis 2005) and thus each PDS was fitted with a power-law plus Lorentzian profile to derive 
the central frequencies and widths of each observed QPO. One has to be careful in rebinning 
the frequency scale (Papadakis and Lawrence 1993) as it may mis-represent the behaviour 
especially at low frequencies. In our case, we rebinned the PDS with a geometrical factor 
of $-1.02$ to have a nearly equispaced logarithmic frequency bins. For  this choice, any 
QPO below $0.0122$ Hz would not be detectable. For the best fitting of the PDS as well as QPO 
profiles we used the least square fit technique. After fitting PDS, we used ``fit err" 
task to calculate +/- error for QPO frequencies and widths. This task calculates the 90\% 
confidence range of any fitted parameter. For the best fit, we occasionally used another 
broad Lorentzian component at the break frequency position. 

\begin{figure}[h]
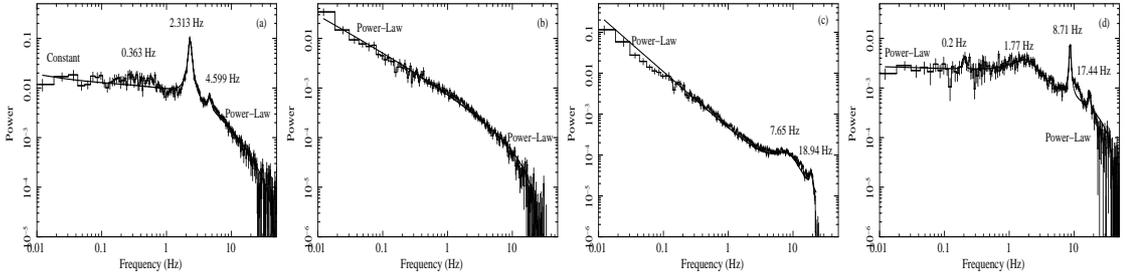

\vskip -0.2cm
\centering
\includegraphics[height=1.4in,width=1.4in,angle=270]{fig4.5a.ps}
\includegraphics[height=1.4in,width=1.4in,angle=270]{fig4.5b.ps}
\includegraphics[height=1.4in,width=1.4in,angle=270]{fig4.5c.ps}
\includegraphics[height=1.4in,width=1.4in,angle=270]{fig4.5d.ps}
\caption{(a-d): (a) The model fitted PDS of $10^{th}$ March, 2005 (ID: 90704-04-01-00), 
a typical day of initial hard state. QPO was found at $2.313$ Hz,  with $0.363$ Hz break 
frequency. (b) The model fitted PDS of $20^{th}$ March, 2005 (ID: 91702-01-08-00). 
No QPO was observed in this case. In this observation day source was at soft/very soft state. 
(c) The model fitted PDS of $17^{th}$ May, 2005 (ID: 91702-01-57-00G), a typical day of 
intermediate state. The QPO was observed at $18.94$ Hz with a bump at $7.65$ Hz. This time 
source was at intermediate state. (d) The model fitted PDS of $17^{th}$ September, 2005 
(ID: 91702-01-79-00). In this observation day source was at final declining hard state. 
QPOs are found at $0.203$ Hz, $8.71$ Hz  with a break frequency at $1.77$ Hz 
(Debnath et al. 2008a).} 
\label{kn : fig4.5}
\end{figure}

In Tab. 4.1, we present a summary of the results, where we put the centroid frequency 
($\nu$) of the QPO, its width ($\Delta\nu$) (both in Hz),  the coherence parameter 
$Q$ (= $\nu$/$\Delta\nu$). The RMS amplitudes $R$ of the fitted QPOs are also included which 
were calculated from $R = 100~\sqrt{(PW\pi/<\phi>)}$, where, $P$, $W$ and $\phi$ are the power, 
half-width ($\Delta\nu/2$) of the Lorentzian fitted QPO and the mean count rate of the source 
respectively. If $Q > 2$, it is considered to be a strong QPO, otherwise it is not strong and 
looks more like a bump on the PDS. Since we are interested only in the QPO properties, namely, 
the frequencies associated with the QPO, bump and the break, only these are included in the 
Table and not the power-law features which may been used for the best fit. In our study for 
the full outburst of total $150$ observations (spreaded over $122$ days), we observed QPOs 
in only $67$ observations (spreaded over $43$ days). We observed these QPOs in three spectral 
states (initial rising phase hard, intermediate and final declining hard), out of four classified 
spectral states. No QPOs were observed in soft/very soft spectral state (see Fig. 4.5b).

\begin{figure}[h]
\vskip -0.2cm
\centering
\includegraphics[height=2.9in,width=2.9in,angle=0]{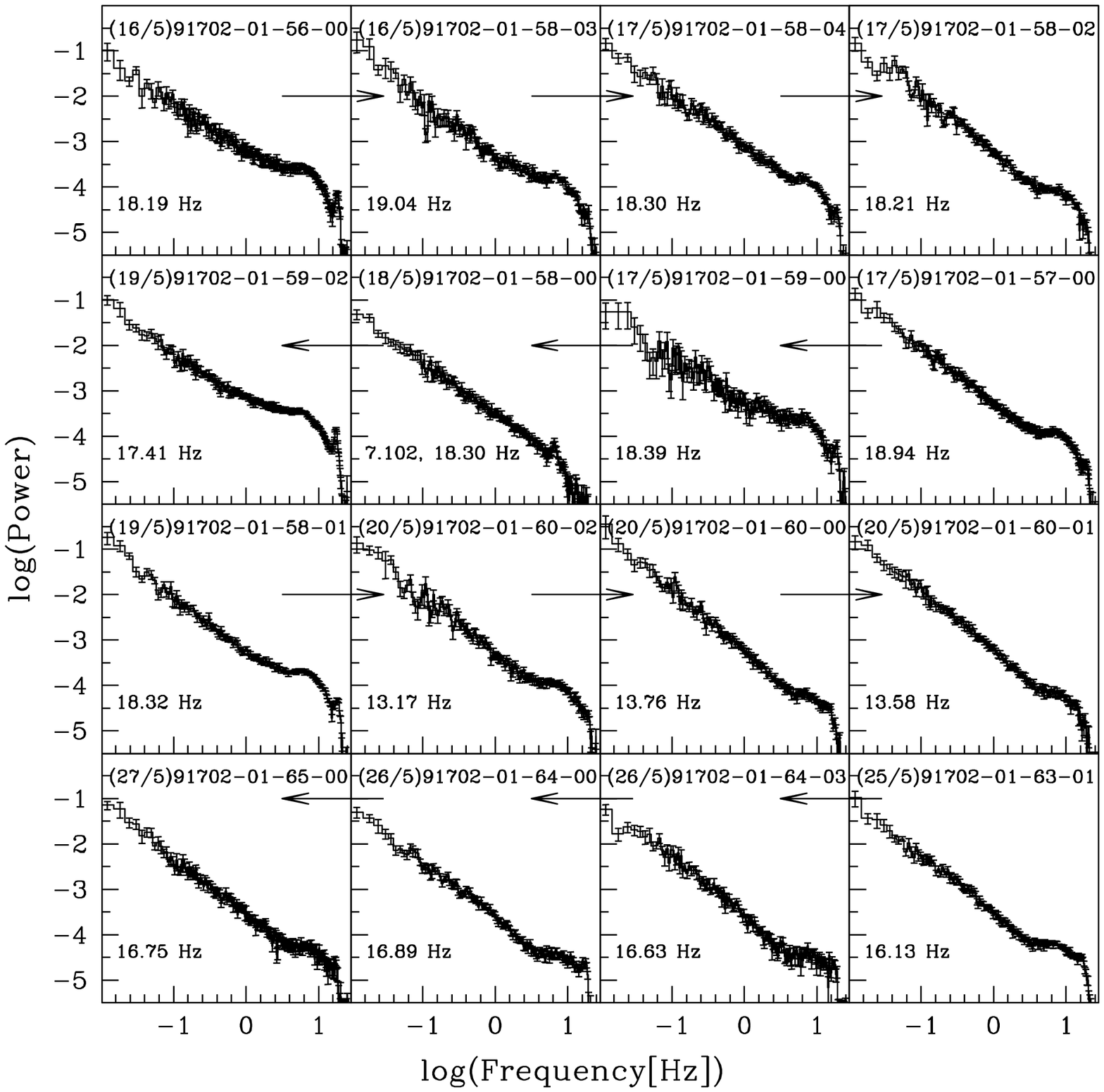}
\includegraphics[height=2.9in,width=2.9in,angle=0]{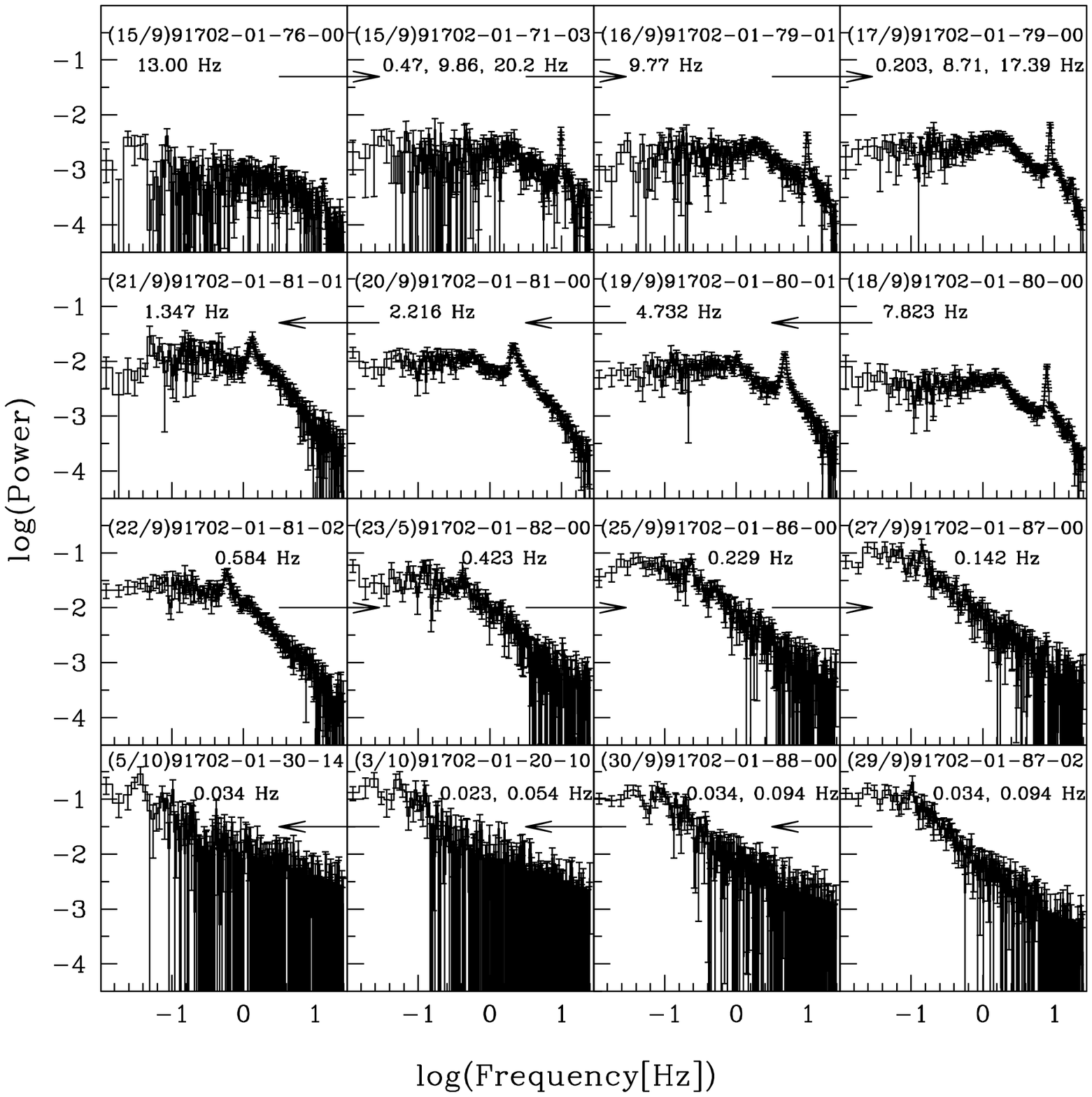}
\caption{(a-b) The variation of the PDS with QPO frequencies marked in the intermediate state 
(a) (from 16$^{th}$ of May, 2005 to 27$^{th}$ of May, 2005) and declining phase hard state (b) 
(from 15$^{th}$ September, 2005 to 5$^{th}$ October, 2005) of the outburst. In the intermediate 
state observed QPO frequencies varied from $13 - 19$ Hz with a bump at $\sim 7$ Hz, whereas 
in the declining hard state, the observed QPO frequencies are seen to decrease monotonically 
from $13.14$ Hz to $0.034$ Hz (Debnath et al. 2008a).}
\label{kn : fig4.6}
\end{figure}

\subsubsection{(i) QPOs of the rising hard state:}

This hard state starts from 25$^{th}$ of February, 2005 (MJD = 53426) (the day when RXTE/PCA 
first observed the source in 2005) and ends at 12$^{th}$ of March, 2005 (MJD = 53441). 
In this phase, QPOs were observed in a wide range of $34$ mHz to $17.78$ Hz. QPOs were observed 
from the very first day of the outburst. The observed QPO frequencies were found to be increased 
monotonically with time (day) from 0.082 Hz to 17.78 Hz (on the first day another QPO at $34$ mHz 
was also seen) (see Figre 4.4b). The increasing nature of the QPO frequency fits well with the 
propagating (towards the black hole) oscillating shock model (Chakrabarti et. al. 2005, 2006a, 
2008a). This theoretical explanation has been discussed in the sub-section follow.

\subsubsection{(ii) QPOs of the intermediate state:}

The intermediate state is seen from the 16$^{th}$ of May, 2005 (MJD = 53506) to 11$^{th}$
of September, 2005 (MJD = 53624). Interestingly, we found QPOs only for $8$ days, 
from 16$^{th}$ of May, 2005 (MJD = 53506) to 20$^{th}$ of May, 2005 (MJD = 53510)
and from 25$^{th}$ of May, 2005 (MJD = 53515) to 27$^{th}$ of May, 2005 (MJD = 53517).
In between, for four days we observed no signature of QPOs (see Fig. 4.5a). The QPO 
frequencies varied from $13.17$ Hz to $19.04$ Hz. In the PDS, we also found one broad 
QPO bump at frequency near $7$ Hz. 

\subsubsection{(iii) QPOs of the declining hard state:}

The final hard state of the declining phase of the outburst was observed from the 12$^{th}$ 
of September, 2005 (MJD = 53625) to 16$^{th}$ of October, 2005 (MJD = 53659). The QPOs of 
$0.023$ Hz to $20.20$ Hz QPOs were observed in this state. If we follow one of the QPO 
frequencies, we find it to decrease monotonically from $13.14$ Hz to $0.034$ Hz within 
$20$ days. The decreasing nature of the QPO frequency fits well with the propagating 
(outwards the black hole) oscillating shock model (Chakrabarti et. al. 2008a). 
The theoretical explanation has been discussed in the next Section.

%\newpage
\begin{table}[t]
%\small
\scriptsize
\centering
\caption{\label{table4.1} Observed QPO fitted parameters for GRO J1655-40 2005 outburst}
\vskip 0.2cm
\begin{tabular}{|lccccc|lccccc|}
\hline
\hline
 & & & & & RMS & & & & & & RMS \\
Obs.&UT Date&$\nu$&$\Delta\nu$&Q&Amp.&Obs.&UT Date&$\nu$&$\Delta\nu$&Q&Amp.\\
\hline

1 &25/02/05&0.082&0.018&4.432&6.410 &  24&10/03/05&2.035&0.243&8.357&19.935 \\
2 &26/02/05&0.034&0.010&3.269&5.021 &  25&10/03/05&2.313&0.298&7.762&19.994 \\
2 &26/02/05&0.106&0.036&2.961&6.852 &  25&10/03/05&4.620&0.450&10.267&5.317 \\
3 &27/02/05&0.116&0.017&6.824&4.908 &  26&11/03/05&3.262&0.989&3.299&2.493  \\
4 &27/02/05&0.051&0.029&1.735&6.048 &  26&11/03/05&6.546&2.305&2.840&6.017  \\  
4 &27/02/05&0.122&0.013&9.760&5.312 &  27&12/03/05&14.54&1.466&9.915&2.146  \\
5 &28/02/05&0.051&0.015&3.446&3.845 &  27&12/03/05&17.78&1.470&12.095&2.149 \\
5 &28/02/05&0.121&0.018&6.612&5.441 &  - -&- - - - - &- - - -&- - - -&- - - -&- - - -\\
6 &02/03/05&0.036&0.011&3.396&4.464 &  57&16/05/05&18.19&0.922&19.735&0.825\\        
6 &02/03/05&0.109&0.028&3.838&5.718 &  58&16/05/05&19.04&1.248&15.256&0.642\\        
7 &02/03/05&0.045&0.026&1.744&7.126 &  59&17/05/05&18.30&0.573&31.954&0.502\\        
7 &02/03/05&0.129&0.012&10.574&6.191&  60&17/05/05&18.21&0.701&25.973&0.420\\        
8 &02/03/05&0.050&0.050&0.998&8.946 &  61&17/05/05&18.94&0.245&77.212&0.317\\        
8 &02/03/05&0.117&0.041&2.875&7.535 &  62&17/05/05&18.65&0.562&33.179&0.722\\        
9 &02/03/05&0.122&0.033&3.754&9.433 &  63&18/05/05&6.566&0.721&9.111&0.767 \\        
10&02/03/05&0.121&0.061&1.977&7.795 &  63&18/05/05&18.30&3.280&5.579&0.753 \\        
11&03/03/05&0.163&0.035&4.644&6.362 &  64&19/05/05&17.41&0.982&17.724&1.459\\        
12&04/03/05&0.160&0.018&8.989&5.673 &  65&19/05/05&18.19&0.817&22.253&0.641\\        
13&05/03/05&0.247&0.088&2.807&8.969 &  66&20/05/05&13.17&3.576&3.683&1.033 \\        
14&05/03/05&0.116&0.028&4.099&4.326 &  67&20/05/05&13.76&3.066&4.488&1.097 \\        
14&05/03/05&0.317&0.076&4.166&9.335 &  68&20/05/05&13.58&4.007&3.389&1.419 \\        
15&06/03/05&0.383&0.102&3.759&10.993&  75&25/05/05&16.13&4.010&4.022&1.442 \\        
16&07/03/05&0.417&0.074&6.608&9.374 &  76&26/05/05&16.63&1.711&9.719&0.696 \\ 
17&07/03/05&0.149&0.015&10.000&3.684&  77&26/05/05&16.89&2.343&7.209&1.102 \\       
17&07/03/05&0.487&0.074&6.608&9.374 &  78&27/05/05&16.75&2.804&5.974&1.049 \\       
18&07/03/05&0.125&0.011&11.792&3.638& - -&- - - - - &- - - -&- - - -&- - - -&- - - -\\ 
18&07/03/05&0.487&0.102&4.761&12.329& 123&15/09/05&13.14&4.007&3.389&1.419  \\       
19&08/03/05&0.119&0.005&23.800&2.023& 124&15/09/05&12.72&1.212&10.495&4.297 \\       
19&08/03/05&0.513&0.107&4.803&12.926& 125&15/09/05&9.863&0.795&12.406&7.878 \\       
20&08/03/05&0.033&0.007&4.783&4.039 & 125&15/09/05&20.20&1.453&13.902&4.192 \\               
20&08/03/05&0.517&0.100&5.149&13.768& 126&15/09/05&10.40&0.783&13.282&6.926 \\       
21&09/03/05&0.886&0.132&6.722&15.537& 127&16/09/05&9.749&0.590&16.529&6.493 \\       
22&09/03/05&1.349&0.213&6.342&19.040& 127&16/09/05&19.40&0.749&25.905&1.680 \\       
23&10/03/05&0.089&0.014&6.449&2.529 & 128&17/09/05&0.207&0.011&19.167&0.783 \\       
23&10/03/05&1.528&0.188&8.132&18.639& 128&17/09/05&8.701&0.457&19.060&7.049 \\       
\hline
\end{tabular}
\end{table}

\begin{table}[h]
%\small
\scriptsize
\centering
\noindent {\small Table 4.1 (Cont'd): Observed QPO fitted parameters for 2005 GRO J1655-40 outburst}
\vskip 0.2cm
\begin{tabular}{|lccccc|lccccc|}
\hline
\hline
 & & & & & RMS & & & & & & RMS \\
Obs.&UT Date&$\nu$&$\Delta\nu$&Q&Amp.&Obs.&UT Date&$\nu$&$\Delta\nu$&Q&Amp.\\
\hline
128&17/09/05&17.39&0.900&19.322&2.711& 138&27/09/05&0.142&0.004&39.444&2.583\\       
129&18/09/05&7.823&0.500&15.663&7.737& 139&28/09/05&0.128&0.042&3.070&7.682 \\       
129&18/09/05&15.20&0.898&16.927&2.760& 140&29/09/05&0.102&0.024&4.232&8.701 \\
130&19/09/05&4.747&0.677&7.007&10.922& 141&30/10/05&0.034&0.021&1.650&6.592 \\
131&20/09/05&2.216&0.821&2.700&15.105& 141&30/10/05&0.094&0.065&1.440&11.400\\
132&21/09/05&1.447&0.579&2.501&13.857& 142&01/10/05&0.073&0.062&1.185&11.902\\
133&21/09/05&1.314&0.316&4.160&10.945& 143&02/10/05&0.060&0.045&1.342&11.677\\
134&22/09/05&0.584&0.140&4.163&9.732 & 144&03/10/05&0.023&0.013&1.729&6.633 \\
135&23/09/05&0.417&0.296&1.409&12.048& 144&03/10/05&0.054&0.023&2.379&8.649 \\
136&25/09/05&0.229&0.114&2.005&10.331& 145&04/10/05&0.048&0.042&1.151&10.512\\
137&26/09/05&0.205&0.144&1.427&11.792& 146&05/10/05&0.034&0.019&1.744&8.707 \\             
\hline
\end{tabular}
\end{table}

%\subsubsection{{\bf A.} Theoretical explanation for the QPO evolution in the rising and the 
%declining phases of the outburst: {\bf Propagatory Shock Oscillation (POS) Model}}

\subsection{Theoretical explanation for the QPO evolution in the rising and the 
declining phases of the outburst: {\it Propagating Oscillatory Shock (POS) Model}}

In our early works (Chakrabarti et al. 2005, 2006a, 2008a) the evolution of the QPO 
frequencies for the initial rising and final decline phases of the outburst with time (day) 
were observed (see, Fig. 4.7). Rapid variation in QPO frequencies in the both rising and 
declining phase of the outburst can be explained by using an oscillating and propagating 
shock model (Chakrabarti et al. 2005, 2006a, 2008a) very satisfactorily. The observational results we describe 
is unique, because with this model we are able to connect the QPO frequency of one 
observation with that of the next by a simple analytical formalism (discussed below). 
To our knowledge no competing model exists which uses a true solution of the flow, 
such as shocks in our case, to explain such a behavior. Other models for the 
QPO generation, such as the rotation of non-axisymmetric blobs cannot explain the smooth rise or 
decrease features of the QPO frequencies. We also successfully showed how a shock wave 
smoothly disappears behind the horizon after $\sim 16$ days of its initial rising phase 
outburst detection. Also with our shock oscillation model, we successfully explained 
monotonically decreasing feature of the QPO frequencies in the declining phase of the 
outburst, which lasts for $\sim 20$ days.

\begin{figure}[h]
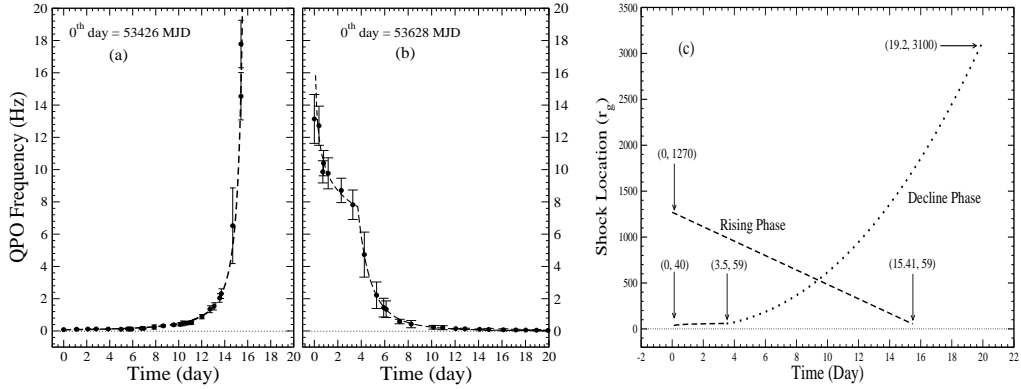

\vskip 0.5cm
\centering
\includegraphics[height=2.0in,width=2.9in,angle=0]{fig4.7a.eps}\hspace{0.2cm}
\includegraphics[height=2.0in,width=2.2in,angle=0]{fig4.7b.eps}
\caption{(a-c) Variation of QPO frequencies and the shock locations with time (in day) 
(a) of the rising phase and (b) of the declining phase. The dotted curves are the solutions 
from oscillating and propagating shocks. While in Fig. (a), the shock appears to be drifting at 
a constant speed towards the black hole, in Fig. (b), the shock initially moves very slowly and 
then runs away at a roughly constant acceleration. According to the fitted solution for the 
rising phase the shock wave goes behind the BH horizon on the $16.14$th day, about $15$ hours 
after the last observed QPO. In Fig. (c), we showed the shock location variation in both rising
and decline phases of the outburst (Chakrabarti et al. 2008a).}
\label{kn : fig4.7}
\end{figure}

In the shock oscillation solution (Molteni, Sponholz \& Chakrabarti 1996, Chakrabarti \&  
Manickam 2000, Chakrabarti, Acharyya \& Molteni 2004) of QPOs, the oscillations take place 
at a frequency inverse to the infall time in the post-shock region (i.e., the region between 
the shock at $r=r_s$ and the horizon). In a shock-free low angular momentum flow, this infall 
time is $t_{infall} \sim r_s/v = r_s(r_s-1)^{1/2}$, where $v=1/(r_s-1)^{1/2}$ is the free-fall 
velocity in a pseudo-Newtonian potential (Paczy\'nski \& Wiita, 1980) $\phi_{PN}=-1/(r_s-1)$. 
Here, distance, velocity and time are measured in units of the Schwarzschild radius 
$r_g=2GM/c^2$, the velocity of light $c$ and $r_g/c$ respectively and where, $G$ and $M$ are 
the universal constant and the mass of the black hole. However, in the presence of a 
significant angular momentum capable of producing centrifugal pressure supported shocks 
around a black hole, the velocity is reduced by a factor of $R$, the compression ratio 
$R=\rho_-/\rho_+$, where, $\rho_-$ and $\rho_+$ are the densities in the pre-shock and 
the post-shock flows, because of the continuity equation $\rho_- v_-= \rho_+ v_+$ across 
a thin shock.

In the presence of a shock, the infall time in the post-shock region is therefore given by
$$
t_{infall}\sim  r_s/v_+ \sim  R~r_s~(r_s-1)^{1/2}
\eqno{(4.1)}
$$
(Chakrabarti \&  Manickam 2000, Chakrabarti et al. 2005). Of course, to trigger the
oscillation, the accretion rate should be such that the cooling time scale roughly match
the infall time scale (Molteni, Sponholz \& Chakrabarti 1996). Thus, the instantaneous
QPO frequency $\nu_{QPO}$ (in $s^{-1}$) is expected to be,
$$
\nu_{QPO} = \nu_{s0}/t_{infall}= \nu_{s0}/[R~r_s~(r_s-1)^{1/2}].
\eqno{(4.2)}
$$
Here, $\nu_{s0}= c/r_g=c^3/2GM$ is the inverse of the light crossing time of the black hole
of mass $M$ in $s^{-1}$ and $c$ is the velocity of light. In a drifting shock scenario,
$r_s=r_s(t)$ is the time-dependent shock location given by,
$$
r_s(t)=r_{s0} \pm v_0~t/r_g.
\eqno{(4.3)}
$$
Here, $r_{s0}$ is the shock location when $t$ is zero and $v_0$ is the shock velocity
(in c.g.s. units) in the laboratory frame. The positive sign in the second term is to be
used for an outgoing shock in the declining phase and the negative sign is to be used for
the in-falling shock in the rising phase. Here, $t$ is measured in seconds from the  first
detection of the QPO.

The physical reason for the oscillation of shocks appears to be a `not-so-sharp' resonance 
between the cooling time scale in the post-shock region and the infall time scale 
(Molteni, Sponholz \& Chakrabarti 1996) or the absence of a steady state solution 
(Ryu, Chakrabarti \& Molteni, 1997). In both the cases, the QPO frequency directly gives 
an estimate of the shock location (Eqn. 4.1). The observed rise of the QPO frequencies 
with luminosity (e.g., Shaposhnikov \& Titarchuk 2006) is explained easily in this model 
since an enhancement of the accretion rate increases the local density and thus the 
cooling rate. The resulting drop of the post-shock pressure reduces the shock location and 
increases the oscillation frequency. In (Chakrabarti \& Manickam 2000,  Rao et al. 2000) 
it was shown that QPOs from the higher energy Comptonized photons, thought to be from the 
post-shock region, (Chakrabarti \& Titarchuk, 1995), have a higher $Q$ value. The latter 
model requires two components, one Keplerian and the other having an angular momentum 
lower than the Keplerian (referred as sub-Keplerian). This model explains a wide variety 
of observations of black hole candidates (Smith, Heindl \& Swank 2002, Smith, 
Dawson \& Swank 2007). As the shocks are the natural solutions of this sub-Keplerian 
component, the explanation of QPOs by shocks is natural. When the {\it Rankine-Hugoniot} 
relation is not exactly satisfied at the shock or the viscous transport rate of the 
angular momentum is different on both sides, the mean shock location would drift slowly 
due to a difference in pressure on both sides. 

Figure 4.7(a-b) shows the variation of the QPO frequencies in the rising (a) and the 
declining (b) phases of the outburst. The full widths at half maxima of the fitted QPOs have 
been used as the error bars. In the rising phase (a), the $0^{th}$ day starts on MJD=53426. 
The fitted curve represents our fit with Eqns. 4.2 \& 4.3, which requires that the shock 
is launched at $r_s=1270$ which drifts slowly at $v_0=1970$cm s$^{-1}$. On the $15^{th}$ 
day after the outburst starts, the noise was high, but we could clearly observe two different 
QPO frequencies with a very short time interval. At the time of the last QPO detection 
($15.41^{th}$ day) at $\nu=17.78$ Hz, the shock was found to be located at $r \approx 59$.
The strength of the shock $R$, which may be strong at the beginning with $R=R_0\sim 4$
should become weaker and ideally $R\sim 1$ at the horizon $r=1$, as it is impossible to 
maintain density gradient on the horizon. If for simplicity we assume the variation of the 
shock strength as $1/R\rightarrow 1/R_0 + \alpha t_d^2$, where $\alpha$ is a very small 
number limited by the time in which the shock disappears (here $t_{ds} \sim 15.5$days).
Thus, the upper limit of $\alpha \sim (1-1/R_0)/t_{ds}^2 =0.75/t_{ds}^2 = 0.003$. We find 
that for a best fit, $\alpha \sim 0.001$ and the reduced $\chi^2=0.96$. However, the fit 
remains generally good ($\chi^2= 1.71$ for $r_{s0}=1245$ and $v_0=1960$cm/s) even with a shock 
of constant strength ($R=R_0$).

In the rising phase of the outburst, it is easy to verify that the QPO frequencies 
(which are inverses of the infall times from the post-shock flow to the black hole) in 
the infalling phase are simply related, as though the shock itself is drifting towards 
the black hole at a slow pace of $\sim 20~m/s$ (Chakrabarti et al. 2005, 2008a).
A combination of the ram pressure of the incoming flow and rapid cooling in the post-shock 
region (which lowers the thermal pressure) pushes the oscillating shock inward, shock location 
decreases and a result of that we observe the increasing nature of QPO frequencies.

During the rising phase of more than two weeks, the disk got sufficient time to transport
angular momentum and a dominant Keplerian disk is formed which made the flow soft or very
soft. The rapid rise of the black body flux after the QPO disappears and almost total absence
of the hard photons testify to the rushing in of the Keplerian disk towards the inner
edge (Chakrabarti \& Titarchuk, 1995, Ebisawa et al. 1996). If we take the two component 
advective flow (TCAF) model one step further and actually fit the
spectra of a few days spreaded during the outburst we observe, using the same procedure
that was followed in Chakrabarti \& Mandal (2006b), we can obtain the accretions rates of
matter in the Keplerian disk and the sub-Keplerian halo. Table 4.5 gives the rates in
units of Eddington rate on various days. It is clear that the Keplerian disk rate
steady increases from the beginning while the halo rate changes in a shorter time scale. 
At the beginning, the halo rate was higher than the disk rate, but in the rest of the 
time, until the very end the disk rate always dominates. In the soft and the very soft states,
the disk rate required to fit the spectra can be high reaching to about two Eddington rates.
The hardness/softness diagrams also give an idea of how the accretion rates in the Keplerian 
and sub-Keplerian components could be changed on a daily basis. 

In the declining phase (Fig. 4.7b), the QPO frequency on the first day ($MJD=53631$) 
corresponds to launching the shock at $ \sim r_s=40$. It evolves as $ \nu_{QPO} \sim 
t_d^{-0.2}$. Since $ \nu_{QPO} \sim r_s^{-2/3}$ (Eqn. 4.1), the shock was found to drift 
very slowly with time ($r_s \sim t_d^{0.13}$) until about $t_d=3.5$ day where the shock 
location was $\sim r=59$. There is a discontinuity in the behavior at this point whose 
possible origin is obtained from spectral studies presented below. After that, it moves 
out roughly at a constant acceleration ($r_s \sim t_d^{2.3}$) and the QPO frequency 
decreases as $\nu_{QPO}\sim r_s^{-2/3} \sim t_d^{-3.5}$. Finally, when the QPO was last 
detected, on $t_d=19.92$th day ($MJD=53648$), the shock went as far as $r_s=3100$ and 
the oscillation could not be detected any longer.

%\subsubsection{{\bf B.} Dynamic PDS: Observation of the QPO evolution in 3D mode}
\subsection{Dynamic PDS: {\it Observation of the QPO evolution in 3D mode}}

\begin{figure}[h]
\vskip -0.5cm
\centering
\includegraphics[height=2.5in,width=2.5in,angle=0]{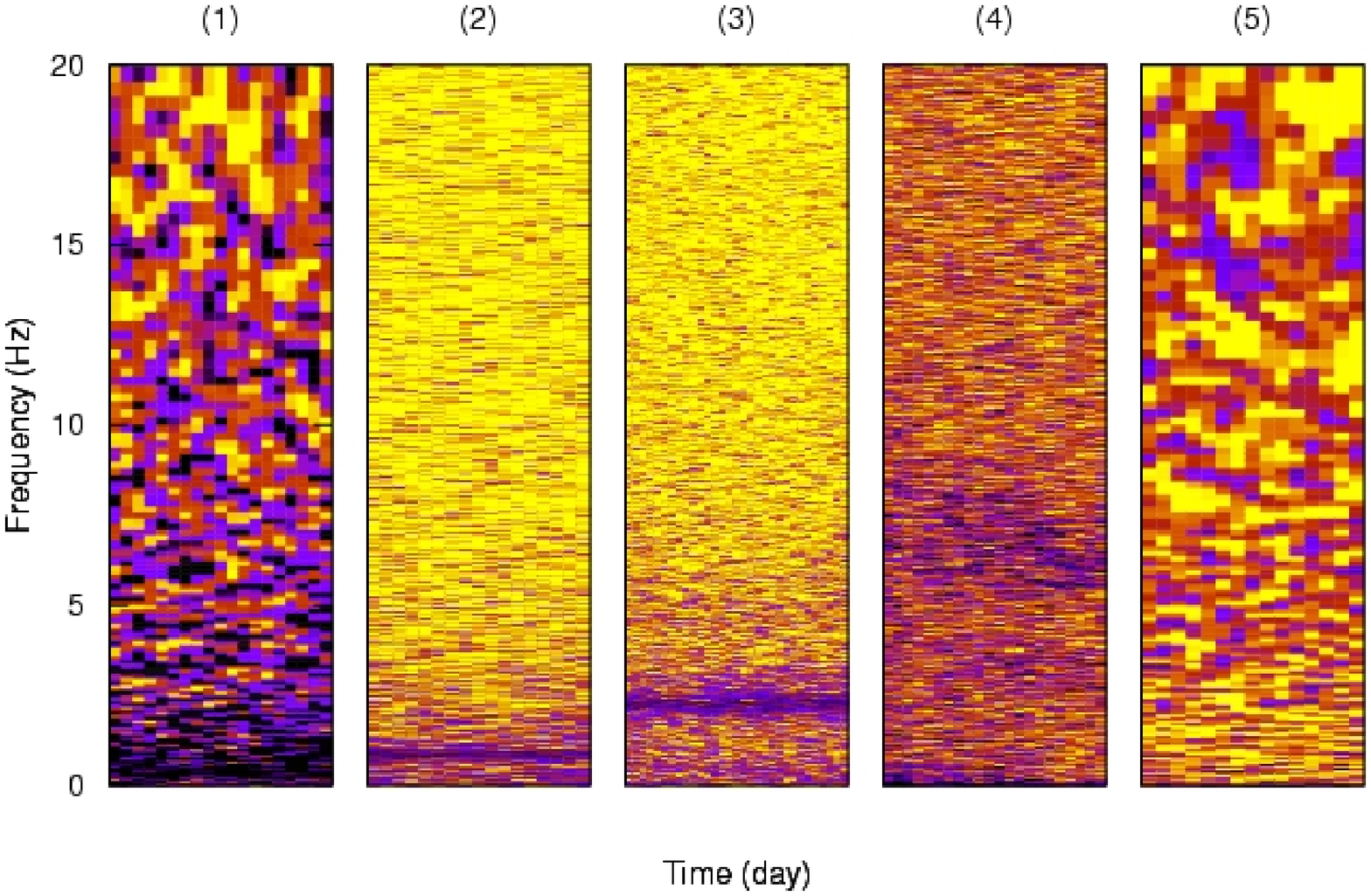}
\includegraphics[height=2.5in,width=2.5in,angle=0]{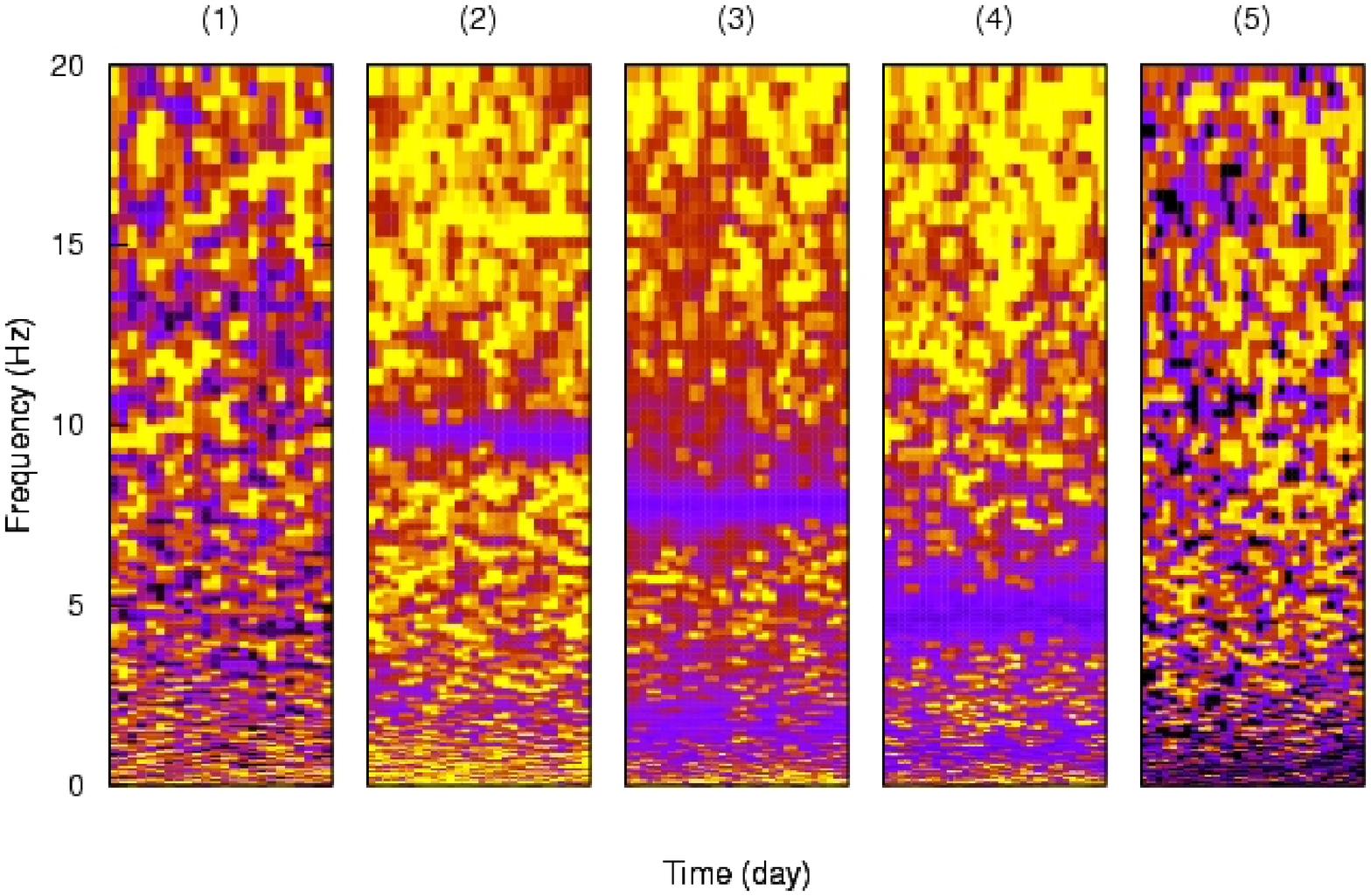}
\caption{(a-b) Left panel (a) shows the dynamic power density spectra over five days in 
the rising phase.
(1) Obs. ID=91404-01-01-01, QPO=0.382 Hz,
(2) Obs. ID=91702-01-01-03, QPO=0.886 Hz,
(3) Obs. ID=90704-04-01-00, QPO=2.3130 Hz,
(4) Obs. ID=91702-01-02-00, QPO=3.45 \& 6.522 Hz with a break frequency
at 0.78 Hz and (5) Obs. ID=91702-01-02-01, QPO=14.54 \& 17.78 Hz.
Right panel (b) shows the dynamic power density spectra over five days in the decline phase. 
(1) Obs. ID=91702-01-76-00,
QPO=13.14 Hz, (2) Obs. ID=91702-01-79-01, QPO=9.77 Hz,
(3) Obs. ID=91702-01-80-00, QPO=7.823 \& 15.2 Hz with a break
frequency at 1.32 Hz, (4) Obs. ID=91702-01-80-01, QPO=4.732 Hz
with a break frequency=0.86 Hz, (5) Obs. ID=91702-01-82-00, QPO=0.423 Hz
(Chakrabarti et al. 2008a).}
\label{kn : fig4.8}
\end{figure}

For the detailed study of the QPO evolution, we made dynamic PDS. Dynamic PDS is the time 
dependent power density spectrum, where the variation QPO frequencies (y-axis) and their 
powers (represented by colours) can be observed with time (x-axis). For making 
these dynamic PDS, we made $1$ sec time averaged lightcurves and PDS.

Figure 4.8(a-b), presents the dynamic PDS of rising and declining phases of the outburst, 
where the vertical direction indicates the QPO frequency. The normalized strength of the 
colours represents the power of the QPOs. It is used to identify the QPO features prominently.
These two Figures also show the same day-wise QPO frequency rise and decrease nature, 
which was observed in Fig. 4.7(a-b). 

%\subsubsection{{\bf C.} Energy dependent QPO signature}
\subsection{Energy dependent QPO signature}

We draw the power density spectra in different energy ranges for the detailed understanding 
of the origin of QPOs, i.e. to know the exact photon energy ranges are responsible for 
generating QPOs (Chakrabarti et. al., 2005 \& Debnath et. al., 2008a). 

\begin{figure}[h]
\vskip -0.5cm
\centering
\hspace{-1.2cm}
\includegraphics[height=3.0in,width=2.1in,angle=0]{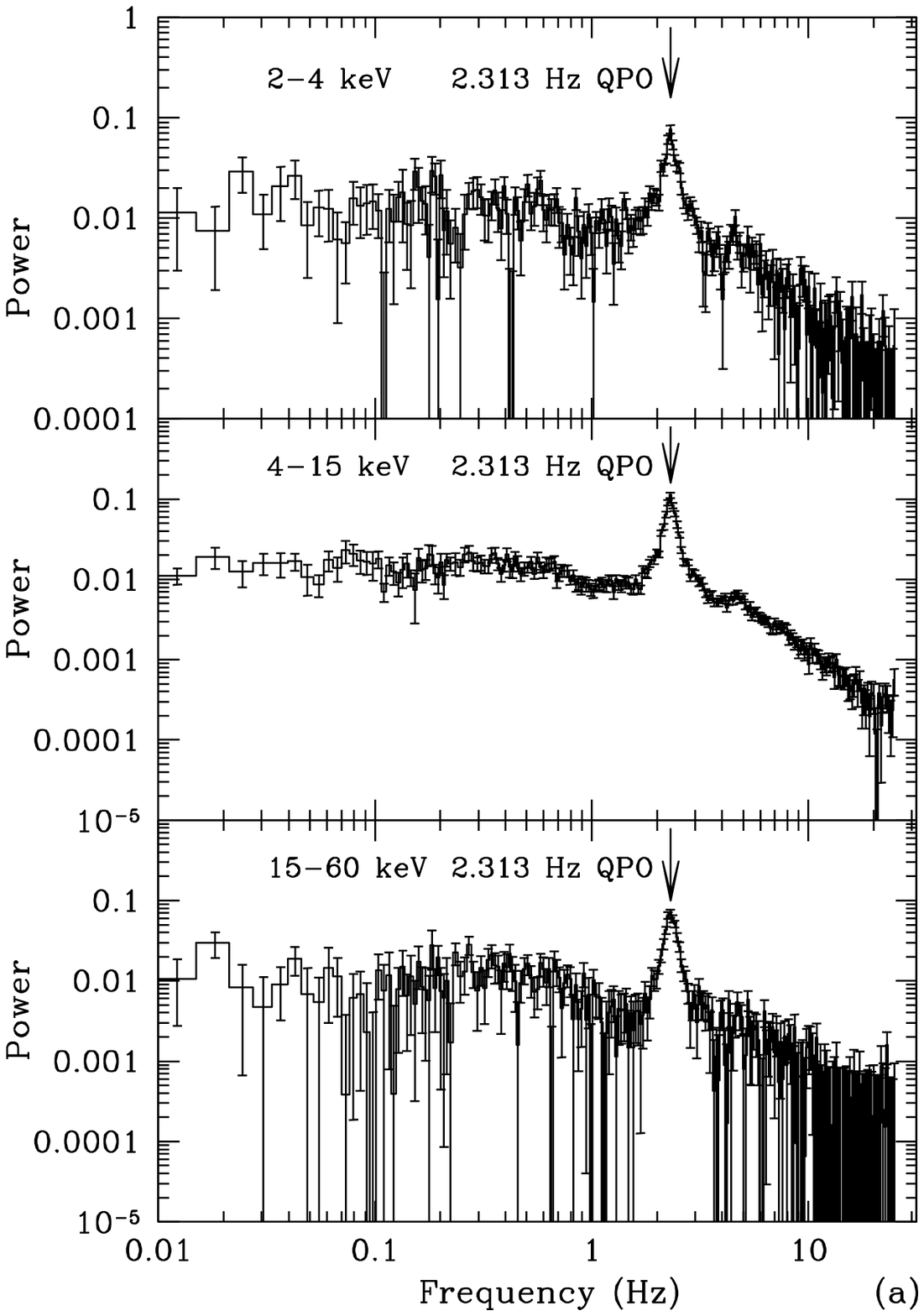}\hspace{-1.9cm}
\includegraphics[height=3.0in,width=2.1in,angle=0]{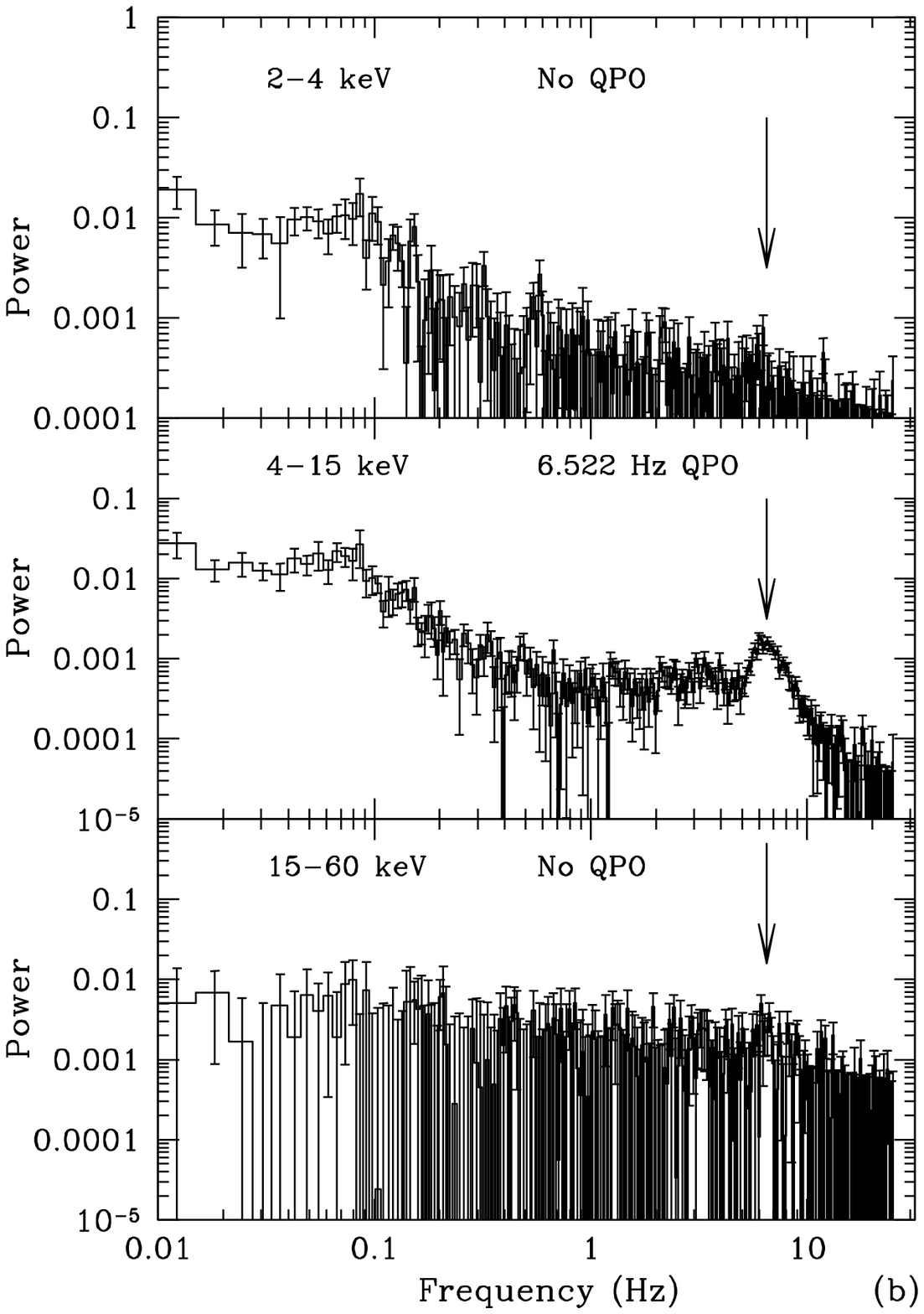}\hspace{-1.9cm}
\includegraphics[height=3.0in,width=2.1in,angle=0]{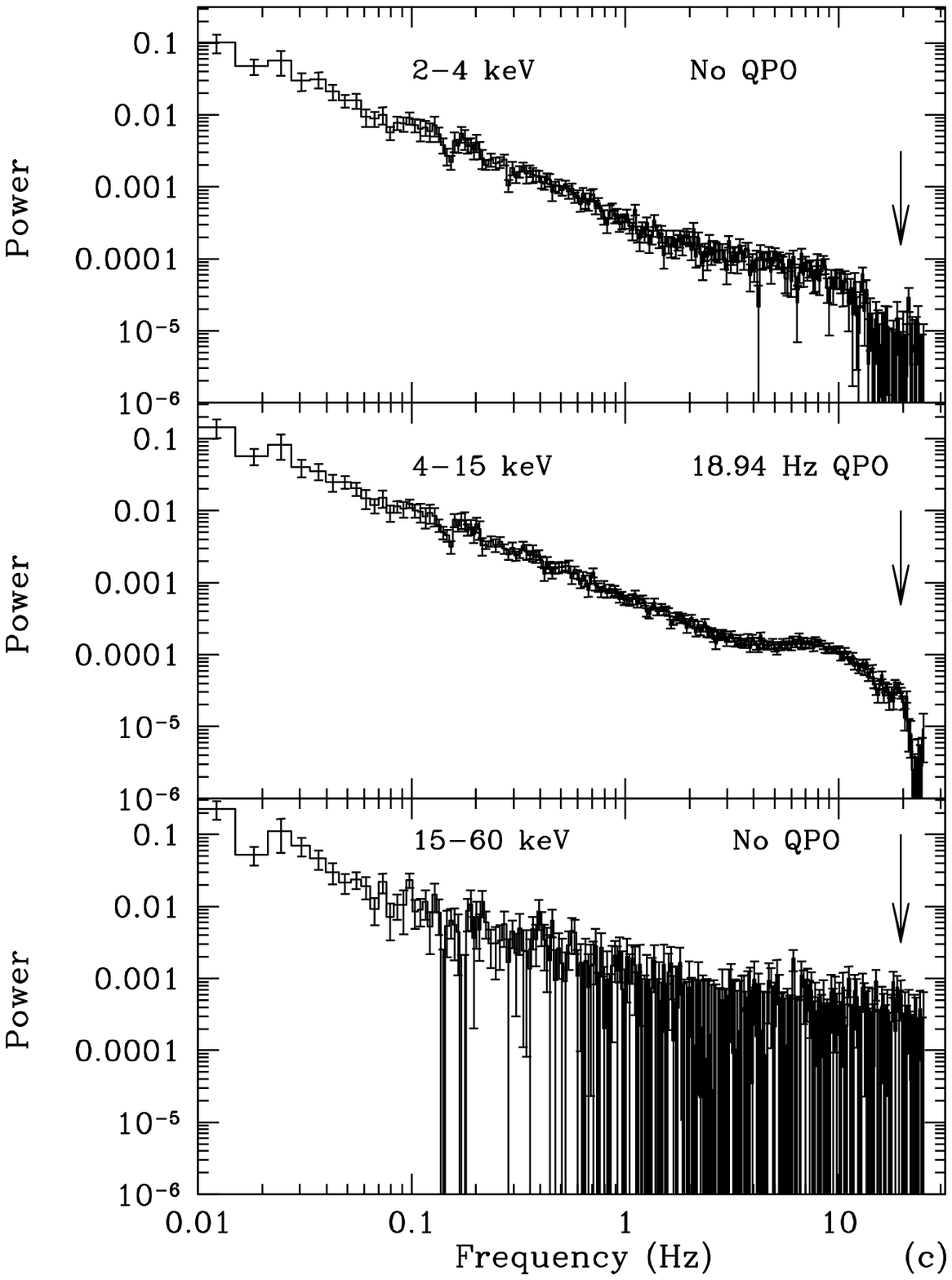}\hspace{-1.9cm}
\includegraphics[height=3.0in,width=2.1in,angle=0]{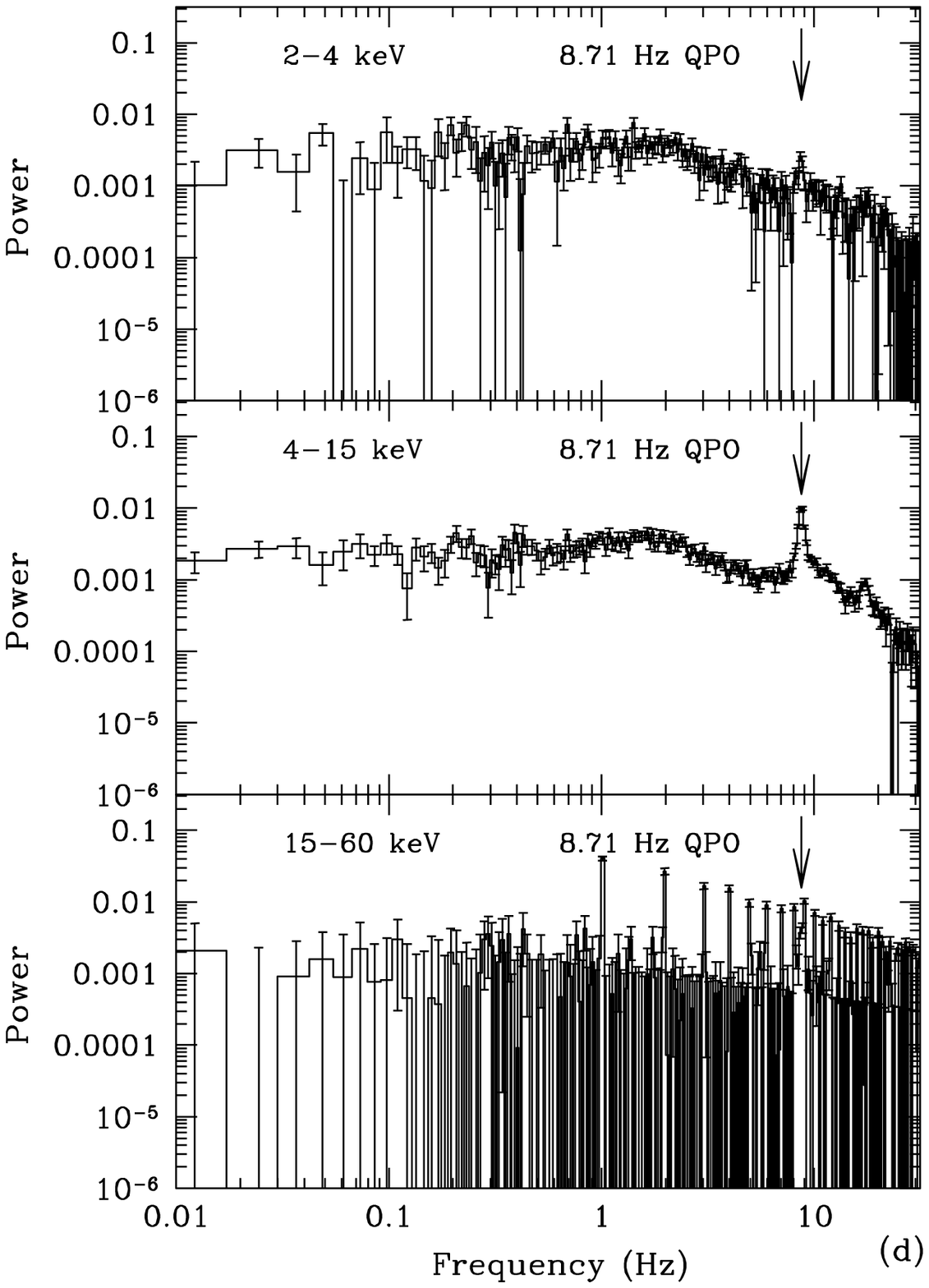}
\caption{(a-d) shows PDS of three different energy bands ($2-4$keV, $4-15$keV and $15-60$keV) for 
different spectral states PCA data. Extreme left panel, (a) shows 3 energy bands PDS 
of 10th March, 2005 (Obs ID:90704040100) observed data. In the same manner (from 2nd left to right) 
(b) shows the same energy bands photons' PDS of March 11th, 2005 (Obs ID:91702-01-02-00G), 
(c) shows the same energy bands photons' PDS of May 17th, 2005 (Obs ID:91702015700), and
(d) shows the same energy bands photons' PDS of September 17th, 2005 (Obs ID:91702017900).
Here in all the Figures data in both X, Y axes are plotted in logarithmic scales 
(Debnath et al. 2008a).}
\label{kn : fig4.9}
\end{figure}

In order to quantify the nature of the energy dependence of the power density spectrum, 
we extracted the lightcurves in three bands with channel numbers 0-8 (2-4 keV), 9-35 (4-15 keV) 
and 36-138 (15-60 keV) respectively in $0.01$ sec time bins and made their PDS using FTOOLS 
task `powspec' with normalization factor -2 (to reduce white noise).
Figure 4.9(a-d), shows PDS in these three energy bands for $10^{th}$ March, $11^{th}$ March, 
$17^{th}$ May, and $17^{th}$ September of 2005 data respectively. $10^{th}$ and $11^{th}$ March 
PDS are of rising hard spectral state, $17^{th}$ May PDS is of intermediate state and $17^{th}$ 
September PDS is of declining hard spectral state respectively.

We found an interesting nature of the energy dependence on the QPO signatures. 
In the 10th March data in Fig. 4.9(a), we see
the 2.313 HZ QPO in its all three energy bands, whereas just after one day 
(on 11th March, Fig. 4.9b) the low (2-4 keV) and high (15-60 keV) energy photons do not show 
any QPO signatures, only intermediate energy (4-15 keV) photons show QPO at frequency 6.522 Hz. 
On the 10th March, the QPO is originated by the both the pre-shock and post-shock photons; where as on 
the 11th March only post-shock photons take part on QPO formation.

Intermediate state data of $17^{th}$ May, Fig. 4.9(c) also shows QPO only in the intermediate 
energy (4-15 keV) band. The observed frequency of the QPO is 18.94 Hz. Here also the post-shock 
photons of high energy band are responsible for the QPO generation.

Declining hard state data of $17^{th}$ September, Fig. 4.9(d) shows 8.71 HZ QPO in the all 
three energy bands like Fig. (a), though powers of the QPO at low (2-4 keV) and high 
(15-60 keV) energy bands are weak and noisy; because in the declining hard state photon count 
rate decreases due to the less availability of the supply matter from the companion star. 
Here also like Fig. 4.9(a), both pre-shock and post-shock photons of all energy bands take part 
in the QPO formation process i.e. oscillates.

%\subsubsection{{\bf D.} Compton Cloud oscillations during QPOs in black hole candidates}
\subsection{Compton Cloud oscillations {\it during} QPOs in black hole candidates: A comparison 
with GRS 1915+105}

In \S 4.2.5, we already discussed the origin of QPOs and their 
evolution under theoretical context. Our POS model is the generalized model for explaining 
QPO evolutions (Chakrabarti, Nandi, Debnath, Sarkar \& Dutta 2005, Chakrabarti, Debnath, 
Nandi \& Pal 2008a and Chakrabarti, Dutta \& Pal 2009a) of the transient black hole candidates, 
those show monotonically rising or decreasing nature of QPO frequencies in their rising or 
declining outburst phases. We believe that QPOs in the stellar mass black hole candidates 
originate due to shock oscillations and the frequency is proportional to the 
inverse of the matter infall time from the location of the shock to the black hole horizon.

In our current work (Debnath, Chakrabarti \&  Pal, 2010c), we show the nature of the oscillations 
of the Compton cloud (i.e., CENBOL) during QPOs. The oscillation of the CENBOL not only changes 
the number of emitted photons causing QPOs (since the intercepted photons change),
their other characteristics, such as the temperature and spectral index also change within 
a single complete oscillation. We demonstrate this by taking the example of QPOs observed 
in GRO J1655-40 and various classes of GRS 1915+105 and demonstrate that the number of photons 
intercepted also oscillate. Our procedure is sufficiently general and is applicable to any 
black hole candidates and also for propagating shocks.

In order to probe what really goes on during a QPO, we consider the lightcurves of GRO 
J1655-40 and various classes of GRS 1915+105 and picked up regions when the QPO frequency 
is fairly constant. If $\nu_{QPO}$ is the frequency of QPO, a chunk of data of duration
$T_{QPO}=1/\nu_{QPO}$ was taken and was divided into 8 sub-chunks corresponding to eight 
average phase bins of the oscillations. For each such sub-chunk with time interval $\Delta$T
equal to 1/8$^{th}$ of $T_{QPO}$, we calculate the count rate and the temperature 
of the black body component of the spectrum. We then plot the variation of the count rate 
and temperature as a function of the mean phase. This directly shows that the 
spectral characteristics are also oscillating along with QPOs. This procedure cannot 
be used for very high frequency QPOs because of low number statistics (i.e., fewer photons 
in each mean phase of the oscillation) unless the source is unusually bright.

\begin{figure}[h]
\vskip -0.0cm
\centering
\hspace{0.0cm}
\includegraphics[height=2.8in,width=4.0in,angle=0]{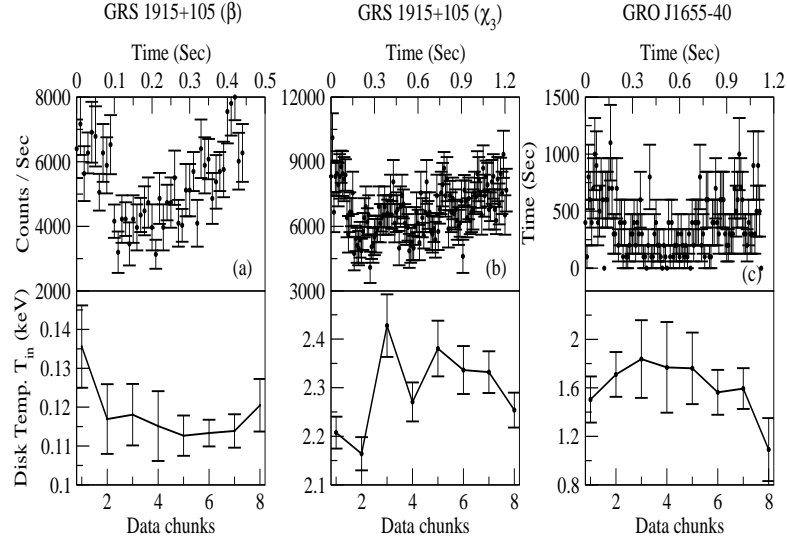}
\caption{(a-c): Variation of the Count rate and disk black body temperature
for (a) GRS 1915+105 $\beta$ class, (b) GRS 1915+105 $\chi_3$ class and (c) GRO J1655-40 
are shown (Debnath et al. 2010c).}
\label{kn : fig4.10}
\end{figure}

Figure 4.10 shows the variation of the photon count rates and disk black body temperature 
for GRS 1915+105 $\beta$ class, GRS 1915+105 $\chi_3$ class and GRO J1655-40 
data. For Fig. 4.10(a), we select GRS 1915+105 data of August 31, 1997 (Obs ID:20402-01-44-00),
where we found QPO at 2.23 Hz. Each chunk contains  data of a time interval ($\Delta$T) 
of [(1/2.23)*(1/8)] = 0.056 sec. Similarly, Fig. 4.10(b), shows the results of October 14, 1997 
GRS 1915+105 data (Obs ID:20402-01-50-00), where we found QPO at 0.849 Hz. Here length of each
data chunk is of 0.152 sec. Fig. 4.10(c) is for the GRO J1655-40 data of 9th March, 2005 
(Obs ID:91702-01-01-03). Here we observe QPO at 0.886 Hz, so each time chunk duration is 
0.14 sec. The QPO fitted results and parameters are summarized in Tab. 4.2.
In each of the three plots of the Figure, we observe count rate variation by about 
$\sim 50\%$. At the same time, we also observe variations in disk black body temperatures 
and photon indices. These clearly show that sometimes depending on the accretion rate, the 
count rate may decrease even when the temperature decreases (i.e., the shock recedes away 
from the black hole).

\begin{table}[h]
%\small
\scriptsize
\centering
\caption{\label{table4.2} QPO fitted results for 3 data sets} %of Figure 4.10
\vskip 0.2cm
\begin{tabular}{|l|c|c|c|c|c|c|c|c|}
\hline
Obs. & Obs. Id & UT Date & MJD &$\nu_{QPO}$&$\Delta\nu_{QPO}$&Q&RMS Amp.&$\Delta$T\\
\hline
1 & 20402-01-44-00&31/08/1997&50691.2557&2.300&0.328&7.01&9.35&0.056\\
2 & 20402-01-50-00&24/11/1997&50776.5543&0.849&0.130&6.53&11.6&0.152\\
3 & 91702-01-01-03&09/03/2005&53446.0653&0.886&0.132&6.71&15.5&0.140\\
\hline
\end{tabular}
\end{table}

\section{Spectral Analysis}

For the spectral analysis we mainly used 3 - 25 keV ``{\bf Standard 2}" mode data from 
RXTE Proportional Counter Unit 2  (PCU2). In general, the black hole energy spectra (2-25 keV)
are modeled with `diskbb' and `power-law' components, though sometimes the best fit can be 
obtained by adding a Gaussian component $\sim 6.5$~keV (for Iron-line). 

\begin{figure}[h]
\vskip -0.5cm
\centering
\hspace{0.0cm}
\includegraphics[height=2.8in,width=2.8in,angle=0]{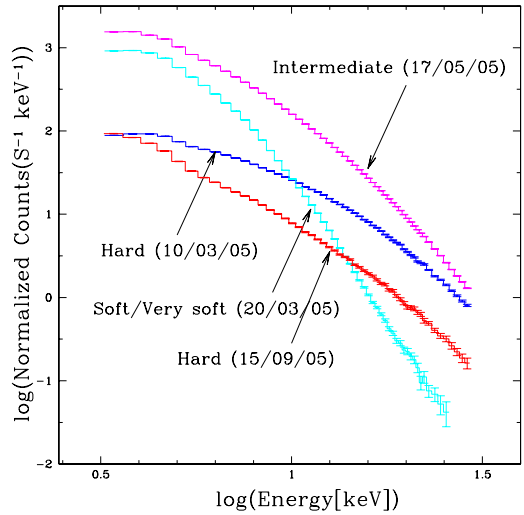}
\caption{RXTE/PCA 3-25 keV spectra of 2005 GRO J1655-40 outburst. Four states are seen.}
\label{kn : fig4.11}
\end{figure}

In our communication (Debnath et al. 2008a) on 2005 GRO J1655-40 outburst, we made 
a detailed spectral analysis of the $150$ PCA observations of $123$ days 
in mainly 3 - 25 keV range. Based on the degree of importance of the black body and the 
power-law components, we divide the entire outburst in four spectral states, namely, 
{\it hard, soft, very soft} and {\it intermediate} in the sequence: hard $\rightarrow$ 
soft/very soft $\rightarrow$ intermediate $\rightarrow$ hard. In our previous 
communications (Chakrabarti et al. 2005, 2006a, 2008a) we published some significant
spectral results as well.

\subsection{Spectral data reduction and fitting}

For the spectral study we use PCU2 `standard 2' mode (FS\_4a*) data of RTXE/PCA instrument.
It is to be noted that for the spectral study background subtraction is very important. 
For generating background file from the raw data we need to run FTOOLS task `runpcabackest'. 
At first we generate spectral file (*.pha) from raw data and background files independently.
Then we need to rebin these pha files using FTOOLS task `rbnpha'. Then these rebinned data
and background pha files along with response file (*.rsp) can be loaded in FTOOLS XRONOS 
package (XSPEC) and fitted with physical models. 

Black hole energy spectra in low energy X-ray band ($<$ 30 keV), can be fitted with 
multi-colour disk black body (diskbb) and power-law models. Low energy spectra ($<$ 10 keV) 
are normally dominated by diskbb component and high energy spectra ($>$ 10 keV) spectra are
dominated by power-law component. In our spectral analysis, most of the cases, we used 
``diskbb + power-law" models. Sometimes for the best-fit, we added an extra component 
in the form of `Gaussian' (for $Fe$-line) at $\sim 6.5$~keV or `CompST' (for Compton cloud). 
The best-fit of the data, can be defined as when the value of the fitted reduced $\chi^2$ is 
$\sim 1$. Also we checked the requirement of extra model component for the fitting with 
the {\it `ftest'} task. Table 4.3 summarized `ftest' results for the spectral fittings. 
We choose the combination of the model components for which the F-test probability is the lowest.

\begin{table}[h]
%\small
\scriptsize
\centering
\caption{\label{table4.3} F-test results for the 4 set of spectra of Fig. 4.12}% {\Large CHECK???}}
\vskip 0.2cm
\begin{tabular}{|l|cc|cc|c|c|}
\hline
\hline
Obs. Id${^**}$&Set 1$^*$& $\chi^2_{red.}$ \& d.o.f.$^*$ &Set 2$^*$&$\chi^2_{red.}$ \& d.o.f.$^*$ &ftest results$^\dagger$&Remarks\\
\hline
I &diskbb+po&4.009, 50&diskbb+CompST+po&2.582, 47&8.690, 1.078e-04& \\
I &diskbb+po&4.009, 50&diskbb+ga+po&1.019, 48&70.438, 5.266e-15&Set2 used\\
\hline
II &diskbb+po&3.621, 45&diskbb+CompST+po&2.716, 33&0.916, 0.542& \\
II &diskbb+po&3.621, 45&diskbb+ga+po&2.210, 43&10.043, 2.636e-04&Set2 used\\
\hline
III &diskbb+po&2.778, 50&diskbb+CompST+po&0.936, 47&30.811, 3.644e-11& \\
III &diskbb+po&2.778, 50&diskbb+ga+po&0.934, 48&49.804, 1.955e-12&Set2 used \\
\hline
IV &diskbb+po&8.172, 50&diskbb+CompST+po&1.179, 47&92.924, 8.981e-20&Set1 used\\
IV &diskbb+po&8.172, 50&diskbb+ga+po&4.99, 49&51.358, 3.665e-09& \\
\hline
\end{tabular}
\noindent{
\leftline {$^*$diskbb: disk black body, po: power-law, ga: Gaussian,
CompST: Sunyaev-Titarchuk Comptonization model,} 
\leftline {~d.o.f.:degrees of freedom.}
\leftline {$^{**}$Observation Ids, I: 90704-04-01-00, II: 91702-01-08-00, III: 91702-01-57-00G, 
IV: 91702-01-79-00}
\leftline { $^\dagger$F-test was done between Set1 and Set2 model fitted results and
F statistic value \& probability values are given.}}
\end{table}

\subsection{Model fitted spectral results}

Initial rising hard state spectra (from $25^{th}$ February 2005, MJD = 53426 to $12^{th}$ 
March 2005, MJD = 53441) were fitted with ``diskbb + power-law" and 
``diskbb + Gaussian + power-law" models. From the spectral fittings, we can conclude 
that at the very beginning days of this outburst the sub-Keplerian disk component 
was dominated, but as the day progresses, the Keplerian component became stronger. 
Another evidence for this is that in the beginning (rising hard state) 
we found monotonically increasing QPO frequencies in the timing data. 

From the spectral fitting we found hard to soft state transition on $13^{th}$ March, 2005 
(MJD = 43442). This soft/very soft state (from $13^{th}$ March 2005 to $15^{th}$ May 2005, 
MJD = 53505) was completely dominated by black body photons (diskbb component). The power-law 
signature in this state was very little. So, we fitted this state spectra by the standard  and simple 
``diskbb + power-law" model. 

\begin{figure}[h]
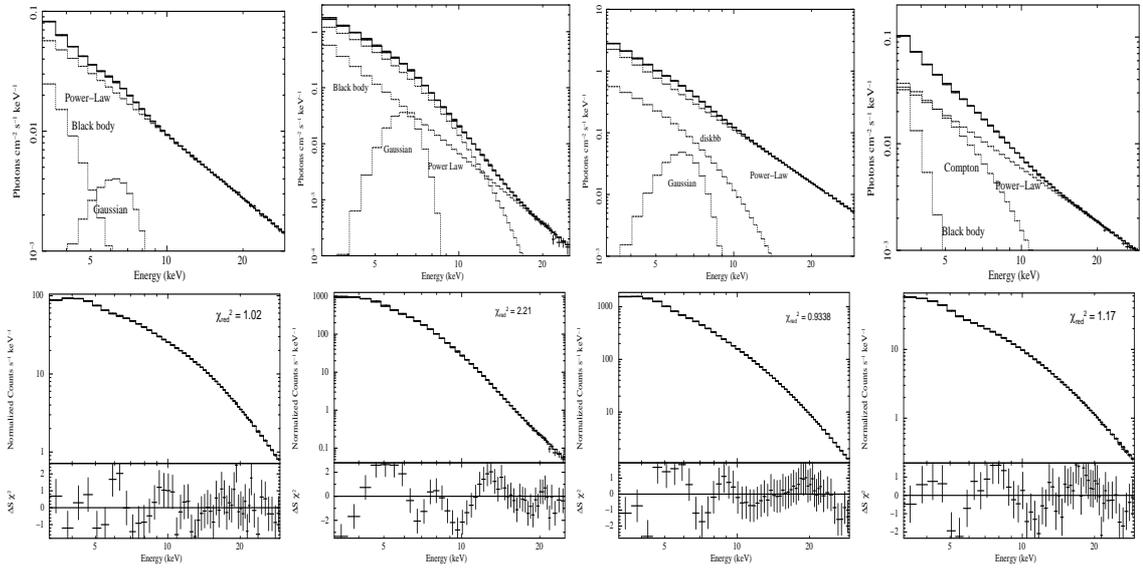

\vskip -0.0cm
\centering
\hspace{0.0cm}
\includegraphics[height=1.42in,width=1.45in,angle=270]{fig4.12al.ps}
\includegraphics[height=1.42in,width=1.45in,angle=270]{fig4.12bl.ps}
\includegraphics[height=1.42in,width=1.45in,angle=270]{fig4.12cl.ps}
\includegraphics[height=1.42in,width=1.45in,angle=270]{fig4.12dl.ps}\vspace{0.1cm}
\includegraphics[height=1.42in,width=1.42in,angle=270]{fig4.12ar.ps}
\includegraphics[height=1.42in,width=1.42in,angle=270]{fig4.12br.ps}
\includegraphics[height=1.42in,width=1.42in,angle=270]{fig4.12cr.ps}
\includegraphics[height=1.42in,width=1.42in,angle=270]{fig4.12dr.ps}
\caption{(a-d) The Figure shows four sets of 3-25 keV PCA fitted spectra. In the top panels we show the
model fitted spectra with various components and in the bottom panels we show corresponding 
reduced $\chi^2$ plots. The diskbb, Gaussian and power-law model fitted spectra 
from the left to the right (a) of 10th March, 2005 (Obs ID:90704040100), (b) of March 20th, 2005 (Obs ID:91702-01-08-00), 
(c) of May 17th, 2005 (Obs ID:91702015700), and (d) show diskbb, CompST and power-law models 
fitted spectra of September 17th, 2005 (Obs ID:91702017900) (Debnath et al. 2008a).}
\label{kn : fig4.12}
\end{figure}

We observed the soft to intermediate state transition on $16^{th}$ May, 2005 (MJD = 53506). This 
intermediate state (from $16^{th}$ May 2005 to $11^{th}$ September 2005, MJD = 53624) was
dominated by ``diskbb + power-law", sometimes for the best-fit an extra component Gaussian
was added. In this state,  we observed QPOs in the timing data, due to local shock oscillations.

From the spectral fitting we observed another state transition from intermediate to 
hard state on $12^{th}$ September, 2005 (MJD = 53625) during declining state. 
This hard state (from $12^{th}$ 
September 2005 to $16^{th}$ October 2005, MJD = 53659) shows interesting spectral features.
Initial few days of this state were fitted by ``diskbb + CompST + power-law" models. The extra 
component CompST (Sunyaev-Titarchuk Comptonization model) was needed because of the presence of 
Compton cloud in the disk, which was different from the cloud generated by the power-law. After 
this the spectra were fitted with  normal ``diskbb + power-law" or ``diskbb + gaussian + power-law" 
models.

\begin{table}[h]
%\small
\scriptsize
\centering
\caption{\label{table4.4} Spectral fit results for the 4 sets of spectra of Fig. 4.12}
\vskip 0.0cm
\begin{tabular}{|l|c|c|c|c|c|}
\hline
Obs. Id & UT Date &Photon & \multicolumn{2}{|c|}{Flux (in $10^{-9} ergs~cm^{-2}~s^{-1}$)}&$\chi^2_{red}$ \\
\cline{4-5}
       &         &Index($\Gamma$)&~~~~3-10keV~~~~ & 10-25keV& \\
\hline
90704-04-01-00 &2005-03-10&1.766&1.849&1.533&1.019 \\
91702-01-08-00 &2005-03-20&6.164&2.416&0.804&2.210 \\
91702-01-57-00G&2005-05-17&2.711&46.86&11.75&0.934 \\
91702-01-79-00G&2005-09-17&1.691&1.838&1.065&1.179 \\
\hline
\end{tabular}
\end{table}

Figure 4.12 shows 4 sets of spectral plots with model components and their 
corresponding $\chi^2_{red}$ plots. These four sets of the spectra were taken from 4 different 
spectral states' observational data. A summary of the spectral fitted parameters
of the four sets of spectra are presented in Tab. 4.4.

We also fitted the spectra of a few days with TCAF model (using the same procedure as 
in Chakrabarti \& Mandal, 2006b) and from our fittings we obtained accretion matter flow rates 
for the two components of accretion flows (the Keplerian or disk rate and the sub-Keplerian 
or halo rate). These results are summarized in the following Tab. 4.5.

\begin{table}[h]
%\small
\scriptsize
\centering
\caption{\label{table4.5} Theoretically fitted GRO J1655-40 outburst data with TCAF model}
\vskip 0.2cm
\begin{tabular}{|l|c|c|c|}
\hline
\hline
Date& MJD & Halo rate & Disk rate \\
\hline
Mar. 3, 2005 & 53432.7930& 0.015& 0.0100\\
Mar. 9, 2005 & 53438.0539& 0.004& 0.0400\\
Mar. 10, 2005& 53439.7400& 0.009& 0.0650\\
Mar. 11, 2005& 53440.6780& 0.0043& 0.5100\\
Mar. 12, 2005& 53441.5109& 0.0100& 0.7800\\
May 16, 2005& 53506.9531& 0.0550& 1.5640\\
May 20, 2005& 53510.0994& 0.0560& 1.7400\\
Sep. 16, 2005& 53629.3761& 0.0038& 0.0700\\
Sep. 19, 2005& 53632.4557& 0.0029& 0.0300\\
Sep. 22, 2005& 53635.4696& 0.0065& 0.0140\\
Sep. 29, 2005& 53641.0383& 0.009&0.0100\\
\hline
\end{tabular}
\end{table}

\subsection{Daily variation of fluxes and spectral parameters}

\begin{figure}[h]
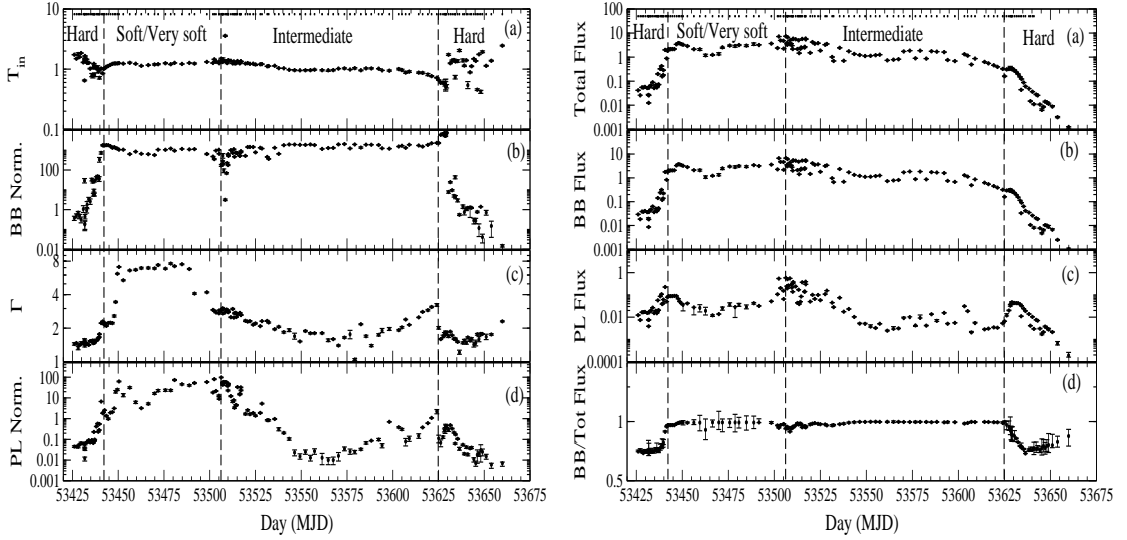

\vskip 0.6cm
\centering
\includegraphics[height=2.8in,width=2.8in,angle=0]{fig4.13a.eps}\hspace{0.2cm}
\includegraphics[height=2.8in,width=2.8in,angle=0]{fig4.13b.eps}
\caption{(A-B): Left panel (A) shows the fitted parameters of RXTE 3 - 25 keV PCA spectra
are plotted with time (MJD). The panels are: (a) disk black body temperature (T$_{in}$) in keV,
(b) disk black body normalization, (c) power-law photon index ($\Gamma$) and (d) power-law 
normalization plotted with day. Logarithmic scale is used in the y-axis and the error bars 
are at $1 \sigma$ level. Right panel (B) shows the daily flux variations. 
The panels are: (a) 3 - 25 keV total flux, (b) 3-10 keV bolometric disk black body flux, 
(c) 10-25 keV power-law flux and (d) the ratio of the total and power-law fluxes. In the 
soft/very soft and intermediate states the total flux is dominated by the black body flux. 
Only in the hard states of the rising and declining phases the ratio is less than unity. 
Here we also use logarithmic scales along the y-axis (Debnath et al. 2008a).}
\label{kn : fig4.13}
\end{figure}

To get a clear idea about the spectral evolution in the GRO J1655-40 
outburst of 2005, we fitted PCU2 data of RXTE/PCA instrument in 3 - 25 keV energy range for the 
total of $150$ observations of $123$ days data. From the spectral fits we found disk black 
body temperature ($T_{in}$) in keV, normalization factors for black body, power-law photon 
index ($\Gamma$) and its normalization factor components. Also we found Gaussian and 
CompST fitted parameters, where they were needed for the fitting. We also calculated 
disk black body (mainly in 3-10 keV energy range), power-law (mainly in 10-25 keV energy 
range) and total (in 3-25 keV energy range) fluxes.

From the nature of the variation of the power-law indices and the disk black body 
components of Fig. 4.13(A), we classified the full outburst into four spectral states: 
(i) Hard state from from 25$^{th}$ of February, 2005 (MJD = 53426) to 
12$^{th}$ of March, 2005 (MJD = 53441), (ii) Soft/Very soft state from 13$^{th}$ of March, 
2005 (MJD = 53442) to 15$^{th}$ of May, 2005 (MJD = 53505). (iii) Intermediate  state is 
from 16$^{th}$ of May, 2005 (MJD = 53506) to 11$^{th}$ of September, 2005 (MJD = 53624) 
and finally (iv) Hard state from the 12$^{th}$ of September, 2005 (MJD = 53625) till 
16$^{th}$ of October, 2005 (MJD = 53659). We kept the hydrogen column density (N$_{H}$) 
fixed at 7.5$\times$ 10$^{21}$ atoms cm$^{-2}$ and the systematics at $0.01$.

Daily variations of the fitted parameters are plotted in Fig. 4.13(A), which clearly reveals 
the justification of separating the full outburst in the above mentioned four states. The 
panels (a-d) are respectively the black body temperature $T_{in}$ in keV, the black body 
normalization factor, the photon index $\Gamma$ and the power-law normalization (plotted 
in the log scale along Y-axis).

\begin{figure}[h]
\vskip -0.2cm
\centering
\hspace{0.0cm}
\includegraphics[height=3.0in,width=3.2in,angle=0]{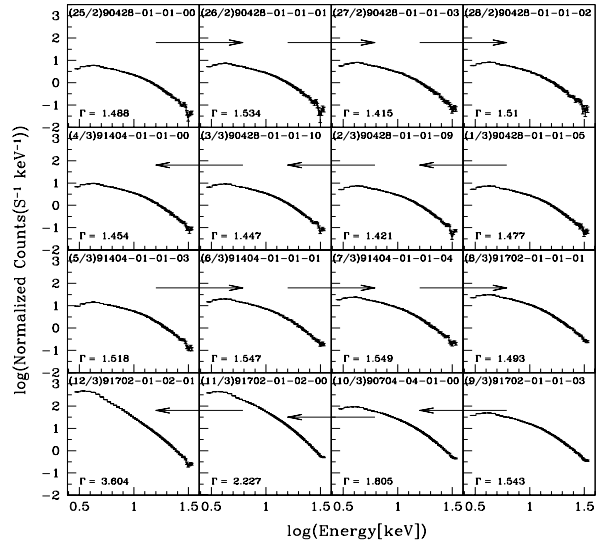}
\caption{ The spectral evolution for the hard state in the rising phase from 25$^{th}$ of February, 2005 
to 12$^{th}$ of March, 2005. Though the spectral slope remains more or less constant for the 
first twelve days or  so, the intensity of radiation increase steadily. Subsequently, the 
spectral slope (i.e., photon index $\Gamma$) and total flux both increase monotonically 
(Chakrabarti et al. 2006a).}
\label{kn : fig4.14}
\end{figure}

Daily variations of the total flux (panel a), black body flux (panel b) and the power-law flux 
(panel c) are shown in Fig. 4.13(B). The panel (d) shows how the ratio of the black body to 
total flux changes daily. Generally in the soft and very soft states the ratio is almost unity, 
indicating the dominance of the soft component in both of these divisions. However, in Fig. 4.13(A), 
we observe a distinct difference in power-law normalization and power-law index in 
these two states. The QPOs are observed only in certain days of the intermediate state. 
It may be noted that $\Gamma$ obtained right in the middle of the soft/very soft state is 
unphysically high ($>4$). We believe that this is due very poor statistics (e.g., only one 
good PCU of RXTE, PCU2 was working properly and photon energy was $>20$keV) rather than 
any unusual absorption at high energies. We find that other workers (Saito et al. 2006) 
also reported a high photon index for these observations.

In Fig. 4.14, the spectral variations for the initial hard state are shown. From the Figure it 
is observed that though in the initial few days spectral slopes and fluxes were remained 
more or less constant, but these parameters were increased monotonically in the subsequent 
days. This is another indication that the Compton cloud moved inward during the rising phase 
of the outburst.

	\reseteqn
	\resetsec
	\resetfig
	\resettab
\alpheqn
\resec
\refig
\retab
\def\k{{\bf k}}
\def\aug{{\tilde{\cal H}}}
\def\ahat{{\mathaccent "7E {\cal A}}}
\def\bhat{{\mathaccent "7E {\cal B}}}
\def\chat{{\mathaccent "7E {\cal F}}}
\def\dhat{{\mathaccent "7E {\cal D}}}
\def\shat{{\mathaccent "7E {\cal S}}}
\def\proj{{\cal P}}
\def\ket{\vert \vert  \{ \emptyset \} \rangle}
\def\ket2{\vert \vert \otimes \{ R \} \rangle}
\def\sqr{$^{2}$}
\def\T{$T$}
\def\P{$P$}
\def\Q{$Q$}
\def\G{$G$}
\def\M{{\bf M}}
\def\H{{\bf H}}
\def\I{{\bf I}}
\def\Pr{{\bf P}}
\def\Tr{{\bf T}}
\def\eq{\enskip =\enskip}
\def\pls{\enskip+\enskip}
\def\mns{\enskip -\enskip}
\def\oper{O^{\bf k}_{R}}
\def\aug{\tilde{\cal H}}
\def\envfn{\chi^{\alpha}_{RL}}
\def\struc{S^{\alpha}_{RL,R^{\prime}L^{\prime}}}
\def\rR{r_{R}}

\newpage
\markboth{\it GX 339-4: A Transient Low-mass X-ray Binary}
{\it GX 339-4: A Transient Low-mass X-ray Binary}
\chapter{GX 339-4: A Transient Low-mass X-ray Binary}

%**********************extra*******************

%**********************extra*******************
GX 339-4 is a well known stellar-mass Galactic black hole candidate. This bright variable 
X-ray source was first observed by MIT X-ray detector aboard OSO-7 Satellite (Markert 
et al. 1973) in the energy range of 1-60 keV. GX 339-4, a transient Low Mass X-ray Binary 
(LMXB) system is located at $(l,b) = (338^\circ.93 \pm 0.20,-4^\circ.27\pm 0.20)$ (Markert 
et al. 1973) with R.A.=$16^h58.8^m\pm 0.8^m$ and Dec.= $-48^\circ48' \pm 12'$. Optical 
spectroscopic study indicates mass function of the source is $M$ = $5.8\pm0.5~M_\odot$ 
and distance $D$ = $6$~kpc (Hynes et al. 2003, 2004).

\section{The recent outburst}

Since its discovery, GX 339-4 has undergone several outburst phases during which the source 
was observed in energy bands from radio to high energy X-ray to reveal its 
multi wavelength nature (Makshima et al. 1986, Liu et al. 2001, Homan et al. 2005). During the 
RXTE era, this source exhibits frequent X-ray outbursts (1998, 2002/2003, 2004/2005, 2006/2007) 
at 2-3 years of interval with very low luminosity states in between each episode. The complex 
outburst profile in each epoch generally begins and ends in the low/hard state, which is quiet 
common in other outbursting black hole  candidates (e.g., GRO J1655-40, XTE J1550-564). 
This general behaviour is understood to be due to sudden variation of viscosity in the system 
(Mandal \& Chakrabarti 2010) which causes the Keplerian disk to rush in. The temporal 
and spectral state evolution during the previous outbursts are well studied (Nowak et al. 1999, 
Belloni et al. 2005, Motta, Belloni \& Homan 2009) but still not clearly understood due to the 
complex nature of the accretion process in the system. Miller et al. (2004) has reported 
this black hole candidate to be a rotating Kerr-black hole with a low spin value of around zero.

Recently, after 3 years of remaining in the quiescent state, the source GX 339-4 became again 
X-ray active on January 03, 2010 with the first detection by MAXI/GSC onboard ISI (Yamaoka et al.
2010). Immediately after the announcement of X-ray trigger, RXTE has started monitoring the
transient source from Jan 12, 2010 (Tomsick, 2010). Other scientific space satellites (Swift,
INTEGRAL, etc.) are also observing this source in regular intervals.
We studied timing and spectral properties of the source in its initial outburst phase, 
using RXTE archival data (Debnath et al. 2010a). Now we present these results obtained from 
the current rising phase of the GX 339-4 outburst using RXTE data of PCA and ASM instruments 
for an analysis of a total of $30$ observational data which are spreaded over the period 
of $5{th}$ March, 2010 (MJD = 55260) to $29{th}$ April, 2010 (MJD = 55315). From our detailed 
timing and spectral analysis of the initial phase of the outburst, we classified the initial 
rising phase of the GX 339-4 outburst mainly in three spectral states, hard, hard/intermediate 
and soft/intermediate. Spectra up to $9^{th}$ of April 2010 were at 'pure' hard state. After that 
up to $17^{th}$ of April it is in hard/intermediate state and then finally after April $18$ 
it moved to soft/intermediate state. During this outburst phase, on $10^{th}$ of April 2010, 
we observe a state transition between hard-hard/intermediate states and on $18^{th}$ 
of April 2010 another state transition between hard/intermediate-soft/intermediate spectral 
states. The results are given in Debnath et al. (2010a).

\section{Results of the timing analysis}

For studying timing properties of the recent GX 339-4 outburst, we use archival data of 
RXTE PCA and ASM instruments. For the PCA data analysis, we mainly use most stable and well 
conditioned proportional counter unit 2 (PCU2) data. NASA's standard analysis software package 
(Heasoft 6.8) is used for the detailed timing as well as spectral analysis of the current 
GX 339-4 outburst. In the study of the temporal properties of any black hole candidate, finding 
quasi-periodic oscillations (QPOs) in power density spectra (PDS) is as important as observation 
of the photon count/flux variations. Generally, most of the black hole candidates (BHCs) show low 
frequency QPOs in the frequency range between $0.01$ to  $30$ Hz (Remillard et al. 1999, 
Debnath et al. 2008a). Some BHCs (GRO J1655-40, GRS 1915+105, XTE J1859+226, H1743-322 etc.) 
also show the signatures of dual high frequency QPOs in 3:2 ratio (Remillard et al. 1999, 
Cui et al. 2000, Strohmayer 2001, Belloni et al. 2006, Remillard et al. 2006).

We studied temporal properties of the initial phase of the GX 339-4 outburst in a 
systematic manner. First we extracted lightcurves from the `Event mode' data using 
FTOOLs task `sefilter' in both $1$ sec and $0.01$ sec time bins. Later $0.01$ sec time bin 
lightcurves are used to make PDS using FTOOLs task `powspec', which actually do FFTs of the 
lightcurves. From the PDS we tried to find QPO signatures and further it was fitted with the 
model `Lorentzian' line profile to find centroid QPO frequency, FWHM and its highest power.

During initial outburst phase, initially the source was in the hard state without 
any signature of QPO in the PDS up-to 3rd week of March, 2010. We first observed $102$~mHz 
QPO on March $22$, 2010. After that the QPO frequency monotonically increased up to $5.69$~Hz 
(observed on April $17$, 2010) within $\sim 26$ days. Afterwards, QPOs were on and off 
sporadically (e.g., $5.739$ Hz on April $18$, 2010, $5.677$ Hz on April $22$, 2010 and 
$5.913$ Hz on April $29$, 2010), always remaining roughly at the same value.
In the soft/intermediate state, we have not seen any signature of QPOs from 
April $19$th - $21$st, 2010 and from April $23$th - $28$th, 2010. In the PDS, we 
also interestingly found $3$ harmonics of primary node $2.42$ Hz QPO on April $13$, 2010 
(Yu,  2010), generally $2$ harmonics of a QPO frequency are seen.

The monotonically rising nature of the QPO frequencies (from $22$nd of March 2010 to $17$th 
of April 2010) in the initial outburst phase was fitted with theoretical model of the 
propagating oscillatory shock (POS). We followed the same procedure, what we had used to 
fit the QPO evolution of GRO J1655-40 initial rising phase of 2005 outburst (Chapter 4), 
where we also observed monotonically increasing nature of QPO frequencies. 
From the model fitting, we calculated initial shock location at $1500~r_g$ 
(on $5{th}$ March, 2010, where we first observed QPO at $0.102$ Hz), where $r_g = 2GM/c^2$, 
Schwarzschild radius and final shock location at $172~r_g$ on $17{th}$ April, 2010 and 
initial and final compression ratios, $R$ ($4$ \& $0.99$), which are inverse of the shock 
strengths, $\beta$ ($= 1/R$). The compression ratio (R) is ratio between the pre-shock 
and the post-shock densities, i.e., $R=\rho_-/\rho_+$, where, $\rho_-$ and $\rho_+$ are the 
densities in the pre-shock and the post-shock flows. We also found shock wave moves towards 
the black hole in at a roughly constant speed of $\sim 10~m~s^{-1}$. 

\begin{figure}[t]
\vskip 0.5cm
\centering
\hspace{0.0cm}
\includegraphics[height=2.3in,width=2.7in,angle=0]{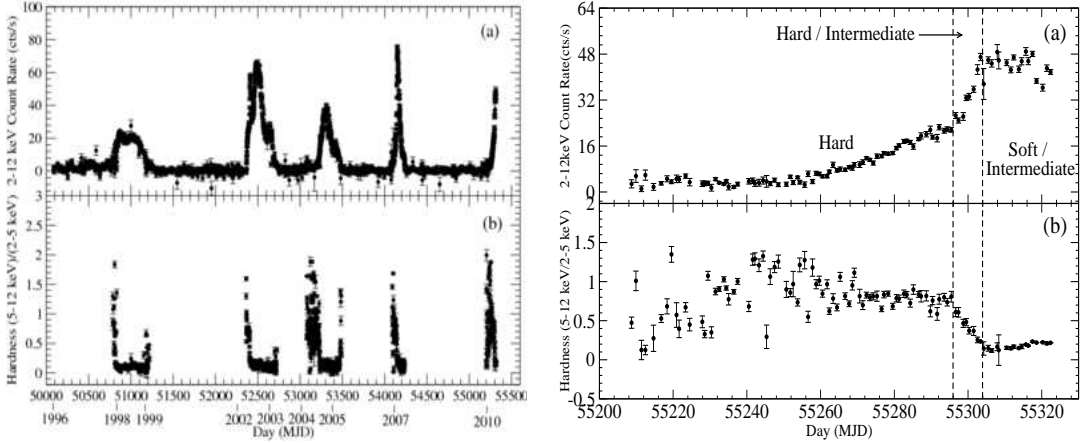}\hskip 0.4cm
\includegraphics[height=2.3in,width=2.7in,angle=0]{fig5.1b.eps}
\vspace{0.0cm}
\caption{Left panel (Fig. A) shows the RXTE ASM lightcurve and hardness ratio plots for the 
black hole candidate GX 339-4 during the RXTE era (1996 to April, 2010) and right panel (Fig. B) 
shows the same thing in recent 2010 GX 339-4 outburst. In each Figure, the top panel (a) is the 
2-12 keV ASM lightcurve and (b) is the hardness ratio (5-12 keV vs. 2-5 keV count ratio) as a 
function of the MJD of the event are plotted. The vertical dashed line at MJD $55296$ in (B), 
indicates the day of state transition from hard to hard/intermediate spectral states. The
other vertical dashed line at MJD $55304$ indicates the state transition from
hard/intermediate to soft/intermediate (Fig.B: Debnath et al. 2010a).}
\label{kn : fig5.1}
\end{figure}

Since on the $18^{th}$, $22^{nd}$ and $29^{th}$ of April, $2010$, we observed QPOs in 
$\sim 5.7$~Hz, nearly same QPO frequency as was observed on the $17^{th}$ April, last day 
of our theoretical model fit, we can conclude that the shock wave became steady and oscillated 
nearly at the same locations in the soft/intermediate state. According to our model in 
these three observation days, we found that the value of the compression ratio (R) to become 
less than unity, which is unphysical for a stable propagating shock. As $R$ became less 
than unity at $\sim 175 r_g$, quite before the black hole horizon, we can conclude that the shock 
was not able to fall into the black hole. This was also observed during the initial 
rising phase of 2005 GRO J1655-40 outburst (Chakrabarti et al. 2008a, Debnath et al. 2008a).

\subsection{ASM Light Curve}

RXTE ASM data (Levine et al. 1998) were downloaded and analyzed in different energy bands 
(2-3, 3-5, 5-12 \& 2-12 keV). In Fig. 5.1, 2-12 keV ASM lightcurves (counts/sec) and the ASM 
hardness ratios (ratio between the photon count rates in 5-12 keV  and 2-5 keV bands) are plotted 
for the RXTE ASM data (Fig. 5.1A) from 6th January, 1996 (MJD = 50088) to 6th May, 2010 
(MJD = 55322) for the BHC GX 339-4 and for its ASM data of the current 2010 outburst 
(Fig. 5.1B) (from 12th January 2010 (MJD = 55208) to 6th of May 2010). In the RXTE 
era so far the transient BHC GX 339-4 has shown five outbursts including the present one.

\subsection{Hardness-Intensity Diagram}

We extracted and analyzed the PCA (Jahoda et al., 1996) data of the initial rising phase of 
the outburst from point $A$, $5^{th}$ of March 2010 (MJD = 55260) to April 29, 2010 (MJD = 55315). 
Figure 5.2 shows RXTE PCA hardness-intensity diagram (HID) (Mandal \& Chakrabarti 2010). 
In Fig. 5.2, the PCA 3-20 keV count rate of the above mentioned period against the X-ray 
color (PCA count ratio between 6-20 keV and 3-6 keV energy bands) are plotted. It is evident 
from the Figure that initially the source was at 'pure' hard state up to $9^{th}$ of April 2010. Subsequently,
it went to the hard/intermediate state for a short duration of period from April $10$, 2010 
(point B) to April $17$, 2010. Finally from $18^{th}$ of April 2010 (point $C$) it went to the 
soft/intermediate state. We observed low-frequency QPOs in all the three spectral states, 
which are most common phenomena for low-mass X-ray black hole binaries.

\begin{figure}[h]
\centering
\includegraphics[height=2.0in,width=2.5in,angle=0]{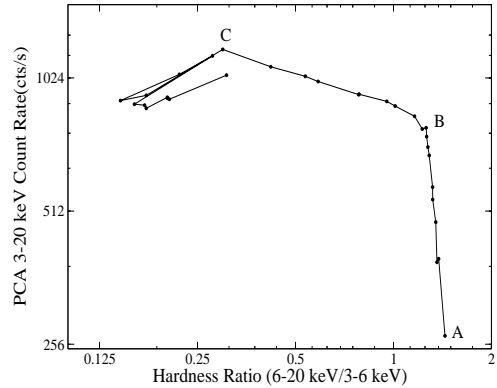}
\caption{The Hardness Intensity Diagram (HID) observed with RXTE PCA. The y-axis of the plot 
represents the PCA count rates in 3-20 keV energy band and the x-axis represents the hardness 
ratio between the count rates of the energy bands of 6-20 keV and 3-6 keV bands. 
Both axes are in logarithmic scales. Point $A$ means MJD 55260, start date of our observation,
$B$ means MJD 55296 and $C$ means MJD 55304. Source was in a 'pure' hard state between the days of 
point A \& B, was at a hard/intermediate state between the days of point B \& C and finally after 
point C, it moved to the soft/intermediate state (Debnath et al. 2010a).}
\label{kn : fig5.2}
\end{figure}

\subsection{Light curves and Hardness, Softness diagrams}

For the detailed study of the X-ray photon count variations over the initial rising phase 
of the outburst, we extracted 2-15 keV (0-35 Channels) PCA lightcurves with time bin of 1 sec. 
Also, we studied both the hardness and softness ratio variations. These lightcurves were 
extracted using FTOOLS taskes ``sefilter" \& ``seextrct" on Event mode data. To plot the 
lightcurves ``fplot" task was used. The hardness and the softness ratio diagrams were made by 
extracting lightcurves for three energy bands: $A:0-8$ channels ($2-4$ keV), $B:9-35$ channels 
($4-15$ keV) and $C:36-138$ channels ($15-60$ keV). A hardness diagram is the plot between $C/A$ 
(HR2) vs. $B/A$ (HR1), while the softness diagram is the plot between $B/C$ (HR3) vs. $A/C$ (HR4). 

\begin{figure}[h]
\vskip -0.2cm
\centering
\includegraphics[height=1.45in,width=1.44in,angle=0]{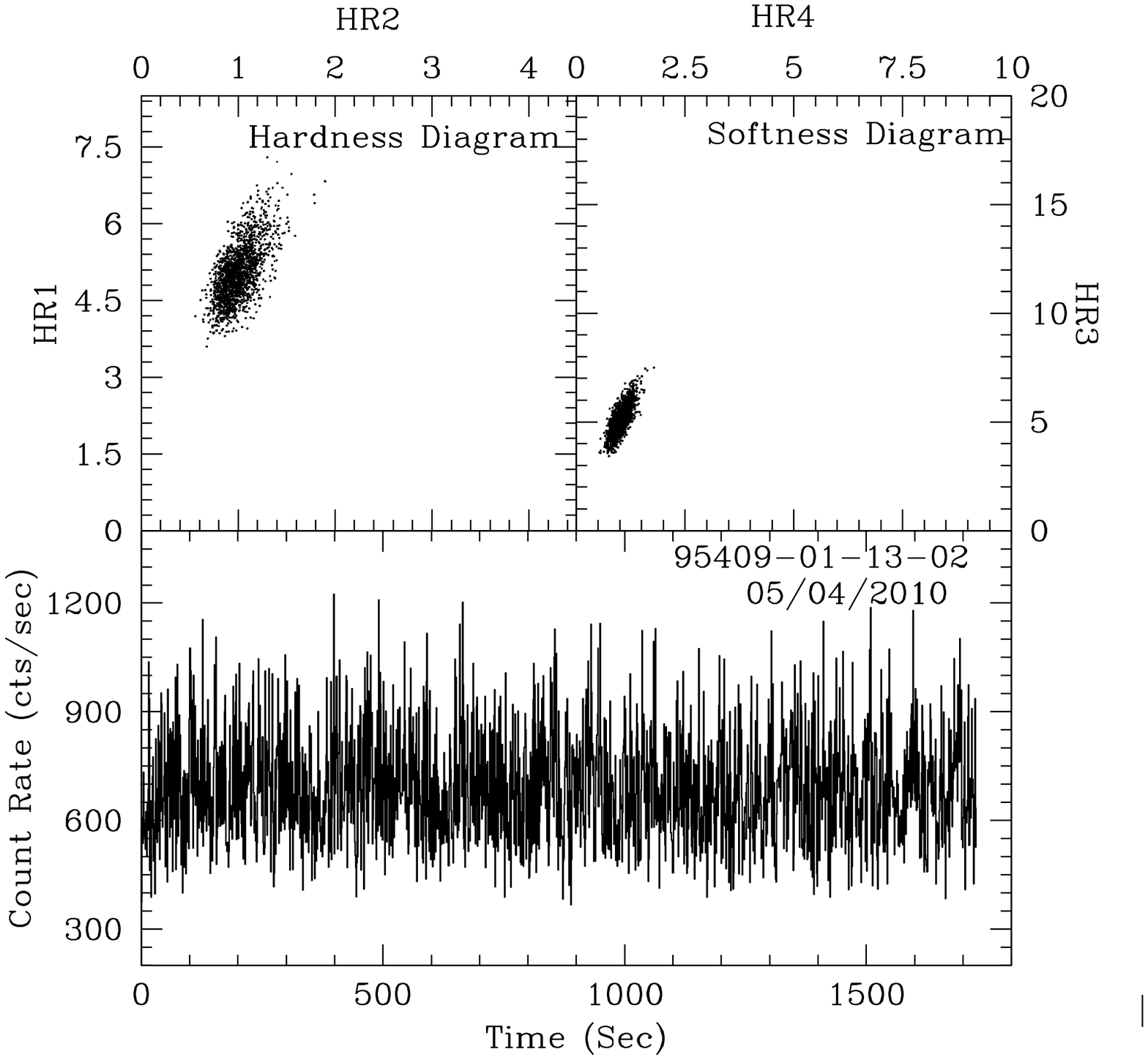}
\includegraphics[height=1.45in,width=1.44in,angle=0]{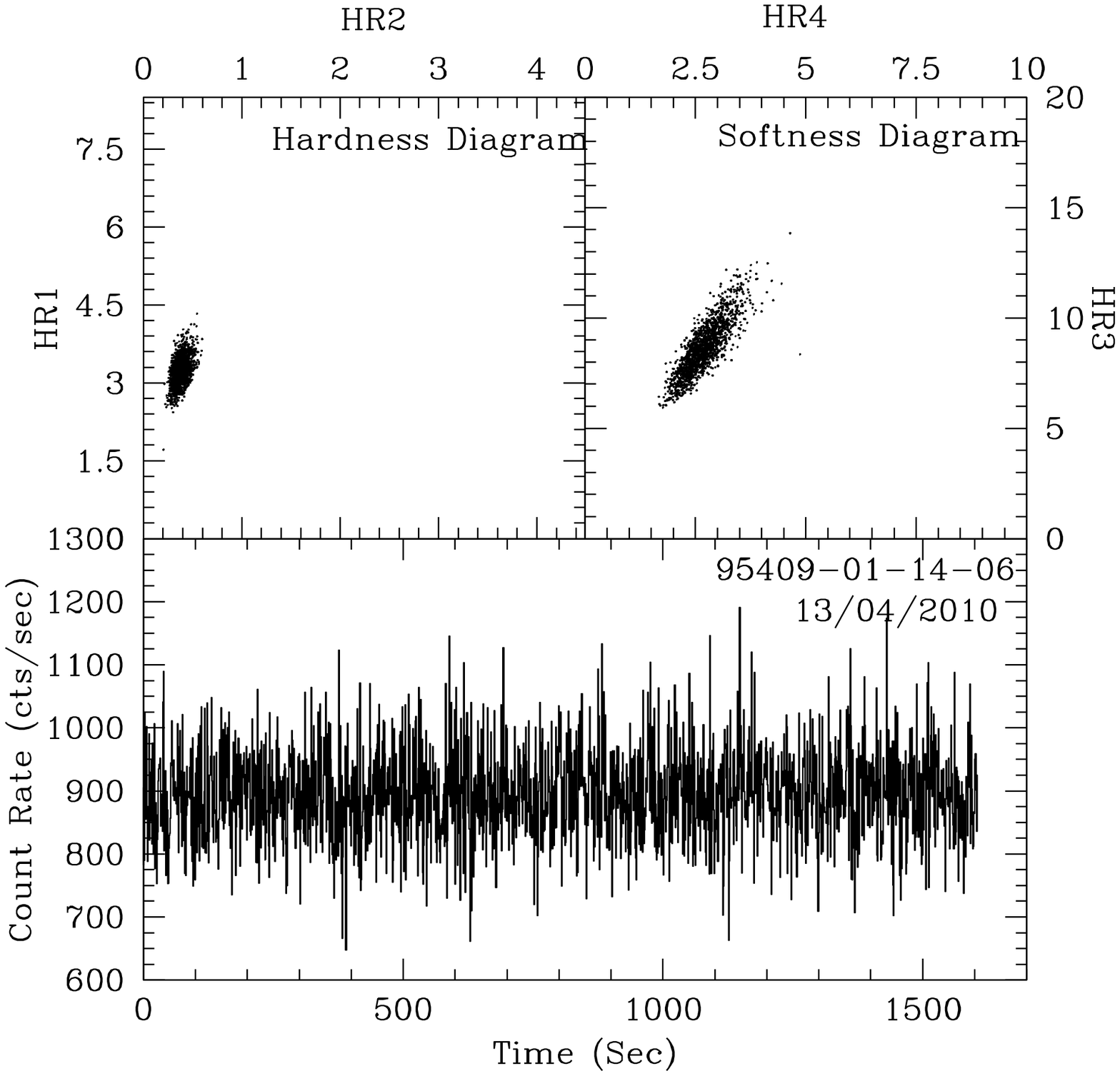}
\includegraphics[height=1.45in,width=1.44in,angle=0]{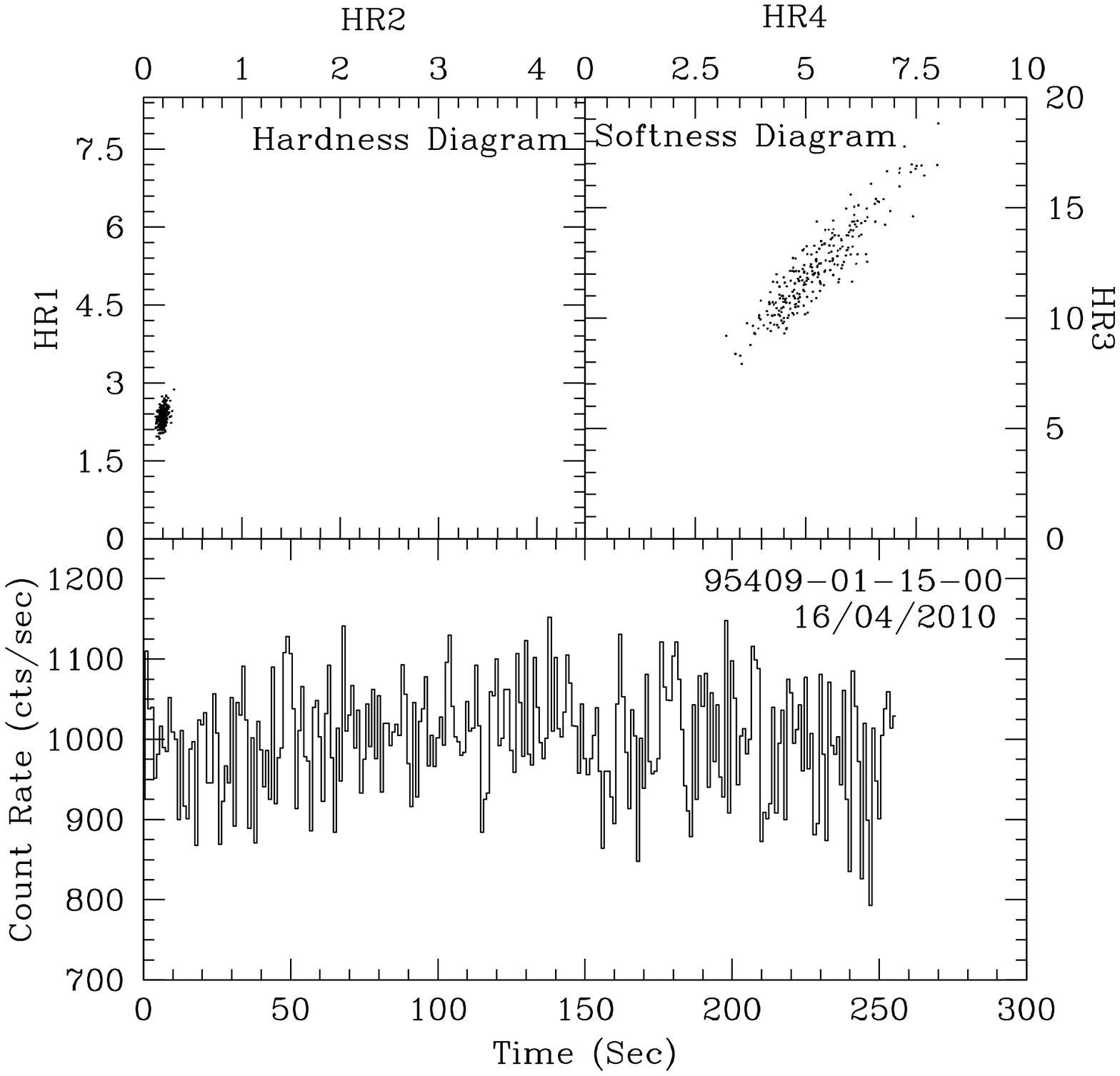}
\includegraphics[height=1.45in,width=1.44in,angle=0]{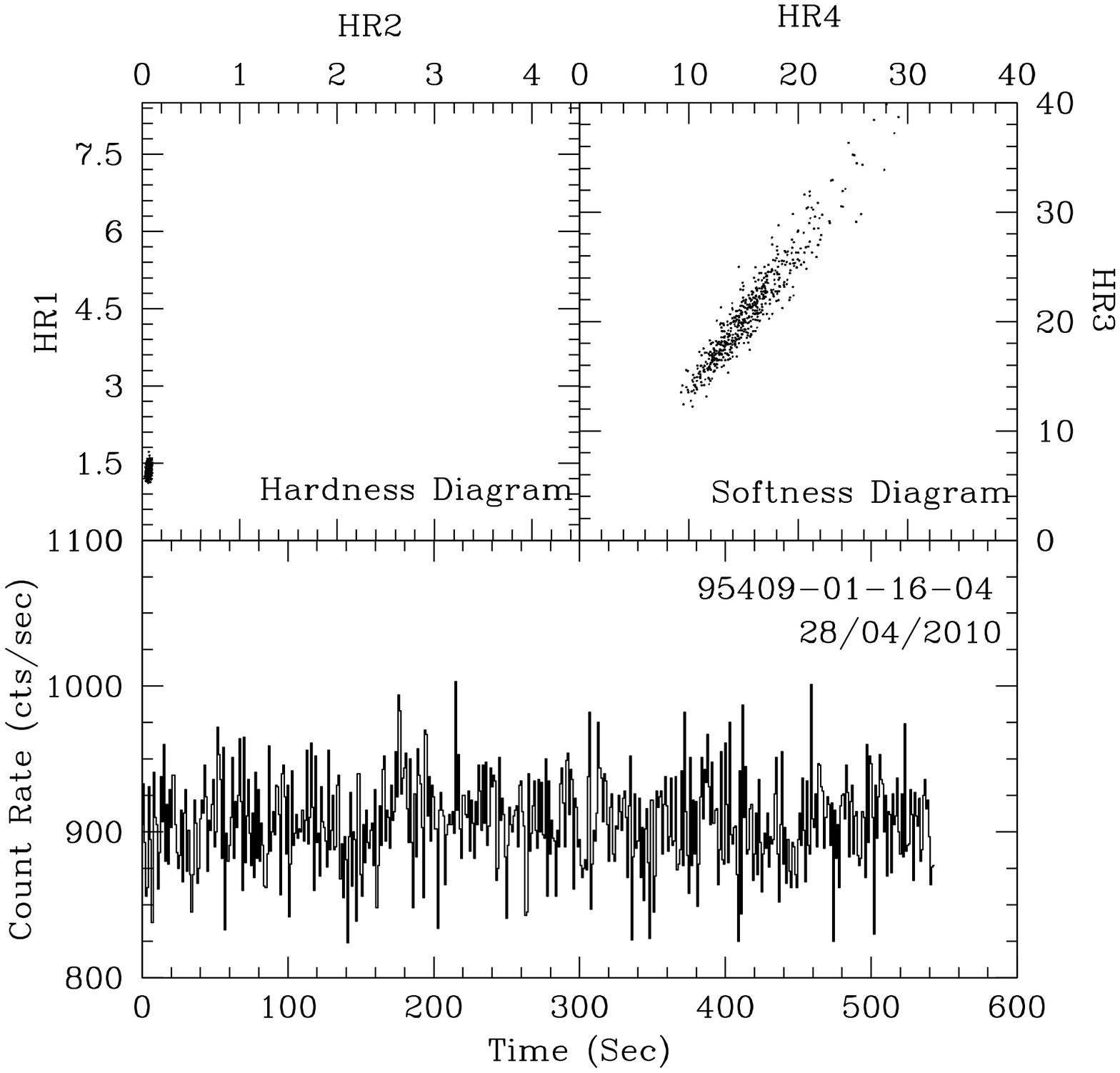}
\caption{(a-d):In the lower panel, 2-15 keV (0-35 Channels) PCA lightcurve and in the
upper panel the hardness and softness diagrams are plotted. In hardness diagrams,
the dashed horizontal ($B=A$), vertical ($C=A$) and the dotted line ($B=C$) are for
reference purpose. In softness diagram they represent $C=A$, $B=C$ and $A=B$
respectively. Observation dates are: (left to right) (a) on 5th of April,
2010 (Obs ID:95409-01-13-02), (b) on 13th April, 2010 (Obs ID:95409-01-14-06), 
(c) on 16th of April, 2010 (Obs ID:95409-01-15-00), and 
(d) on 28th of April, 2010 (Obs ID:95409-01-16-04). Fig. (a) is taken from the 'pure' 
hard state, where as Figs. (b) \& (c) are taken from the hard/intermediate state and remaining 
Fig. (d) is taken from soft/intermediate states. In Fig. (d) x, y scale for the hardness and 
softness diagrams are changed.}
\label{kn : fig5.3}
\end{figure}

Our motivation of splitting the lightcurves in above mentioned energy bands is to separate 
the low energy Keplerian disk photons ($\lsim 4$ Kev) from the high energy sub-Keplerian 
disk photons. Thus, $A$ would be emitted mostly from the Keplerian component. The component 
$B$ would be emitted from the region where the moderate thermal Comptonization of the Keplerian
photons took place. The component $C$ would be emitted from the region which was definitely
depleted or enhanced during state transitions as it was represented at the higher energy side
of the pivotal energy [$\sim 15$ keV] in the spectrum. Thus, these diagrams are not directly
connected to the spectral states -- rather, they are connected to the geometry, i.e., the
number of soft photons produced by the Keplerian disk ($\sim A$) and the seed photons 
intercepted by the `Compton cloud' [$\sim (B + C)$] and the number of scatterings they
undergo ($\sim B$ or $\sim C$).

From the Fig. 5.3(a-d), we got a rough idea about the evolution of accretion rates of high 
energy, the sub-Keplerian (halo) and low energy, the Keplerian (disk) seed photons during 
the initial rising phase of the outburst.

\subsection{Power Density Spectra}

Power Density Spectra (PDS) were generated by using ``powspec" task of XRONOS package 
with a normalization factor of `$-2$' to reduce `white' noise from the rms fractional
variability. We also used a geometrical factor of $-1.02$ to get a nearly equispaced 
logarithmic frequency bins. The power has been obtained in units of $rms^2/Hz$. The 
lightcurve of X-ray variability from which PDS was obtained were binned at $0.01$ sec 
time resolution so that the Nyquist frequency (for higher limit) becomes $50$ Hz. 
It general practice for studying PDS, is to find QPO. QPOs are generally Lorentzian 
type (Nowak 2000, van der Klis 2005) and thus we fitted each PDS by zooming QPO 
peak with Lorentzian profile to find the centroid frequency and width of each observed 
QPO. Table 5.1 shows the observed QPO frequencies and their fitted parameters.

\begin{figure}[h]
\vskip -0.2cm
\centering
\hskip -0.5cm
\includegraphics[height=1.50in,width=1.55in,angle=0]{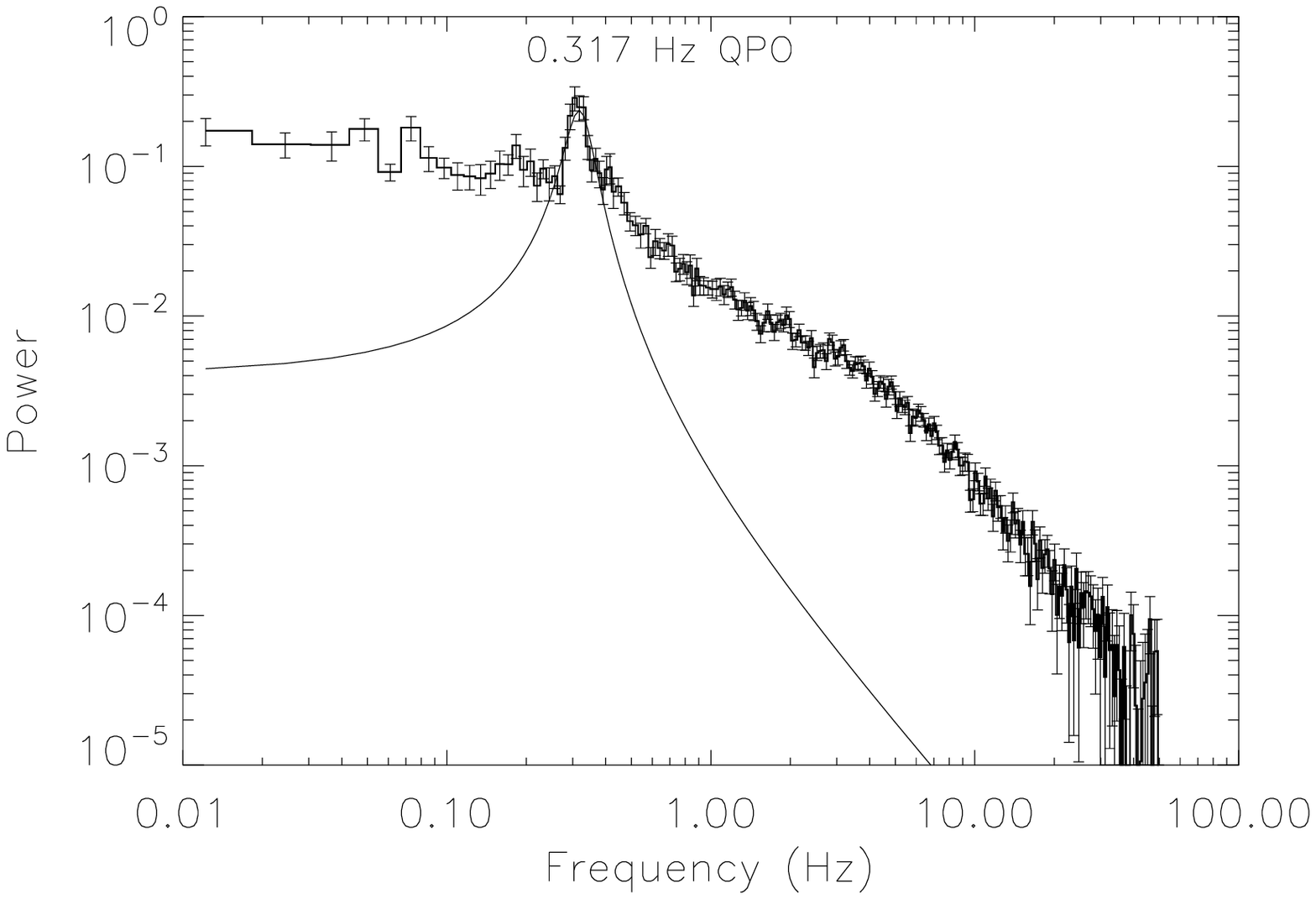}\hskip -0.3cm
\includegraphics[height=1.50in,width=1.55in,angle=0]{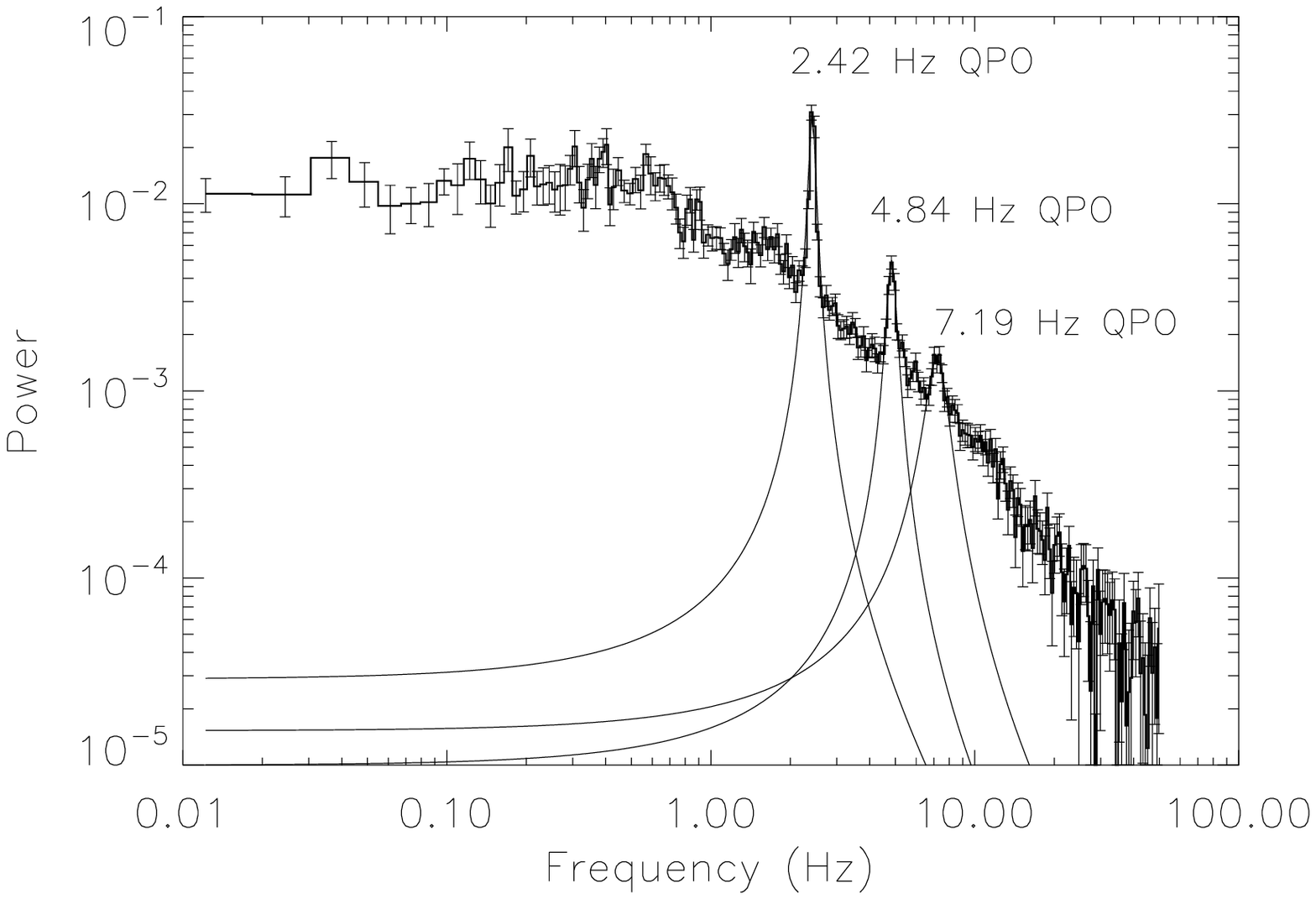}\hskip -0.3cm
\includegraphics[height=1.50in,width=1.55in,angle=0]{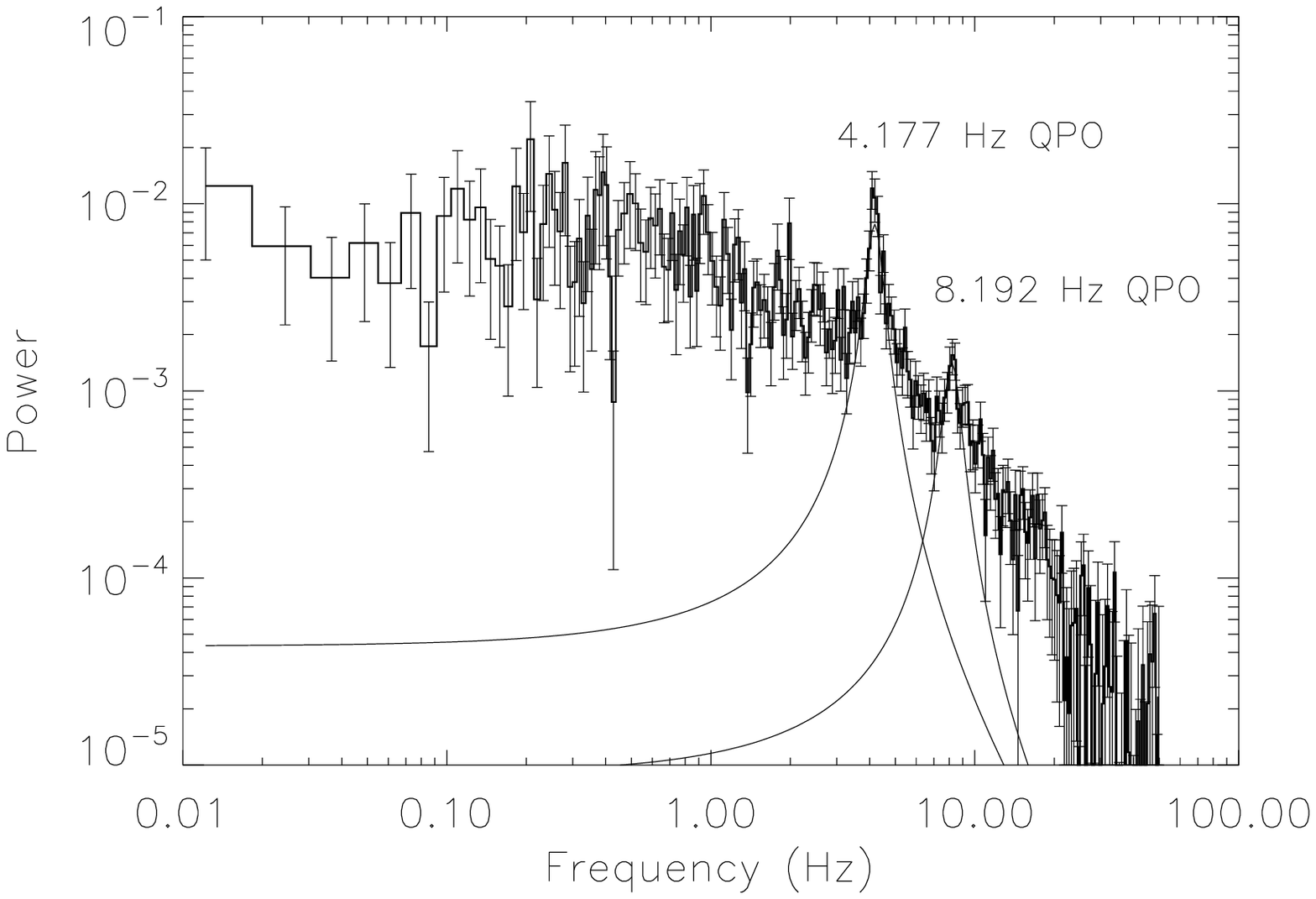}\hskip -0.3cm
\includegraphics[height=1.50in,width=1.55in,angle=0]{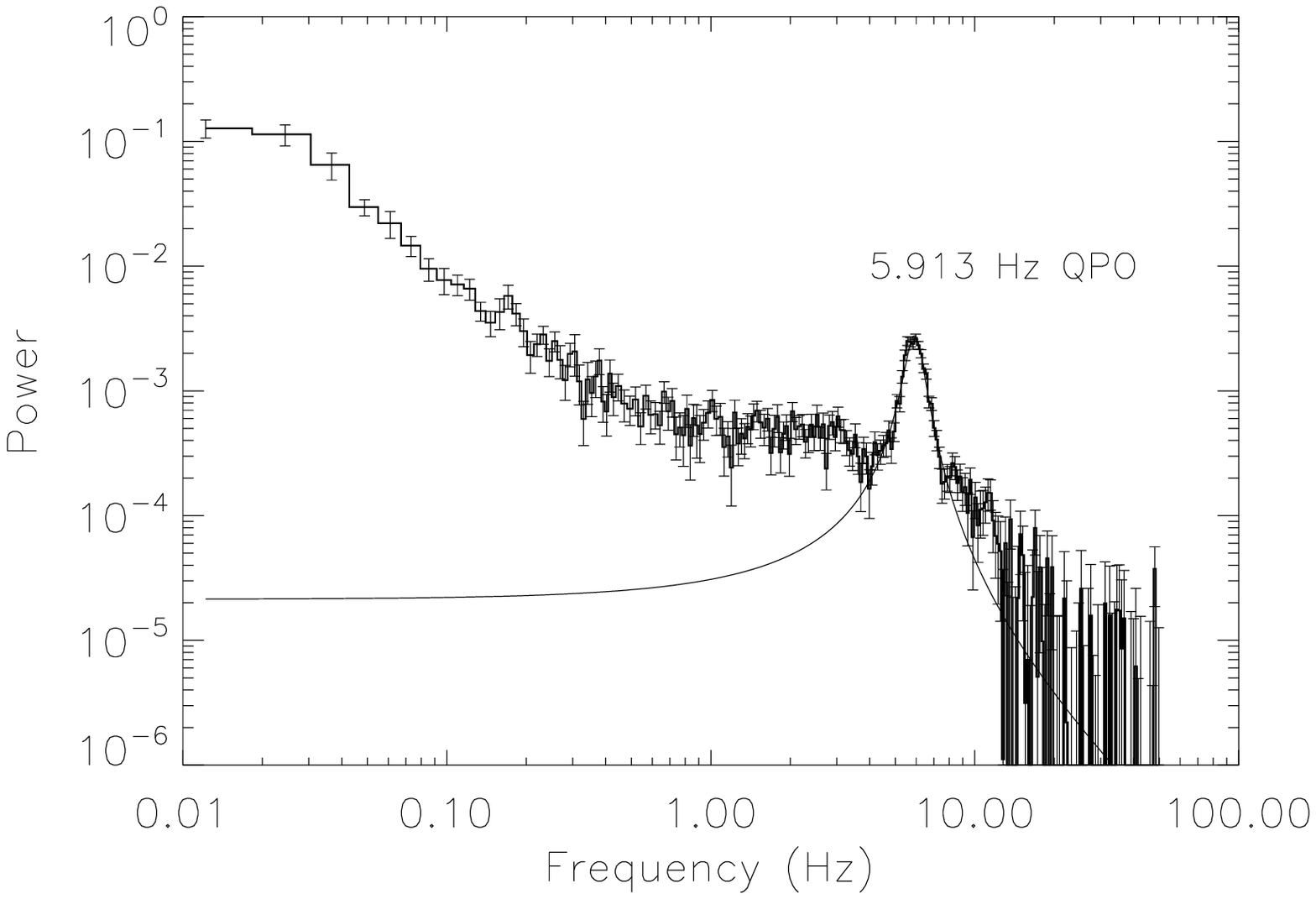}
\caption{(a-d): (a) The model fitted PDS of $5^{th}$ April, 2010 (Obs. ID:95409-01-13-02),
example of a typical hard state. QPO was found at $0.317$ Hz. (b) The model fitted PDS 
of $13^{th}$ April, 2010 (Obs. ID:95409-01-14-06). Here we observed 3 harmonics of the
primary $2.42$ Hz QPO (other two harmonics are $4.84$ Hz and $7.19$ Hz).
(c) The model fitted PDS of $16^{th}$ April, 2010 (Obs. ID:95409-01-15-00), a typical 
day just 1 day before hard-soft/intermediate state transition. The QPOs were observed at 
frequencies $4.177$ and $8.192$ Hz. (d) The model fitted PDS of $29^{th}$ April, 2010 
(Obs. ID:95409-01-16-05), a typical day of soft/intermediate state. Here we observed 
strong signature of intermediate state QPO at $5.913$~Hz frequency.}
\label{kn : fig5.4}
\end{figure}

In Tab. 5.1, we present a summary of the QPO fitted results, where we put the centroid 
frequency ($\nu$) of the QPO, its width ($\Delta\nu$) (both in Hz), the coherence parameter
or Quality factor $Q$ (= $\nu$/$\Delta\nu$) and percentage of RMS amplitude. The RMS amplitude 
($R$) of the fitted QPO was calculated from the relation $R = 100~\sqrt{(I/mean~count~rate)}$, 
where, $I$ is the integral of the Lorentzian profile of the fitted QPO. Our selection criterion
for defining a peak at PDS as QPO was the value $Q$ parameter, for QPO it should have to be more 
than 2 ($Q > 2$).

Figure 5.4(a-d) shows the QPO fitted PDS of the initial rising phase of the outburst, taken from 
different spectral states. First Figure is taken from the 'pure' hard state data, next two are 
taken from the hard/intermediate state data and the remaining last Figure is taken from the 
soft/intermediate state data. In the intermediate state, sometimes we observe QPOs in PDS and 
sometimes we do not. This may be because the generation of the shock due to collision of incoming 
Keplerian matter and receding sub-Keplerian matter.

\begin{figure}[h]
\vskip -0.2cm
\centering
\includegraphics[height=4.0in,width=4.0in,angle=0]{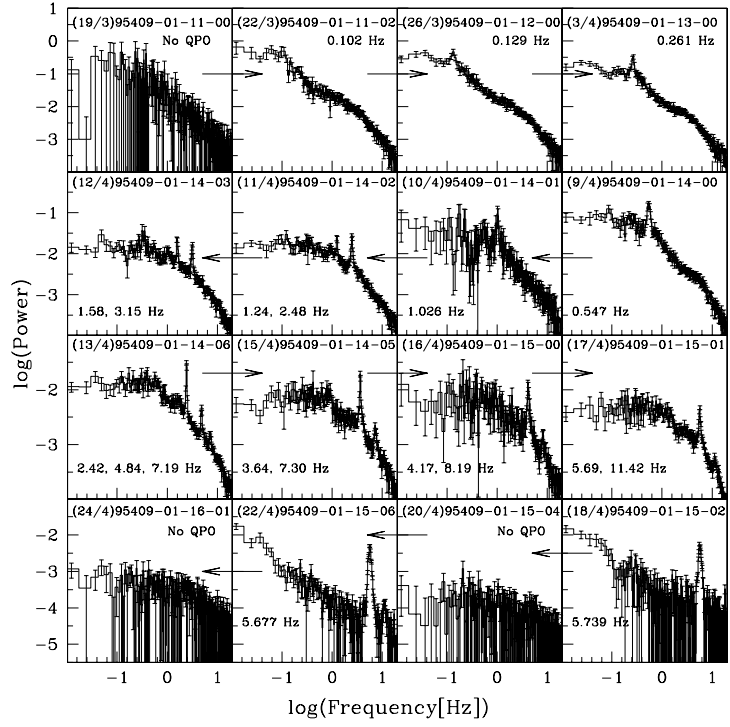}
\caption{The variation of the PDS with QPO frequencies of the initial rising phase of the 
outburst. The dates (dd/mm), the observation ID and the frequency of the QPOs are in the inset. 
Arrows indicate the direction in which the dates are increasing.}
\label{kn : fig5.5}
\end{figure}

Figure 5.5 shows the date-wise QPO evolution of the initial rising phase of the most recent GX 339-4 
outburst. $16$ PDS plots are drawn. In the insets of each PDS observation ID, 
observation date and QPO frequency/frequencies if observed are noted. Arrows indicate the direction 
in which the dates are increasing. Our first plot is of $19^{th}$ March, 2010 data, three days 
before first observed QPO and second plot is of $22^{nd}$ March, 2010 data, where we observed
QPO at $102$~mHz (our first observed QPO). We fitted the QPO evolution up to $17^{th}$ April, 2010
data (where we observed QPO at $5.692$~Hz). After that we also found QPOs at frequencies 
$5.739$~Hz on April 18, 2010, $5.677$~Hz on April 22, 2010 and $5.913$~Hz on April 29, 
2010 (not shown in Figure). These last three observed QPOs are of soft/intermediate state.

%\newpage
\begin{table}[h]
%\small
\scriptsize
\centering
\caption{\label{table5.1} Observed QPO fitted parameters for recent GX 339-4 outburst}
\vskip 0.2cm
\begin{tabular}{|l|c|c|c|c|c|}
%\begin{tabular}{|lccccc|lccccc|}
\hline
\hline
%Obs. & & & & & RMS\\% &Obs. & & & & & RMS \\
%ID$^\dagger$&UT Date&$\nu$&$\Delta\nu$&Q& Amp.\\%&ID$^\dagger$&UT Date&$\nu$&$\Delta\nu$&Q&Amp.\\
Obs. ID&UT Date&$\nu$&$\Delta\nu$&Q& RMS Amp.\\
\hline
95409-01-11-02&2010-03-22&0.102&0.0318&3.207&17.438\\
95409-01-11-03&2010-03-24&0.113&0.0452&2.500&18.879\\
95409-01-12-00&2010-03-26&0.133&0.0443&3.002&18.276\\
95409-01-12-03&2010-04-01&0.202&0.0624&3.237&19.044\\
95409-01-13-00&2010-04-03&0.261&0.0802&3.254&17.158\\
95409-01-13-02&2010-04-05&0.317&0.0845&3.751&16.952\\
95409-01-13-05&2010-04-06&0.363&0.0377&9.628&15.391\\
95409-01-14-00&2010-04-09&0.547&0.1088&5.027&16.346\\
95409-01-14-01&2010-04-10&1.026&0.1261&8.136&14.074\\
95409-01-14-02&2010-04-11&1.241&0.2192&5.661& 9.278\\
95409-01-14-02&2010-04-11&2.488&0.3328&7.475&12.098\\
95409-01-14-03&2010-04-12&1.588&0.2509&6.329& 9.095\\
95409-01-14-03&2010-04-12&3.156&0.3433&9.193& 9.575\\
95409-01-14-06&2010-04-13&2.420&0.1502&16.11& 8.412\\
95409-01-14-06&2010-04-13&4.842&0.4634&10.45& 5.824\\
95409-01-14-06&2010-04-13&7.193&1.4608&4.924& 5.847\\
95409-01-14-04&2010-04-14&2.374&0.1787&13.28& 9.175\\
95409-01-14-04&2010-04-14&4.727&0.7466&6.331& 6.434\\
95409-01-14-05&2010-04-15&3.643&0.2659&13.70& 9.137\\
95409-01-14-05&2010-04-15&7.308&1.1724&6.233& 5.962\\
95409-01-15-00&2010-04-16&4.177&0.6248&6.685& 8.738\\
95409-01-15-00&2010-04-16&8.192&1.3115&6.246& 5.390\\
95409-01-15-01&2010-04-17&5.692&0.9941&5.725& 7.297\\
95409-01-15-01&2010-04-17&11.42&2.9427&3.880& 4.950\\
95409-01-15-02&2010-04-18&5.739&0.5257&10.92& 6.794\\
95409-01-15-05&2010-04-21& - - - - &- - - - &- - - - &- - - - \\
95409-01-15-06&2010-04-22&5.677&0.4053&14.01& 6.150\\
95409-01-15-06&2010-04-22&11.09&1.4958&7.414& 1.814\\
95409-01-16-04&2010-04-28& - - - - &- - - - &- - - - &- - - - \\
95409-01-16-05&2010-04-29&5.913&1.0369&5.703& 6.741\\
\hline
\end{tabular}
\end{table}

\subsection{QPO evolution in the rising phase of the outburst and POS Model}

We fitted monotonically day-wise increasing nature of the QPO frequencies with our 
propagating oscillatory shock (POS) model. We followed the same methodology as was
followed for fitting of the QPO evolution of the 
initial rising phase of 2005 GRO J1655-40 outburst. For the present QPO evolution, we take 
observed QPOs from $22^{nd}$ of March, 2010 to $17^{th}$ of April, 2010. In this 
period, we saw that QPO frequencies were increased monotonically from $102$~mHz to $5.69$~Hz 
with in a period of $\sim 26$ days. Spectral analysis (discussed in the next Section) 
indicates during the above mentioned evolution period photon indices of the spectra were 
less than $2$, i.e., source was at 'pure' hard state (up to $9^{th}$ of April) and hard/intermediate 
state ($10^{th}$ April to $17^{th}$ April). After that on the $18^{th}$ of April 2010, photon 
index became $\sim 2.5$, the source was moved to the soft/intermediate spectral state. 
From the analysis, we found hard to hard/intermediate transition on $10^{th}$ of April 2010 
and hard/intermediate to soft/intermediate spectral state transition on $18^{th}$ of April 
2010. In this soft/intermediate state, we also observed QPOs at frequencies $5.739$~Hz, 
$5.677$~Hz \& $5.913$~Hz respectively on 18th, 22nd \& 29th of April 2010.

\begin{figure}[h]
\vskip 0.6cm
\centering
\includegraphics[height=2.8in,width=5.0in,angle=0]{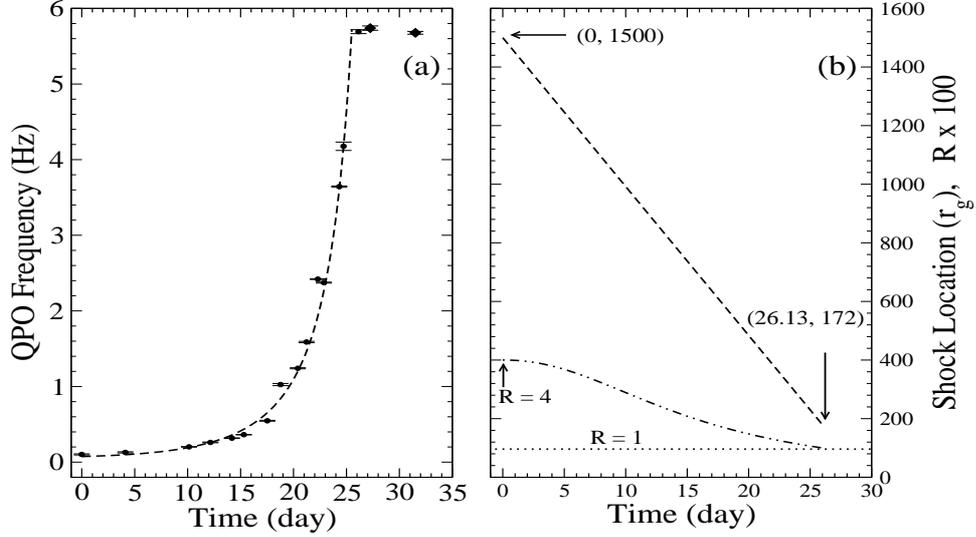}
\caption{(a-b): (a) Variation of QPO frequency with time (in day) of the initial rising phase of 
the outburst with fitted POS model (dotted curve). The diamonds indicate the last 
two observed QPOs (on April 18 \& 22, 2010) in the soft/intermediate spectral state, not included 
in our model fitting, since the shock already achieved its weakest strength with compression ratio 
(inverse of shock strength $\beta$) $R\sim 1$, on earlier day, April $17^{th}$. (b) Variation of the 
shock location (in $r_g$) and compression ratio ($R$) multiplied by $100$ are plotted (Debnath et al. 2010a).}
\label{kn : fig5.6}
\end{figure}

The governing equations for our theoretical POS model fitting for the present GX 339-4 QPO 
evolution are the same as mentioned in the previous Chapter (Eqns. 4.1-4.3) for GRO J1655-40 
evolution, except that here the shock strength ($\beta = 1/R$) 
is assumed to vary by the relation $1/R\rightarrow 
1/R_0 + \alpha t_d^{2.15}$, where $R_0$ is the initial compression ratio, $t_d$ is the time 
in day (taking first observed QPO frequency day as $0^{th}$ day) and $\alpha$ is a very small 
number ($6.8 \times 10^{-4}$), limited by the time in which the shock disappears. We started 
with strong shock ($R_0 = 4$), which becomes unity in $\sim 26$ days on $17^{th}$ of April, 
2010. Physically, shock strength $\beta = 1$, i.e., $R = 1$ means, the system is in equilibrium 
condition with matter densities of pre- and post-shock are same; no propagating shock oscillation 
is possible. Although after $17^{th}$ of April, we found QPOs near around $5.7$~Hz in the 
soft/intermediate state, this is because of local oscillation near about same location due 
to the generation of the shock wave by the incoming Keplerian matter and receding sub-Keplerian 
matter. This type of behaviour is also observed in 2005 GRO J1655-40 outburst, where we found QPOs 
in the intermediate state at frequencies around $19$, $17$ and $13$~Hz.

From the fitting, we calculated initial shock location at $1500~r_g$ (on $5{th}$ March, 2010, 
where we first observe QPO at $0.102$~Hz), where $r_g = 2GM/c^2$, Schwarzschild radius and 
final shock location at $172~r_g$ on $17{th}$ April, 2010 and initial and final compression 
ratios ($4$ \& $0.99$). Since in this current rising phase of GX 339-4 outburst,  
shock strength ($\beta$) reaches its weakest value unity at $\sim 175~r_g$, much before the black 
hole horizon, we can conclude that the shock was not able to fall into the black hole. This was 
also observed in the rising phase of 2005 GRO J1655-40 outburst (Chakrabarti et al. 2008a). From 
our theoretical POS model fitting, we also found that the shock wave was moved towards the black 
hole with a constant velocity ($\sim 10~m~s^{-1}$). 

\subsection{Energy dependent QPO signature}

In order to quantify the energy dependent nature of the power density spectra, we extracted 
the lightcurves using PCU2 data of RXTE PCA instrument in three different energy bands with 
channel numbers 0-8 (2-4 keV), 9-35 (4-15 keV) and 36-71 (15-30 keV) respectively in $0.01$ 
sec time bins and made their PDS using FTOOLS task `powspec' with normalization factor -2 
(to reduce white noise) and geometric area factor -1.02 (to get a nearly equispaced logarithmic 
frequency bins). Our motivation of splitting the 2-30 keV PDS in above mentioned energy bands 
is to separate the low energy Keplerian disk photons ($\lsim 4$ Kev) from the high energy 
sub-Keplerian disk photons. Thus, 2-4 keV PDS would be mostly dominated by the Keplerian 
component, 4-15 keV PDS would be emitted from the region where the moderate thermal 
Comptonization of the Keplerian photons took place and higher energy band (15-30 keV) 
PDS would be dominated by the non-thermal sub-Keplerian component.

\begin{figure}[h]
\vskip -0.2cm
\centering
\hskip -1.0cm
\includegraphics[height=2.8in,width=2.6in,angle=0]{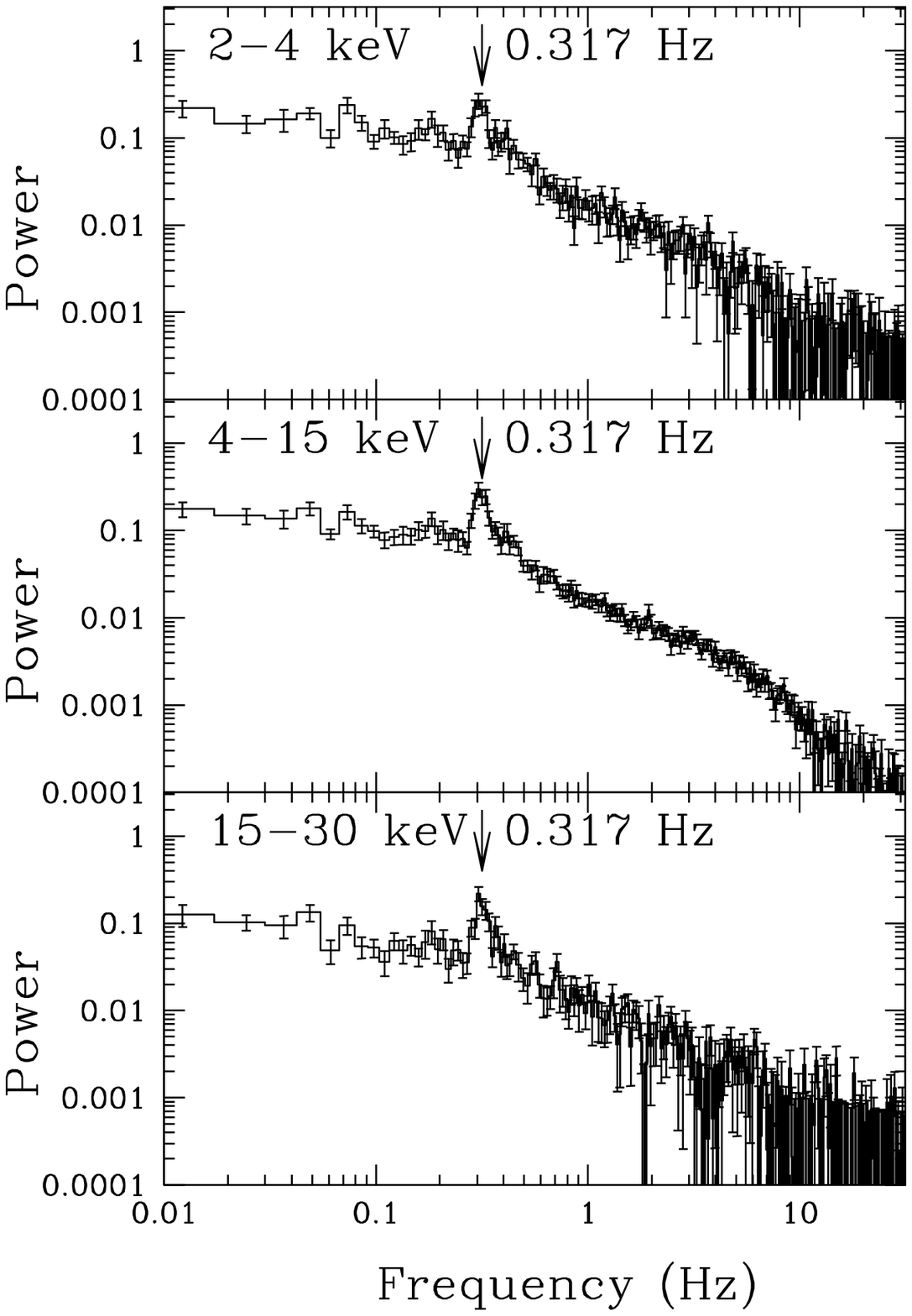}\hskip -2.0cm
\includegraphics[height=2.8in,width=2.6in,angle=0]{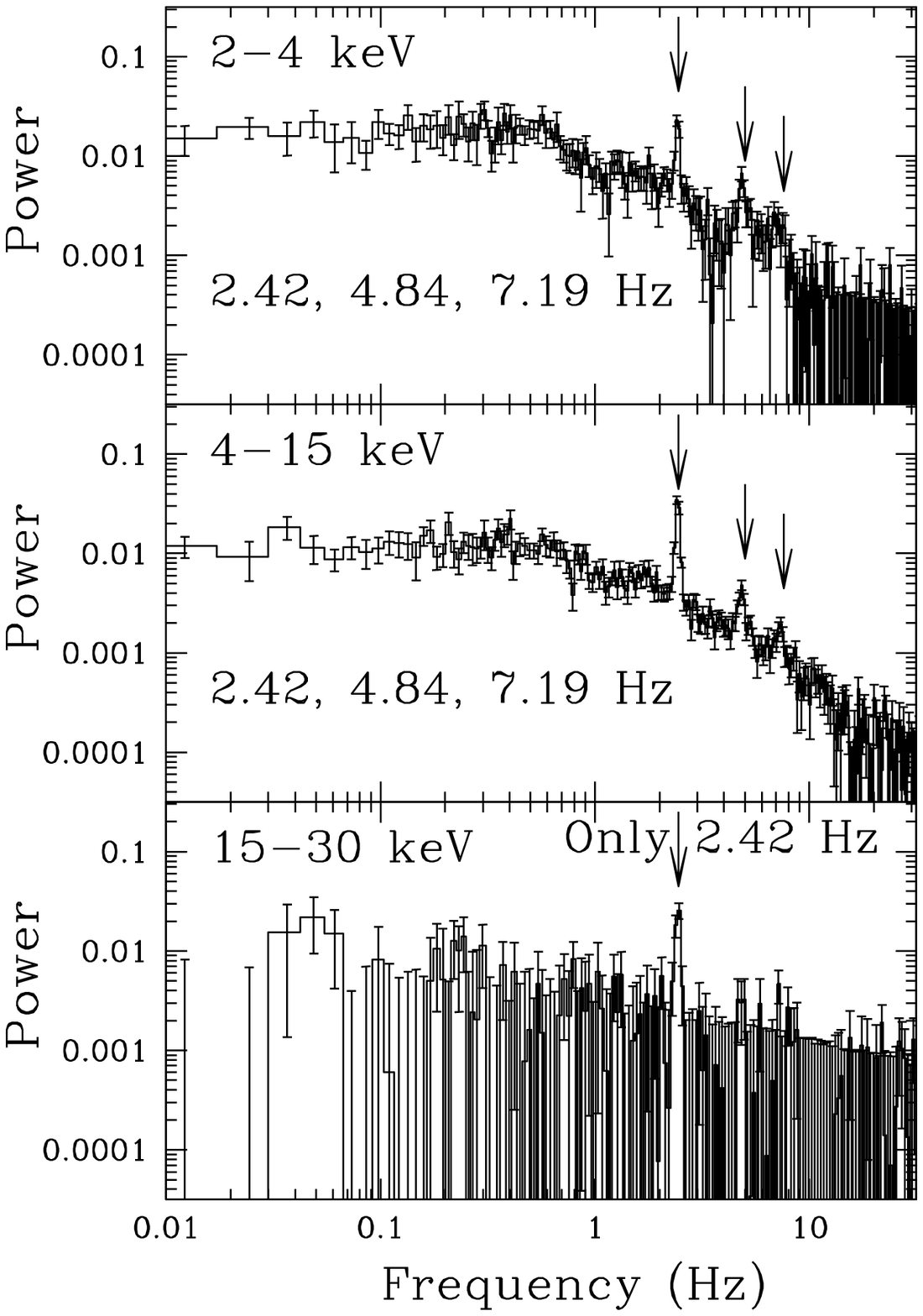}\hskip -2.0cm
\includegraphics[height=2.8in,width=2.6in,angle=0]{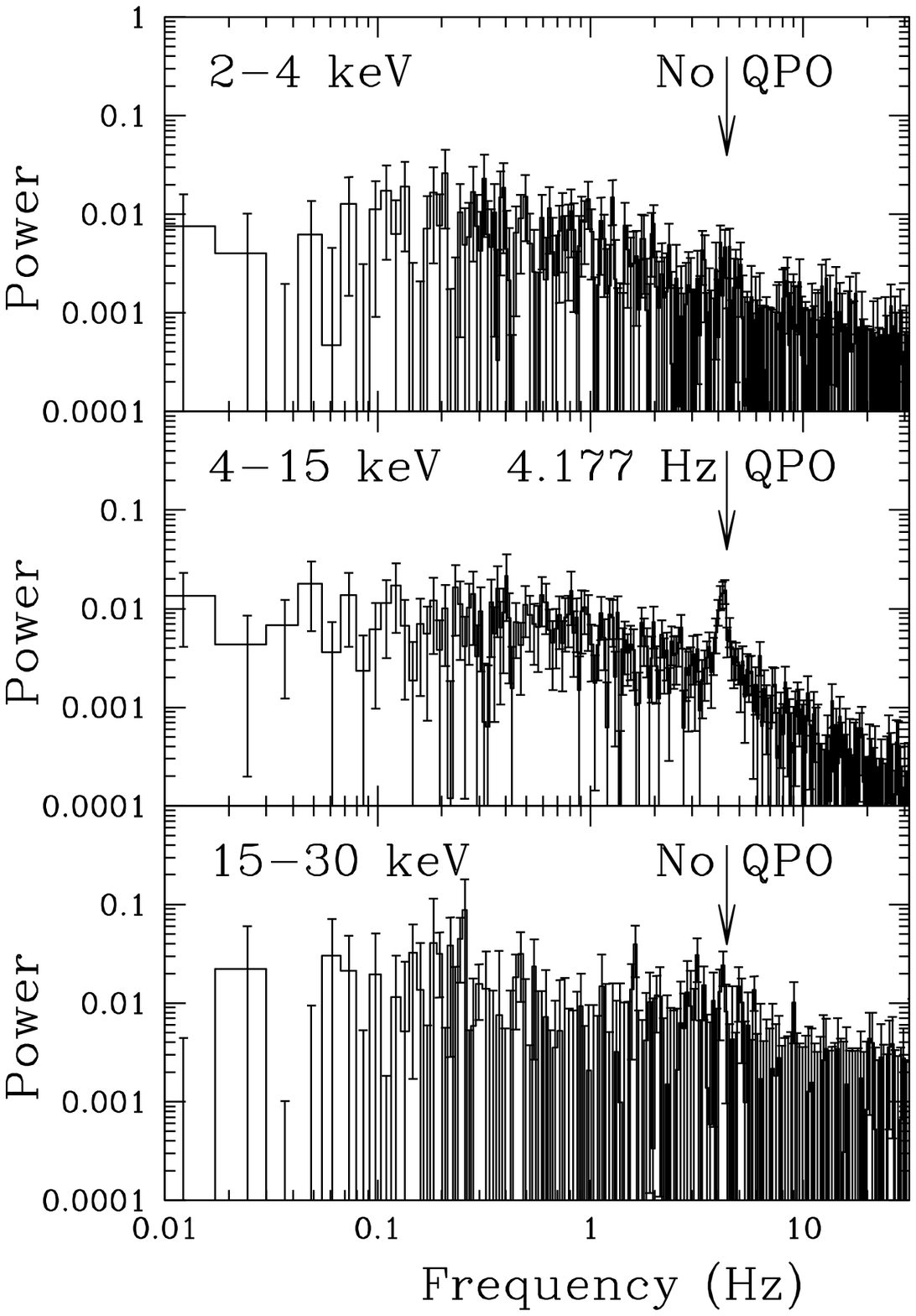}
\caption{(a-c): PDS of three different energy bands of three different observation IDs, taken from 
initial, middle and final stages of the initial rising phase outburst. In each Fig. top panel is 
for 2-4 keV, middle panel is for 4-15 keV and bottom panel is for 15-30 keV bands. (Left) Fig. 
(a) is for Obs ID:95409-01-13-02 (observed on 5th April, 2010), (middle) Fig. (b) is for 
Obs ID:95409-01-14-06 (observed on 13th April, 2010) and (right) Fig. (c) is for Obs ID:95409-01-15-00 
(observed on 16th April, 2010). Fig. (a) is for initial QPO evolution phase, where all the three 
energy bands show QPO signature, whereas (c) is for final QPO evolution phase, only the intermediate 
energy photons take part into QPO formation. Fig. (b) is a special PDS, where we found 3 QPO harmonics 
(Yu  2010) of $2.42$~Hz primary node frequency QPO. Here, the higher energy band photons do not show 
other two harmonics (Debnath et al. 2010a).}
\label{kn : fig5.7}
\end{figure}

Figure 5.7(a-c) shows energy dependent PDS of the three different observation IDs data, taken from 
very beginning, middle and final stages of GX 339-4 initial rising phase of the QPO evolution process. 
Figure 5.7(a) represents the energy dependent PDS of 95409-01-13-02 observation data (observed on 
$5^{th}$ of April 2010). Here, the X-ray photons of all the three energy bands take part into the 
generation of $0.317$~Hz QPO. Similarly, in Figure 5.7(c), the energy dependent PDS of April $16^{th}$, 
2010 (observation ID: 95409-01-15-00) are plotted. Here only intermediate energy photons take 
part into the generation process of $4.177$~Hz QPO. In Figure 5.7(b), the energy dependent PDS of 
observation ID: 95409-01-14-06 (observed on $13^{th}$ of April 2010) is shown. In this particular 
observation we found two more harmonics of fundamental $2.42$~Hz. From the Figure, we observe X-ray 
photons of all the three energy bands taking part into the generation of $2.42$~Hz QPO, but higher 
energy photons do not take part into the generation of other two harmonics ($4.84$ \& $7.19$~Hz QPO).

%\newpage
\section{Results of spectral analysis}

For the spectral study we use $3-30$~keV PCU2 `standard 2' mode (FS\_4a*) data of RTXE/PCA 
instrument. It is to be noted that for the spectral study the background subtraction is very 
important. For generating background file from the raw data we ran FTOOLS task `runpcabackest'.
At first we generate spectral pulse height analysis (pha) files from the both raw data and 
the background files independently. Then we rebinned these pha files using FTOOLS task 
`rbnpha' and finally these rebinned source and and background pha files along with 
response file (*.rsp) were loaded in FTOOLS XRONOS package (XSPEC) and fitted with physical models. 

Black hole energy spectra in low energy X-ray band ($<$ 30 keV), can be fitted with
multi-colour disk black body (diskbb) and power-law models. Low energy spectra ($<$ 10 keV)
are normally dominate by thermal diskbb component and high energy spectra ($>$ 10 keV) are 
dominated by non-thermal power-law component.

In the current rising phase of the outburst, we made detailed spectral analysis of 
a total of  $30$ observational IDs data which are spreaded over the period of $5{th}$ March, 
2010 (MJD = 55260) to $29{th}$ April, 2010 (MJD = 55315). 
For the spectral data analysis, XSPEC package of version 12.5 were used, and for all 
observations we kept the hydrogen column density ($N_H$) fixed at $5 \times 10^{21}$ 
(Motta et al. 2009) using absorption model wabs. We fitted the spectra of each day, using 
standard diskbb and power-law models with a Gaussian line (for Fe emission line at $6.5$~keV). 

\begin{figure}[h] \vskip 0.7cm
\centering
\includegraphics[height=2.8in,width=3.6in,angle=0]{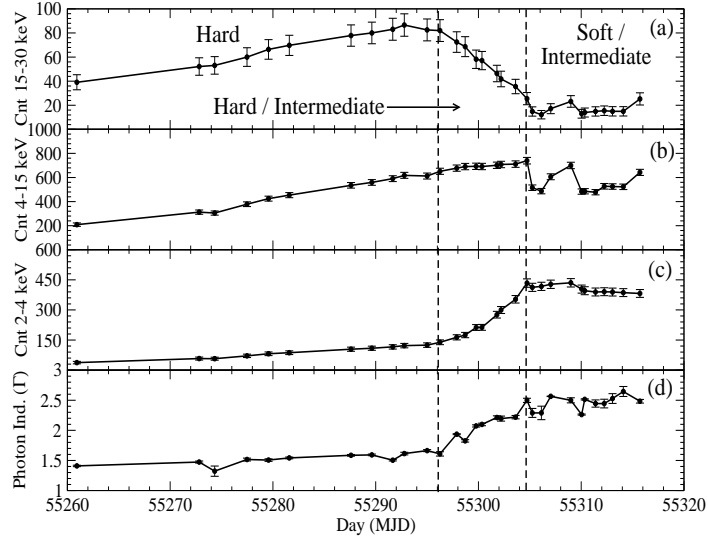}
\caption{Variation of RXTE PCA count rates in (a) 15-30 keV, (b) 4-15 keV and (c) 2-4 keV energy 
bands are plotted in top three panels and in the bottom panel power-law photon index variation 
with day (in MJD) are plotted. The vertical dashed line at MJDs $55296$ and $55304$ indicates 
the state transitions between hard to hard/intermediate to soft/intermediate states 
(Debnath et al. 2010a).}
\label{kn : fig5.8}
\end{figure}

From the detailed spectral analysis we can conclude that the source was initially in the 'pure' 
hard state up to $9^{th}$ of April, 2010 (MJD = 55295), completely dominated by the sub-Keplerian 
flow. After that it started becoming softer (as more Keplerian flow matter started coming, 
which flows in dynamic viscous time scale) and spectral state became hard/intermediate for 
a short duration of time (from $10^{th}$ of April, MJD = 55296 to $17^{th}$ of April, 2010, 
MJD = 55303). Finally the spectra became soft on $18^{th}$ of April 2010 (MJD = 55304) and remained 
in the soft/intermediate state. So, we observed hard-hard/intermediate and 
hard/intermediate-soft/intermediate spectral state transition on $10^{th}$ of April and $18^{th}$ 
of April 2010 respectively. Figure 5.8, clearly shows this signature. Here, the RXTE PCA (PCU2) 
photon counts in three different energy bands ($2-4$ keV, $4-15$ keV and $15-30$ keV) along with 
spectral photon index are plotted. Our motivation for finding PCA photon counts in the above 
mentioned energy bands is to separate the Keplerian and the sub-Keplerian photons. Lower energy 
band ($2-4$ keV) photons come from Keplerian flow component, whereas $15-30$ keV energy band 
photons mainly come from the Comptonized sub-Keplerian flow component. From the Fig. 5.8, it is 
clear that low energy count rates were more or less constant up to $10^{th}$ of April 2010 
(MJD = 55296), where as during this period $15-30$ keV photon count rates were also increased. 
After that more Keplerian component photons started coming, as a result day-by-day $2-4$ keV 
count rates were increased and also in the same time sub-Keplerian supply started decreasing. 
These processes were seen up to $18^{th}$ of April 2010 (MJD = 55304). After that we found the 
steady flow rates from both the Keplerian and the sub-Keplerian flows and the spectra becomes 
soft/intermediate. We observe spectral photon indices at $\sim 1.50$ up to $9^{th}$ of April 2010, 
after that it started increasing (as the spectra started becoming softer) and on $17^{th}$ of 
April 2010, (last day in the hard state) it was $2.22\pm 0.028$. Finally, on $18^{th}$ of 
April 2010, we found the photon index at $2.49\pm 0.035$, 'pure' soft/intermediate state. 
Also, in the photon index plot, we observe a `kink' on MJD = 55274, immediately after that, 
we started observing QPOs in PDS.

\begin{figure}[h]
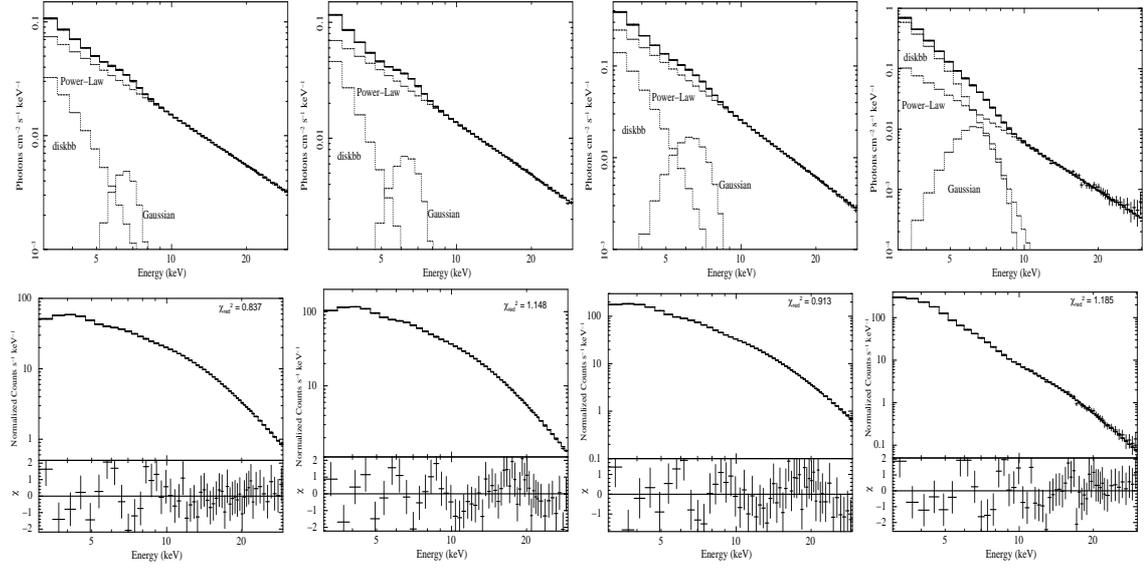

\vskip -0.0cm
\centering
\hspace{0.0cm}
\includegraphics[height=1.42in,width=1.45in,angle=270]{fig5.9al.ps}
\includegraphics[height=1.42in,width=1.45in,angle=270]{fig5.9bl.ps}
\includegraphics[height=1.42in,width=1.45in,angle=270]{fig5.9cl.ps}
\includegraphics[height=1.42in,width=1.45in,angle=270]{fig5.9dl.ps}\vspace{0.1cm}
\includegraphics[height=1.42in,width=1.42in,angle=270]{fig5.9ar.ps}
\includegraphics[height=1.42in,width=1.42in,angle=270]{fig5.9br.ps}
\includegraphics[height=1.42in,width=1.42in,angle=270]{fig5.9cr.ps}
\includegraphics[height=1.42in,width=1.42in,angle=270]{fig5.9dr.ps}
\caption{(a-d) show four sets of 3-30 keV PCA fitted spectra. In the top panels it shows model 
fitted spectra with various fitted components and in the bottom it shows corresponding reduced 
$\chi^2$ plots. All the spectra are fitted with standard diskbb, power-law and Gaussian 
(for Fe-line emission line at $\sim 6.5$~keV) models. From left to right: the model fitted spectra of
(a) 17th March, 2010 (Obs ID:90409-01-10-05), (b) 5th April, 2010 (Obs ID:95409-01-13-02), 
(c) 13th April, 2010 (Obs ID:95409-01-14-06), and (d) 28th April, 2010 (Obs ID:95409-01-16-04) 
are plotted. First spectrum is taken from 'pure' hard state, next two are taken from the hard/intermediate 
state and the remaining one is taken from the soft/intermediate state data.}
\label{kn : fig5.9}
\end{figure}

\begin{table}[h]
%\small
\scriptsize
\centering
\caption{\label{table5.2} Spectral fitted results for the 4 sets of spectra of Fig. 5.9}
\vskip 0.0cm
\begin{tabular}{|l|c|c|c|c|c|}
\hline
Obs. Id & UT Date &Photon & \multicolumn{2}{|c|}{Flux (in $10^{-9} ergs~cm^{-2}~s^{-1}$)}&$\chi^2_{red}$ \\
\cline{4-5}
       &         &Index($\Gamma$)&~~~~3-10keV~~~~ & 10-30keV& \\
\hline
95409-01-10-05&2010-03-17&1.473&2.533&3.626&0.872 \\
95409-01-13-02&2010-04-05&1.504&2.405&3.206&1.148 \\
95409-01-14-06&2010-04-13&2.073&6.495&4.333&0.913 \\
95409-01-16-04&2010-04-28&2.613&6.238&0.750&1.185 \\
\hline
\end{tabular}
\end{table}

Figure 5.9 shows model-fitted 4 sets of spectral plots with model components and their 
corresponding $\chi^2_{red}$ plots. First two spectra of obs. ID: 95409-01-10-05 ($17^{th}$ of 
March, 2010) and 95409-01-13-02 ($5^{th}$ of April, 2010) produce 'pure' hard states.
Obs. ID: 95409-01-14-06 ($13^{th}$ of April, 2010) produces the hard/intermediate 
state and the remaining one Obs. ID: 95409-01-16-04 ($28^{th}$ April, 2010) produces
the soft/intermediate state data. All the spectra are fitted with combination of diskbb, 
power-law and Gaussian models with a multiplicative absorption model wabs. The summary of 
the spectral fittings of these 4 sets of spectra are presented in Tab. 5.2.

In this Chapter, so far, we have discussed initial analysis results of the current GX 339-4 
outburst (Debnath et al. 2010a). The results are mainly from the initial rising phase of the 
outbursts. The detailed timing and spectral analysis for the full period of the outburst are 
going on and results will be published in near future (Debnath et al. 2010e).

	\reseteqn
	\resetsec
	\resetfig
	\resettab
\alpheqn
\resec
\refig
\retab
\def\k{{\bf k}}
\def\aug{{\tilde{\cal H}}}
\def\ahat{{\mathaccent "7E {\cal A}}}
\def\bhat{{\mathaccent "7E {\cal B}}}
\def\chat{{\mathaccent "7E {\cal F}}}
\def\dhat{{\mathaccent "7E {\cal D}}}
\def\shat{{\mathaccent "7E {\cal S}}}
\def\proj{{\cal P}}
\def\ket{\vert \vert  \{ \emptyset \} \rangle}
\def\ket2{\vert \vert \otimes \{ R \} \rangle}
\def\sqr{$^{2}$}
\def\T{$T$}
\def\P{$P$}
\def\Q{$Q$}
\def\G{$G$}
\def\M{{\bf M}}
\def\H{{\bf H}}
\def\I{{\bf I}}
\def\Pr{{\bf P}}
\def\Tr{{\bf T}}
\def\eq{\enskip =\enskip}
\def\pls{\enskip+\enskip}
\def\mns{\enskip -\enskip}
\def\oper{O^{\bf k}_{R}}
\def\aug{\tilde{\cal H}}
\def\envfn{\chi^{\alpha}_{RL}}
\def\struc{S^{\alpha}_{RL,R^{\prime}L^{\prime}}}
\def\rR{r_{R}}

\newpage
\markboth{\it Conclusions and Future Works}
{\it Conclusions and Future Works}
\chapter{Conclusions and Future Works}

%**********************extra*******************

%**********************extra*******************
My Thesis contains the study of the X-ray properties of the two major stages of the life 
cycle of the stars: one is the normal life of a lighter mass star (Sun) and another is the 
collapsed state (black hole) of a star (black hole candidates GRO J1655-40, GX 339-4 and GRBs). 
I am lucky to be a team member for developing X-ray solar space 
instruments RT-2 (S, G and CZT) which observed both the Sun and GRBs from space. 
A part of my Thesis contains development of 
RT-2 instruments, characterization of CZT \& CMOS imaging detectors (used in RT-2/CZT 
instrument), some observational results of solar flares. 
My Thesis also contains the detailed timing \& spectral properties of the 2005 outburst 
of the well known Galactic black hole candidate GRO J1655-40 and on-going 2010 outburst of the 
transient Galactic steller mass black hole candidate GX 339-4.

In Chapter 1, we gave a general introduction of the subject of Astronomy and Astrophysics. 
In \S 1.2, we briefly discussed the life cycle of a star from birth (at nebula) to death. Here 
we also discussed intermediate stages of a star depending upon mass. In \S 1.3 we gave a detailed 
description and some physical properties of our nearest star, namely, the Sun. Also in this 
Section, we discussed important high energy space astronomy missions, dedicated to the study 
of the solar properties. \S 1.4 is devoted to the introduction of compact objects (specially 
black holes). Here, we started with X-ray binary systems: white dwarf, neutron star and black 
hole and their properties. Then we concentrated on the black hole properties: radiative processes 
associated with the matter around a black hole and physical processes responsible for the generation 
of electromagnetic radiations. In this Section, we also discussed the development of accretion 
flow models. Also this Section contains the description of the major X-ray or $\gamma$-ray astronomy 
missions, so far launched in space for the black hole study. In \S 1.5, we discuss the properties 
of GRBs, their origin and classifications. Here we also discussed the commonly accepted two 
(Fireball and Cannonball) models, for describing the origin of GRBs.
In \S 1.6, we briefly discussed the analysis methods (specially timing and spectral) for 
observational study, which we used in a major part of the Thesis work. 

Chapter 2 is dedicated for space instruments and their data acquisition and analysis procedures. 
We used Indo-Russian collaborative RT-2 instruments for the solar X-ray study and NASA 
satellite RXTE for the black hole study. In \S 2.1 - \S 2.4, we discussed
the RT-2 instruments and in \S 2.5 - 2.8, we discussed the RXTE instruments and data
analysis procedures. In \S 2.1, we gave brief introduction of RT-2 mission and in \S 2.2
we discussed about the features of the RT-2 scientific payloads. In \S 2.3, we discussed
the details of the RT-2 data acquisition processes and \S 2.4, RT-2 data analysis procedures 
are discussed. In \S 2.5 and \S 2.6, we gave a brief introduction of the RXTE satellite and 
description of its scientific payloads. In \S 2.7, RXTE data analysis procedures are 
discussed. In \S 2.8, we discussed about the theoretical fitting models used for 
black hole study using RXTE archival data.

Chapter 3, contains the observational results of the solar flares. These solar flares were observed 
by our RT-2 instruments. On $5^{th}$ July 2009, hard X-ray solar flare, we observed quasi-periodic 
pulsations of period $\sim$ 12 seconds. The detailed analysis results of this particular flare has
been discussed in \S 3.2.1 and in \S 3.2.2, we discussed some preliminary results of the low
energetic hard X-ray flares, observed on $26^{th}$ of October, 2009.

\index{NASA}Chapter 4, contains analysis of detailed timing and spectral results of the 2005 
outburst of Galactic black hole candidate GRO J1655-40. We found some important results. 
We started the Chapter by giving a brief description of the source GRO J1655-40 and discussed major results 
obtained from its two outbursts occurred in 1996-97 and in 2005. On the basis of 
our detailed timing and spectral analysis of the 2005 GRO J1655-40 outburst, we  
classified the total outburst into four spectral states: {\it hard, soft, very soft and 
intermediate}, in the sequence of {\it hard $\rightarrow$ soft/very soft $\rightarrow$ 
intermediate $\rightarrow$ hard}.
\S 5.2 is dedicated to the detailed timing analysis results of the 2005  
GRO J1655-40 outburst. In this Section, we discussed the ASM lightcurve, ASM \& PCA 
hardness-intensity diagrams, pca lightcurves, hardness and softness 
diagrams. Also in this Section, we discussed power density spectra and observed QPOs 
in details. We found monotonically increasing and monotonically decreasing nature of the 
QPO frequencies in the initial rising and final declining hard states respectively. 
This nature of the QPO evolution and origin of QPOs are discussed in \S 5.2.5. 
Using theoretical propagating oscillatory shock (POS) model these results were
explained. In the next two sub-sections we discussed the results of dynamic PDS (where we  
showed the rising and the declining phases of QPO evolutions) and energy dependent QPO 
signatures. In \S 5.2.8, we discussed the observed Compton cloud 
oscillations during QPOs in black hole candidates (GRO J1655-40 \& GRS 1915+105). 
In \S 5.3, a detailed spectral analysis results of the 2005 GRO J1655-40 
outburst are discussed. In the first sub-section we discussed the spectral data reduction 
and fitting techniques. We used these in our spectral analysis. In 
\S 5.3.2 we discussed model fitted spectral results for the outburst. In the next
sub-section a daily variation of the spectral parameters and fluxes are discussed.

Detailed timing and spectral analysis results for the initial rising phase of the current 
GX 339-4 outburst are discussed in Chapter 5. After the launch of RXTE satellite on 
December 30, 1995, GX 339-4 has shown 5 major X-ray 
outbursts in $\sim 2-3$ years of intervals. After remaining `dormant' for $\sim 3$ years
GX 339-4 has become again X-ray active in first week of January, 2010. Presently it is in 
the rising phase of outburst. RXTE started observing the source in regular intervals from 
$12^{th}$ of January, 2010, after Japanese satellite MAXI/GSC onboard HETE announced its 
X-ray trigger on January 03, 2010. We analyzed a total $30$ of observation ID data of RXTE 
PCA data spreading over a period of $5{th}$ March, 2010 (MJD = 55260) to $29{th}$ April, 
2010. We study the detailed timing and spectral properties of the source during its initial 
phase of the outburst. From our detailed timing and spectral studies, we found that the source 
already passed through three spectral states (hard, hard/intermediate and soft/intermediate) during 
the above mentioned period of the outburst. In the PDS, we found QPOs in all three states. 
We also observed monotonically increasing nature of QPO frequencies during the hard and 
hard/intermediate spectral states, which fit well with our theoretical POS model.
In \S 6.1 and \S 6.2, we discussed detailed timing and spectral analysis results respectively. 
In \S 6.1.1, RXTE ASM lightcurves and corresponding hardness diagrams are plotted. In 6.1.2, 
the hardness intensity diagram using RXTE ASM and PCA data are plotted and 
discussed. In \S 6.1.3, PCA lightcurves and hardness, softness diagrams are plotted for various 
outburst stages data, from where we get an rough idea about variation of the two component 
(Keplerian and sub-Keplerian) advective flow rates. In \S 6.1.4, model fitted PDS and QPO signatures 
are discussed. In \S 6.1.5, monotonically increasing nature of the QPO frequencies with our theoretical 
POS model fitted results are discussed. In \S 6.1.6, energy dependent nature of 
the QPOs during the initial phase of the outburst are discussed. In \S 6.2, detailed 
spectral results using $3-30$ keV RXTE PCA data are discussed. Here we showed the day-wise variation 
of the spectral photon indices with PCA count rates. From there we get a clear idea about the 
variations of the thermal flow component Keplerian and non-thermal flow component sub-Keplerian rates.

So far, I worked on the life of a typical normal star (e.g. Sun) and the end product of a star 
(black hole). In future, I want to work on origin of black holes as is believed to be
through the gamma-ray bursts (GRBs). 
After the successful launch on January 30, 2009, RT-2 observed four GRBs.
In the following sub-sections, we briefly discuss the  nature of the lightcurves of the observed GRBs, 
giving more details for the first GRB (GRB 090618).

\subsubsection{\bf A. GRB 090618}

GRB 090618 was first detected with the {\it Swift} Burst Alert Telescope (BAT) (Schady et al.,
2009) on 18th June, 2009 at 08:28:29 UT with source location RA(J2000): 19h 35m 58.3s
and Dec(J2000): $+78^\circ 21^{'} 25.3^{''}$. Most of the space instruments working in the
high energy X-ray domain has detected this object.

The GRB lightcurve shows a complex profile (Golenetskii et al., 2009a)
of time duration of around $150$ seconds. The burst profile
has three main pulses with the brightest pulse ($\sim 700$ counts/sec) started at
T$_0$+65 sec, the second one at T$_0$+85 sec and the last one at T$_0$+115 sec, where
T$_0$ = 30504.974 s (08:28:24.974) UT.

\begin{figure}[h]
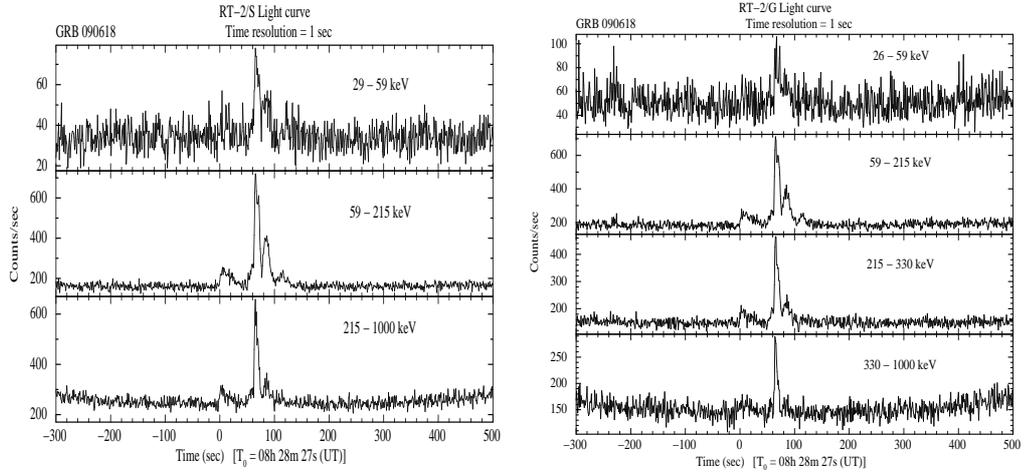

\centering
\includegraphics[height=2.6in,width=2.5in,angle=270]{fig6.1a.ps}\hspace{0.1 cm}
\includegraphics[height=2.6in,width=2.5in,angle=270]{fig6.1b.ps}
\caption{(a-b) Light curves in different energy band (marked) of (a) RT-2/S and (b) RT-2/G
detectors. Four emission peaks along with the {\it ignition (precursor trigger) pulse} at
T$_0$ = 08h 28m 27s UT are detected clearly in the energy band of 59 - 215 keV (Rao et al. 2010b).}
\label{kn : fig6.1}
\end{figure}

RT-2 Experiment onboard CORONAS-PHOTON mission also detected this object at 08h 28m 27s (UT)
(Rao et al., 2009, 2010b). During the GRB event, the satellite was completely in the SHADOW
mode (away from the Sun) started at 08:16:10.207 (UT) and ended at 08:37:35.465 (UT) and the
burst event was detected by the RT-2 instruments with a large off-axis of about $77^\circ$.
Spectroscopic study of afterglow of GRB 090618 suggests that the associated source could
be located at a redshift of z = 0.54 (Cenko et al., 2009).
Both RT-2/S and RT-2/G detectors also have
registered the multi-structured burst profile of the lightcurve in the energy band of
$60 - 215$ keV. In the high-energy band of $330 - 1000$ keV, the profile becomes simple with
a single bright peak at T$_0$+65 sec (see, Figure 6.1), where T$_0$ = 08:28:27 UT.
It is also noted that the burst width decreases with the increase in the energy band,
a quite natural phenomenon of GRB burst profile. It showed the typical band spectrum
with peak energy at about $180$ keV and integrated $20$ keV - $1$ MeV flux of
$2.8 \times 10^{-4}$ ergs/ cm$^2$.
During the GRB time, two RT-2 phoswich detector instruments (RT-2/S \& RT-2/G) were working
properly, but unfortunately the solid-state detector payload (RT-2/CZT) was not completely
in working condition, so it has missed the GRB.

In the following sub-sections we will discuss the detailed timing and spectral results
obtained using our RT-2 data. In our recent communication (Rao et al., 2010b), we gave
a detailed timing and spectral analysis results of the multi-structured burst profile of the
GRB.

\subsubsection{$\bullet$ Timing analysis results for GRB 090618}

GRB 090618 was detected by the three different energy bands of RT-2/S and by the four different
energy bands of RT-2/G detectors (see Figure 6.1). The burst profile is not clearly observable in
the low energy ($<$ 59 keV) band. But in the high energy bands (in between 59 - 1000 keV), light
curves clearly show multi-structured (mainly four peaks) burst profiles. To find the widths
of the peaks, we tried to fit RT-2/G lightcurve data of 59-215 keV energy band with
fast-rise-exponential-decay (FRED) profile developed by Kocevski et al. (2003). According to the
FRED profile GRB photon flux can be defined as:
$$
F(t) = F_m(\frac {t}{t_m})^r[\frac {d}{d+r} + \frac {r}{d+r}
        (\frac {t}{t_m})^{(r+1)}]^{-(r+d)/(r+1)},
\eqno (6.1)
$$
where $F_m$ is the maximum flux at time $t_m$, $r$ and $d$ are the decaying indices for rising
and decaying phases of any pulse (peak) respectively. We first fit individual peaks and then
total lightcurve with the pulse profile of Eqn. (2.1). This model fitted lightcurve for RT-2/G
detector in the energy band of 59 - 215 keV is shown on Figure 6.2. We also fitted this model
in the other energy bands of RT-2/G lightcurves. Finally we calculated widths of each peaks
from our fittings. All the fitted parameters and pulse widths are noted in Table 6.1. We also
compared our fitted results with Swift BAT data (Rao et al. 2010b).

\begin{figure}[h]
\centering
\includegraphics[height=2.4in,width=2.8in,angle=000]{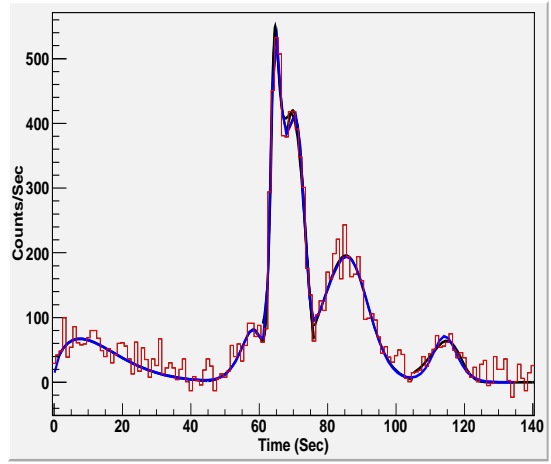}
\caption{Multiple FRED profiles fitted 59-215 keV band RT-2/G background subtracted lightcurve
(Time 0 means $T_0$ = 08h 28m 27s UT) for GRB 090618 (Rao et al. 2010b).}
\label{kn : fig6.2}
\end{figure}

\begin{table}[h]
\small
\centering
\caption{\label{table2.4} FRED profile fitted results for RT-2/G lightcurves of GRB 090618}
\vskip 0.2cm
\begin{tabular}{|l|c|c|c|c|c|c|}
\hline
 Pulse & Energy range (keV) & $F_m$ (s$^{-1}$) & $t_m$ (s) & $r$ (s)  & $d$ (s)  & Width (s) \\
\hline
{1} &        &  532.0 & 65.0 & 81.1 & 13.8 & 6.30 \\
{2} & 59-215 &  422.0 & 70.0 & 34.5 & 34 & 7.15  \\
{3} &        &  210.0 & 85.0 & 20.0 & 20.0 & 14.60 \\
{4} &        &  72.0 & 114.0 & 42.0 & 40.0 & 9.49\\
\hline
{1} &          & 315.0 & 65 & 82.5 & 16.4 & 5.59 \\
{2} & 215-330 & 200.0 & 70.0 & 41.6 & 34.2 & 6.17 \\
{3} &         & 76 & 85.0 & 25.0 & 22.0 & 12.17 \\
\hline
{1} & 330-1000 & 144.0 & 65.0 & 78.3 & 22.5 & 4.77  \\
{2} &        &  64.0 & 70.0 & 88.0 & 84.0 & 2.93  \\
\hline
\end{tabular}
\end{table}

\subsubsection{$\bullet$ Spectral analysis results for GRB 090618}

\begin{figure}[h]
\centering
\includegraphics[height=2.8in,width=2.4in,angle=270]{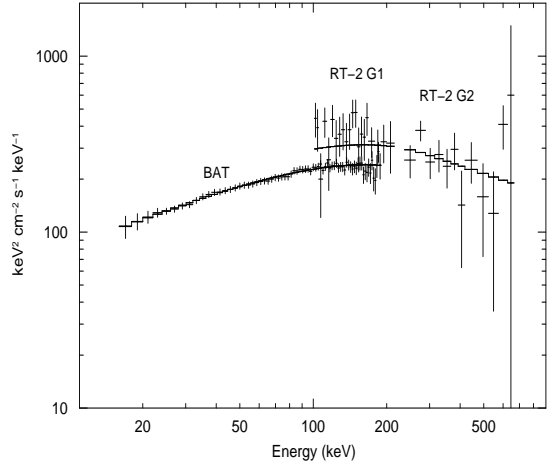}
\caption{The unfolded $Swift$ BAT and $RT$-2/S combined energy spectrum of GRB 090618 
(Rao et al. 2010b).}
\label{kn : fig6.3}
\end{figure}

Spectral analysis for the GRB was done by using the RT-2/S and Swift BAT data. We used
$15-200$ keV BAT, $100-210$ keV $RT$-2/S G1, and $250-650$ keV $RT$-2/S G2 data for
our combined spectral fitting analysis. From our RT-2/S spectral data, we found the GRB
peak energy at $\sim 180$ keV and total flux in $20 - 1000$ keV energy band is equal to
$2.8 \times 10^{-4}$ ergs/cm$^2$.

For RT-2 spectral data fitting we first generated response matrix using FTOOLs {\it genrsp}
task. The GRB spectrum can be well fitted with the double power-law model introduced by Band
et al. (1993). In this model, two power-laws join at break frequency $(\alpha-\beta)$E$_p$,
where $\alpha$ and $\beta$ are the first and second power-law indices respectively.
Figure 6.3 shows the unfolded power-law model fitted $Swift$ BAT and $RT$-2/S combined
energy spectrum of the GRB 090618.

\subsubsection{\bf B. GRB 090820}

The Fermi Gamma-Ray Burst Monitor (GBM) first detected the GRB 090820 by its trigger
number 272421498 (Connaughton V. 2009) on 20th August, 2009 at 00:38:16.19 UT.
The source location was at RA(J2000) = 87.7 degrees (5h 51m) and Dec(J2000) = 27.0
degrees ($+27^\circ 0^{'}$). The lightcurve of this bright GRB shows GBM triggered
on a weak precursor, followed by a bright pulse beginning 30 sec later and lasting a
further 30 seconds.

Both the RT-2 phoswich detectors (RT-2/S \& RT-2/G) have detected this bright
astrophysical event (Chakrabarti et al., 2009d). The satellite was in the
GOOD/LIGHT mode with Earth latitude = $-64.09$ and longitude = $169.11$ with 600 sec
duration of GOOD time observation, starting at 00h 29m 16sec (UT) and ending at
00h 39m 16sec (UT). After 30 sec, the satellite completely entered into the BAD mode
of high charge particle region.

\begin{figure}[h]
\centering
\includegraphics[height=2.6in,width=2.5in,angle=270]{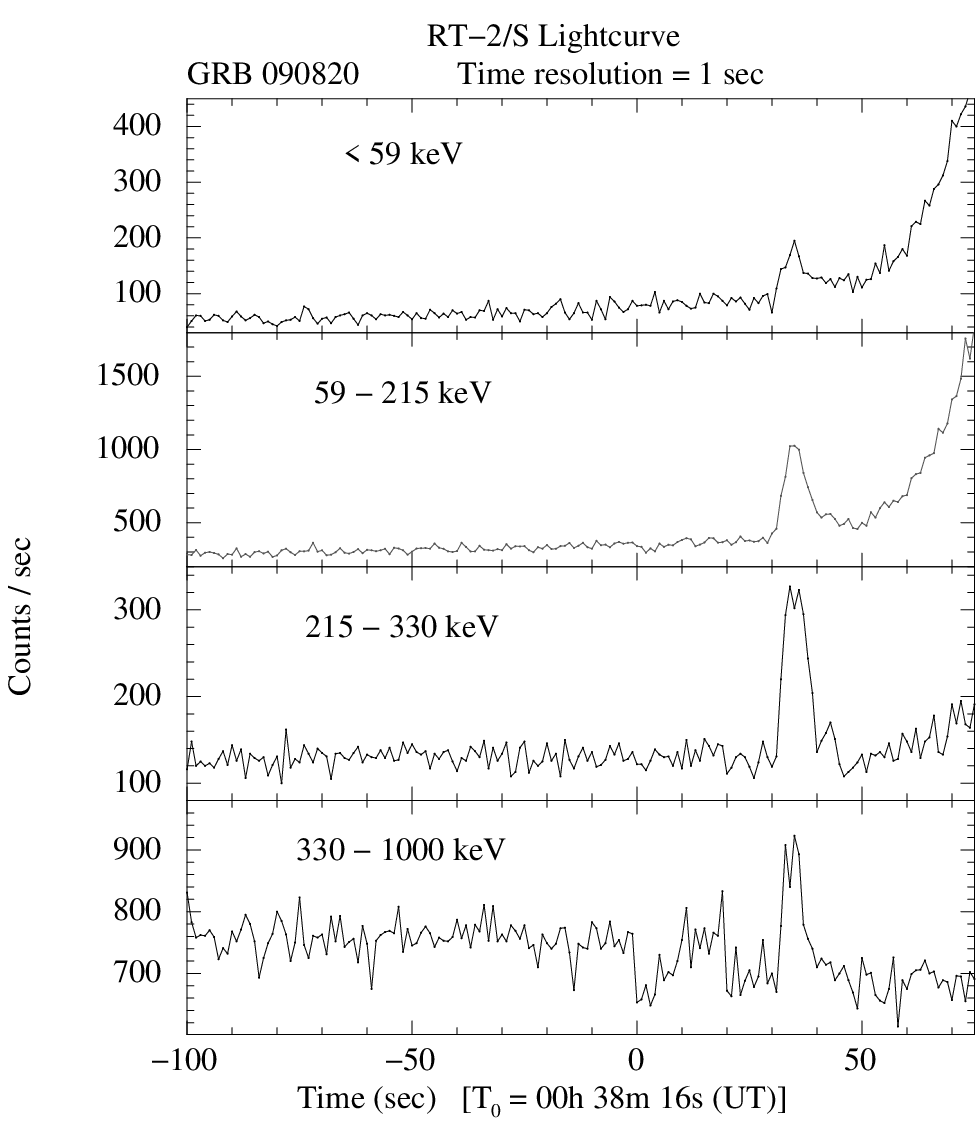}\hspace{0.1 cm}
\includegraphics[height=2.6in,width=2.5in,angle=270]{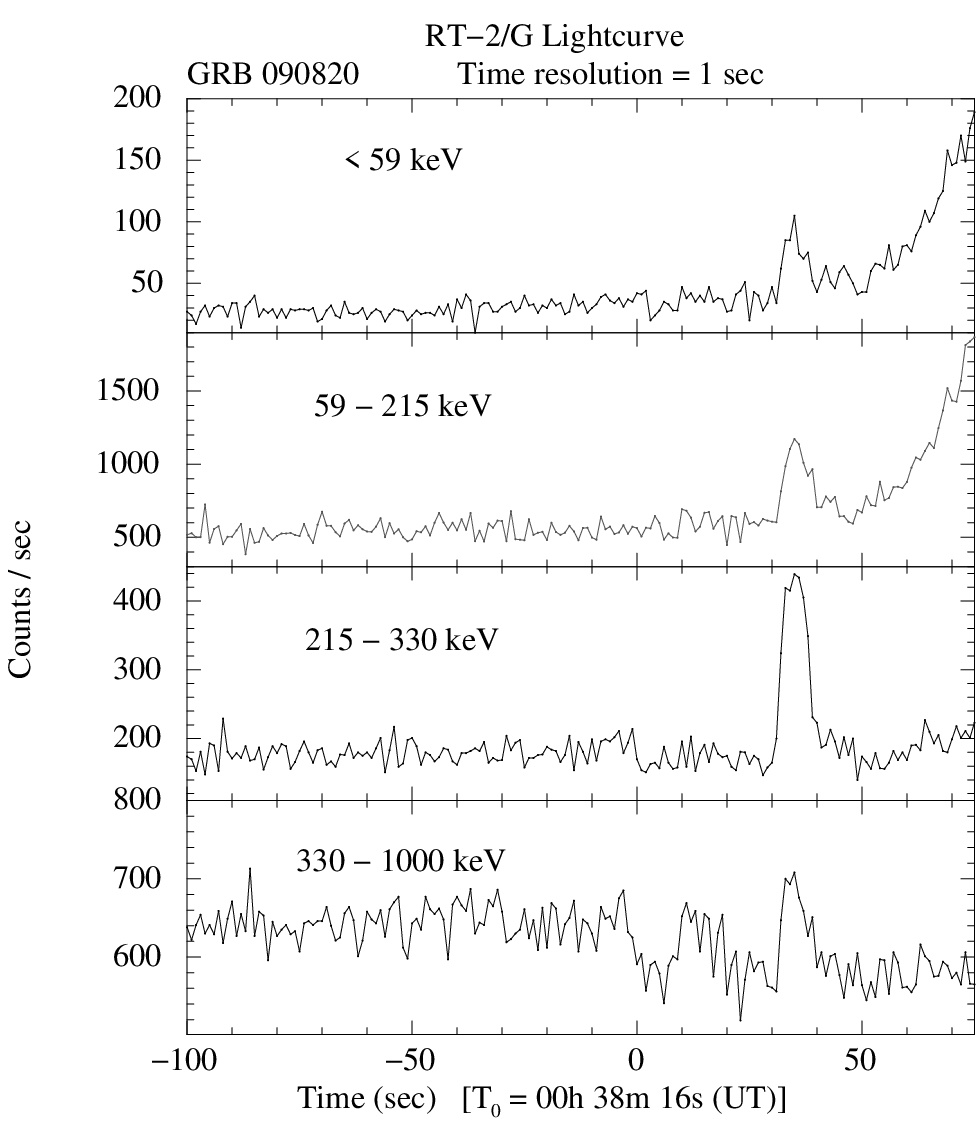}
\caption{(a,b) Light curves in different energy band (marked) of (a) RT-2/S and  (b) RT-2/G
detectors. Emission peaks along with the weak precursor trigger pulse at T$_0$ = 00h 38m 16s UT
are shown (Chakrabarti et al., 2009d).}
\label{kn : fig6.4}
\end{figure}

The GRB lightcurve shows a simple profile of a strong emission peak at T$_0$+34 sec
with a weak precursor (T$_0$ = 00:38:16 UT). The time duration of the brightest
emission is around $19$ sec with $\sim 1100$ counts/sec.

Both RT-2/S and RT-2/G detectors have registered this burst profile in the
energy band of $15 - 1000$ keV with the strongest emission in the energy range of
around $100$ keV to $330$ keV. It is also noted that the burst width decreases
with the increase in the energy band.

\subsubsection{\bf C. GRB 090926A}

The very bright GRB 090926A was first detected by the Fermi-GBM trigger 275631628
(Bissaldi 2009) on 26th September, 2009 at 04:20:26.99 UT. The source location
was at RA(J2000) = $354.5$ degrees (23h 38m), DEC(J2000) = $-64.2$ degrees
($-64^\circ 12^{'}$), with an uncertainty of $1^\circ$. The GBM lightcurve
consists of single pulse with a duration (T90) of 20 s (8-1000 keV).

\begin{figure}[h]
\centering
\includegraphics[height=2.6in,width=2.5in,angle=270]{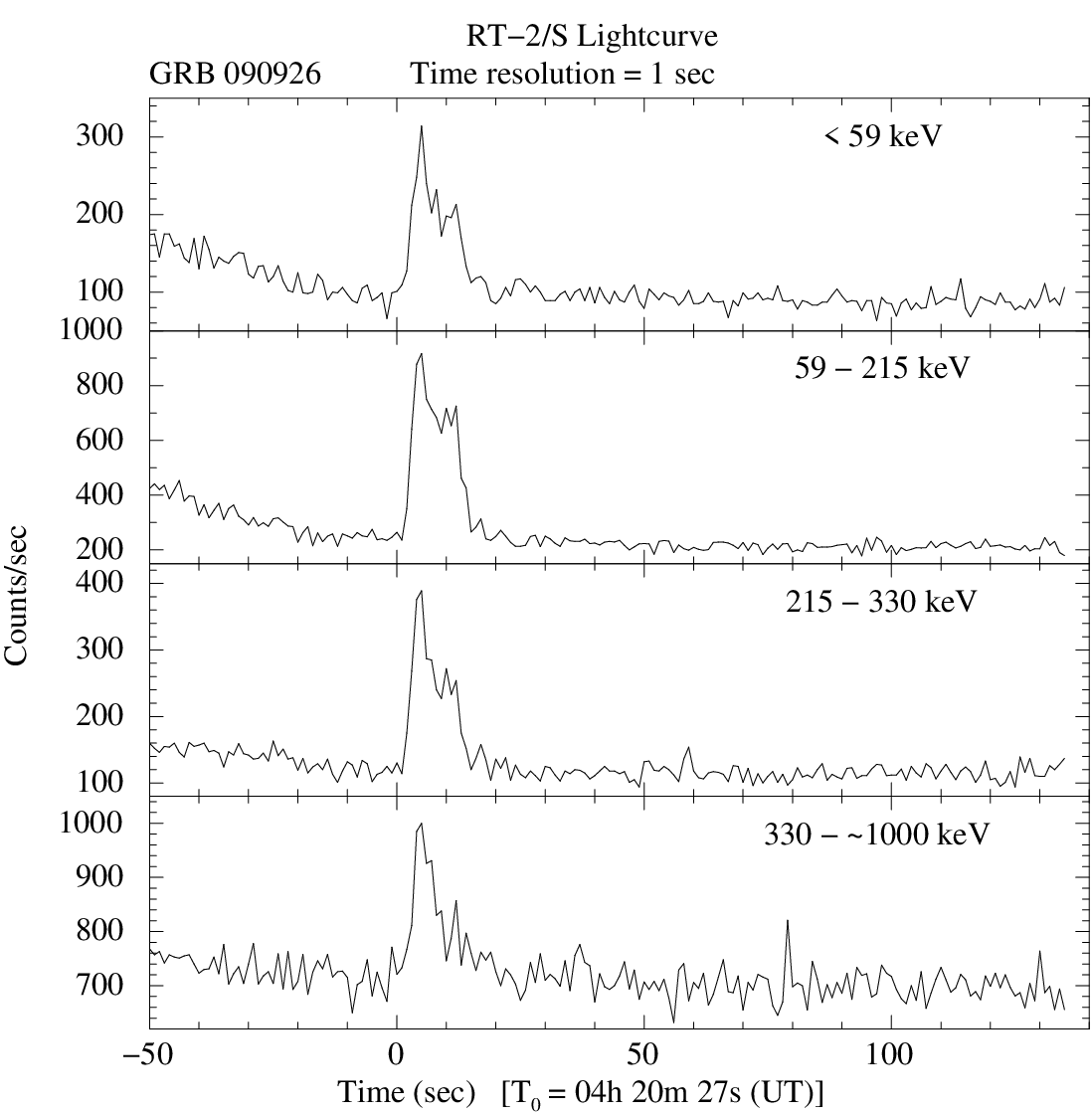}\hspace{0.1 cm}
\includegraphics[height=2.6in,width=2.5in,angle=270]{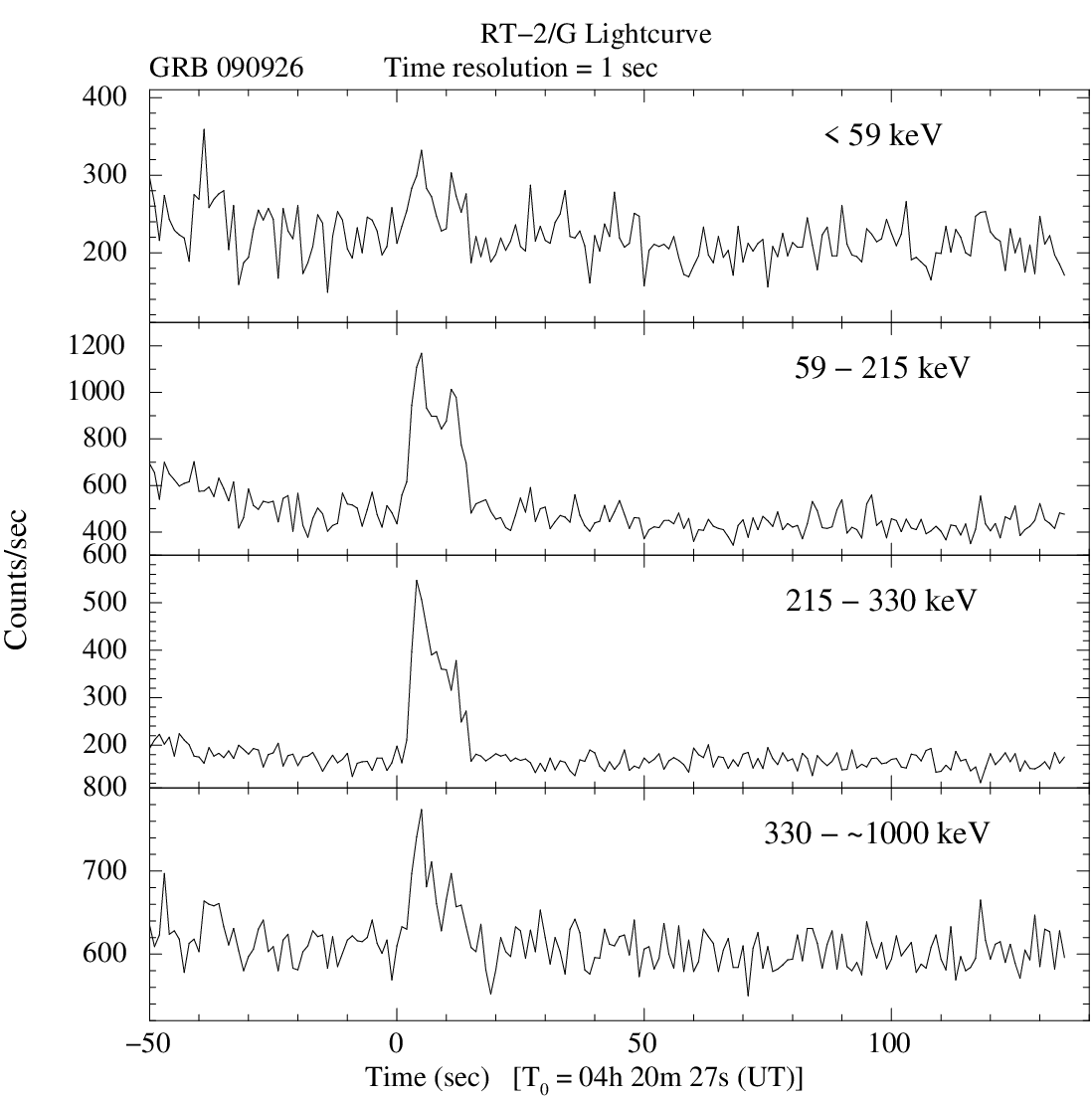}
\caption{(a,b) Light curves in different energy band (marked) of (a) RT-2/S and (b) RT-2/G
detectors. Two emission peaks along with the weak precursor trigger pulse at T$_0$ = 04h 20m 27s UT
are detected in a wide energy band of 15 - 1000 keV (Chakrabarti et al., 2009e).}
\label{kn : fig6.5}
\end{figure}

RT-2 instruments (RT-2/S \& RT-2/G) also has detected this bright GRB at T$_0$ = 04h 20m 27s 
(UT) (Chakrabarti et al., 2009e). The satellite was in the LIGHT mode (pointing towards the Sun) 
for a short duration at a high latitude in its orbit. During this time, the GOOD time (away from 
the polar and SAA regions) observation was for 348 sec starting at 04h 16m 55sec (UT) and ending 
at 04h 22m 43sec (UT). This burst was also detected by KONUS-RF, another instrument onboard C
ORONAS-PHOTON satellite (Golenetskii et al., 2009b).

The burst lightcurve consists of multiple peaks of total duration of $\sim 17$ sec,
followed by a weak tail ending at T$_0$+30 sec. The strongest peak count rate is $\sim 1200$
counts/sec.

Both RT-2/S and RT-2/G detectors have registered the burst profile of this
bright GRB in the energy band of 15 - 1000 keV with strongest emission in
60 - 215 keV energy band.

\subsubsection{\bf D. GRB 090929A}

The Fermi Gamma-Ray Burst Monitor first reported the GRB 090929A by its trigger $275891585$
(Rau, A., 2009). The event occurred at 04:33:03.97 UT on 29th September, 2009
in the sky location of RA(J2000) = 51.7 degrees (03h 27m), DEC(J2000) = -7.3 degrees
($-7^\circ 18^{'}$), with an uncertainty of 1.3 degrees.

\begin{figure}[h]
\centering
\includegraphics[height=2.6in,width=2.5in,angle=270]{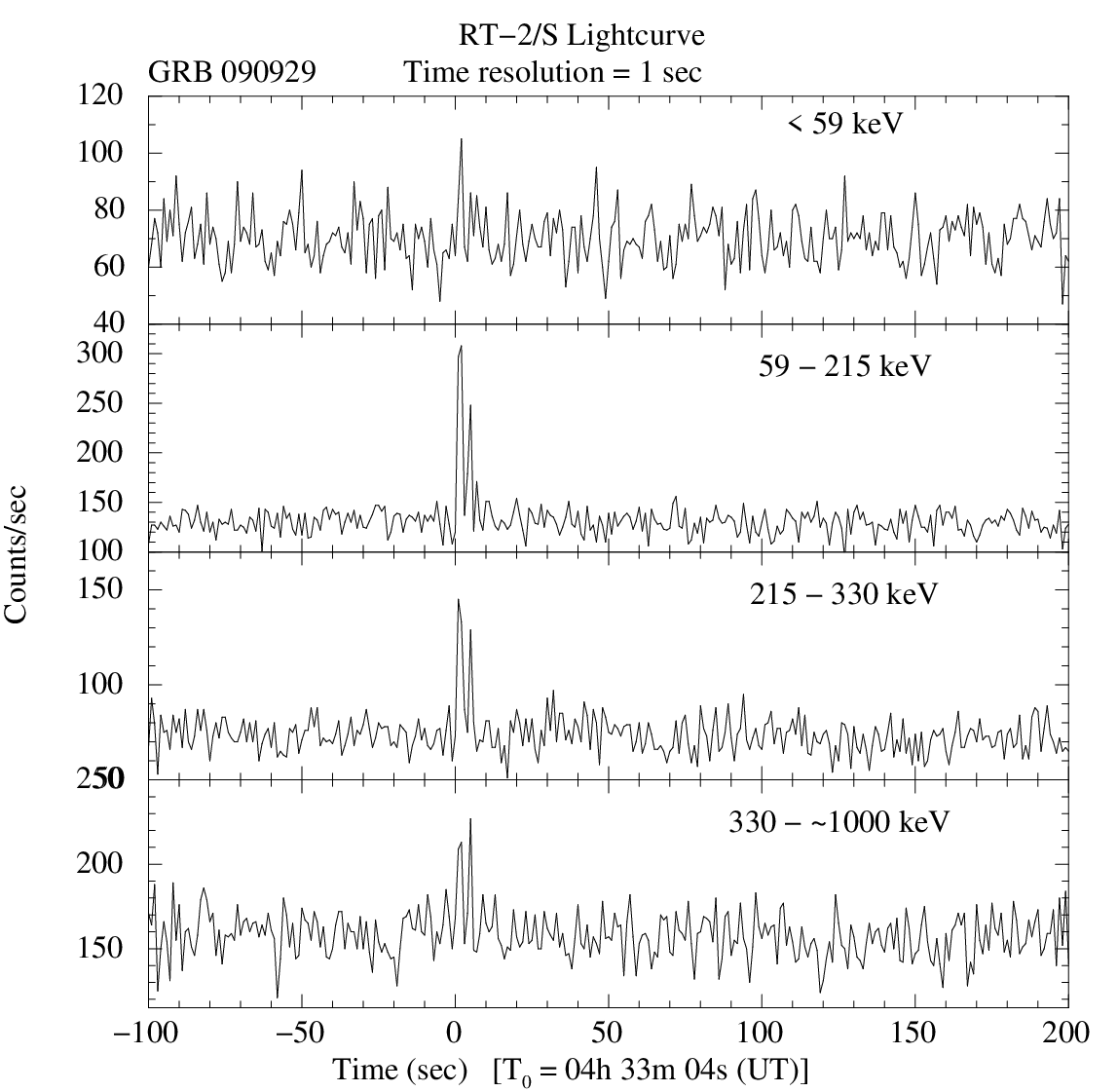}\hspace{0.1 cm}
\includegraphics[height=2.6in,width=2.5in,angle=270]{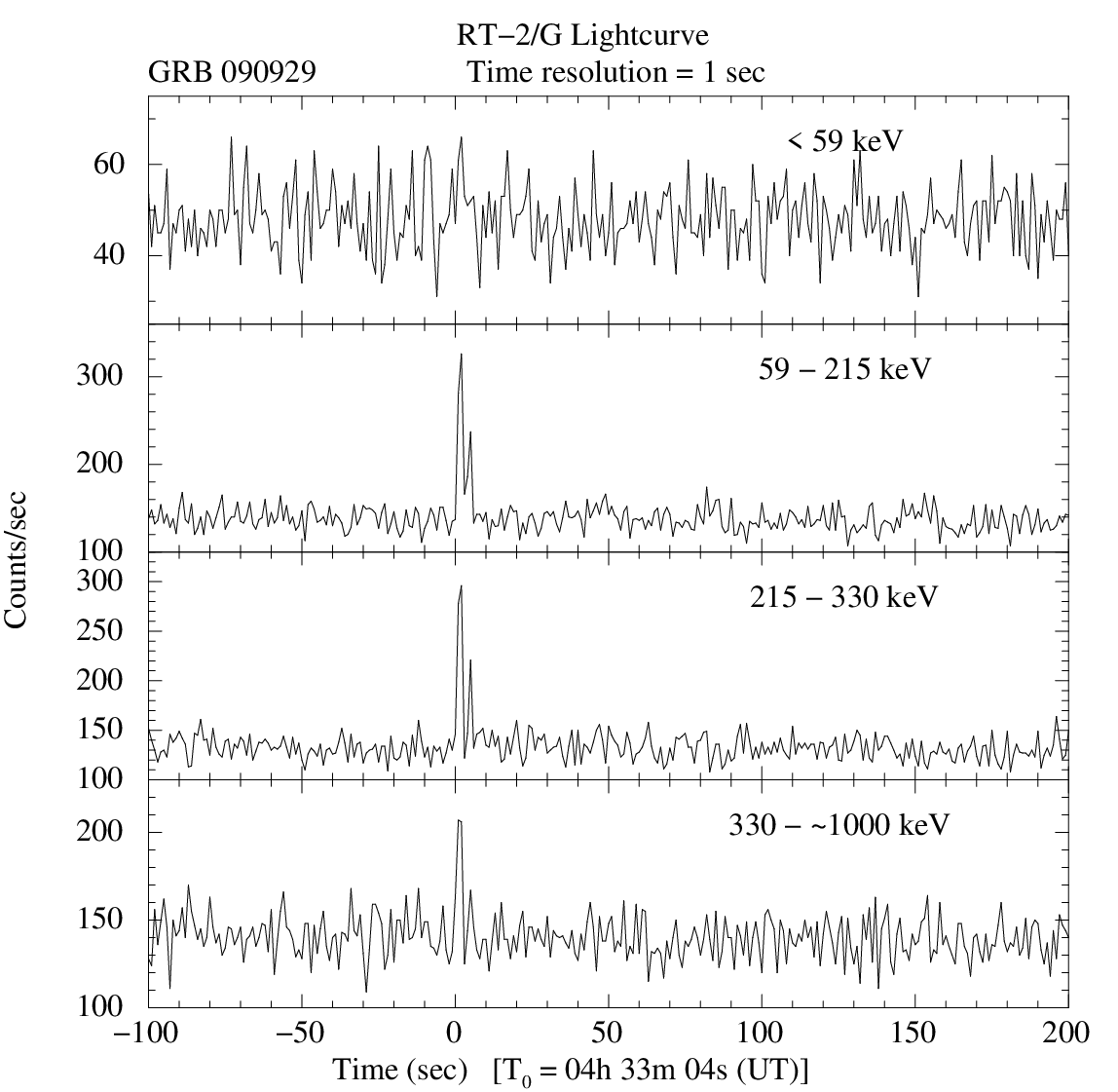}
\caption{(a,b) Light curves in different energy band (marked) of (a) RT-2/S and (b) RT-2/G
detectors. Emission peaks along with the weak precursor trigger pulse at T$_0$ = 04h 33m 04s UT
are detected in a wide energy band (Chakrabarti et al., 2009f).}
\label{kn : fig6.6}
\end{figure}

Both the phoswich detectors (RT-2/S \& RT-2/G) of the RT-2 Experiment, has also detected 
this GRB (Chakrabarti et al., 2009f). During the time of GRB, the satellite was in the LIGHT mode 
(pointing towards the Sun) with GOOD time (i.e., away from the polar charge particle and SAA 
regions) observation of 1537 sec starting at 04h 21m 20sec (UT) and ending at 04h 46m 57sec (UT). 
This burst is also detected by KONUS-RF (Golenetskii et al., 2009c).

The burst lightcurve (1 sec bin) shows double peak profile of a total duration
of just around 5 sec. The strongest peak count rate is $\sim 310$ counts/sec.

Both RT-2/S and RT-2/G detectors have registered the prompt emission from the
GRB090929A in the energy band of 60 - 1000, with strongest emission in
60 - 215 keV energy band.

	\reseteqn
	\resetsec
	\resetfig
	\resettab
%\vfill\eject
%\pagestyle{newheadings}
%\pagenumbering{roman}
%\setcounter{page}{1}

\def\k{{\bf k}}
\def\aug{{\tilde{\cal H}}}
\def\ahat{{\mathaccent "7E {\cal A}}}
\def\bhat{{\mathaccent "7E {\cal B}}}
\def\chat{{\mathaccent "7E {\cal F}}}
\def\dhat{{\mathaccent "7E {\cal D}}}
\def\shat{{\mathaccent "7E {\cal S}}}
\def\proj{{\cal P}}
\def\ket{\vert \vert  \{ \emptyset \} \rangle}
\def\ket2{\vert \vert \otimes \{ R \} \rangle}
\def\sqr{$^{2}$}
\def\T{$T$}
\def\P{$P$}
\def\Q{$Q$}
\def\G{$G$}
\def\M{{\bf M}}
\def\H{{\bf H}}
\def\I{{\bf I}}
\def\Pr{{\bf P}}
\def\Tr{{\bf T}}
\def\eq{\enskip =\enskip}
\def\pls{\enskip+\enskip}
\def\mns{\enskip -\enskip}
\def\oper{O^{\bf k}_{R}}
\def\aug{\tilde{\cal H}}
\def\envfn{\chi^{\alpha}_{RL}}
\def\struc{S^{\alpha}_{RL,R^{\prime}L^{\prime}}}
\def\rR{r_{R}}

\newpage
\appendix
\markboth{\it ABBREVIATIONS}
{\it ABBREVIATIONS}
\chapter{ABBREVIATIONS}
%\notations

%\centerline{\bf \large A}
\begin{quote}
\begin {tabular}{l l}
{2D}&{Two Dimensional}\\
{3D}&{Three Dimensional}\\
{$\dot{A}$}&{$\dot{A}$ngstrom, $1 \times 10^{-8}~cm$}\\
{A \& A}&{Astronomy \& Astrophysics}\\
{ACE}&{Advanced Composition Explorer}\\
{ADC}&{Analog to Digital Converter}\\
{ADAF}&{Advection Dominated Accretion Flow}\\
{AGN}&{Active Galactic Nuclei}\\
{AIP}&{The American Institute of Physics}\\
{al.}&{all}\\
{Al}&{Aluminium}\\
{ALEXIS}&{Array of Low Energy X-ray Imaging Sensor}\\
{ANS}&{Astronomical Netherlands Satellite}\\
{APPIDS}&{APPlication IDS}\\
{APJ}&{Astro-Physical Journal}\\
{AR}&{Angular Resolution}\\
{ASCII}&{American Standard Code for Information Interchange}\\
{ASI}&{Astronomical Society of India}\\
{ASIC}&{Application-Specific Integrated Circuit}\\
{ASM}&{All Sky Monitor}\\
{ASTROSAT}&{Indian Multi-wavelength ASTROnomy SATellite}\\
{AU}&{Astronomical Unit, $1.495 \times 10^{13}~cm$ (Distance between Earth \& Sun)}\\

{bb}&{black body}\\
{BASI}&{Bulletin of Astronomical Society of India}\\
{BAT}&{Burst Alert Telescope}\\
{BH}&{Black Hole}\\

\end {tabular}
\end{quote}
\begin{quote}
\begin {tabular}{l l}
{BHC}&{Black Hole Candidate}\\
%{BATSE}&{}\\

{c}&{Velocity of light in vacuum, $2.99 \times 10^{10}~cm~s^{-1}$}\\
{$^\circ$C}&{degree centigrade}\\
{CAM}&{Coded Aperture Mask}\\
{CB}&{Cannonball}\\
{CCAFS}&{Cape Canaveral Air Force Station}\\
{Cd}&{Cadmium}\\
{CENBOL}&{CENtrifugal pressure supported BOundary Layer}\\
{CERN}&{European Council for Nuclear Research}\\
{CGRO}&{Compton Gamma-Ray Observatory}\\
{cm}&{centimeter}\\
{CME}&{Coronal Mass Ejection}\\
{CMOS}&{Complementary Metal Oxide Semiconductor}\\
{CompST}&{Sunyaev-Titarchuk Comptonization model}\\
{CONFIG}&{CONFIGuration}\\
{CORONAS}&{Complex ORbital Observations Near-earth of Activity of the Sun}\\
{Cs}&{Cesium}\\
{CsI}&{Cesium Iodide}\\
{CSIR}&{Council for Scientific and Industrial Research}\\
{CZT} & {Cadmium Zinc Telluride}\\

{DC}&{Direct Current}\\
{DEC}&{DEClination}\\
{diskbb}&{disk black body}\\
{DOF}&{Degrees of Freedom}\\

{EA}&{Event Analyzer}\\
{EDS}&{Experiment Data System}\\
{e.g.}&{exempli gratia (for example)}\\
{EM}&{Emission Measure}\\
{ESA}&{European Space Agency}\\
{ESRO}&{European Space Research Organization}\\
{et}&{etc., excetera}\\
{EUV}&{Extreme Ultraviolet}\\
{eV}&{electron Volt, $1.6 \times 10^{-12}~erg$}\\
{Exp. Astron.}&{Experimenal Astronomy}\\
%{EXIST}&{}\\

{FB}&{Fireball}\\
{Fe}&{Iron}\\
{f-f}&{free-free}\\
{FFT}&{Fast-Fourier Transformation}\\
{FITS}&{Flexible Image Transport System}\\
{FMI}&{FITS Master Index}\\
{FOM}&{Figure Of Merit}\\
{FOV}&{Field Of View}\\
\end {tabular}
\end{quote}
\begin{quote}
\begin {tabular}{l l}
{FPGA}&{Field Programmable Gate Array}\\
{FRED}&{Fast Rise Exponential Decay}\\
{FWHM}&{Full Width at Half Maximum}\\
{FZP}&{Fresnel Zone Plate}\\

{g}&{gram}\\
{G}&{Gravitational constant, $6.672 \times 10^{-8}~dyn~cm^2~g^{-2}$}\\
{ga}&{gaussian}\\
{GBM}&{Gamma-ray Burst Monitor}\\
{Gd}&{Gadolinium}\\
{GEANT4}&{GEometry ANd Tracking-4 (CERN particle simulation software 4)}\\
{GGS}&{Global Geospace Science}\\
{GOES}&{Geostationary Operational Environmental Satellites}\\
%{GRANAT}&{}\\
{GRB}&{Gamma Ray Burst}\\
{GTI}&{Good Time Intervals}\\

{H}&{Hydrogen}\\
{HEAO}&{High Energy Astrophysical Observatory}\\
{HESS}&{High Energy Stereoscopic System}\\
{HEXTE}&{High-Energy X-ray Timing Experiment}\\
%{HETE}&{}\\
{HID}&{Hardness-Intensity Diagram}\\
{HMXB}&{High Mass X-ray Binary}\\
{HV}&{High Voltage}\\
{HR}&{Hertzsprung-Russell}\\

{I}&{Iodine}\\
{IACT}&{Imaging Air Cherenkov Telescope}\\
{ICSP}&{Indian Centre for Space Physics}\\
{ICTP}&{the Abdus Salam International Centre for Theoretical Physics}\\
{IDL}&{Interactive Data Language}\\
{IJP}&{Indian Journal of Physics}\\
{IMXB}&{Intermediate Mass X-ray Binary}\\
{INFN}&{National Institute of Nuclear Physics (Italy)}\\
{INTEGRAL}&{INTErnational Gamma-Ray Astrophysics Laboratory}\\
{IR}&{Infrared}\\
{ISM}&{Inter-Stellar Medium}\\
{ISRO}&{Indian Space Research Organization}\\

{LabVIEW}&{Laboratory Virtual Instrumentation \& Engineering Workbench}\\
{LC}&{Light Curve}\\
{LEO}&{Low Earth Orbit}\\
{LLD}&{Lower Level Discriminator}\\
{LMXB}&{Low Mass X-ray Binary}\\

\end {tabular}
\end{quote}
\begin{quote}
\begin {tabular}{l l}

{$k$}&{Boltzmann Constant, $1.38 \times 10^{-16}~erg~K^{-1}$}\\
{K}&{Kelvin}\\
{keV}&{kilo electron Volt}\\
{kg}&{kilogram}\\
{km}&{kilo meter}\\
{kpc}&{kilo parsec}\\

{m}&{meter}\\
{$M_\odot$}&{Solar Mass, $1.989 \times 10^{33}$ g}\\
{$M_E$}&{Earth Mass, $5.976 \times 10^{27}$ g}\\
{$M_{Moon}$}&{Moon Mass, $7.347 \times 10^{25}$ g, ($0.0123\times M_E$)}\\
{MC}&{Monte Carlo}\\
{MeV}&{Mega Electron Volt}\\
{MHB}&{Modified Horizontal Bridgman}\\
{$\mu$}&{micro}\\
{MJD}&{Modified Julian Day}\\
{MK}&{Mega Kelvin}\\
{mlab}&{Microprocessor Laboratory (ICTP, Italy)}\\
{mm}&{millimeter}\\
{MPhei}&{Moscow PHysics-Engineering Institute}\\

{Na}&{Sodium}\\
{NaI}&{Sodium Iodide}\\
{NASA}&{National Aeronautics and Space Administration (USA)}\\
{ns}&{nano second}\\
{nT}&{nano-Tesla, $1~Tesla = 10^4~Gauss$}\\

{O}&{Oxygen}\\
{OEBH}&{Observational Evidence of Black Hole}\\

{pc}&{parsec, $3.26$ light years ($9.46\times 10^{17}~cm$), $3.085 \times 10^{18}~cm$}\\
{PCA}&{Proportional Counter Array}\\
{PCU}&{Proportional Counter Unit}\\
{PDS}&{Power Density Spectra}\\
{PHA}&{Pulse Height Analysis}\\
{Ph.D.}&{Doctor of Philosophy}\\
{PL}&{Power-Law}\\
{p-p}&{proton-proton}\\
{PRL}&{Physical Research Laboratory}\\
{PS}&{Pulse Shape}\\
{PSD}&{Pulse Shape Discriminator}\\
{PSF}&{Point Spread Function}\\
{POS}&{Propagating Oscillatory Shock}\\
{PSPC}&{Position Sensitive Proportional Counter}\\
%{PULSAR}&{}\\

{QPO}&{Quasi-Periodic Oscillation}\\
{QPP}&{Quasi-Periodic Pulsation}\\
{QUASAR}&{Quasi-Stellar Radio Source}\\

\end {tabular}
\end{quote}
\begin{quote}
\begin {tabular}{l l}

{$R_\odot$}&{Solar radius, $6.959 \times 10^{10}$ cm}\\
{$R_E$}&{Earth radius, $6.3713 \times 10^8$ cm}\\
{$R_{Moon}$}&{Moon radius, $1.737 \times 10^8$ cm, ($0.273\times R_E$)}\\
{RA}&{Right Ascension}\\
{RHESSI}&{Reuven Ramaty High Energy Solar Spectroscopic Imager}\\
{RMS}&{Root Mean Square}\\
{RXTE}&{Rossi X-ray Timing Explorer}\\
{RT-2}&{R\"{o}ntgen-2}\\

{S}&{Sulphar}\\
{SAA}&{South Atlantic Anomaly}\\
{SAC}&{Satellite Application Centre}\\
{SDD}&{Silicon Drift Detector}\\
{SFM}&{Solar Flare Mode}\\
{SolarMax}&{Solar Maximum Mission satellite}\\
{SOHO}&{Solar and Heliospheric Observatory}\\
{SQM}&{Solar Quiet Mode}\\
{SS disk}&{Shakura-Sunyav disk}\\
{SSC}&{Scanning Shadow Camera}\\
{SSW}&{Solar SoftWare}\\
{STEREO}&{Solar TErrestrial RElations Observatory}\\
%{SWIFT}&{}\\

{TCAF}&{Two Component Advective Flow}\\
{Te}&{Telluride}\\
{TIFR}&{Tata Institute of Fundamental Research}\\
{Tl}&{Thallium}\\
{TRACE}&{Transition Region and Coronal Explorer}\\

{V}&{Volt}\\
{VCO}&{Voltage Control Oscillator}\\
{VSSC}&{Vikram Sarabhai Space Centre}\\

{ULD}&{Upper Level Discriminator}\\
{URA}&{Uniformly Redundant Arrays}\\
{UT}&{Universal Time}\\
{UV}&{Ultraviolet}\\

{W}&{Watt}\\

{XMM-Newton}&{X-ray Multi-Mirror Mission - Newton}\\

{Zn}&{Zinc}\\

\end {tabular}
\end{quote}

        \resetsec
        \resettab
\vfill\eject
\pagestyle{newheadings}
{\baselineskip 15pt\newpage
%\appendix
\def\aug{\vert R,L,\{\emptyset\}\rangle }
  \def\ket{\vert \vert	\{ \emptyset \} \rangle}
  \def\ket2{\vert \vert \otimes \{ R \} \rangle}
  \def\sqr{$^{2}$}
\def\dpr{\prime\prime}
\def\tpr{\prime\prime\prime}
\def\pr#1{ Phys.Rev. {\bf B #1}}
\def\pj#1{\proj_{{\cal #1}}}
\def\barr#1{{\overline{#1}}}
\def\jpc#1{J.Phys. Condensed Matter {\bf #1}}
\def\prl#1{ Phys. Rev. Lett. {\bf #1}}
\def\.#1{\mathaccent 95#1}
\def\^#1{\mathaccent 94 #1}
\def\~#1{\mathaccent "7E #1}
\def\Ir{{\mbox{I}}}
\def\Mr{{\mbox{M}}}
\def\Hr{{\mbox{H}}}
\def\sund{\mathaccent 22{\sigma}}
\def\equal{\enskip =\enskip}
\def\plus{\enskip +\enskip}
\def\minus{\enskip -\enskip}
\def\eq{\enskip =\enskip}
\def\pls{\enskip +\enskip}
\def\mns{\enskip -\enskip}
\def\Gund{\mathaccent 22 {G}}
\def\ul#1{\underline{#1}}
\def\ac#1{\mathaccent 95#1}
\def\td#1{\mathaccent "7E#1}
\def\un#1{\underline{#1}}
\def\nbox{\raisebox{.6ex}{\fbox{{\scriptsize{\phantom{$\sqrt{}$}}}}}$\:$}
\def\ybox{\raisebox{.6ex}{\fbox{{\scriptsize{$\sqrt{}$}}}}$\:$}
\def\pbox#1{\raisebox{.6ex}{\fbox{{#1}}}$\:$}
\def\c#1{\mbox{\bf #1}}
\def\und#1{$\underline{\mbox{\bf #1}}\:$}
\def\unit{{\cal I}}
\def\trans{{\cal T}}
\def\proj{{\cal P}}
\def\T{$T$}
\def\P{$P$}
\def\Q{$Q$}
\def\G{$G$}
\def\M{{\bf M}}
\def\H{{\bf H}}
\def\I{{\bf I}}
\def\Pr{{\bf P}}
\def\Tra{{\bf T}}
\def\diag{\varepsilon_{i}}
  \def\proj{{\cal P}}
  \def\trans{{\cal T}}
  \def\ket{\vert \vert	\{ \emptyset \} \rangle}
  \def\ket2{\vert \vert \otimes \{ R \} \rangle}
  \def\sqr{$^{2}$}
\def\k{{ ( k}}
  \def\ahat{{\mathaccent "7E  A}}
  \def\bhat{{\mathaccent "7E  B}}
  \def\chat{{\mathaccent "7E  C}}
  \def\fhat{{\mathaccent "7E  F}}
  \def\dhat{{\mathaccent "7E  D}}
  \def\shat{{\mathaccent "7E  S}}
  \def\phat{{\mathaccent "7E  P}}
  \def\jhat{{\mathaccent "7E  J}}
  \def\khat{{\mathaccent "7E  K}}
  \def\ohat{{\mathaccent "7E o}}
\def\ve{\varepsilon}
\def\car{\{{\cal C}\}}
\def\gt{\; > \;}
\def\lt{\: < \:}

%\end{References}

}

\printindex %Make an index AUTOMATICALLY

\end{document}